\pdfoutput=1
\documentclass[11pt]{article}

\usepackage[a4paper,margin=1in]{geometry}

\usepackage[T1]{fontenc}
\usepackage[utf8]{inputenc}

\usepackage{amsmath,amssymb,amsfonts}
\usepackage{dsfont}   % \mathds{1}: indicator function
\usepackage{mhchem}
\usepackage{siunitx}

\usepackage{booktabs}
\usepackage{array}
\usepackage{graphicx}
\usepackage{caption}
\usepackage{subcaption}
\usepackage{longtable}
\usepackage{multirow}
\usepackage{float}      % [H] for the one-page-per-MOF blocks of Section S7
\usepackage{pdflscape}  % landscape table of all strategies
\usepackage[section]{placeins}   % floats never cross a \section; \FloatBarrier used inside S2

\usepackage{enumitem}

\usepackage{titlesec}
\titleformat{\section}{\normalfont\Large\bfseries}{Section \thesection.}{0.6em}{}
\titleformat{\subsection}{\normalfont\large\bfseries}{\thesubsection.}{0.6em}{}
\titleformat{\subsubsection}{\normalfont\normalsize\bfseries}{\thesubsubsection.}{0.6em}{}

\usepackage{fvextra}
\usepackage[most]{tcolorbox}
\newtcolorbox[auto counter]{sibox}[2]{breakable, enhanced, title={Box S\thetcbcounter. #1}, label={#2},
  colback=white, colframe=black!55, coltitle=white, fonttitle=\bfseries\small, boxrule=0.5pt, arc=1pt,
  left=4pt, right=4pt, top=3pt, bottom=3pt}
\definecolor{vsgreen}{HTML}{008000}

\newcommand{\siVerb}[1]{\VerbatimInput[fontsize=\scriptsize, breaklines, breakanywhere, breaksymbolleft={}, commandchars=\|\@\~]{SI_boxes/#1}}

\usepackage[numbers,sort&compress]{natbib}
\usepackage{xurl}

\usepackage[colorlinks=true,linkcolor=blue,citecolor=blue,urlcolor=blue]{hyperref}

\usepackage{xcolor}
\newcommand{\todo}[1]{\textcolor{red}{\textbf{[TODO: #1]}}}
\newcommand{\qTopPointZeroOne}{144{,}556}   % tier thresholds (main Fig. 2); define once here
\newcommand{\qTopPointOne}{100{,}288}
\newcommand{\qTopOne}{42{,}463}
\newcommand{\qTopFive}{4{,}167}
\newcommand{\siFig}[2][\linewidth]{\IfFileExists{SI_figures/#2.pdf}{\includegraphics[width=#1]{SI_figures/#2.pdf}}{\IfFileExists{SI_figures/#2.png}{\includegraphics[width=#1]{SI_figures/#2.png}}{\placeholderfig[3.5cm]{SI\_figures/\detokenize{#2}}}}}
\newcommand{\placeholderfig}[2][8cm]{%
  \begin{center}
    \fbox{\parbox[c][#1][c]{0.85\linewidth}{\centering\todo{#2}}}
  \end{center}
}



\begin{document}

% =====================================================================
%                              MAIN TEXT
% =====================================================================
\setcitestyle{super,open={},close={}}
\renewcommand{\bibnumfmt}[1]{(#1)}
\begin{center}
{\LARGE\bfseries ActMOF: Mapping Reticular Synthesis through Active Learning\par}
\end{center}
\vspace{0.5em}

\begin{center}
Hongrui~Sheng$^{1,2}$, Shutian~Liang$^{3}$, Reid~A.~Coyle$^{1}$, Shyam~Chand~Pal$^{1}$, Peter~Walther$^{1}$, Bin~Feng$^{1,4}$, Xinyu~Ding$^{1,5}$, Zhiling~Zheng$^{1,4,\ast}$

\bigskip
\begin{minipage}{0.95\linewidth}
\small\raggedright
\noindent
$^{1}$Department of Chemistry, Washington University, St.\ Louis, Missouri 63130, United States\\
$^{2}$Department of Chemistry, Fudan University, Shanghai 200438, China\\
$^{3}$Department of Energy, Environmental and Chemical Engineering, Washington University, St.\ Louis, Missouri 63130, United States\\
$^{4}$Institute of Materials Science and Engineering, Washington University, St.\ Louis, Missouri 63130, United States\\
$^{5}$Department of Chemistry, Imperial College London, London W12 0BZ, United Kingdom

\medskip
\noindent
$^{\ast}$Corresponding author. Email: \href{mailto:z.z@wustl.edu}{z.z@wustl.edu}
\end{minipage}
\end{center}

\vspace{1em}
\begin{abstract}
\noindent Reticular chemistry allows metal--organic frameworks (MOFs) to be designed from their building units, yet their synthesis remains constrained by sparse experimental data, noisy crystallization outcomes, and the absence of benchmarks for comparing active-learning strategies. While data-driven approaches such as Bayesian optimization (BO) can improve synthesis success rates, selecting appropriate optimizers beforehand for mapping MOF crystallization landscapes remains largely empirical. Here we present ActMOF, an experiment-grounded benchmark for comparing search strategies in aluminum-carboxylate MOF synthesis. ActMOF is emulated from experimentally measured microwave syntheses with each encoded condition assigned a crystallinity index calculated based on diffraction peak intensity and full width at half-maximum. By benchmarking policies guided by Gaussian processes (GP), tree ensembles and large language models (LLM), ActMOF placed the GP family among the leading policies for active learning in sparse synthesis landscapes. It then demonstrated transferability by guiding the microwave synthesis of six reticularly related aluminum MOFs. All six campaigns gave crystalline products within a fixed budget of 12 experiments each, and established a reusable testbed for AI-driven MOF synthesis.
\end{abstract}


\section*{Introduction}

The discovery of crystalline reticular materials is an experimental science governed by many variables that are individually simple yet collectively nontrivial.\cite{m1,m2,m3,m4,m5,m6} Reticular synthesis, the assembly of molecular building units into predetermined ordered frameworks held together by strong bonds,\cite{m6} makes the structure of a metal--organic framework (MOF) a matter of design.\cite{m7,m8,m9,m10,m11,m12} In its isoreticular form, the net and the secondary building unit (SBU) are kept while the linker is lengthened or functionalized, as first shown for the IRMOF series and later extended to the large-pore IRMOF-74 family;\cite{m13,m14} in its multivariate (MTV) form,\cite{m15,m16} several linkers share one underlying topology.\cite{m15,m17} Despite progress in rational design, the synthesis of the intended crystalline phase depends on concentration, modulator identity and amount, solvent environment, time, temperature, and nucleation kinetics,\cite{m18,m19,m20,m21,m22,m23,m24} and as reticular space expands, how to navigate synthesis becomes as pressing as what to synthesize. In light of this complexity, the most consequential decisions in MOF synthesis are often made not at the selection of framework based on properties alone,\cite{m25,m26} but at the level of synthesizable materials verification and reaction condition discovery.\cite{m27,m28,m29,m30,m31,m32,m33,m34} Hitherto, this condition-search problem has been addressed by chemical intuition\cite{m24,m31,m35}, literature analogy\cite{m13,m36}, and basic design-of-experiments\cite{m37,m38} campaigns.\cite{m2,m39,m40,m41} A chemist can rationally design the net, node, and linker, yet still face a narrow crystallization window in which a small change in acidity, dilution, or thermal dose transforms a crystalline product into amorphous solid, side phase, or unreacted precursor.\cite{m32,m34,m42,m43,m44,m45}

A human-centered machine learning (ML) strategy for synthesis should therefore be judged in a manner aligned with how chemists actually make decisions and provide early confidence for navigating vast synthesis space.\cite{m46,m47,m48} Active learning (AL) offers an iterative strategy that mimics how chemists explore chemical space by refining experimental choices in response to prior outcomes.\cite{m49,m50,m51,m52,m53} Concretely, active learning proceeds as an iterative predict--synthesize--measure--update loop: a surrogate model, a computationally cheap predictive model, proposes the condition or batch of conditions expected to be most informative or most likely to improve crystallinity, that proposal is then evaluated, either against the experiment-grounded emulative landscape in the dry lab or by real microwave synthesis and powder X-ray diffraction (PXRD) measurement in the wet lab, and the observed outcome is fed back to update the surrogate before the next round of proposals (\textbf{Figure 1a}). Active learning therefore operates on both sides of this work: dry-lab active learning is used to benchmark the search behavior of candidate policies against the emulative landscape, while wet-lab active learning subsequently guides and revises real microwave synthesis campaigns for MOFs whose crystallization behavior lies outside the original benchmark. In lieu of being outcome-oriented, MOF synthesis screening is inherently process-driven, wherein chemists' confidence must be accumulated throughout the search itself.\cite{m31,m54} A practical closed-loop workflow should accordingly provide (i) early promising signs, as each iteration consumes reagents, instrument time, and human attention\cite{m55,m56,m57}; (ii) steady improvement, because a modest but visible change in crystallinity can keep a researcher willing to continue the campaign\cite{m52,m53,m58}; and (iii) interpretable suggestions\cite{m59,m60,m61}, as a proposed reaction condition that conflicts strongly with chemical intuition will often be deprioritized by synthetic chemists unless supported by compelling chemical or statistical rationale.

Herein, we report ActMOF as an experiment-grounded benchmark emulated from 95 measured microwave syntheses spanning various synthesis parameters, on which synthesis hypotheses can be quantitatively compared. Built on a prior microwave synthesis campaign,\cite{m33} this dataset expanded into a hypothetical reaction landscape of 6,101,172 encoded synthesis conditions by combining experimental reaction parameters, PXRD-derived outputs, machine-learning emulation, and local experimental-neighbor rules (Results and Supporting Information, Section~\hyperref[si:benchmark]{S2}). We then investigated Gaussian processes (GPs),\cite{m52,m53,m62,m63} statistical ensembles,\cite{m33,m52,m64,m65,m66} large language model (LLM)-only suggestions,\cite{m67,m68,m69,m70} and hybrid GP-LLM workflows\cite{m67,m71,m72,m73} under an identical active-learning protocol to identify which model suggestions of reaction conditions best navigate the sparse crystallization landscapes of MOFs. To this end, the selected strategies are subsequently transferred, following the isoreticular principle of keeping the rod-shaped Al--OH SBU while changing the linker,\cite{m6,m13,m74,m75,m76,m77} to CAU-23\cite{m75}, MIL-160\cite{m78}, CAU-10\cite{m79,m80}, MIL-53-muc\cite{m81}, MTV-MOF-303\cite{m82,m83}, and MTV-MIL-160\cite{m74,m84,m85} whose microwave synthesis has not been systematically optimized. This application therefore supports the transferability of benchmark-guided active learning across related reticular synthesis systems and sheds light on how statistical optimization and chemical intuition can be synergistically integrated for actual MOF crystallinity discovery.

\begin{figure}[!t]
\centering
\includegraphics[width=\linewidth]{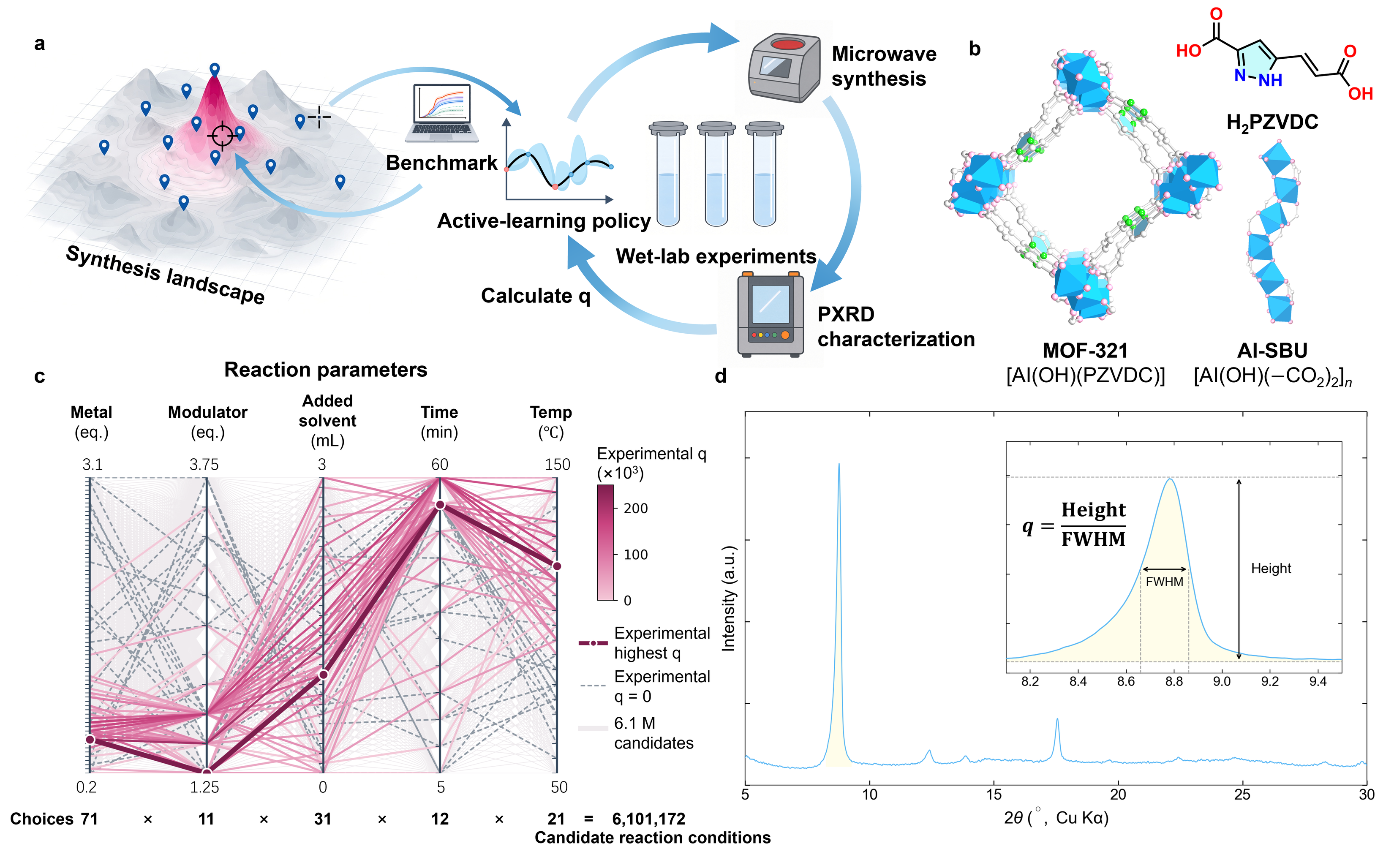}
\caption{Overview of the ActMOF active-learning benchmark. (a) Closed-loop active-learning workflow, in which proposed conditions are evaluated on the synthesis landscape (dry lab) or by microwave synthesis and PXRD (wet lab). (b) Structure of MOF-321, [Al(OH)(PZVDC)], with its linker H\textsubscript{2}PZVDC and rod-shaped Al-SBU. (c) The 95 measured MOF-321 syntheses across the five encoded reaction parameters, colored by experimental \textit{q}. (d) Definition of \textit{q} as peak height over full width at half-maximum (FWHM) of the diagnostic reflection, shown on a representative PXRD pattern. For illustrative clarity: Al is represented by deep blue octahedra; O, pink; C, light gray; N, green. H atoms are omitted.}
\label{fig:main-1}
\end{figure}

\section*{Results and Discussion}

At the onset of our study, we selected aluminum-based MOF-321\cite{m33} (\textbf{Figure 1b}) as the model reaction system because it adopts reproducible and programmable microwave synthesis,\cite{m86,m87,m88} has strong relevance to rich carboxylate-based reticular chemistry,\cite{m7,m9,m89,m90} and carries a reaction space too vast for manual brute-force screening. In particular, the encoded synthesis space contains five variables (\textbf{Figure 1c} and Supporting Information, Table~\hyperref[tab:encoding]{S5}), which include aluminum precursor amount, modulator amount, added solvent volume, microwave reaction time, and microwave reaction temperature, adding up to 95 deduplicated data entries.

The primary experimental observable is PXRD crystallinity. In this study, we define the quality score \textit{q} as the ratio of the intensity (I) of a diagnostic diffraction peak to its full width at half-maximum (FWHM, \textbf{Figure 1d}). The enhanced crystalline formation is associated with stronger and sharper diffraction features, especially at low angles where framework periodicity is readily visible,\cite{m2,m4,m91,m92,m93} and, for MOF chemists, the crystallinity index \textit{q} is usually considered as a first-stage synthesis metric, suitable for ranking crystallization outcomes during screening, whereas stability and sorption performance remain necessary downstream validations (Supporting Information, Table~\hyperref[tab:q-bet]{S3}).\cite{m2,m9,m41,m94,m95}

Starting from a set of 95 unique measured reaction outcomes, we developed a rule-guided statistical emulator (Supporting Information, Section~\hyperref[si:emulator]{S2.4}), in which crystallization neighborhoods were defined using a Chebyshev-distance rule over the five encoded synthesis variables,\cite{m96,m97,m98,m99,m100} to construct a grid of 6,101,172 candidate conditions. In essence, we note that ActMOF is not intended to declare the experimental outcome of every unmeasured condition (Supporting Information, Section~\hyperref[si:loo-andgate]{S2.5}). Inherently, one should consider it as an experiment-grounded reaction landscape designed to preserve the qualitative structure of the synthesis campaign, including sparse success, local neighborhoods of useful conditions, and broad low-quality regions, which are necessary for comparing AL search strategies. The emulator itself is built from random forests, so tree-based policies, which share this model family, might find the landscape unusually easy; in practice, they gained no measurable advantage over the Gaussian-process policies (Supporting Information, Section~\hyperref[si:landscape-robustness]{S4.4}).

The global structure of ActMOF is shown in \textbf{Figure 2}. High-\textit{q} regions, where top 1\% \textit{q} has a threshold of 42,463 and 0.1\% \textit{q} is higher than 100,288, were enriched in comparatively low metal loading, low modulator amount, long microwave heating time, and elevated temperature, with added water mattering less significantly (see \textbf{Figure 2b}). In particular, in a chemically interpretable sense, the resulting landscape suggests an insufficient base amount may leave the linker protonated, moderate dilution may improve nucleation, and an inappropriate thermal dose may produce amorphous solid or side phases. In fact, the sparse-success landscape and highly skewed distribution\cite{m101} of crystallinity index \textit{q} is consistent with the nature of MOF crystallization where most conditions are poor and the experimenter often encounters abrupt transitions between no product, poorly crystalline product, side phases, and the desired crystalline phase. In contrast to computational property-optimization tasks,\cite{m102,m103,m104,m105,m106,m107} where smooth movement along a continuous surface gradually improves performance, such difference reveals the type of rugged basins of crystalline products that AL needs to navigate and has direct consequences for which optimizer should be adopted in the laboratory (Supporting Information, Section~\hyperref[si:landscape]{S2.6}).

\begin{figure}[!t]
\centering
\includegraphics[width=\linewidth]{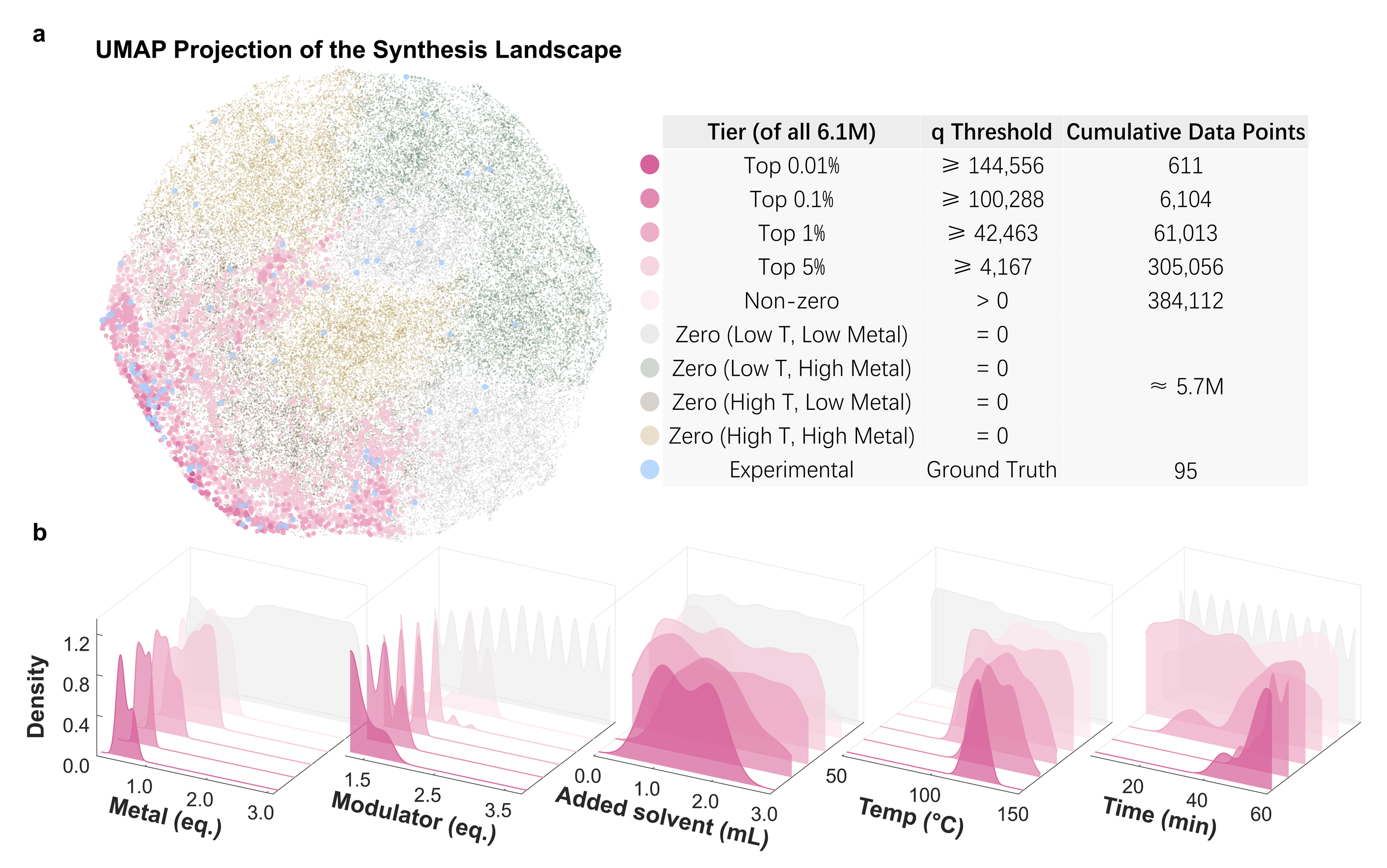}
\caption{Synthesis landscape and parameter distributions across \textit{q} tiers. (a) Uniform Manifold Approximation and Projection (UMAP) of a 1\% stratified random subsample of $\sim$6.1M enumerated synthesis conditions, colored by \textit{q} percentile tier. Zero-crystallinity conditions (\textit{q} = 0) are shaded by temperature--metal quadrant, split at the dataset medians: T \textless{} 100 \textdegree{}C (Low T) or $\geq$ 100 \textdegree{}C (High T), and metal \textless{} 1.66 equiv (Low Metal) or $\geq$ 1.66 equiv (High Metal). Blue points mark the 95 experimental ground-truth conditions. (b) Ridgeline kernel density estimates (KDE) of each \textit{q} tier along the five synthesis parameters, from a 2\% random subsample within each tier; each curve is normalized to its maximum. The favorable window narrows with increasing tier stringency. Metal and modulator equivalents are mmol of AlCl\textsubscript{3}\textperiodcentered{}6H\textsubscript{2}O and NaOH, respectively, per 1 mmol of linker; added solvent is deionized water. UMAP and KDE settings are given in Supporting Information, Section~\hyperref[si:landscape]{S2.6}.}
\label{fig:main-2}
\end{figure}

With the dataset in hand, we compared 39 policies (Supporting Information, Table~\hyperref[tab:strategies]{S13}) under conditions that mimic practical synthesis screening. Each run was initialized with three starting experiments drawn at random and different for each method and run, followed by iterative batches of three proposed experiments (Supporting Information, Section~\hyperref[si:protocol]{S3.1}). The total budget was 120 experiments, corresponding to 39 active-learning iterations after initialization. The objective was the cumulative best \textit{q} found by the search (\textbf{Figure 3}), and the surrogate models were trained on logarithmic scale to reduce the undue influence of rare high-\textit{q} values and emphasize relative crystalline enhancement across the synthesis landscape. In addition, we evaluated the trajectory of improvement, hit rates for the top 1\% and top 0.1\% of the landscape, and normalized regret (Supporting Information, Section~\hyperref[si:cum-best]{S4.1}).

Notably, in practical laboratory MOF screening, a bench chemist rarely continues a campaign out of faith in asymptotic convergence after over a hundred experiments; rather, confidence must be accumulated from the first few batches of PXRD patterns. To this end, cumulative best-\textit{q} trajectories in ActMOF distinguish the approaches showing early confidence with those demonstrating high terminal performance. Once \textit{q} exceeds top 0.1\% threshold ($\sim$100,000), the product already corresponds to sharp and intense diffraction features that many synthetic chemists would regard as operationally successful, even if subsequent optimization can still refine crystal quality.

In the first five iterations, probability of improvement (PI), which scores a candidate by its chance of exceeding the current best, behaved slightly more exploitatively than expected improvement (EI), which scores it by the size of the gain it is expected to deliver. In the benchmark runs, the best-\textit{q}-so-far of PI led that of EI at each of these iterations. This lead, however, came from the starting batches rather than from the acquisition rule: 7 of the 20 PI starting batches already contained a crystalline product against 1 of 20 for EI, and when both rules were rerun from identical starts, their early trajectories overlapped (Supporting Information, Figure~\hyperref[fig:si-pi-ei-early]{S19}). Conceptually, in the early stage of MOF microwave synthesis, uncertainty alone is not a reliable proxy for reaction promise in a sparse crystallinity landscape. A condition far from previous experiments may be statistically uncertain yet chemically unproductive, because it lies outside the window for metal hydrolysis, linker deprotonation and framework nucleation. For MOF synthesis, where the first goal is to locate a robust crystalline window before fine optimization, acquisition rules that weight predicted crystallinity over predictive variance are therefore advantageous, a property PI and EI share and one that separates both from uncertainty-driven sampling. The two nonetheless differ in what they reward. PI treats every improvement over the current best as equal, however small; EI asks how large that improvement is expected to be. On the certain map, where consulting a point is free, this difference barely matters. When each query is a microwave synthesis whose \textit{q} carries a $\sim$30\% repeat-loading scatter, it does. With noise added, PI lost 8.8\% of its final best \textit{q} and EI 0.3\% (Supporting Information, Figure~\hyperref[fig:si-noise]{S20}), and EI, the criterion least affected by noise, was taken forward.

\begin{figure}[!t]
\centering
\includegraphics[width=\linewidth]{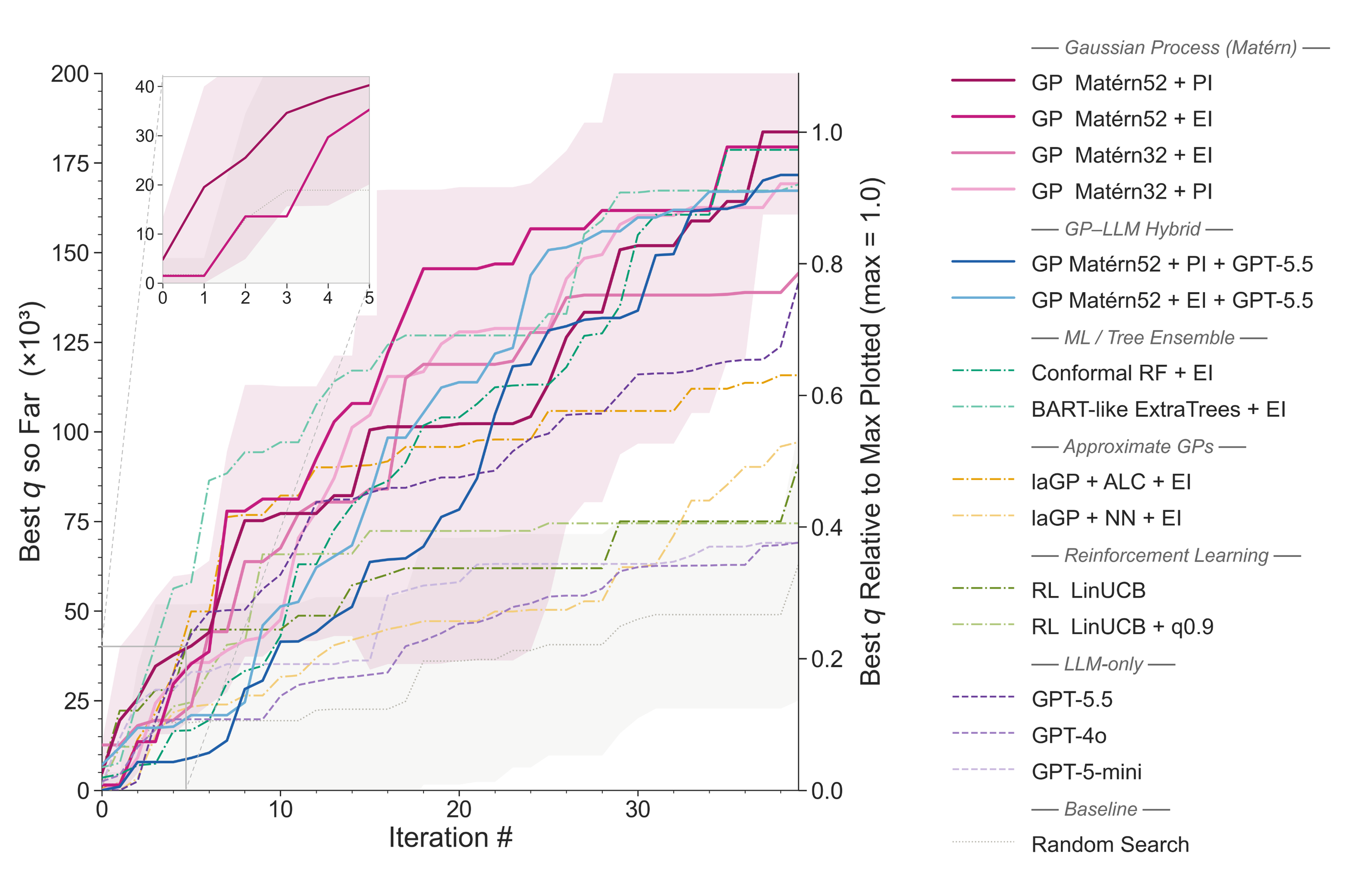}
\caption{Benchmark comparison of active-learning policies on the ActMOF landscape. Cumulative best \textit{q} is plotted against iteration (three random starting experiments, then 39 batches of three; 120 experiments in total) and averaged over five runs per policy, each run starting from its own random experiments. Shaded bands are 95\% confidence intervals of the mean for random search and Matérn-5/2 PI. Shown are 16 of the 39 policies: Gaussian processes (Matérn-5/2 or Matérn-3/2 kernel; PI or EI), GP--LLM hybrids, tree ensembles (conformal random forest; Bayesian additive regression tree (BART)-like ExtraTrees), local approximate GPs (laGP), linear bandits (LinUCB), LLM-only proposers (GPT-5.5, GPT-4o, GPT-5-mini) and random search (Supporting Information, Table~\hyperref[tab:strategies]{S13}). The right axis normalizes each curve to the highest plotted \textit{q}. Inset: iterations 0--5 for Matérn-5/2 PI and EI.}
\label{fig:main-3}
\end{figure}

The ActMOF benchmark also supports the continued use of common GP kernels for MOF synthesis. Matérn-based GPs,\cite{m108,m109,m110} especially those coupled to PI or EI acquisition functions, remained among the most reliable AL performers in ActMOF, which is encouraging because these kernels are already widely used in vanilla Bayesian optimization.\cite{m111} This outcome is chemically reasonable because the ActMOF landscape contains local neighborhoods of related crystallization behavior, while still retaining abrupt changes caused by coupled concentration, base, solvent, time, and temperature effects. In this regard, a Matérn kernel is well matched to a landscape that is neither fully smooth nor purely discontinuous, although radial basis function (RBF) and rational-quadratic kernels performed comparably (Supporting Information, Table~\hyperref[tab:n5-all]{S15}). \textbf{Figure 4} follows the Matérn-5/2 EI policy as it settles on long hold times and high temperatures; its learned length scales rank the metal amount as the most sensitive variable and added water as nearly irrelevant (Supporting Information, Figure~\hyperref[fig:si-lengthscales]{S10}). At the same time, tree-based uncertainty models provided a complementary advantage. Paired with EI, a conformal random forest matched the Matérn-5/2 GP hit rate for the top-1\% region within the first 7 iterations, and a tree-ensemble surrogate (BART-like ExtraTrees) reached the top-1\% region in a median of 4 iterations (Supporting Information, Section~\hyperref[si:n5-all]{S4.2}). This is consistent with their ability to capture threshold-like rules such as minimum temperature for framework formation, moderate dilution improving nucleation, or narrow metal/base windows. An alternating GP--LLM proposer stayed with the leading GP curves without overtaking them, and LLM-only proposals lagged both in mean best \textit{q}; language models are therefore kept as a control in \textbf{Figure 3} (Supporting Information, Table~\hyperref[tab:n5-all]{S15}). Thus, ActMOF shows that statistical models and rule-like ensemble models can both be useful, but for different chemical reasons.

\begin{figure}[!t]
\centering
\includegraphics[width=\linewidth]{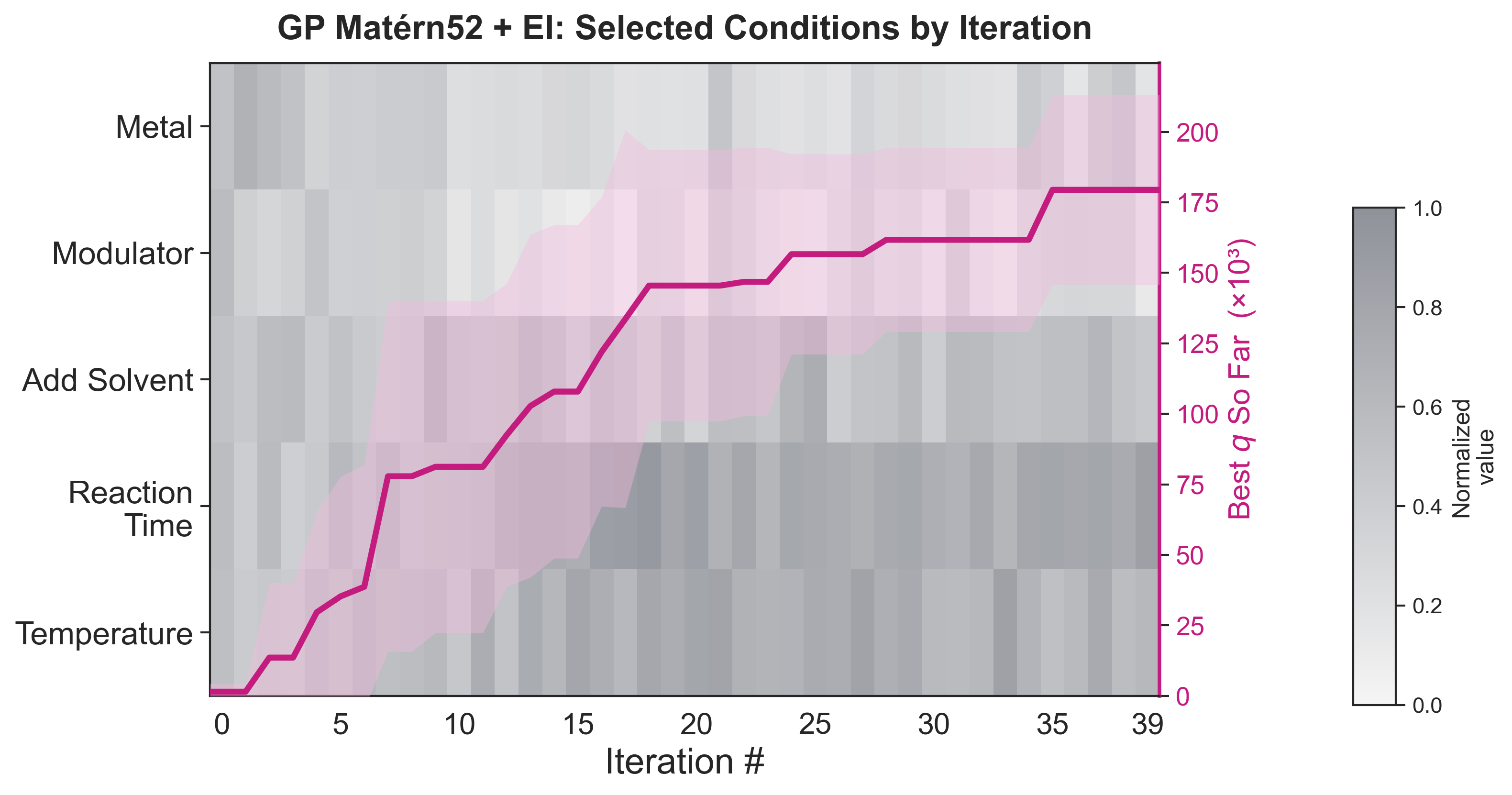}
\caption{Selected experimental conditions and cumulative best \textit{q} for a Matérn-5/2 Gaussian process with expected improvement (batch size = 3; mean of five runs). Heatmap rows correspond to the five decision variables (metal, modulator, additional solvent, reaction time, temperature); color encodes the normalized selected value at each iteration. The pink curve with right axis shows best \textit{q} so far; the shaded band is the 95\% confidence interval of the mean.}
\label{fig:main-4}
\end{figure}

Having identified a leading search behavior in ActMOF, we next asked whether the benchmark-guided strategy could be transferred to new MOF syntheses. To this end, we selected expected-improvement Matérn GP (Supporting Information, Section~\hyperref[si:app]{S5.1}) for experimental microwave synthesis of CAU-10, CAU-23, MIL-160, MIL-53-muc, MTV-MIL-160, and MTV-MOF-303 (\textbf{Figure 5}). None of the six reticularly related MOFs has been systematically optimized under microwave conditions.\cite{m75,m78,m112} These materials were further chosen because they preserve a common aluminum-based reticular chemistry while introducing chemically meaningful differences in organic linker solubility, hydrophilicity, and carboxylate angularity.\cite{m74,m82,m83,m84,m85} Accordingly, they pose the isoreticular question for synthesis that with the SBU kept and the linker changed, whether a benchmark-guided optimizer can transfer without assuming that the same reaction coordinate remains optimal.

Notably, rather than starting each new MOF crystallinity optimization from scratch, the validation workflow used the MOF-321 data behind ActMOF as a calibrated transfer prior, where a reference GP first learns synthesis trends from the 95 measured MOF-321 syntheses, including the 28 without product, then a residual GP adapts those trends to the measured PXRD responses of the MOF candidate (Supporting Information, Section~\hyperref[si:residual-gp]{S5.2}). Even the three conditions used to initialize each MOF's campaign were themselves proposed from this MOF-321-anchored prior rather than drawn at random, and each subsequent round refits the recalibration between the reference prediction and the accumulating real measurements for that specific MOF, reshaping the local landscape as evidence arrives (Supporting Information, Section~\hyperref[si:calibration]{S5.3}). On the MOF-321 landscape, starting from these three prior conditions placed every run in the top 1\%; from there the leading policies climbed to the top 0.1\% in 65--100\% of runs by iteration 13, while random search from the same start remained at its starting \textit{q} in the median (Supporting Information, Table~\hyperref[tab:warmstart]{S17}). This design directly tests whether ActMOF teaches a transferable search principle across various MOFs with underexplored microwave reaction conditions rather than a memorized numerical crystallinity surface of one particular MOF. We note that such transferability is particularly important for the two multivariate reticular systems, MTV-MIL-160 and MTV-MOF-303, because the coexistence of multiple linker environments inherently amplifies the complexity of nucleation and crystal growth, and microwave reaction of multivariate reticular materials remains far less established than that of their single-linker analogues. The ability to initialize such a campaign from a calibrated prior, while allowing new experimental residuals to rapidly overwrite system-specific errors, provides a practical route for reticular synthesis in which linker variation does not require a completely \textit{de novo} search.

Indeed, the transfer-learning workflow proved effective for MTV-MOF-303 (\textbf{Figure 5}), which represents one of the two multivariate cases among the six validation systems. Starting from three prior-informed experiments, the search reached a phase-matched MTV-MOF-303 pattern within the 12-experiment budget, while best \textit{q} then rose only modestly (1.23-fold) over the remaining rounds. Importantly, the search did not merely reproduce the optimal MOF-321 synthesis condition: its best condition lies inside the MOF-321 top-1\% tier but at no MOF-321 optimum (Supporting Information, Table~\hyperref[tab:six-mof-localization]{S11}). Rather, the transferred prior accelerated identification of productive crystallization neighborhoods, whereas the residual adaptation layer corrected for framework-specific linker behavior and nucleation kinetics from that prior. Each proposal is returned as a laboratory understandable prescription, including metal precursor amount, modulator equivalents, solvent volume, reaction time and temperature, so a chemist can accept and annotate it before the run (Supporting Information, Tables~\hyperref[tab:al-cau10]{S20}--\hyperref[tab:al-mtvmof303]{S25}). A parallel campaign on MTV-MIL-160 followed the same budget and likewise ended in agreement with the simulated phase, further supporting transfer of the benchmark-guided policy to multivariate reticular chemistry.

To further demonstrate the broader applicability of benchmark-guided AL, the same transfer strategy was then applied to CAU-10, CAU-23, MIL-160, and MIL-53-muc. Gain within each campaign is the ratio of the best \textit{q} in the 12-run budget to the best \textit{q} among that campaign's own three starting experiments. Because those starts were already prior-informed, this is a conservative denominator; the average increase ratio across the six frameworks is 2.2-fold (Supporting Information, Table~\hyperref[tab:transfer-summary]{S19}). CAU-23 never exceeded its starting best \textit{q}, and that starting solid already matched literature quality, whose Brunauer--Emmett--Teller (BET) surface area has reached 1173 m\textsuperscript{2}/g, within 6\% of the reported value for this framework\cite{m75} (Supporting Information, Table~\hyperref[tab:bet]{S26}). The remaining campaigns still moved the needle, about 1.8-fold for MIL-160, MIL-53-muc and MTV-MIL-160, and 5.4-fold for CAU-10. The self-correction had a visible price once: for CAU-10 and CAU-23, the first recalibration inverted the MOF-321 trends, and the round-2 batch at 60--65 \textdegree{}C gave little or no product before the calibration recovered in the next round (Supporting Information, Section~\hyperref[si:calibration]{S5.3}). In every case the final PXRD matched the simulated target phase (\textbf{Figure 5}). Early confidence therefore arrives with the first batch, including the multivariate starts: all 18 starting experiments crystallized, as did 67 of the 72 experiments overall (Supporting Information, Figure~\hyperref[fig:si-transfer-traj]{S24}); steady improvement is visible on the next few PXRD patterns rather than only at the end of the budget.

Chemically, all six best recipes stayed in the window that MOF-321 had mapped, with little metal (0.66--0.83 equiv AlCl\textsubscript{3}\textperiodcentered{}6H\textsubscript{2}O), little base (1.5--1.75 equiv NaOH) and long holds of 50--60 min, so the general rules of Al-carboxylate crystallization carried over (Supporting Information, Table~\hyperref[tab:transfer-summary]{S19}). Notably, the two frameworks containing isophthalate, CAU-10 and MTV-MIL-160, converged on the same recipe at 140 \textdegree{}C, whereas those built from five-membered heteroaromatic linkers (MIL-160, CAU-23 and MTV-MOF-303) and from linear muconate (MIL-53-muc) peaked at 100--110 \textdegree{}C, on either side of the 120 \textdegree{}C reported for MOF-321.\cite{m33,m76} This split follows the linker rather than the rod-shaped SBU. The cis-only chains of CAU-10 and MIL-160 fall into opposite groups, whereas the trans-only and alternating cis--trans chains of MIL-53-muc and CAU-23 fall together.\cite{m75} Merely copying the MOF-321 optimum would likely have missed both groups; the transfer prior instead carried over the shared stoichiometry and left each campaign to find its own temperature. These optima were reached with 50--60 min of microwave heating, compared with 12 h of solvothermal synthesis for CAU-10\cite{m79} and 24 h and 6 h of reflux for MIL-160\cite{m80} and CAU-23,\cite{m75} respectively (Supporting Information, Section~\hyperref[si:synthesis]{S6}).

Because the policy was chosen on a single benchmark landscape, these campaigns serve as a falsifiable test of transfer. A policy that had merely memorized the MOF-321 optimum would have missed frameworks whose best temperatures lie on either side of it. That the same policy instead located linker-specific conditions for six targets under one encoded reaction scheme supports the central premise of this work, that an experiment-grounded benchmark can select model suggestions that transfer across related reticular syntheses, provided the synthesis variables keep a comparable chemical meaning.

\begin{figure}[!t]
\centering
\includegraphics[width=\linewidth]{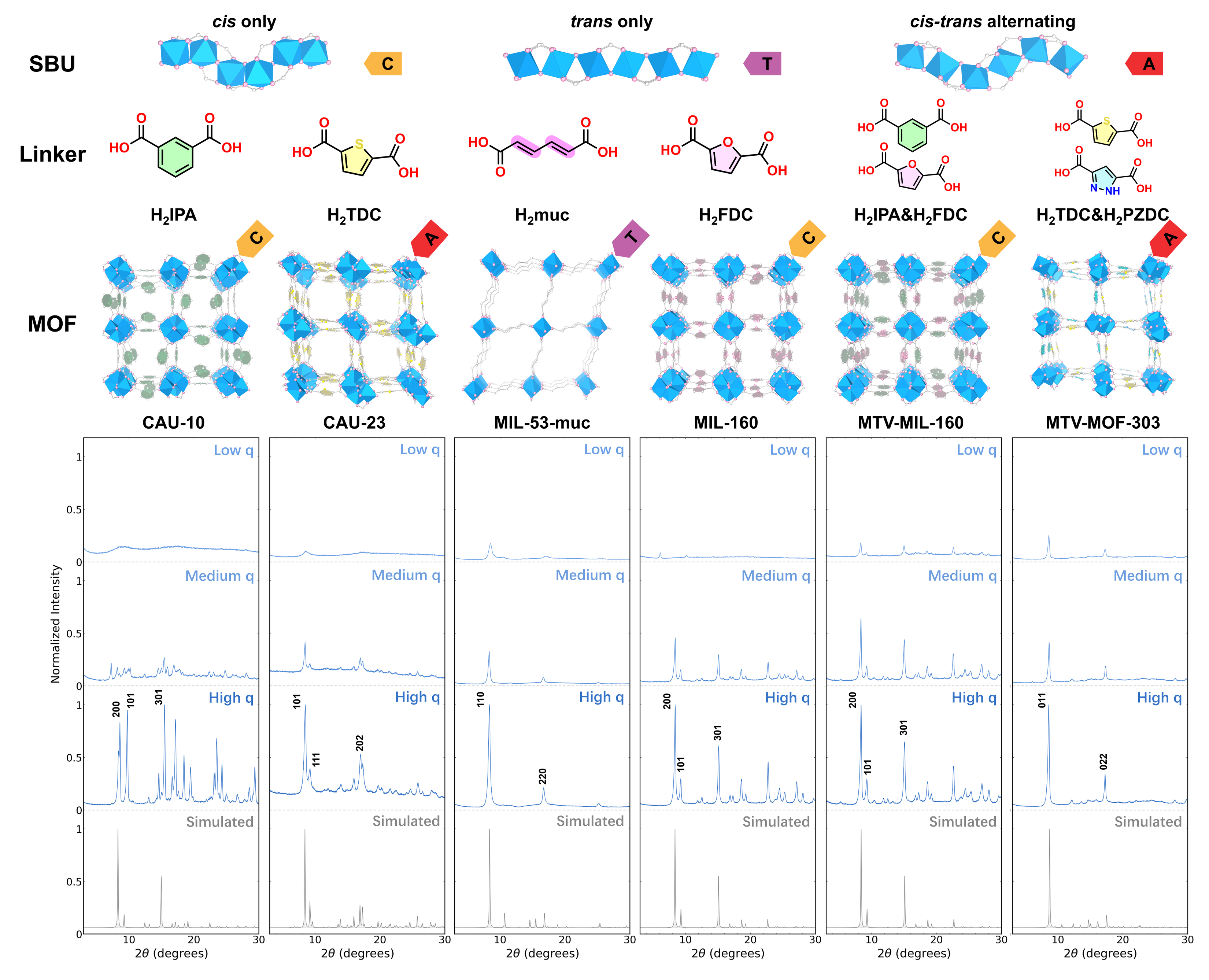}
\caption{PXRD patterns for the six MOF systems used to validate transfer of the benchmark-guided active-learning policy (CAU-10, CAU-23, MIL-53-muc, MIL-160, MTV-MIL-160, and MTV-MOF-303). Crystal structures are shown above each column. For every system, representative experimental patterns from low-, medium-, and high-\textit{q} outcomes encountered during the search are compared with the corresponding simulated reference pattern shown at the bottom of each column. Selected reflections in the high-\textit{q} traces are indexed. For illustrative clarity: Al is represented by deep blue octahedra; O, pink; C, light gray; N, green; S, yellow. H atoms are omitted.}
\label{fig:main-5}
\end{figure}

\section*{Concluding Remarks}

Considering the continued expansion of automated experimentation, foundation models, and AI-guided decision-making in chemistry, it is increasingly important to move beyond isolated trial-and-error attempts toward the conversion of crystallization outcomes into transferable synthetic knowledge in a quantitative and interpretable manner. ActMOF advances this direction by introducing a crystallinity-guided reaction benchmark in which encoded synthesis conditions are derived from experimentally measured PXRD outcomes of an aluminum carboxylate-based MOF, thereby providing a reproducible platform for studying how optimization algorithms navigate sparse-success reticular synthesis spaces. In this regard, our work shows that efficient MOF discovery is facilitated by (i) early confidence within the first few experimental batches, (ii) controlled exploitation through Matérn GP models with expected-improvement acquisition, and (iii) suggestions returned as inspectable laboratory prescriptions rather than latent indices. It is envisioned that such benchmarks will move synthetic chemistry from reporting optimal recipes toward reporting transferable search principles. The transferable element is therefore not a fixed recipe but an adaptive prior as framework-specific evidence accumulates. The validation across CAU-10, CAU-23, MIL-160, MIL-53-muc, MTV-MIL-160 and MTV-MOF-303 suggests a future in which multivariate and isoreticular structure discovery can be pursued through adaptive, benchmark-guided synthesis intelligence.

\section*{Associated Content}

\smallskip\noindent \textbf{Data and Software Availability.} The ActMOF dataset and benchmark code are available at \url{https://github.com/deep-synthesis/ActMOF}. An interactive web demonstration for recommending synthesis conditions is available at \url{https://actmof.streamlit.app/}.

\smallskip\noindent\textbf{Supporting Information.} The Supporting Information follows the main text of this document (from page~\pageref{si:start}). It contains materials and methods, construction and validation of the ActMOF benchmark, benchmarking protocols and statistics, transfer-learning framework, synthesis details, PXRD patterns and N$_2$ sorption data.

\section*{Author Information}

\medskip\noindent\textbf{Corresponding Author}\par\nopagebreak\smallskip

{\raggedright\noindent\hangindent=1.5em \textbf{Zhiling Zheng} \textit{$-$ Department of Chemistry; Institute of Materials Science and Engineering, Washington University, St.\ Louis, Missouri 63130, United States;} orcid.org/0000-0001-6090-2258. Email: z.z@wustl.edu\par}\smallskip

\medskip\noindent\textbf{Authors}\par\nopagebreak\smallskip

{\raggedright\noindent\hangindent=1.5em \textbf{Hongrui Sheng} -- \textit{Department of Chemistry, Washington University, St.\ Louis, Missouri 63130, United States; Department of Chemistry, Fudan University, Shanghai 200438, China;} orcid.org/0009-0008-7444-2692\par}\smallskip

{\raggedright\noindent\hangindent=1.5em \textbf{Shutian Liang} -- \textit{Department of Energy, Environmental and Chemical Engineering, Washington University, St.\ Louis, Missouri 63130, United States}\par}\smallskip

{\raggedright\noindent\hangindent=1.5em \textbf{Reid A. Coyle} -- \textit{Department of Chemistry, Washington University, St.\ Louis, Missouri 63130, United States}\par}\smallskip

{\raggedright\noindent\hangindent=1.5em \textbf{Shyam Chand Pal} -- \textit{Department of Chemistry, Washington University, St.\ Louis, Missouri 63130, United States}\par}\smallskip

{\raggedright\noindent\hangindent=1.5em \textbf{Peter Walther} -- \textit{Department of Chemistry, Washington University, St.\ Louis, Missouri 63130, United States}\par}\smallskip

{\raggedright\noindent\hangindent=1.5em \textbf{Bin Feng} -- \textit{Department of Chemistry, Washington University, St.\ Louis, Missouri 63130, United States; Institute of Materials Science and Engineering, Washington University, St.\ Louis, Missouri 63130, United States}\par}\smallskip

{\raggedright\noindent\hangindent=1.5em \textbf{Xinyu Ding} -- \textit{Department of Chemistry, Washington University, St.\ Louis, Missouri 63130, United States; Department of Chemistry, Imperial College London, London W12 0BZ, United Kingdom}\par}\smallskip

\medskip\noindent\textbf{Author Contributions}\par\nopagebreak\smallskip

\smallskip\noindent Z.Z. conceptualized the study and designed the methodology. H.S. performed sample preparation, PXRD and BET measurements, sorted the benchmark code, carried out data analysis, and prepared the manuscript and Supporting Information. S.L. prepared samples, collected PXRD data, and contributed to the Supporting Information. R.C. assisted with sample preparation, PXRD measurements, and code review. S.C.P. assisted with BET measurements and analysis. P.W. assisted with benchmark code and sample preparation. B.F. participated in discussions and contributed to data visualization. X.D. assisted with sample preparation. Z.Z. provided critical review and discussion and finalized the manuscript.

\smallskip\noindent The manuscript was written through contributions of all authors. All authors have given approval to the final version of the manuscript.

\medskip\noindent\textbf{Conflicts of interest}\par\nopagebreak\smallskip

\smallskip\noindent There are no conflicts to declare.

\section*{Acknowledgment}

\smallskip\noindent Z.Z. acknowledges funding and extends thanks for the support provided by the EQT Foundation through the Breakthrough Science Grant (No. 059994), Thinking Machines through the Tinker Research Grant, the Royal Society of Chemistry through the Researcher Collaborations grant (GR0042769), and Washington University in St.\ Louis through the Startup Grant. The authors thank Prof. Nan Lin (Department of Statistics and Data Science, Washington University in St.\ Louis) for insightful discussions on strengthening the statistical basis of the ActMOF landscape. H.S. thanks Chenglin Li (Washington University in St.\ Louis) for building the initial version of the interactive web application and Zeyuan Li (Tsinghua University) for testing the reliability of the ActMOF landscape.

\renewcommand{\refname}{References}

% =====================================================================
%                        SUPPORTING INFORMATION
% =====================================================================
\clearpage
\pagenumbering{arabic}
\renewcommand{\thepage}{S\arabic{page}}
\setcounter{section}{0}\setcounter{figure}{0}\setcounter{table}{0}\setcounter{equation}{0}
\renewcommand{\thesection}{S\arabic{section}}
\renewcommand{\thesubsection}{\thesection.\arabic{subsection}}
\renewcommand{\thesubsubsection}{\thesubsection.\arabic{subsubsection}}
\renewcommand{\thefigure}{S\arabic{figure}}
\renewcommand{\thetable}{S\arabic{table}}
\renewcommand{\theequation}{S\arabic{equation}}
% distinct link targets for the "S" numbers
\renewcommand{\theHsection}{S\arabic{section}}
\renewcommand{\theHfigure}{S\arabic{figure}}
\renewcommand{\theHtable}{S\arabic{table}}
\renewcommand{\theHequation}{S\arabic{equation}}
\setcitestyle{numbers,square,open={[},close={]}}
\renewcommand{\bibnumfmt}[1]{[#1]}
\phantomsection\label{si:start}
\begin{center}
{\Huge\bfseries Supporting Information}\\[1.2em]
{\LARGE ActMOF: Mapping Reticular Synthesis through Active Learning}
\end{center}
\vspace{1em}
% =====================================================================
% Author block (manual, multiple affiliations per author)
% =====================================================================
\begin{center}
Hongrui~Sheng$^{1,2}$, Shutian~Liang$^{3}$, Reid~A.~Coyle$^{1}$, Shyam~Chand~Pal$^{1}$, Peter~Walther$^{1}$, Bin~Feng$^{1,4}$, Xinyu~Ding$^{1,5}$, Zhiling~Zheng$^{1,4,\ast}$

\bigskip
\begin{minipage}{0.95\linewidth}
\small\raggedright
\noindent
$^{1}$Department of Chemistry, Washington University, St.\ Louis, Missouri 63130, United States\\
$^{2}$Department of Chemistry, Fudan University, Shanghai 200438, China\\
$^{3}$Department of Energy, Environmental and Chemical Engineering, Washington University, St.\ Louis, Missouri 63130, United States\\
$^{4}$Institute of Materials Science and Engineering, Washington University, St.\ Louis, Missouri 63130, United States\\
$^{5}$Department of Chemistry, Imperial College London, London W12 0BZ, United Kingdom

\medskip
\noindent
$^{\ast}$Corresponding author. Email: \href{mailto:z.z@wustl.edu}{z.z@wustl.edu}
\end{minipage}
\end{center}

\vspace{1.5em}
\setcounter{tocdepth}{2}
\tableofcontents
\newpage

% =====================================================================
\section{Materials and Methods}
\label{si:methods}

% =====================================================================

\paragraph{Chemicals.}\label{si:chemicals} All chemicals were purchased commercially and used without further purification (Table~\ref{tab:chemicals}). The MOF-321 syntheses of Section~\ref{si:seed-data} are taken from ref.~\cite{zheng2023acscs} and were not repeated here.

\begin{table}[htbp]
\centering
\caption{Chemicals used for the validation syntheses, with molar masses $M$. Deionized (DI) water was used throughout.}
\label{tab:chemicals}
\small
\setlength{\tabcolsep}{4pt}
\begin{tabular}{@{}p{0.31\linewidth}rp{0.23\linewidth}p{0.12\linewidth}p{0.11\linewidth}@{}}
\toprule
Chemical & $M$ (g\,mol$^{-1}$) & Use & Supplier & Purity \\
\midrule
Aluminium chloride hexahydrate, \ce{AlCl3.6H2O} & 241.43 & Al source (all MOFs) & AK Scientific & $\ge$99\% \\
Isophthalic acid & 166.13 & linker, CAU-10, MTV-MIL-160 & Ambeed & $\ge$99.99\% \\
Thiophene-2,5-dicarboxylic acid & 172.16 & linker, CAU-23, MTV-MOF-303 & Ambeed & $\ge$97\% \\
\textit{trans,trans}-Muconic acid & 142.11 & linker, MIL-53-muc & A2B & $\ge$97\% \\
Furan-2,5-dicarboxylic acid & 156.09 & linker, MIL-160, MTV-MIL-160 & Ambeed & $\ge$99.30\% \\
1\textit{H}-Pyrazole-3,5-dicarboxylic acid hydrate & 174.11 & linker, MTV-MOF-303 & Ambeed & $\ge$99.75\% \\
Sodium hydroxide & 40.00 & base (modulator) & A2B & $\ge$97\% \\
Nitrogen & & cooling of the microwave vial & Linde & Grade 4.8 \\
\bottomrule
\end{tabular}
\end{table}

\paragraph{Microwave reactor.}\label{si:microwave} All syntheses were carried out in an Anton Paar Monowave 400 microwave reactor in G10 (``Glass 10'') vials with a magnetic stir bar, at 850~W and 600~rpm stirring, at the temperature and hold time of each experiment (Section~\ref{si:encoding}). After the hold time the vial was cooled to 50~$^{\circ}$C with nitrogen. The reaction mixture was transferred to a centrifuge tube, and the solid was isolated by centrifugation (6800~rpm, acceleration setting 9) and washed with deionized (DI) water (twice), ethanol (twice) and acetone (once). The washed solid was measured by PXRD directly (Sections~\ref{si:q-mount} and~\ref{si:pxrd-instrument}); the samples for N$_2$ sorption were activated (Section~\ref{si:sorption}). No unexpected or unusually high safety hazards were encountered.

% =====================================================================
\section{The ActMOF Landscape}
\label{si:benchmark}

% =====================================================================

ActMOF is built in three steps: a crystallinity index $q$ read from PXRD (Section~\ref{si:q-index}); 95 measured MOF-321 syntheses (Section~\ref{si:seed-data}) placed in a five-dimensional encoded condition space (Section~\ref{si:encoding}); and a rule-guided emulator that fills the remaining 6.1 million conditions (Section~\ref{si:emulator}). Sections~\ref{si:loo-andgate}--\ref{si:six-mof-localization} examine where the emulated landscape can be trusted, what it looks like and where the six validation optima lie on it.

% =====================================================================
\FloatBarrier
\subsection{Crystallinity Index \textit{q}}
\label{si:q-index}

\paragraph{(i) Definition.}
\label{si:q-def}

Following ref.~\cite{zheng2023acscs}, $q$ is the intensity $I$ of the diagnostic PXRD reflection divided by its full width at half maximum,
\begin{equation}
q = \frac{I}{\mathrm{FWHM}}.
\label{eq:q-definition}
\end{equation}
$I$ is the intensity at the maximum of the diagnostic reflection (Table~\ref{tab:q-peaks}), and $\mathrm{FWHM} = x_2 - x_1$, where $x_1$ and $x_2$ are the $2\theta$ positions on either side of the maximum at which the intensity has fallen to $I/2$. A pattern without this reflection gives $q=0$ (FWHM $=30^{\circ}$ is entered as a placeholder that marks $q=0$; $q$ is not computed from it). Because the intensity depends on the structure factor of the reflection, $q$ is compared only between products of the same phase measured under the same protocol. The diagnostic reflection is the strongest reflection of the simulated pattern below $10^{\circ}$ $2\theta$ (Table~\ref{tab:q-peaks}). Because it arises from the largest lattice spacing of the framework (about 10~\AA), it reports the long-range order of the porous lattice.

\begin{table}[htbp]
\centering
\caption{Strongest simulated reflection below $10^{\circ}$ $2\theta$ (pymatgen, Cu K$\alpha$, $\lambda=1.5406$~\AA), used as the diagnostic reflection. MTV-MIL-160 and MTV-MOF-303 are simulated from the MIL-160 and MOF-303 structures.}
\label{tab:q-peaks}
\small
\begin{tabular}{@{}lccc@{}}
\toprule
Framework & $2\theta$ ($^{\circ}$) & $d$ (\AA) & $hkl$ \\
\midrule
CAU-10 & 8.30 & 10.65 & (200) \\
CAU-23 & 8.43 & 10.48 & (101) \\
MIL-53-muc & 8.40 & 10.52 & (110) \\
MIL-160 & 8.39 & 10.54 & (200) \\
MTV-MIL-160 & 8.39 & 10.54 & (200) \\
MTV-MOF-303 & 8.73 & 10.12 & (011) \\
\bottomrule
\end{tabular}
\end{table}

\paragraph{(ii) Sample mounting.}
\label{si:q-mount}

\begin{figure}[htbp]
\centering
\siFig[0.95\linewidth]{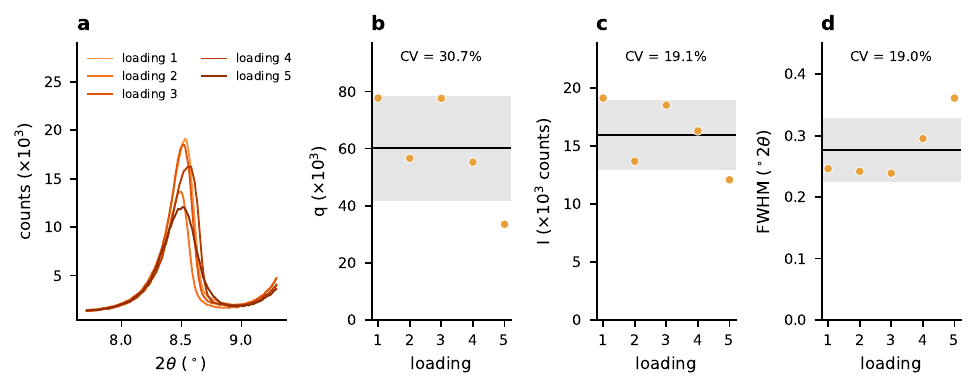}
\caption{Repeated loading of one MIL-160 powder (reflux, 4~h): loaded, measured, emptied and loaded again, five loadings. (a) Diagnostic reflection; (b) $q$, (c) intensity $I$, (d) FWHM for each loading. Line: mean; band: $\pm1$ s.d.; CV = s.d./mean.}
\label{fig:si-repack}
\end{figure}

All samples were prepared and measured by one procedure. About one spatula of the washed powder was placed in a 1.5~mL Eppendorf tube, acetone was added to the 1~mL mark and the suspension was mixed. Four drops were placed with a dropper on the center of a Rigaku silicon sample holder (Si(510), 5~mm $\times$ 0.2~mm well, part no.\ 1016621) and left to dry; the solid was then scraped into the well, filling it completely, and pressed with the spatula. Patterns were recorded as described in Section~\ref{si:pxrd-instrument}. To measure the mounting error directly, one powder (MIL-160, reflux 4~h) was loaded and measured, emptied from the holder and loaded again, five times in all: $q$ varied with a coefficient of variation of 30.7\% ($I$ 19.1\%, FWHM 19.0\%; $n=5$; Figure~\ref{fig:si-repack}). A single $q$ value is therefore resolved to about $\pm$30\%, whereas $q$ of the crystalline anchors spans three orders of magnitude (200--247{,}107).

\paragraph{(iii) Choice of reflection: CAU-23.}
\label{si:q-bet}

\begin{figure}[htbp]
\centering
\siFig[0.95\linewidth]{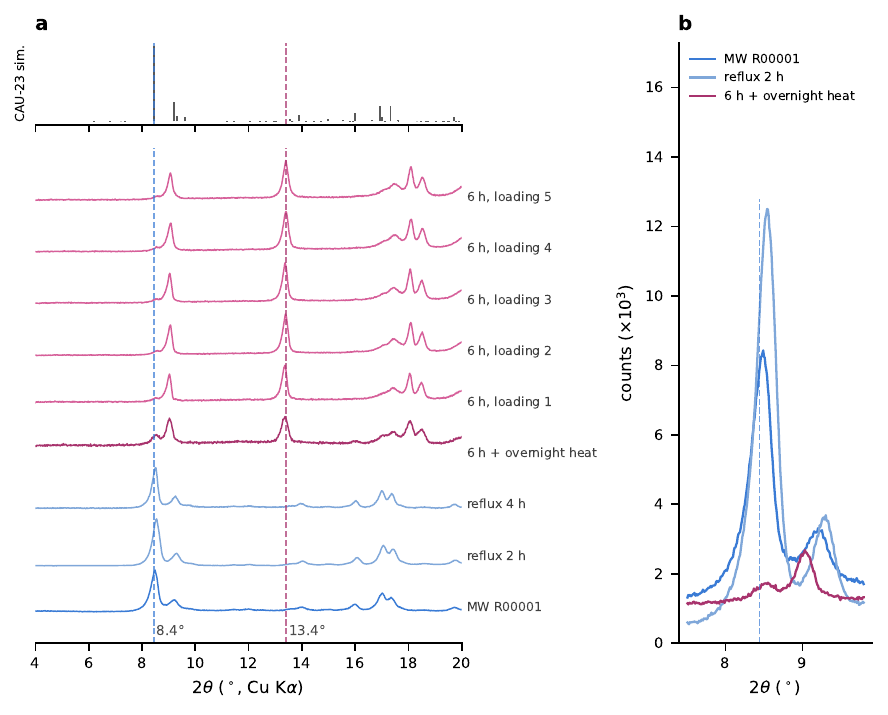}
\caption{CAU-23 products. (a) Normalized PXRD patterns, offset: microwave optimum R00001 (MW), reflux products after 2 and 4~h, and the 6~h reflux product that was afterwards heated overnight (one scan and five loadings), below the pattern simulated from the CAU-23 structure (CCDC 1878820). Dashed lines: position of the diagnostic reflection in the simulated CAU-23 pattern (8.4$^{\circ}$) and the 13.4$^{\circ}$ peak of the phase formed on overnight heating. (b) Diagnostic region on a common absolute scale. $q$ and $S_{\mathrm{BET}}$ of these samples: Section~\ref{si:q-bet} and Table~\ref{tab:bet}.}
\label{fig:si-bet}
\end{figure}

The highest-$q$ CAU-23 product (R00001, $q=22{,}254$) is phase-pure CAU-23 and porous, with a BET area of 1173~m$^2$\,g$^{-1}$, 6\% below the 1250~m$^2$\,g$^{-1}$ reported for CAU-23 made by reflux\cite{lenzen2019} (Section~\ref{si:sorption}). Our own CAU-23 reflux product (6~h) was mistakenly left heating overnight afterwards and converted to another phase (strongest peak at 13.4$^{\circ}$, $S_{\mathrm{BET}} = 56$~m$^2$\,g$^{-1}$; Figure~\ref{fig:si-bet}). Read at that peak, its $q$ (20{,}354) would equal that of the microwave optimum, so $q$ is read only after the target phase is confirmed, at the reflection listed in Table~\ref{tab:q-peaks}.

\paragraph{(iv) Reference scale: MIL-160.}
\label{si:q-reflux}

\begin{figure}[htbp]
\centering
\siFig[0.6\linewidth]{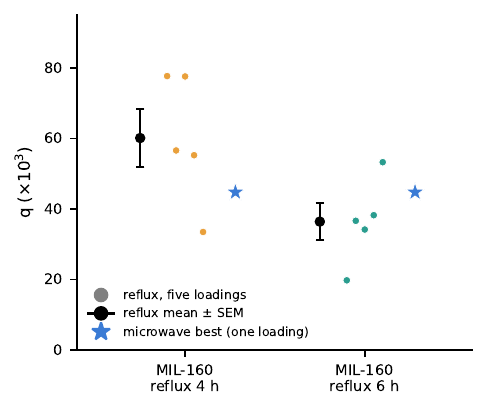}
\caption{$q$ of the MIL-160 microwave optimum R00007 (star, one loading) against reflux products after 4 and 6~h (dots, five loadings each; black, mean $\pm$ s.e.m.). The microwave value lies inside the range of the five loadings of both reflux products.}
\label{fig:si-mil160}
\end{figure}

To place $q$ on a known scale, MIL-160 was also made by the conventional reflux route\cite{tannert2019} (4 and 6~h; five loadings each). The best microwave product of the campaign (R00007, 55~min, 100~$^{\circ}$C, $q = 44{,}395$) lies within the range of both reflux products ($q = 60{,}106 \pm 8{,}246$ and $36{,}401 \pm 5{,}337$, mean $\pm$ s.e.m.; Figure~\ref{fig:si-mil160}), and its BET area is 80\% of that of the 6~h product (Section~\ref{si:sorption}).

\paragraph{(v) $q$ and porosity.} Table~\ref{tab:q-bet} collects $q$ and the BET area of the four samples for which both were measured. The two microwave optima, which have the highest $q$ of their campaigns, are porous, with 94\% (CAU-23) and 63\% (MIL-160) of the literature BET area, and the MIL-160 sample of higher BET area (reflux) has a $q$ in the same range as the microwave sample. The CAU-23 product that had converted to another phase has almost no CAU-23 reflection left and is almost non-porous. Within one phase, a high $q$ at the diagnostic reflection therefore goes with a porous product; $q$ ranks samples of one phase and is not a measure of surface area.

\begin{table}[htbp]
\centering
\caption{$q$ at the diagnostic reflection (Table~\ref{tab:q-peaks}) and BET area of the samples measured by both methods. $S_{\mathrm{BET}}$ in m$^2$\,g$^{-1}$ ($p/p_0$ 0.05--0.10, Section~\ref{si:sorption}); $q$ of the reflux products, mean of five loadings.}
\label{tab:q-bet}
\small
\begin{tabular}{@{}llrcc@{}}
\toprule
Framework & Sample & $q$ & $S_{\mathrm{BET}}$ & Lit.\ $S_{\mathrm{BET}}$ \\
\midrule
CAU-23 & microwave optimum, R00001 & 22{,}254 & 1173 & 1250\cite{lenzen2019} \\
       & reflux 6~h, mistakenly heated overnight & 1{,}722 & 56 & \\
MIL-160 & microwave optimum, R00007 & 44{,}395 & 729 & 1150\cite{permyakova2017} \\
        & reflux, 6~h & 36{,}401 & 908 & \\
\bottomrule
\end{tabular}
\end{table}

% =====================================================================
\FloatBarrier
\subsection{Anchor Experiments}
\label{si:seed-data}

The measured data are the microwave syntheses of MOF-321 (MOF-LA2-1; Al$^{3+}$, H$_2$PZVDC, NaOH) reported in ref.~\cite{zheng2023acscs}, reused here. We call these measured conditions \emph{anchor experiments}, or anchors: the emulator is trained on them and keeps their measured values (Section~\ref{si:emulator}), so they anchor the emulated landscape to experiment. Table~\ref{tab:seed-raw} lists all 96 recorded entries. Entries 26 and 31 are identical in condition and outcome, so 95 unique conditions remain: 67 crystalline ($q>0$) and 28 without product ($q=0$), with $q$ up to 247{,}107 (entry 73). The entries and their metadata (recording order, conditions, measured $I$ and FWHM) are also provided as a data file in the \href{https://github.com/deep-synthesis/ActMOF}{GitHub repository}.

\begingroup
\footnotesize
\setlength{\tabcolsep}{4pt}
\begin{longtable}{@{}rcrrrrrrrr@{}}
\caption{All 96 recorded MOF-321 entries. Encoded vector $\mathbf{x}$ as in Table~\ref{tab:encoding}. $^{\dagger}$Entries 26 and 31 are identical; the data set was deduplicated to 95 unique conditions.}
\label{tab:seed-raw}\\
\toprule
No. & Encoded $\mathbf{x}$ & Al (mL) & NaOH (mL) & \ce{H2O} (mL) & $t$ (min) & $T$ ($^{\circ}$C) & $I$ & FWHM ($^{\circ}$) & $q$ \\
\midrule
\endfirsthead
\multicolumn{10}{l}{\small\itshape Table~\ref{tab:seed-raw} (continued)}\\
\toprule
No. & Encoded $\mathbf{x}$ & Al (mL) & NaOH (mL) & \ce{H2O} (mL) & $t$ (min) & $T$ ($^{\circ}$C) & $I$ & FWHM ($^{\circ}$) & $q$ \\
\midrule
\endhead
\bottomrule
\endfoot
1 & (35, 5, 22, 9, 14) & 3.5 & 0.5 & 2.2 & 45 & 70 & 0 & 30 & 0\\
2 & (68, 7, 15, 12, 24) & 6.8 & 0.7 & 1.5 & 60 & 120 & 0 & 30 & 0\\
3 & (12, 9, 1, 7, 29) & 1.2 & 0.9 & 0.1 & 35 & 145 & 785 & 2.2 & 357\\
4 & (36, 10, 24, 3, 18) & 3.6 & 1.0 & 2.4 & 15 & 90 & 0 & 30 & 0\\
5 & (55, 11, 8, 2, 15) & 5.5 & 1.1 & 0.8 & 10 & 75 & 0 & 30 & 0\\
6 & (69, 6, 23, 1, 20) & 6.9 & 0.6 & 2.3 & 5 & 100 & 605 & 30 & 0\\
7 & (14, 12, 2, 11, 15) & 1.4 & 1.2 & 0.2 & 55 & 75 & 0 & 30 & 0\\
8 & (42, 7, 27, 6, 17) & 4.2 & 0.7 & 2.7 & 30 & 85 & 0 & 30 & 0\\
9 & (9, 14, 7, 4, 11) & 0.9 & 1.4 & 0.7 & 20 & 55 & 182 & 30 & 0\\
10 & (45, 6, 11, 8, 28) & 4.5 & 0.6 & 1.1 & 40 & 140 & 0 & 30 & 0\\
11 & (27, 12, 1, 2, 25) & 2.7 & 1.2 & 0.1 & 10 & 125 & 880 & 3.9 & 226\\
12 & (41, 14, 5, 3, 16) & 4.1 & 1.4 & 0.5 & 15 & 80 & 0 & 30 & 0\\
13 & (5, 5, 0, 1, 30) & 0.5 & 0.5 & 0.0 & 5 & 150 & 1{,}457 & 0.32 & 4{,}553\\
14 & (5, 15, 0, 1, 30) & 0.5 & 1.5 & 0.0 & 5 & 150 & 0 & 30 & 0\\
15 & (75, 15, 0, 1, 30) & 7.5 & 1.5 & 0.0 & 5 & 150 & 0 & 30 & 0\\
16 & (16, 7, 22, 3, 20) & 1.6 & 0.7 & 2.2 & 15 & 100 & 2{,}345 & 0.37 & 6{,}338\\
17 & (16, 9, 23, 4, 21) & 1.6 & 0.9 & 2.3 & 20 & 105 & 960 & 1.1 & 873\\
18 & (15, 7, 12, 11, 19) & 1.5 & 0.7 & 1.2 & 55 & 95 & 16{,}751 & 0.22 & 76{,}141\\
19 & (6, 5, 29, 11, 27) & 0.6 & 0.5 & 2.9 & 55 & 135 & 7{,}575 & 0.18 & 42{,}083\\
20 & (62, 7, 26, 11, 26) & 6.2 & 0.7 & 2.6 & 55 & 130 & 0 & 30 & 0\\
21 & (5, 15, 2, 11, 16) & 0.5 & 1.5 & 0.2 & 55 & 80 & 0 & 30 & 0\\
22 & (33, 10, 7, 11, 18) & 3.3 & 1.0 & 0.7 & 55 & 90 & 8{,}060 & 0.25 & 32{,}240\\
23 & (9, 6, 1, 12, 19) & 0.9 & 0.6 & 0.1 & 60 & 95 & 1{,}318 & 0.4 & 3{,}295\\
24 & (48, 14, 3, 12, 23) & 4.8 & 1.4 & 0.3 & 60 & 115 & 920 & 4.6 & 200\\
25 & (19, 8, 12, 11, 16) & 1.9 & 0.8 & 1.2 & 55 & 80 & 0 & 30 & 0\\
26$^{\dagger}$ & (15, 7, 14, 11, 24) & 1.5 & 0.7 & 1.4 & 55 & 120 & 23{,}969 & 0.21 & 114{,}138 \\
27 & (12, 8, 10, 11, 18) & 1.2 & 0.8 & 1.0 & 55 & 90 & 7{,}305 & 0.15 & 48{,}700\\
28 & (15, 7, 13, 11, 18) & 1.5 & 0.7 & 1.3 & 55 & 90 & 7{,}703 & 0.27 & 28{,}530\\
29 & (14, 7, 11, 11, 20) & 1.4 & 0.7 & 1.1 & 55 & 100 & 12{,}434 & 0.24 & 51{,}808\\
30 & (7, 6, 20, 11, 15) & 0.7 & 0.6 & 2.0 & 55 & 75 & 0 & 30 & 0\\
\textit{31}$^{\dagger}$ & \textit{(15, 7, 14, 11, 24)} & 1.5 & 0.7 & 1.4 & 55 & 120 & 23{,}969 & 0.21 & 114{,}138 \\
32 & (15, 6, 10, 11, 14) & 1.5 & 0.6 & 1.0 & 55 & 70 & 0 & 30 & 0\\
33 & (15, 7, 10, 11, 12) & 1.5 & 0.7 & 1.0 & 55 & 60 & 0 & 30 & 0\\
34 & (15, 8, 12, 11, 19) & 1.5 & 0.8 & 1.2 & 55 & 95 & 1{,}103 & 0.8 & 1{,}379\\
35 & (9, 8, 13, 7, 20) & 0.9 & 0.8 & 1.3 & 35 & 100 & 0 & 30 & 0\\
36 & (21, 6, 13, 12, 19) & 2.1 & 0.6 & 1.3 & 60 & 95 & 17{,}269 & 0.21 & 82{,}233\\
37 & (6, 6, 12, 11, 20) & 0.6 & 0.6 & 1.2 & 55 & 100 & 1{,}077 & 0.48 & 2{,}244\\
38 & (15, 8, 9, 12, 19) & 1.5 & 0.8 & 0.9 & 60 & 95 & 903 & 2 & 452\\
39 & (75, 12, 6, 11, 18) & 7.5 & 1.2 & 0.6 & 55 & 90 & 0 & 30 & 0\\
40 & (25, 13, 14, 7, 24) & 2.5 & 1.3 & 1.4 & 35 & 120 & 0 & 30 & 0\\
41 & (17, 13, 16, 12, 24) & 1.7 & 1.3 & 1.6 & 60 & 120 & 0 & 30 & 0\\
42 & (38, 13, 16, 8, 24) & 3.8 & 1.3 & 1.6 & 40 & 120 & 0 & 30 & 0\\
43 & (65, 5, 0, 12, 24) & 6.5 & 0.5 & 0.0 & 60 & 120 & 0 & 30 & 0\\
44 & (74, 7, 23, 1, 16) & 7.4 & 0.7 & 2.3 & 5 & 80 & 0 & 30 & 0\\
45 & (5, 7, 21, 12, 20) & 0.5 & 0.7 & 2.1 & 60 & 100 & 0 & 30 & 0\\
46 & (15, 7, 16, 12, 29) & 1.5 & 0.7 & 1.6 & 60 & 145 & 31{,}082 & 0.2 & 155{,}410\\
47 & (16, 7, 15, 11, 18) & 1.6 & 0.7 & 1.5 & 55 & 90 & 12{,}791 & 0.3 & 42{,}637\\
48 & (62, 10, 12, 10, 17) & 6.2 & 1.0 & 1.2 & 50 & 85 & 0 & 30 & 0\\
49 & (11, 7, 12, 11, 24) & 1.1 & 0.7 & 1.2 & 55 & 120 & 28{,}179 & 0.26 & 108{,}381\\
50 & (17, 7, 16, 11, 27) & 1.7 & 0.7 & 1.6 & 55 & 135 & 29{,}258 & 0.29 & 100{,}890\\
51 & (13, 7, 10, 12, 25) & 1.3 & 0.7 & 1.0 & 60 & 125 & 24{,}240 & 0.24 & 101{,}000\\
52 & (13, 6, 13, 11, 24) & 1.3 & 0.6 & 1.3 & 55 & 120 & 37{,}307 & 0.18 & 207{,}261\\
53 & (15, 5, 0, 12, 24) & 1.5 & 0.5 & 0.0 & 60 & 120 & 15{,}489 & 0.28 & 55{,}318\\
54 & (21, 5, 26, 11, 23) & 2.1 & 0.5 & 2.6 & 55 & 115 & 9{,}906 & 0.26 & 38{,}100\\
55 & (16, 7, 8, 11, 24) & 1.6 & 0.7 & 0.8 & 55 & 120 & 21{,}878 & 0.21 & 104{,}181\\
56 & (28, 7, 0, 12, 30) & 2.8 & 0.7 & 0.0 & 60 & 150 & 11{,}411 & 0.22 & 51{,}868\\
57 & (10, 7, 0, 12, 24) & 1.0 & 0.7 & 0.0 & 60 & 120 & 688 & 0.6 & 1{,}147\\
58 & (23, 7, 30, 11, 24) & 2.3 & 0.7 & 3.0 & 55 & 120 & 18{,}169 & 0.2 & 90{,}845\\
59 & (20, 7, 30, 11, 29) & 2.0 & 0.7 & 3.0 & 55 & 145 & 10{,}048 & 0.22 & 45{,}673\\
60 & (23, 7, 14, 11, 25) & 2.3 & 0.7 & 1.4 & 55 & 125 & 12{,}005 & 0.24 & 50{,}021\\
61 & (16, 7, 18, 12, 25) & 1.6 & 0.7 & 1.8 & 60 & 125 & 35{,}900 & 0.22 & 163{,}182\\
62 & (17, 7, 18, 12, 24) & 1.7 & 0.7 & 1.8 & 60 & 120 & 35{,}967 & 0.21 & 171{,}271\\
63 & (24, 7, 18, 12, 24) & 2.4 & 0.7 & 1.8 & 60 & 120 & 13{,}410 & 0.22 & 60{,}955\\
64 & (16, 7, 14, 11, 30) & 1.6 & 0.7 & 1.4 & 55 & 150 & 2{,}331 & 0.3 & 7{,}770\\
65 & (15, 7, 13, 9, 23) & 1.5 & 0.7 & 1.3 & 45 & 115 & 23{,}301 & 0.29 & 80{,}348\\
66 & (15, 7, 7, 7, 26) & 1.5 & 0.7 & 0.7 & 35 & 130 & 11{,}663 & 0.23 & 50{,}709\\
67 & (13, 6, 25, 12, 24) & 1.3 & 0.6 & 2.5 & 60 & 120 & 23{,}658 & 0.28 & 84{,}493\\
68 & (13, 7, 11, 11, 23) & 1.3 & 0.7 & 1.1 & 55 & 115 & 17{,}558 & 0.26 & 67{,}531\\
69 & (14, 7, 7, 11, 26) & 1.4 & 0.7 & 0.7 & 55 & 130 & 7{,}418 & 0.29 & 25{,}579\\
70 & (15, 6, 23, 12, 24) & 1.5 & 0.6 & 2.3 & 60 & 120 & 32{,}110 & 0.3 & 107{,}033\\
71 & (15, 5, 8, 12, 23) & 1.5 & 0.5 & 0.8 & 60 & 115 & 20{,}121 & 0.21 & 95{,}814\\
72 & (15, 6, 14, 12, 21) & 1.5 & 0.6 & 1.4 & 60 & 105 & 19{,}118 & 0.25 & 76{,}472\\
73 & (13, 5, 10, 11, 24) & 1.3 & 0.5 & 1.0 & 55 & 120 & 34{,}595 & 0.14 & 247{,}107\\
74 & (17, 6, 11, 11, 22) & 1.7 & 0.6 & 1.1 & 55 & 110 & 17{,}880 & 0.22 & 81{,}273\\
75 & (19, 5, 2, 12, 24) & 1.9 & 0.5 & 0.2 & 60 & 120 & 6{,}978 & 0.16 & 43{,}613\\
76 & (6, 6, 19, 12, 30) & 0.6 & 0.6 & 1.9 & 60 & 150 & 0 & 30 & 0\\
77 & (25, 7, 18, 11, 24) & 2.5 & 0.7 & 1.8 & 55 & 120 & 8{,}241 & 0.26 & 31{,}696\\
78 & (27, 7, 27, 12, 26) & 2.7 & 0.7 & 2.7 & 60 & 130 & 7{,}910 & 0.23 & 34{,}391\\
79 & (15, 5, 8, 11, 23) & 1.5 & 0.5 & 0.8 & 55 & 115 & 26{,}277 & 0.22 & 119{,}441\\
80 & (15, 5, 11, 10, 21) & 1.5 & 0.5 & 1.1 & 50 & 105 & 18{,}112 & 0.29 & 62{,}455\\
81 & (24, 5, 30, 12, 24) & 2.4 & 0.5 & 3.0 & 60 & 120 & 4{,}394 & 0.31 & 14{,}174\\
82 & (13, 7, 15, 12, 25) & 1.3 & 0.7 & 1.5 & 60 & 125 & 27{,}452 & 0.24 & 114{,}383\\
83 & (16, 7, 11, 12, 24) & 1.6 & 0.7 & 1.1 & 60 & 120 & 16{,}348 & 0.28 & 58{,}386\\
84 & (14, 6, 15, 12, 26) & 1.4 & 0.6 & 1.5 & 60 & 130 & 30{,}033 & 0.23 & 130{,}578\\
85 & (14, 7, 15, 11, 26) & 1.4 & 0.7 & 1.5 & 55 & 130 & 28{,}764 & 0.29 & 99{,}186\\
86 & (18, 7, 10, 12, 23) & 1.8 & 0.7 & 1.0 & 60 & 115 & 28{,}299 & 0.22 & 128{,}632\\
87 & (10, 7, 13, 12, 29) & 1.0 & 0.7 & 1.3 & 60 & 145 & 1{,}082 & 0.41 & 2{,}639\\
88 & (23, 7, 2, 12, 19) & 2.3 & 0.7 & 0.2 & 60 & 95 & 20{,}334 & 0.27 & 75{,}311\\
89 & (18, 7, 1, 4, 19) & 1.8 & 0.7 & 0.1 & 20 & 95 & 19{,}300 & 0.32 & 60{,}313\\
90 & (19, 6, 4, 12, 25) & 1.9 & 0.6 & 0.4 & 60 & 125 & 12{,}874 & 0.21 & 61{,}305\\
91 & (17, 7, 10, 12, 25) & 1.7 & 0.7 & 1.0 & 60 & 125 & 28{,}836 & 0.26 & 110{,}908\\
92 & (16, 6, 6, 5, 23) & 1.6 & 0.6 & 0.6 & 25 & 115 & 16{,}918 & 0.31 & 54{,}574\\
93 & (20, 6, 0, 12, 23) & 2.0 & 0.6 & 0.0 & 60 & 115 & 11{,}668 & 0.3 & 38{,}893\\
94 & (17, 7, 20, 11, 21) & 1.7 & 0.7 & 2.0 & 55 & 105 & 26{,}346 & 0.28 & 94{,}093\\
95 & (24, 7, 22, 8, 21) & 2.4 & 0.7 & 2.2 & 40 & 105 & 17{,}895 & 0.31 & 57{,}726\\
96 & (25, 7, 19, 10, 27) & 2.5 & 0.7 & 1.9 & 50 & 135 & 11{,}657 & 0.34 & 34{,}285\\
\end{longtable}
\endgroup

% =====================================================================
\FloatBarrier
\subsection{Variable Space}
\label{si:encoding}

\begin{figure}[htbp]
\centering
\siFig[0.92\linewidth]{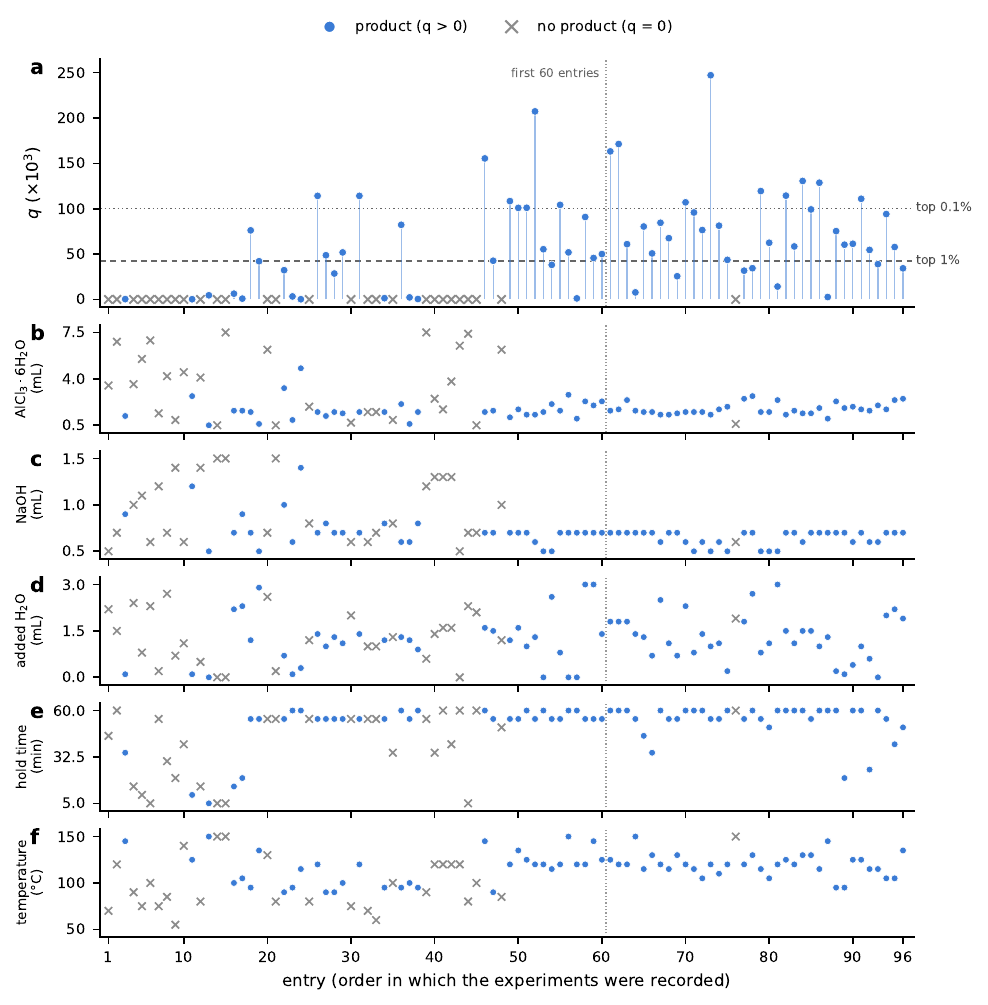}
\caption{The anchor experiments in the order in which they were recorded (entries 1--96, Table~\ref{tab:seed-raw}). (a) $q$ of every entry, with the top-1\% and top-0.1\% thresholds of the landscape (Table~\ref{tab:tiers}). (b--f) The five synthesis variables of every entry; the vertical axes span the full range of the grid. Dotted line: end of the first 60 entries, used in Sections~\ref{si:emulator-validation} and~\ref{si:landscape-robustness}.}
\label{fig:si-seed-coverage}
\end{figure}

Every condition is a vector of five integers $\mathbf{x}=(x_1,\dots,x_5)$ converted to laboratory quantities by
\begin{align}
V_{\mathrm{Al}} &= 0.1\,x_1~\mathrm{mL}, & x_1 &\in \{5, \dots, 75\}, \label{eq:enc-metal}\\
V_{\mathrm{NaOH}} &= 0.1\,x_2~\mathrm{mL}, & x_2 &\in \{5, \dots, 15\}, \label{eq:enc-mod}\\
V_{\mathrm{H_2O}} &= 0.1\,x_3~\mathrm{mL}, & x_3 &\in \{0, \dots, 30\}, \label{eq:enc-solvent}\\
t &= 5\,x_4~\mathrm{min}, & x_4 &\in \{1, \dots, 12\}, \label{eq:enc-time}\\
T &= 5\,x_5~^{\circ}\mathrm{C}, & x_5 &\in \{10, \dots, 30\}, \label{eq:enc-temp}
\end{align}
where $V_{\mathrm{Al}}$ and $V_{\mathrm{NaOH}}$ are volumes of \ce{AlCl3.6H2O} and NaOH stock solutions, both 100~mg\,mL$^{-1}$ in DI water, and $V_{\mathrm{H_2O}}$ is additional DI water (Table~\ref{tab:encoding}). One step of $x_1$ or $x_2$ is therefore 10~mg of \ce{AlCl3.6H2O} (0.041~mmol) or 10~mg of NaOH (0.25~mmol). The variables are coded as integers because the syntheses are set up in fixed increments (0.1~mL of a stock solution, 5~min, 5~$^{\circ}$C): with integer codes, every point of the grid is a recipe that can be dispensed exactly, all search methods choose from the same finite set of conditions, and two conditions are the same experiment exactly when their codes are equal. The full Cartesian product,
\begin{equation}
71 \times 11 \times 31 \times 12 \times 21 = 6{,}101{,}172,
\label{eq:grid-size}
\end{equation}
is the ActMOF candidate set. Every search policy and every wet-lab campaign uses the same encoding, so any proposal is directly a recipe. The anchors cover only part of this space (Figure~\ref{fig:si-seed-coverage}): the first entries are spread over the whole range of every variable and mostly give no product, whereas the later ones concentrate at low \ce{AlCl3.6H2O} and NaOH volumes, long hold times and 90--150~$^{\circ}$C, where the original campaign found its products. Of the 71 \ce{AlCl3.6H2O} volumes on the grid, 36 were used at least once.

\begin{table}[htbp]
\centering
\caption{Encoded synthesis variables.}
\label{tab:encoding}
\small
\setlength{\tabcolsep}{4pt}
\begin{tabular}{@{}cllccc@{}}
\toprule
Index & Program variable & Laboratory quantity & Integer range & Step & Physical range \\
\midrule
$x_1$ & \texttt{metal\_amount} & \ce{AlCl3.6H2O}, 100 mg\,mL$^{-1}$ & 5--75 & 0.1~mL & 0.5--7.5~mL \\
$x_2$ & \texttt{modulator} & NaOH, 100 mg\,mL$^{-1}$ & 5--15 & 0.1~mL & 0.5--1.5~mL \\
$x_3$ & \texttt{add\_solvent} & Additional DI water & 0--30 & 0.1~mL & 0--3.0~mL \\
$x_4$ & \texttt{reaction\_time} & Microwave hold time & 1--12 & 5~min & 5--60~min \\
$x_5$ & \texttt{reaction\_temperature} & Microwave temperature & 10--30 & 5~$^{\circ}$C & 50--150~$^{\circ}$C \\
\bottomrule
\end{tabular}
\end{table}

% =====================================================================
\FloatBarrier
\subsection{Emulator}
\label{si:emulator}

\begin{figure}[htbp]
\centering
\siFig{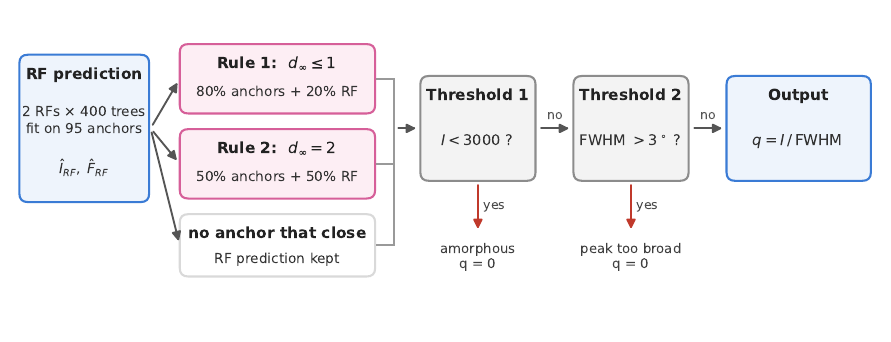}
\caption{Construction of the emulator in three stages: random-forest prediction, blending with nearby anchors, hard thresholds. $d_\infty$, Chebyshev distance (grid steps) to the nearest anchor; Rule~2 applies when one or two variables are two steps away and the others within one step.}
\label{fig:si-emulator}
\end{figure}

The emulator predicts $I$ and FWHM for any $\mathbf{x}$ in three stages (Figure~\ref{fig:si-emulator}).

\paragraph{(i) Random forests.} Two independent \texttt{RandomForestRegressor} models (scikit-learn), one for the measured peak height and one for the FWHM, take the five encoded variables as input and give $\hat I_{\mathrm{RF}}(\mathbf{x})$ and $\hat F_{\mathrm{RF}}(\mathbf{x})$; $q$ itself is not a training target. Training samples without product carry the placeholder FWHM $= 30^{\circ}$ (Table~\ref{tab:seed-raw}). Settings and five-fold cross-validation are given in Table~\ref{tab:rf}. The FWHM model has two tasks, and its overall error mixes them: the 28 samples without product sit at $30^{\circ}$ and the 67 crystalline ones mostly at 0.2--0.4$^{\circ}$ (median 0.26$^{\circ}$), so the RMSE of $7.3^{\circ}$ mainly reflects how well the two groups are told apart. Taken separately, all 28 held-out samples without product are recognized as such ($\hat I < 3000$ or $\hat F > 3^{\circ}$), 55 of the 67 held-out crystalline samples are predicted as products, and for these 55 the median error of the FWHM is 0.06$^{\circ}$ (21\% of the measured value).

\begin{table}[htbp]
\centering
\caption{Random-forest settings (scikit-learn defaults unless stated) and five-fold cross-validation (\texttt{KFold}, shuffled, \texttt{random\_state}~=~42; mean $\pm$ s.d.\ over folds).}
\label{tab:rf}
\small
\begin{tabular}{@{}ll@{\hspace{2.5em}}lcc@{}}
\toprule
Setting & Value & & $R^2$ & RMSE \\
\midrule
\texttt{n\_estimators} & 400 & Peak height & $0.64\pm0.13$ & $6{,}685\pm1{,}535$ \\
\texttt{max\_depth} & None & FWHM & $0.66\pm0.20$ & $7.3\pm2.1^{\circ}$\,$^{a}$ \\
\texttt{min\_samples\_split} / \texttt{leaf} & 2 / 1 & & & \\
\texttt{max\_features} & 1.0 (all five) & & & \\
\texttt{bootstrap} & True & & & \\
\texttt{random\_state} & 42 & & & \\
\bottomrule
\end{tabular}
\\[2pt]{\footnotesize $^{a}$Dominated by the samples without product (placeholder FWHM $=30^{\circ}$); for crystalline samples predicted as products, the median error is 0.06$^{\circ}$ (see text).}
\end{table}

\paragraph{(ii) Local neighbor blending.} Close to an anchor, the forest prediction is pulled toward the measured values. Distances between conditions are counted in grid steps, as the largest difference in any one variable (Chebyshev distance):
\begin{equation}
d_{\infty}(\mathbf{x}, \mathbf{x}_i) = \max_{k} \left| x_k - x_{i,k} \right|.
\label{eq:chebyshev}
\end{equation}
$\mathcal{N}_1(\mathbf{x})$ denotes the set of anchors within one step of $\mathbf{x}$ in every variable ($d_\infty\le1$), and $\overline{I}_{\mathcal{N}_1}$ their mean measured intensity. If this set contains at least one anchor (Rule~1), the prediction is 80\% anchor mean and 20\% forest. Otherwise, $\mathcal{N}_2(\mathbf{x})$ denotes the set of anchors two steps away in one or two variables and within one step in the others, and the prediction is a 50:50 blend (Rule~2):
\begin{equation}
\hat I = 0.8\,\overline{I}_{\mathcal{N}_1} + 0.2\,\hat I_{\mathrm{RF}}
\quad\text{or}\quad
\hat I = 0.5\,\overline{I}_{\mathcal{N}_2} + 0.5\,\hat I_{\mathrm{RF}},
\label{eq:rule1}
\end{equation}
and likewise for FWHM. If neither set contains an anchor, the forest prediction is kept.

\paragraph{(iii) Hard boundaries.}
\begin{equation}
q_{\mathrm{final}} =
\begin{cases}
0, & \hat I < 3000 \ \text{or}\ \hat F > 3^{\circ} \\
\hat I / \hat F, & \text{otherwise.}
\end{cases}
\label{eq:hard-thresh}
\end{equation}
These boundaries apply to emulator output only, not to measured data.

% =====================================================================
\FloatBarrier
\subsection{Reliability}
\label{si:loo-andgate}
\label{si:emulator-validation}

\textbf{The emulator is not used to predict any single reaction in one shot.} A search benchmark needs two things from it: that conditions are ranked as experiments would rank them, and that the productive region is not inflated. We therefore judge the search strategies by their behavior on the landscape and by six wet-lab campaigns, rather than treating the fitted landscape as a definitive physical model of MOF-321 crystallization. Both requirements were tested on the measured data.

\paragraph{Leave-one-out test.} Each anchor was removed in turn, the emulator was refit on the other 94 (five forest random states), and the removed condition was predicted. The error is measured on the logarithmic scale and averaged over the five forest random states $s$,
\begin{equation}
e_{\log}(i) = \frac{1}{5}\sum_{s=1}^{5}\left| \log(1+q_{\mathrm{pred},i,s}) - \log(1+q_{\mathrm{true},i}) \right|;
\label{eq:eloge}
\end{equation}
$e_{\log} = 0$ is a perfect prediction; a prediction of 20{,}000 for a measured $q$ of 10{,}000 (or the reverse) gives $e_{\log} = 0.69$, and a prediction of no product for a measured $q$ of 10{,}000 gives $e_{\log} = 9.2$. Table~\ref{tab:loo} lists the result for every anchor.
All 28 anchors without product were predicted as $q = 0$. For the 67 crystalline anchors, predicted and measured $q$ rank alike (Spearman $\rho = 0.78$), 57\% are predicted between half and twice the measured $q$ ($e_{\log} \le 0.69$), and 9 are wrongly predicted as no product. \textbf{The errors are one-sided: the emulator can miss part of the productive region, which makes the benchmark harder, but it never predicted a product where none was measured.}

\paragraph{Occurrence of prediction errors.} Two quantities tell whether a prediction is backed by data. The first is $d_{\min}$, the Chebyshev distance to the nearest other anchor (Eq.~\ref{eq:chebyshev}). The second measures how the five forests (fitted with five random states) vote at this condition: $B_I$ is the fraction of them whose predicted intensity lies below the intensity threshold ($\hat I < 3000$), $B_F$ the fraction whose predicted FWHM lies above the width threshold ($\hat F > 3^{\circ}$; Eq.~\ref{eq:hard-thresh}), and $B = \max(B_I, B_F)$. $B = 0$ means that all five forests predict a product, $B = 1$ that all five predict none. A prediction is flagged when it is far from data and most forests predict no product:
\begin{equation}
\mathrm{flag}(\mathbf{x}) = \mathds{1}\!\left[d_{\min}(\mathbf{x}) > 3\right] \cdot \mathds{1}\!\left[B(\mathbf{x}) > 0.5\right],
\label{eq:andgate-si}
\end{equation}
where $\mathds{1}[\cdot]$ is 1 if the condition in brackets holds and 0 otherwise. A flag is a warning, not an error: it marks a prediction of ``no product'' that no nearby measurement supports. The flagged anchors without product (Table~\ref{tab:loo}) were all predicted correctly.

\begin{figure}[htbp]
\centering
\siFig[0.75\linewidth]{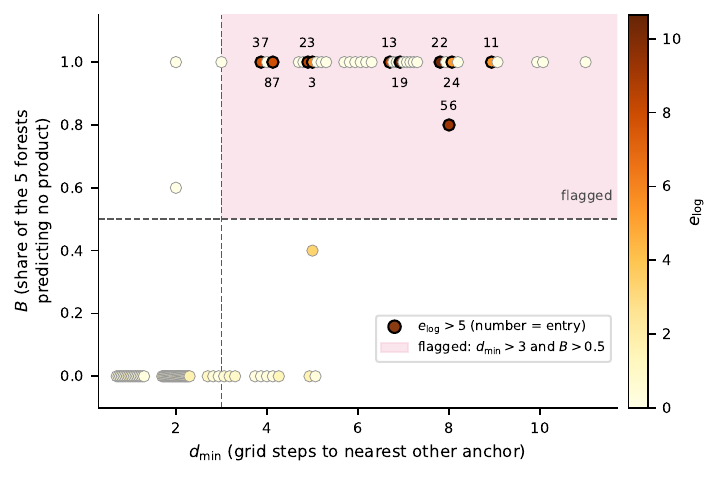}
\caption{Leave-one-out predictions of the 95 anchors. Horizontal axis, $d_{\min}$, grid steps to the nearest other anchor; vertical axis, $B$, the share of the five forests that predict no product; color, error $e_{\log}$ (light, small; dark, large). Shaded: flagged region of Eq.~\ref{eq:andgate-si} ($d_{\min} > 3$ and $B > 0.5$), i.e.\ far from data and predicted as no product by most forests. Outlined and numbered (entry of Table~\ref{tab:seed-raw}): the ten predictions with $e_{\log} > 5$, all of them in the flagged region. $d_{\min}$ and $B$ take only a few values, so many anchors fall on the same position; they are drawn side by side to keep every one visible.}
\label{fig:si-andgate-interaction}
\end{figure}

The large errors are confined to the flagged predictions (Figure~\ref{fig:si-andgate-interaction}, shaded region). The 49 crystalline anchors within three steps of another anchor have a median $e_{\log}$ of 0.51, and none has $e_{\log} > 5$. Of the 18 that lie farther away, the 8 for which the forests predict a product are predicted as well (median 0.32); the 10 for which most forests predict no product are the 10 flagged ones, and all of them are large errors (median 7.99): these are isolated products that the emulator cannot know about once they are removed.

\begingroup
\footnotesize
\setlength{\tabcolsep}{4pt}
\begin{longtable}{@{}rrrr@{\hspace{2.5em}}rrrr@{}}
\caption{Leave-one-out prediction of every anchor. No., entry of Table~\ref{tab:seed-raw}; $q$, measured; $\hat q$, mean prediction of the five forests refitted without this anchor; $e_{\log}$, Eq.~\ref{eq:eloge} (lower is better). $^{\ast}$Flagged by Eq.~\ref{eq:andgate-si}: far from the other anchors and predicted as no product by most forests. For an anchor without product ($q = 0$, $\hat q = 0$), the flag means that this correct prediction is not supported by a nearby measurement.}\label{tab:loo}\\
\toprule
No. & $q$ & $\hat q$ & $e_{\log}$ & No. & $q$ & $\hat q$ & $e_{\log}$ \\
\midrule
\endfirsthead
\toprule
No. & $q$ & $\hat q$ & $e_{\log}$ & No. & $q$ & $\hat q$ & $e_{\log}$ \\
\midrule
\endhead
\bottomrule
\endlastfoot
1$^{\ast}$ & 0 & 0 & 0.00 & 50 & 100{,}890 & 127{,}790 & 0.24 \\
2$^{\ast}$ & 0 & 0 & 0.00 & 51 & 101{,}000 & 117{,}006 & 0.15 \\
3$^{\ast}$ & 357 & 0 & 5.88 & 52 & 207{,}261 & 104{,}582 & 0.68 \\
4$^{\ast}$ & 0 & 0 & 0.00 & 53 & 55{,}318 & 71{,}696 & 0.26 \\
5$^{\ast}$ & 0 & 0 & 0.00 & 54 & 38{,}100 & 43{,}835 & 0.17 \\
6$^{\ast}$ & 0 & 0 & 0.00 & 55 & 104{,}181 & 88{,}542 & 0.16 \\
7$^{\ast}$ & 0 & 0 & 0.00 & 56$^{\ast}$ & 51{,}868 & 610 & 9.25 \\
8$^{\ast}$ & 0 & 0 & 0.00 & 57 & 1{,}147 & 1{,}612 & 3.33 \\
9$^{\ast}$ & 0 & 0 & 0.00 & 58 & 90{,}845 & 25{,}054 & 1.29 \\
10$^{\ast}$ & 0 & 0 & 0.00 & 59 & 45{,}673 & 33{,}407 & 0.32 \\
11$^{\ast}$ & 226 & 0 & 5.42 & 60 & 50{,}021 & 64{,}823 & 0.26 \\
12$^{\ast}$ & 0 & 0 & 0.00 & 61 & 163{,}182 & 157{,}476 & 0.04 \\
13$^{\ast}$ & 4{,}553 & 0 & 8.42 & 62 & 171{,}271 & 157{,}035 & 0.09 \\
14$^{\ast}$ & 0 & 0 & 0.00 & 63 & 60{,}955 & 34{,}698 & 0.56 \\
15$^{\ast}$ & 0 & 0 & 0.00 & 64 & 7{,}770 & 12{,}987 & 0.51 \\
16 & 6{,}338 & 6{,}922 & 0.09 & 65 & 80{,}348 & 90{,}515 & 0.14 \\
17 & 873 & 2{,}066 & 0.86 & 66 & 50{,}709 & 15{,}857 & 1.16 \\
18 & 76{,}141 & 19{,}113 & 1.38 & 67 & 84{,}493 & 114{,}560 & 0.30 \\
19$^{\ast}$ & 42{,}083 & 0 & 10.65 & 68 & 67{,}531 & 129{,}292 & 0.65 \\
20$^{\ast}$ & 0 & 0 & 0.00 & 69 & 25{,}579 & 93{,}364 & 1.29 \\
21$^{\ast}$ & 0 & 0 & 0.00 & 70 & 107{,}033 & 93{,}728 & 0.13 \\
22$^{\ast}$ & 32{,}240 & 0 & 10.38 & 71 & 95{,}814 & 116{,}891 & 0.20 \\
23$^{\ast}$ & 3{,}295 & 0 & 8.10 & 72 & 76{,}472 & 88{,}473 & 0.15 \\
24$^{\ast}$ & 200 & 0 & 5.30 & 73 & 247{,}107 & 106{,}067 & 0.85 \\
25 & 0 & 0 & 0.00 & 74 & 81{,}273 & 122{,}612 & 0.41 \\
26 & 114{,}138 & 111{,}644 & 0.02 & 75 & 43{,}612 & 49{,}088 & 0.12 \\
27 & 48{,}700 & 3{,}761 & 2.57 & 76$^{\ast}$ & 0 & 0 & 0.00 \\
28 & 28{,}530 & 15{,}484 & 0.61 & 77 & 31{,}696 & 40{,}010 & 0.23 \\
29 & 51{,}808 & 20{,}383 & 0.93 & 78 & 34{,}391 & 13{,}856 & 0.94 \\
30$^{\ast}$ & 0 & 0 & 0.00 & 79 & 119{,}441 & 95{,}194 & 0.23 \\
32 & 0 & 0 & 0.00 & 80 & 62{,}455 & 71{,}463 & 0.13 \\
33 & 0 & 0 & 0.00 & 81 & 14{,}174 & 78{,}046 & 1.70 \\
34 & 1{,}379 & 11{,}958 & 2.16 & 82 & 114{,}383 & 112{,}761 & 0.01 \\
35$^{\ast}$ & 0 & 0 & 0.00 & 83 & 58{,}386 & 112{,}287 & 0.65 \\
36 & 82{,}233 & 44{,}972 & 0.64 & 84 & 130{,}578 & 108{,}447 & 0.19 \\
37$^{\ast}$ & 2{,}244 & 0 & 7.72 & 85 & 99{,}186 & 116{,}373 & 0.16 \\
38 & 452 & 12{,}821 & 3.34 & 86 & 128{,}632 & 80{,}176 & 0.47 \\
39$^{\ast}$ & 0 & 0 & 0.00 & 87$^{\ast}$ & 2{,}639 & 0 & 7.88 \\
40$^{\ast}$ & 0 & 0 & 0.00 & 88 & 75{,}311 & 55{,}192 & 0.32 \\
41$^{\ast}$ & 0 & 0 & 0.00 & 89 & 60{,}312 & 11{,}694 & 1.64 \\
42$^{\ast}$ & 0 & 0 & 0.00 & 90 & 61{,}305 & 48{,}592 & 0.23 \\
43$^{\ast}$ & 0 & 0 & 0.00 & 91 & 110{,}908 & 64{,}697 & 0.54 \\
44$^{\ast}$ & 0 & 0 & 0.00 & 92 & 54{,}574 & 32{,}235 & 0.55 \\
45$^{\ast}$ & 0 & 0 & 0.00 & 93 & 38{,}893 & 45{,}767 & 0.16 \\
46 & 155{,}410 & 45{,}962 & 1.23 & 94 & 94{,}093 & 81{,}203 & 0.15 \\
47 & 42{,}637 & 16{,}230 & 1.00 & 95 & 57{,}726 & 14{,}716 & 1.38 \\
48$^{\ast}$ & 0 & 0 & 0.00 & 96 & 34{,}285 & 12{,}227 & 1.07 \\
49 & 108{,}381 & 39{,}681 & 1.01 & & & & \\

\end{longtable}
\endgroup

\paragraph{Data support across the landscape.} Table~\ref{tab:dmin-boxstats} applies the same two quantities to the whole landscape. The productive region is well supported: every top-1\% condition lies within 7 steps of an anchor (74\% within 3), all five forests predict a product for every one of them, and none is flagged. The no-product region is different. It makes up 94\% of the landscape, and 93\% of it is flagged: these conditions are predicted as no product by most forests, but no measurement lies within three steps. \textbf{The zeros of the landscape are therefore mostly extrapolations, not observations.} They are consistent with the 28 measured failures, which are spread over the whole range of every variable (Figure~\ref{fig:si-seed-coverage}) and were all predicted correctly when left out, but the exact extent of the no-product region depends on the emulator. For this reason the comparison of the strategies was repeated on 41 landscapes rebuilt with other choices (Section~\ref{si:landscape-robustness}).

\begin{table}[tb]
\centering
\caption{Support of the emulated landscape by the anchors, by region. $d_{\min}$, grid steps to the nearest anchor; Within 3 steps, share of the conditions with $d_{\min} \le 3$; Forests agree, share for which all five forests give the same answer (product or no product); Flagged, Eq.~\ref{eq:andgate-si}. Top $x$\%: the $x$\% of emulated conditions with the highest $q$ (Table~\ref{tab:tiers}).}
\label{tab:dmin-boxstats}
\small
\begin{tabular}{@{}lrcccc@{}}
\toprule
Region of the landscape & Conditions & Median $d_{\min}$ & Within 3 steps & Forests agree & Flagged \\
\midrule
Top 0.1\% & 6{,}104 & 2 & 92\% & 100\% & 0\% \\
Top 1\% & 61{,}013 & 3 & 74\% & 100\% & 0\% \\
Top 5\% & 305{,}056 & 4 & 36\% & 99.8\% & 0\% \\
All with product ($q > 0$) & 384{,}112 & 4 & 36\% & 96\% & 0.6\% \\
No product ($q = 0$) & 5{,}716{,}965 & 6 & 6\% & 93\% & 93\% \\
Full landscape & 6{,}101{,}077 & 6 & 8\% & 93\% & 87\% \\
\bottomrule
\end{tabular}
\end{table}

\paragraph{Prediction of later experiments.} A map built from only the first 60 anchors in recording order (Table~\ref{tab:seed-raw}) predicted the 35 later ones, which it had not seen. It ranked them better than the $q$ of the nearest earlier anchor did (Spearman $\rho = 0.61$ vs.\ 0.43) and again made only false zeros (7 of 34 crystalline anchors), never a false product. With more anchors the ranking improved steadily, from $\rho = 0.48$ at 40 anchors to 0.83 at 80 (Table~\ref{tab:temporal}). In absolute magnitude the nearest earlier anchor was closer (median $e_{\log}$ 0.37 vs.\ 0.78), because the campaign stayed near earlier experiments; \textbf{the benchmark relies on the ranking, not on the magnitude}.

\begin{table}[htbp]
\centering
\caption{Prediction of later anchors from the first $N$ anchors (recording order). Emulator: the landscape rebuilt from the first $N$ anchors only, read at the condition of each later anchor. Nearest anchor: a simple comparison that assigns each later anchor the measured $q$ of the closest of the first $N$ anchors (Chebyshev distance; mean $q$ if several are equally close). $\rho$, Spearman correlation between predicted and measured $q$ (1, same order). False zeros: crystalline later anchors predicted as $q = 0$ by the emulator.}
\label{tab:temporal}
\small
\begin{tabular}{@{}rcccc@{}}
\toprule
$N$ & Later anchors (crystalline) & $\rho$, emulator & $\rho$, nearest anchor & False zeros \\
\midrule
40 & 55 (49) & 0.48 & 0.56 & 23/49 \\
50 & 45 (44) & 0.52 & 0.37 & 20/44 \\
60 & 35 (34) & 0.61 & 0.43 & 7/34 \\
70 & 25 (24) & 0.75 & 0.38 & 3/24 \\
80 & 15 (15) & 0.83 & 0.60 & 1/15 \\
\bottomrule
\end{tabular}
\end{table}

% =====================================================================
\subsection{Structure}
\label{si:landscape}

\paragraph{Sparsity.} Only 6.3\% of the landscape yields a crystalline product, and high $q$ is rare even among these conditions: 1\% of all conditions reach $q \ge \qTopOne$ and 0.1\% reach $q \ge \qTopPointOne$ (Table~\ref{tab:tiers}). The anchors look far less sparse (29.5\% zero, 46\% above the top-1\% threshold) because they come from an optimization campaign concentrated near the productive region (Figure~\ref{fig:si-seed-coverage}).

\begin{table}[H]
\centering
\caption{Quality tiers of the landscape and of the 95 anchors. Top $x$\%: the $x$\% of the 6{,}101{,}077 emulated conditions (anchor positions excluded) with the highest $q$; the threshold is the lowest $q$ in the tier. The same thresholds are used throughout the SI and the main text.}
\label{tab:tiers}
\small
\begin{tabular}{@{}lrrrr@{}}
\toprule
Tier & $q$ threshold & Landscape & Fraction & Anchors \\
\midrule
Top 0.01\% & $\ge\qTopPointZeroOne$ & 611 & 0.01\% & 5 \\
Top 0.1\% & $\ge\qTopPointOne$ & 6{,}104 & 0.10\% & 16 \\
Top 1\% & $\ge\qTopOne$ & 61{,}013 & 1.0\% & 44 \\
Top 5\% & $\ge\qTopFive$ & 305{,}056 & 5.0\% & 57 \\
Non-zero & $>0$ & 384{,}112 & 6.3\% & 67 \\
Zero & $=0$ & 5{,}716{,}965 & 93.7\% & 28 \\
\bottomrule
\end{tabular}
\end{table}

\begin{figure}[htbp]
\centering
\siFig{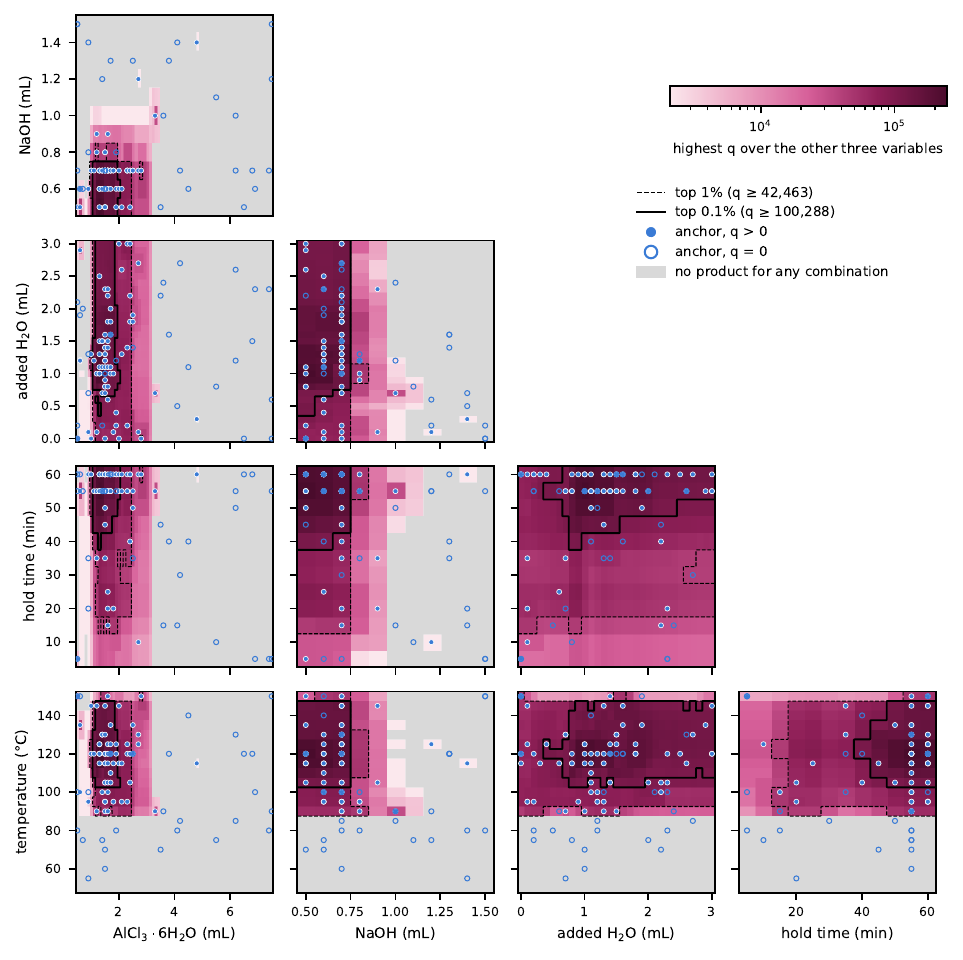}
\caption{The landscape on all ten two-variable projections: in each panel, the highest $q$ over the other three variables, with the outlines of the top-1\% (dashed) and top-0.1\% (solid) tiers, drawn along the edges of the grid cells; gray, no product for any combination of the other three variables. Filled and open circles: anchors with $q>0$ and $q=0$.}
\label{fig:si-heatmaps}
\end{figure}

\paragraph{Two-dimensional projections.} Figure~\ref{fig:si-heatmaps} shows, for all ten pairs of the five variables, the best $q$ reachable when the other three variables are chosen freely. The crystalline region is one block: about 1--3~mL \ce{AlCl3.6H2O}, less than about 1~mL NaOH and at least 90~$^{\circ}$C, at any amount of added DI water and hold time. The top-0.1\% tier is a smaller block inside it, at 1.1--1.9~mL \ce{AlCl3.6H2O}, 0.5--0.7~mL NaOH, hold times of 40~min or more and 100--145~$^{\circ}$C, again over most of the range of added DI water.

\paragraph{Chemical trends.} High-$q$ conditions combine low \ce{AlCl3.6H2O} (1.1--1.8~mL for all 16 top-0.1\% anchors), low NaOH (0.5--0.7~mL), long hold times (55--60~min) and 115--145~$^{\circ}$C, with added DI water mattering least (Figure~\ref{fig:si-heatmaps}). The base window is narrow: too little base leaves the linker protonated, while excess base is clearly detrimental, with 13 of 16 anchors at 1.0~mL NaOH or more giving no product and a single step from 0.7 to 0.8~mL (entries 18 and 34) lowering $q$ from 76{,}141 to 1{,}379. All anchors at or below 75~$^{\circ}$C fail.

The top-0.1\% tier is essentially one connected region (6{,}100 of its 6{,}104 grid points). Its surroundings are steep: of all pairs of neighboring conditions (one grid step apart in one variable) that include a top-1\% condition, 29\% belong to two different quality tiers, so a single grid step often changes the tier.

\paragraph{Settings of the embedding and density plots.} The two-dimensional embedding of the landscape shown in the main text was computed with UMAP (\texttt{n\_neighbors} = 10, \texttt{min\_dist} = 0.2, Euclidean metric, \texttt{random\_state} = 42) on standardized variables, for 1\% of the landscape sampled separately from $q=0$ and $q>0$ (61{,}012 conditions) plus the 95 anchors. The no-product conditions are split at the medians of the embedded set (100~$^{\circ}$C; 4.0~mL \ce{AlCl3.6H2O}). The density curves are Gaussian kernel density estimates (Scott's rule) on a 2\% random sample of each tier.

% =====================================================================
\FloatBarrier
\subsection{Validation Optima}
\label{si:six-mof-localization}

\begin{figure}[htbp]
\centering
\siFig{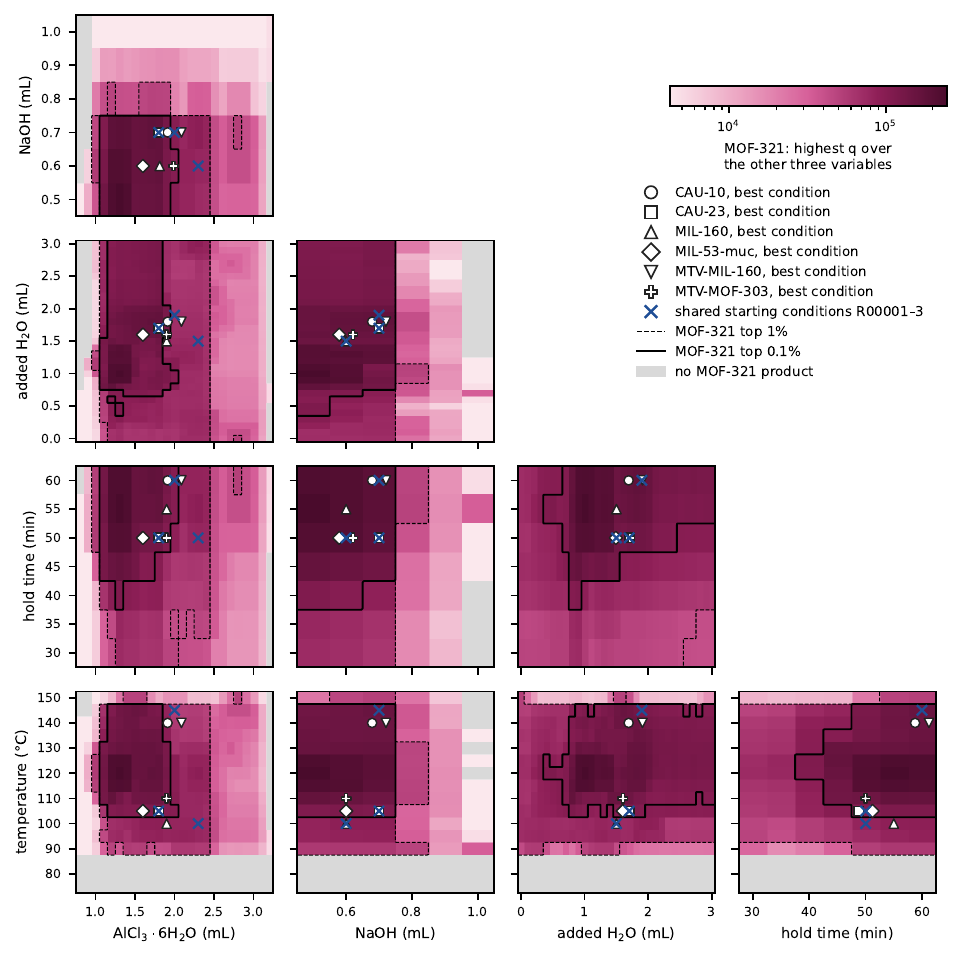}
\caption{Best-found conditions of the six validation MOFs (open symbols) and the three shared starting conditions (crosses) on all ten two-variable projections of the MOF-321 landscape (highest $q$ over the other three variables, with the outlines of the top-1\% (dashed) and top-0.1\% (solid) tiers), zoomed to the productive part of the grid. Symbols in the same grid cell are spread horizontally; the best CAU-23 condition is starting condition R00001.}
\label{fig:si-six-mof}
\end{figure}

Six further Al MOFs were optimized in the wet lab, each in 12 experiments that started from the MOF-321 data; the transfer-learning method and the campaigns are described in Section~\ref{si:transfer}, the syntheses in Sections~\ref{si:synthesis} and~\ref{si:pxrd}. Here we ask only where their best conditions lie on the MOF-321 landscape. We compare positions, not values: $q$ is comparable only within one phase (Section~\ref{si:q-def}), so the $q$ that the landscape predicts for MOF-321 at a condition is not compared with the $q$ measured there for another framework. For each validation MOF, the best-found condition $\mathbf{x}^\ast$ was located on the MOF-321 landscape by its Chebyshev distance to the nearest anchor ($d_{95}$) and to the top-1\% and top-0.1\% tiers (Figure~\ref{fig:si-six-mof}, Table~\ref{tab:six-mof-localization}), measured to the nearest condition of the tier and to its centroid, the mean of the integer codes of all conditions in the tier (a point that need not lie on the grid). All six fall on grid points inside MOF-321's own top-1\% tier and within 1--2 steps of its top-0.1\% tier, whereas a condition drawn at random from the grid lies a median of 16 and 20 steps from the nearest condition of these tiers (20{,}000 random draws). The three shared starting conditions already sit in the top 1\% (Table~\ref{tab:six-mof-localization}, bottom), so the transferred prior places each search in MOF-321's productive region from the start; the searches stayed there while re-learning $q$ for each framework, and MTV-MOF-303 ended closer to the top-0.1\% centroid than any starting point.

\begin{table}[H]
\centering
\caption{Distance of each validation MOF's best condition, and of the three shared starting conditions, to MOF-321 anchors and top tiers (Chebyshev units).}
\label{tab:six-mof-localization}
\small
\begin{tabular}{@{}lccrrrrr@{}}
\toprule
 & & & & \multicolumn{2}{c}{Top 1\%} & \multicolumn{2}{c}{Top 0.1\%} \\
\cmidrule(lr){5-6}\cmidrule(lr){7-8}
MOF & Best ID & Best $q$ & $d_{95}$ & nearest & centroid & nearest & centroid \\
\midrule
CAU-10 & R00012 & 32{,}259 & 3 & 0 & 4.1 & 2 & 4.9 \\
CAU-23 & R00001 & 22{,}254 & 3 & 0 & 2.9 & 1 & 4.2 \\
MIL-53-muc & R00008 & 58{,}064 & 2 & 0 & 2.9 & 1 & 4.2 \\
MIL-160 & R00007 & 44{,}395 & 2 & 0 & 3.9 & 1 & 5.2 \\
MTV-MIL-160 & R00007 & 30{,}990 & 3 & 0 & 4.1 & 2 & 4.9 \\
MTV-MOF-303 & R00009 & 53{,}648 & 2 & 0 & 2.2 & 1 & 3.9 \\
\midrule
\multicolumn{8}{@{}l}{\textit{Shared starting conditions} (landscape percentile)} \\
R00001 (18,7,17,10,21) & & 99.78\% & 3 & 0 & 2.9 & 1 & 4.2 \\
R00002 (23,6,15,10,20) & & 99.30\% & 2 & 0 & 6.2 & 4 & 7.9 \\
R00003 (20,7,19,12,29) & & 99.44\% & 3 & 0 & 5.1 & 2 & 4.9 \\
\bottomrule
\end{tabular}
\end{table}

% =====================================================================
\section{Search Strategies}
\label{si:policies}

% =====================================================================

This section describes how the search policies explore the ActMOF landscape: the shared protocol (Section~\ref{si:protocol}), Gaussian processes (Section~\ref{si:gp}) and the acquisition rules (Section~\ref{si:acquisition}), the LLM and hybrid policies (Sections~\ref{si:llm}--\ref{si:hybrid}) and a list of all 39 strategies that were run, with a short description of the other model types (Section~\ref{si:strategy-table}). The code is in the \href{https://github.com/deep-synthesis/ActMOF}{GitHub repository} (\texttt{benchmark/families/}). The quantities used to compare strategies (quality tiers, iterations, hit rates) are defined in Table~\ref{tab:metric-defs} (Section~\ref{si:cum-best}).

\subsection{Protocol}
\label{si:protocol}

\paragraph{Search loop.} Each run starts from three experiments and then proposes three experiments per iteration. The landscape (Section~\ref{si:landscape}) returns the $q$ of each proposed condition, which is added to the history before the next iteration. A run has a budget of 120 experiments ($3 + 39 \times 3$, iterations 0--39). Every strategy was run five times ($n=5$); all 39 strategies are shown in Figure~\ref{fig:si-overview}. Table~\ref{tab:protocol} lists the settings.

\begin{figure}[htbp]
\centering
\siFig[0.6\linewidth]{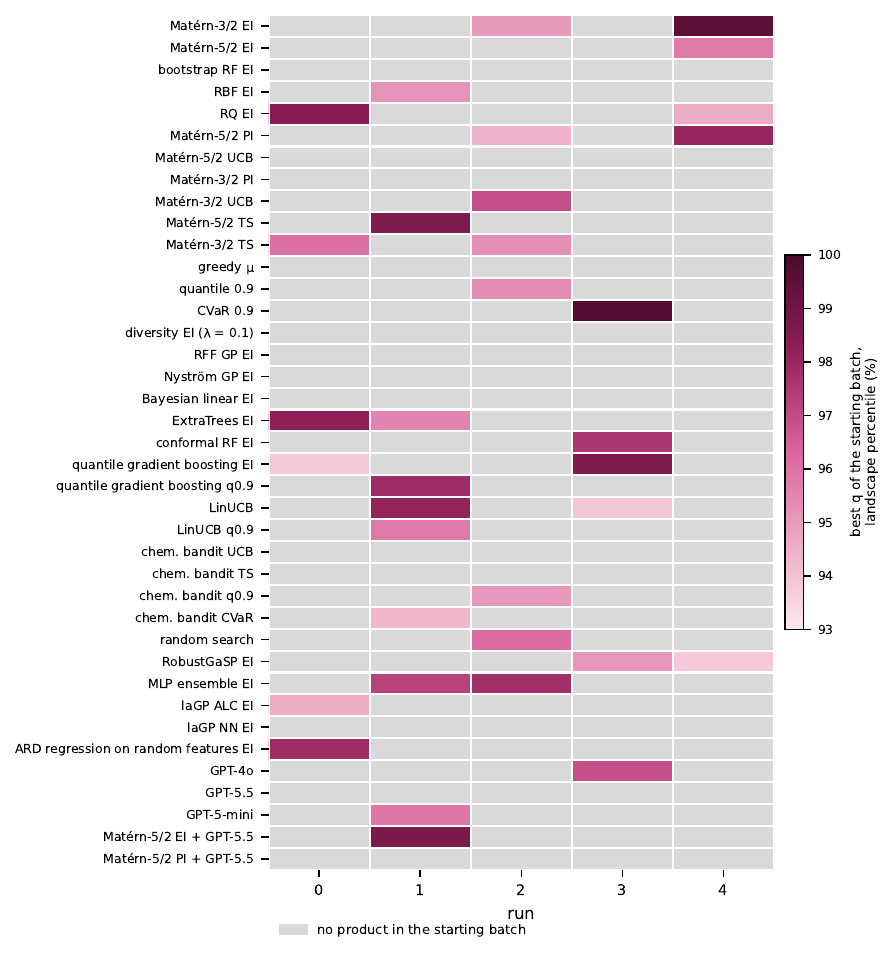}
\caption{Starting batches of all 39 strategies. Each cell is one run; runs 0--4; color gives the landscape percentile of the best of the three starting experiments; gray, no crystalline product in the starting batch.}
\label{fig:si-starts}
\end{figure}

\paragraph{Starting experiments.} The three starting experiments are drawn uniformly at random from the grid of Eq.~\ref{eq:grid-size} (function \texttt{run\_single\_method}, stage \texttt{initial\_random}). The random generator of run $r$ of method $m$ is seeded with
\begin{equation}
s_{m,r} = 42 + 10{,}000\,r + h(m),
\label{eq:seed}
\end{equation}
where $h(m)$ is a hash of the method name: a fixed integer computed from the name by a hash function (\texttt{stable\_int} in the code). Because $h(m)$ differs between methods, \emph{the starting experiments are not shared between methods}, even within the same run index; this holds for all methods, including the LLM and hybrid policies. Each run of each strategy therefore starts from its own random conditions; 35 of the 195 starting batches (39 strategies, five runs each) already contain a crystalline product (Figure~\ref{fig:si-starts}).

\paragraph{Independent and shared starting batches.} Independent starts are the usual protocol of optimization benchmarks. Each strategy is averaged over its own set of starting situations, so no strategy is favored or penalized by one particular set of starts, and the spread over runs includes the effect of the start, as in a real campaign whose first experiments are not known in advance. The price is that, with few runs, two strategies can differ only because one of them drew luckier starts. Where this matters, in the early iterations and in the comparison of PI and EI, the runs were repeated with the same three starting conditions for every strategy (shared starts; Sections~\ref{si:pi-vs-ei}, \ref{si:warmstart} and~\ref{si:feasibility}), which removes this source of difference.

\paragraph{Before the first product.} Because 94\% of the landscape gives no product, a run usually begins with several experiments that all return $q=0$. As long as all results of a run are the same, a model has nothing to learn from: it cannot tell one condition from another. In this phase every candidate receives the same score, and the three conditions of a batch are in effect drawn at random; only the rule of the next paragraph, which keeps them apart, still acts. A model is fitted from the first iteration in which two different results exist.

This phase matters for two reasons. First, until the first product appears, every model-based strategy behaves like random search, whatever its model and acquisition rule, so that differences between strategies in the first iterations mostly reflect luck (Section~\ref{si:pi-vs-ei}). Second, the phase can use up a large part of a short campaign: in the shared-start runs of Mat\'ern-5/2 EI, half of the 20 runs needed more than 12 experiments to find their first product (Section~\ref{si:feasibility}), and in iterations 1--5 of the five Mat\'ern-5/2 runs, 57 of the 75 EI queries and 24 of the 75 PI queries were chosen in this way. This is why the wet-lab campaigns, which have only 12 experiments, start from a prior and not from random conditions (Section~\ref{si:warmstart}).

\paragraph{Candidate sampling and batch selection.} Scoring all 6.1 million conditions in every iteration would be slow, so each iteration scores a random subset: 15{,}000 conditions drawn uniformly from the grid, excluding those already tested in the run. The two local approximate GPs (laGP) score 4{,}000 candidates to reduce the computing cost. Random search draws its three conditions directly; the LLM policies write their conditions without a candidate list (Section~\ref{si:llm}); the hybrid policies use the candidate list only in their GP iterations. Each candidate $\mathbf{x}$ receives a score $a(\mathbf{x})$ from the acquisition function of the strategy (Section~\ref{si:acquisition}), which measures how useful it would be to test this condition next. The three best-scored candidates are often almost the same condition, because neighboring conditions get similar scores, and testing all three would waste two experiments. The batch is therefore built one condition at a time: the first is the best-scored candidate, and each further one maximizes the score plus a bonus for its distance from the conditions already chosen,
\begin{equation}
\tilde{a}(\mathbf{x}) + \lambda \min_{\mathbf{x}' \in B} \lVert \mathbf{x}-\mathbf{x}' \rVert_2, \qquad \lambda = 0.03 .
\label{eq:diverse-batch}
\end{equation}
Here $\tilde{a}$ is the standardized score,
\begin{equation}
\tilde{a}(\mathbf{x}) = \frac{a(\mathbf{x}) - \bar{a}}{s_a},
\label{eq:zscore}
\end{equation}
where $\bar{a}$ and $s_a$ are the mean and the standard deviation of $a$ over the candidates of the iteration; the standardization gives $\lambda$ the same meaning for every acquisition function. $B$ is the set of conditions already in the batch, and distances are measured with every variable scaled to $[0,1]$ over its range. With the small value $\lambda = 0.03$, the bonus decides only between candidates of nearly equal score. One of the 39 strategies, ``diversity EI'' (Table~\ref{tab:strategies}), uses $\lambda = 0.10$ to test the effect of a stronger bonus. Before the batch is built, a random number of order $10^{-12}$ is added to every score. It is far too small to change the order of different scores, but it puts candidates with exactly equal scores (for example, all candidates before the first product) in a random order, so that they are not chosen by their position in the list.

\begin{table}[htbp]
\centering
\caption{Benchmarking protocol.}
\label{tab:protocol}
\small
\begin{tabular}{@{}p{0.3\linewidth}p{0.62\linewidth}@{}}
\toprule
Setting & Value \\
\midrule
Starting experiments & 3, uniform random on the grid; random seed Eq.~\ref{eq:seed}, differs between methods \\
Batch size per iteration & 3 \\
Total budget & 120 experiments ($3 + 39\times3$), iterations 0--39 \\
Replicate runs & $n=5$ for all 39 strategies; the leading group was extended to $n=20$ (Section~\ref{si:leading-n20}) \\
Model target & $y=\log(1+q)$; incumbent $y^{+}=\max y$ observed \\
Input scaling & each variable scaled to $[0,1]$ over its grid range \\
Candidates per iteration & 15{,}000 untried grid points (laGP: 4{,}000) \\
Batch construction & one condition at a time: standardized score plus distance bonus, Eq.~\ref{eq:diverse-batch}, $\lambda = 0.03$ \\
Exploration settings & EI, PI: smallest improvement that counts, $\xi = 0.01$ on the $\log(1+q)$ scale (about 1\% in $q$); UCB: weight of the uncertainty, $\beta = 2$; quantile and CVaR: 90\% level, $\tau = 0.9$ (Section~\ref{si:acquisition}) \\
No product yet & no model until two different results exist; all candidates scored equally \\
Oracle & noise-free landscape $q$ (Section~\ref{si:landscape}); noisy variant in Section~\ref{si:pi-vs-ei} \\
\bottomrule
\end{tabular}
\end{table}

\FloatBarrier
\subsection{Gaussian Processes}
\label{si:gp}

A Gaussian process (GP) is a flexible regression model that gives, for every untested condition, a predicted value $\mu(\mathbf{x})$ and an uncertainty $\sigma(\mathbf{x})$ of that prediction. It assumes only that conditions close to each other give similar results, and it learns from the data how close is close, separately for every variable. Near tested conditions $\sigma$ is small; far from them it is large. The acquisition functions (Section~\ref{si:acquisition}) use $\mu$ and $\sigma$ to weigh refining a good region against exploring an unknown one.

The GP policies use scikit-learn's\cite{pedregosa2011} \texttt{GaussianProcessRegressor}. The similarity of two conditions is given by a Mat\'ern kernel $k_\nu(r)$ of their scaled distance
\begin{equation}
r^2 = \sum_{j=1}^{5} \frac{(x_j - x'_j)^2}{\ell_j^2},
\label{eq:matern}
\end{equation}
plus a noise term that lets the GP fit data that are not exactly smooth. The kernel parameter $\nu$ sets how smooth the GP assumes the response to be: $\nu = 5/2$ gives a smoother response, $\nu = 3/2$ allows sharper changes, as at the edge of a narrow crystallization window (RBF and rational-quadratic kernels were also run; Table~\ref{tab:strategies}). The length scale $\ell_j$ is the change of variable $j$ (as a fraction of its range) over which $q$ is expected to change appreciably: a short length scale means that $q$ is sensitive to that variable, a long one that it hardly matters. All parameters are refitted to the data in every iteration (fitting bounds and optimizer settings in the repository). The GP is fitted to $y=\log(1+q)$: the logarithm compresses the range of $q$ (0 to $2.5\times10^{5}$), so that one very high value does not dominate the fit, and $\log(1+q)$ keeps $q=0$ finite. RobustGaSP (Mat\'ern-5/2 with estimated noise) and the local approximate GPs of the laGP package\cite{gramacy2015lagp}, which fit a small GP to the 10--60 tested conditions nearest to each candidate, were run through R.

\begin{figure}[htbp]
\centering
\siFig[0.95\linewidth]{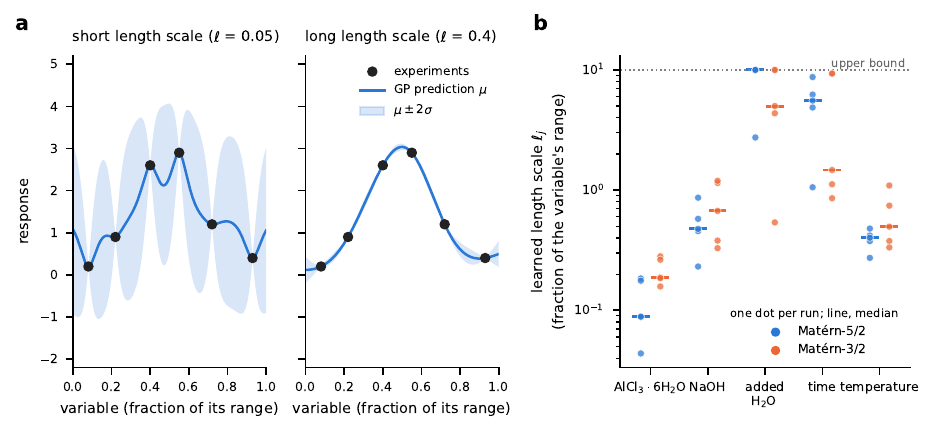}
\caption{Length scales. (a) Illustration with one variable: the same six experiments fitted by a GP with a short and with a long length scale $\ell$; line, prediction $\mu$; band, $\mu \pm 2\sigma$. (b) Length scales learned by the Mat\'ern-5/2 and Mat\'ern-3/2 GPs with EI from the 120 experiments of each run ($n=5$), as a fraction of the range of each variable; one dot per run; line, median; dotted line, upper bound of the fit.}
\label{fig:si-lengthscales}
\end{figure}

Figure~\ref{fig:si-lengthscales}a illustrates the length scale with the same six experiments fitted twice. With a short length scale, the GP expects the response to change over short distances: it follows every point closely and is very uncertain between the points. With a long length scale, it draws one smooth curve through the points and is confident between them. Because one length scale is fitted for each of the five variables, the fitted values show which variables $q$ is sensitive to (Figure~\ref{fig:si-lengthscales}b). After 120 experiments, \ce{AlCl3.6H2O} has the shortest length scale (median 0.09 of its range for Mat\'ern-5/2), followed by temperature (0.40) and NaOH (0.48), whereas added DI water and hold time have length scales longer than their whole range; for added DI water the length scale reaches the upper bound of the fit in four of five Mat\'ern-5/2 runs. This agrees with the landscape (Figure~\ref{fig:si-heatmaps}): the metal and base amounts and the temperature decide whether a product forms; DI water hardly matters.

\FloatBarrier
\subsection{Acquisition Functions}
\label{si:acquisition}

With GP mean $\mu(\mathbf{x})$ and standard deviation $\sigma(\mathbf{x})$ on the $\log(1+q)$ scale,
\begin{equation}
\mathrm{PI}(\mathbf{x}) = \Phi(z), \qquad z = \frac{\mu(\mathbf{x}) - y^{+} - \xi}{\sigma(\mathbf{x})},
\label{eq:pi}
\end{equation}
\begin{equation}
\mathrm{EI}(\mathbf{x}) = \big(\mu(\mathbf{x}) - y^{+} - \xi\big)\,\Phi(z) + \sigma(\mathbf{x})\,\phi(z),
\label{eq:ei}
\end{equation}
as in efficient global optimization\cite{jones1998ego}, where $y^{+}$ is the largest $\log(1+q)$ observed so far in the run, $\xi = 0.01$, and $\Phi$ and $\phi$ are the cumulative distribution function and the density of the standard normal distribution. PI is the probability that a candidate beats the best result so far by at least $\xi$; EI is the expected size of that improvement, counting a result below the best as zero. The other rules in Table~\ref{tab:strategies} score candidates by an optimistic estimate of their result: UCB by $\mu + \beta\sigma$ with $\beta = 2$; the upper quantile by the value that the result exceeds with 10\% probability, $\mu + \Phi^{-1}(\tau)\sigma$ with $\tau = 0.9$; and the upper conditional value at risk (CVaR) by the mean of the best 10\% of possible results, $\mu + \sigma\phi(\Phi^{-1}(\tau))/(1-\tau)$. Thompson sampling draws, for every candidate, one random plausible result from its predicted distribution $\mathcal{N}(\mu,\sigma^2)$ and picks the highest draws, so that uncertain candidates are picked from time to time. Greedy $\mu$ picks the highest predictions and ignores the uncertainty.

\paragraph{What PI and EI reward.} When $\mu$ exceeds $y^{+}$, PI is highest for small $\sigma$, so it rewards any improvement, however small; EI always increases with $\sigma$ at fixed $\mu$, because it also rewards the size of the possible gain. Section~\ref{si:pi-vs-ei} compares the two rules.

\begin{figure}[htbp]
\centering
\siFig[0.92\linewidth]{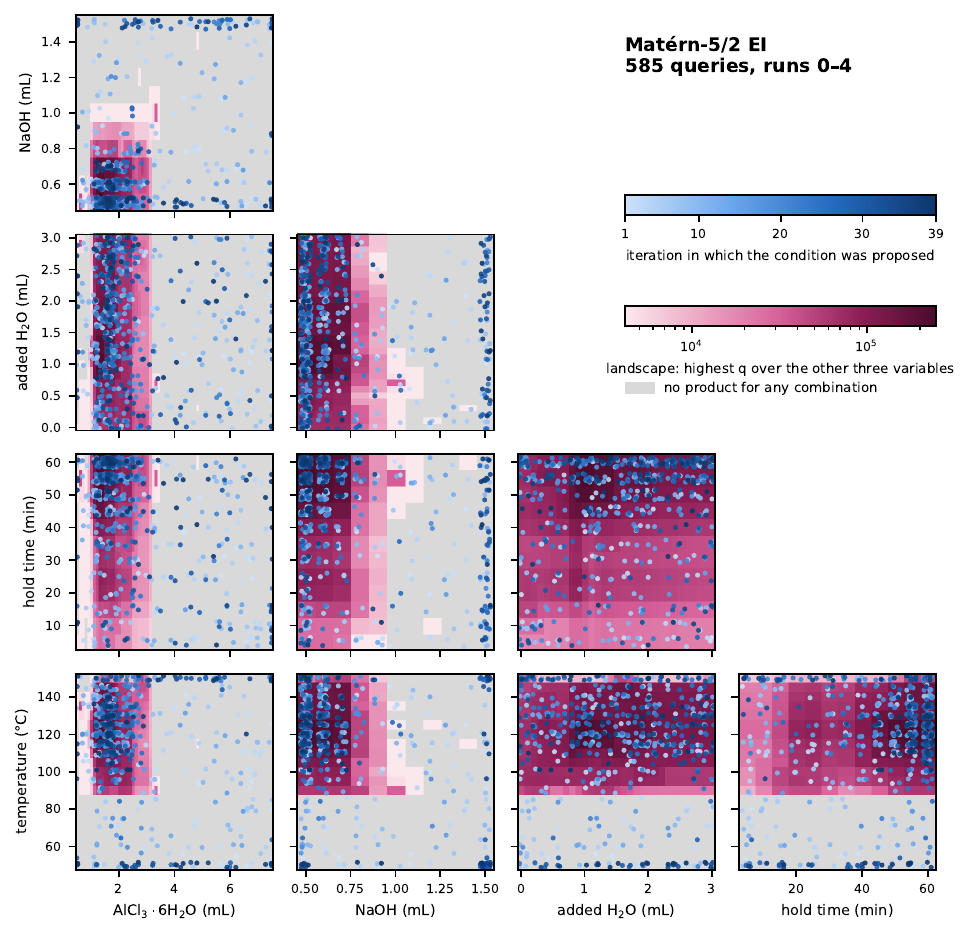}
\caption{Conditions proposed by Mat\'ern-5/2 EI, runs 0--4, iterations 1--39, on all ten two-variable projections of the landscape (highest $q$ over the other three variables; gray, no product). Color gives the iteration in which a condition was proposed, light (iteration 1) to dark (iteration 39); later iterations are drawn on top. Points are jittered within their grid cell.}
\label{fig:si-gp-proposals}
\end{figure}

\paragraph{GP-EI proposals.} Figure~\ref{fig:si-gp-proposals} shows the 585 conditions proposed by Mat\'ern-5/2 EI in runs 0--4. The early proposals (light) are spread over the whole grid; the later ones (dark) concentrate on the productive region. Overall, 59\% of the proposals give a product, 39\% reach the top-1\% tier and 16\% the top-0.1\% tier.

\FloatBarrier
\subsection{LLM Policies}
\label{si:llm}

{\sloppy\paragraph{Models and settings.} Three OpenAI models were used through the Chat Completions API with structured output: \texttt{gpt-4o}, \texttt{gpt-5.5} and \texttt{gpt-5-mini}. The snapshots were those of August 2026: \texttt{gpt-4o-2024-08-06}, \texttt{gpt-5.5-2026-04-23} and \texttt{gpt-5-mini-2025-08-07}. gpt-4o was run at temperature 0.2. The GPT-5 models do not accept a temperature and were run with the lowest reasoning effort (\texttt{none} for gpt-5.5, \texttt{minimal} for gpt-5-mini). A fixed seed was sent with each call: the seed of the run (Eq.~\ref{eq:seed}) plus the number of the iteration. The iteration number is added because the seed of the run alone would give all 39 calls of a run the same seed; with it, every call has its own fixed seed and can be repeated on its own. Each model was run for 5 runs of 39 iterations.\par}
\paragraph{Prompt.} Every request has two messages. The system message is the same in every call and gives the chemistry, the objective and the integer encoding of the five variables (Box~S\ref{box:system}). The user message is a JSON object built from the history of the run (Box~S\ref{box:user}): the objective, the allowed ranges, the last 45 experiments of the run (conditions and $q$, oldest first) and the instructions to propose exactly three new experiments, ``one exploitation condition, one local perturbation, and one exploration condition''. The same messages were used for all three models and for the LLM iterations of the hybrids. Boxes~S\ref{box:system} and~S\ref{box:user} show the two messages in full; lines in green are our comments and are not part of the messages. They describe the set-up of the original MOF-321 syntheses (1~mmol linker, 35~mL vessel), not the validation syntheses. The model is given no list of candidates: it writes the integer codes of three new conditions itself, guided by the previous experiments and the allowed ranges in the user message.

\begin{sibox}{System message.}{box:system}
\siVerb{box1_system_prompt.txt}
\end{sibox}

\begin{sibox}{User message: the JSON object sent with every request. Angle brackets mark the values filled in from the run; \texttt{previous\_experiments} holds the 45 most recent experiments of the run, \texttt{repair\_context} is empty except on a retry (One iteration of the loop).}{box:user}
\siVerb{box2_user_template.txt}
\end{sibox}

\paragraph{Structure of one iteration.} (i)~\emph{Input.} The model receives nothing but the two messages: no landscape, no candidate list and no GP output. The API keeps no state between calls; every call is a new request, so everything the model knows about the run reaches it through \texttt{previous\_experiments}. Each entry holds the five integer codes of an experiment and its $q$ as a plain number (0 for no product), the same value every other strategy observes. The first request contains the three starting experiments; from iteration 16 on, the history exceeds 45 experiments and the oldest drop out of the message. This list is the model's only memory of the run: Box~S\ref{box:example} shows a complete request and the answer to it.
(ii)~\emph{Proposal.} The response is constrained by structured output to a fixed schema: a short strategy summary and exactly three conditions, each with the five integer codes and a rationale. The conditions are then checked by our program, not by the model. The program keeps the complete list of the conditions tested in the run, so it can also detect a repeat of an experiment that is no longer among the 45 sent to the model. A batch is rejected if a condition is out of range, repeats any earlier experiment of the run or is duplicated within the batch; the model is then called again with the same messages and the error in \texttt{repair\_context} (``Your previous answer was invalid: \ldots''; Box~S\ref{box:repair}). After three failed attempts, three random untried conditions are used.
(iii)~\emph{Feedback.} The three accepted conditions are looked up on the landscape, as for every other strategy, and appended with their $q$ to the history of the run; the next request is built from the updated history. Only conditions and $q$ are sent back. The model's earlier rationales and strategy summaries are logged but not returned to it, and rejected answers do not enter the history. Random fallback conditions enter it like any other experiment. In the hybrids, GP and LLM iterations append to the same history, so the LLM sees the GP's experiments in the same form as its own.

\begin{sibox}{Example request and response: gpt-5.5, run 0, iteration 5. The user message contains the 15 experiments of the run so far (3 starting experiments and the 12 of iterations 1--4); the response gives the three conditions with the model's rationales.}{box:example}
\siVerb{box4_example.txt}
\end{sibox}

\begin{sibox}{Example of a retry: gpt-5.5, run 2, iteration 30, from the log of the run.}{box:repair}
\siVerb{box5_repair_example.txt}
\end{sibox}

\paragraph{Validity of the answers.} gpt-5.5 returned a valid batch at the first attempt in 191 of 195 iterations and needed the random fallback once; in the hybrids, gpt-5.5 needed it in 2 (EI hybrid) and 0 (PI hybrid) of 95 LLM iterations. gpt-4o needed the fallback in 54 of 195 iterations (28\%) and gpt-5-mini in 70 (36\%), so part of their queries was random. Almost all rejected answers repeated an earlier experiment of the run (249 of 259 rejections for gpt-4o, all 379 for gpt-5-mini). A repeat can occur even when the earlier experiment is in the message: in Box~S\ref{box:repair}, the rejected condition had been tested in the previous iteration.

\paragraph{Reproducibility.} A language model writes its answer by sampling one word after another from a probability distribution, so the same request can give a different answer each time, even at the lowest temperature or reasoning effort and with a fixed seed (the seed is honored by the provider on a best-effort basis). This variation can be reduced but not removed. We therefore (i)~recorded the snapshot behind each alias; (ii)~fixed every sampling setting the models accept (temperature or reasoning effort, as above; at most 1{,}500 completion tokens) and sent a fixed seed with each call; (iii)~started each campaign with a test call that stopped the run if the API had silently dropped the seed, the temperature or the reasoning effort, or if a GPT-5 model used more than 200 reasoning tokens; (iv)~used the same system message and the same code to build the user message in the LLM-only and hybrid scripts; (v)~drew the starting experiments from the fixed seed of the run (Eq.~\ref{eq:seed}); and (vi)~logged every iteration: the accepted conditions with their rationales, the strategy summary, every attempt with its error, whether the fallback was used, the best-so-far and the cost. Because the user message is a deterministic function of the history, every request can be rebuilt from these logs, and the benchmark curves can be recomputed from the logged conditions without calling the API. Two gaps remain: the \texttt{system\_fingerprint} of the responses was not recorded, so a change of the serving backend within a snapshot cannot be excluded, and the random fallback was drawn without a fixed seed, so its conditions are known only from the logs. A rerun with the same snapshots and seeds should therefore give similar but not necessarily identical proposals; all reported numbers come from the logged runs.

\begin{figure}[htbp]
\centering
\siFig[0.92\linewidth]{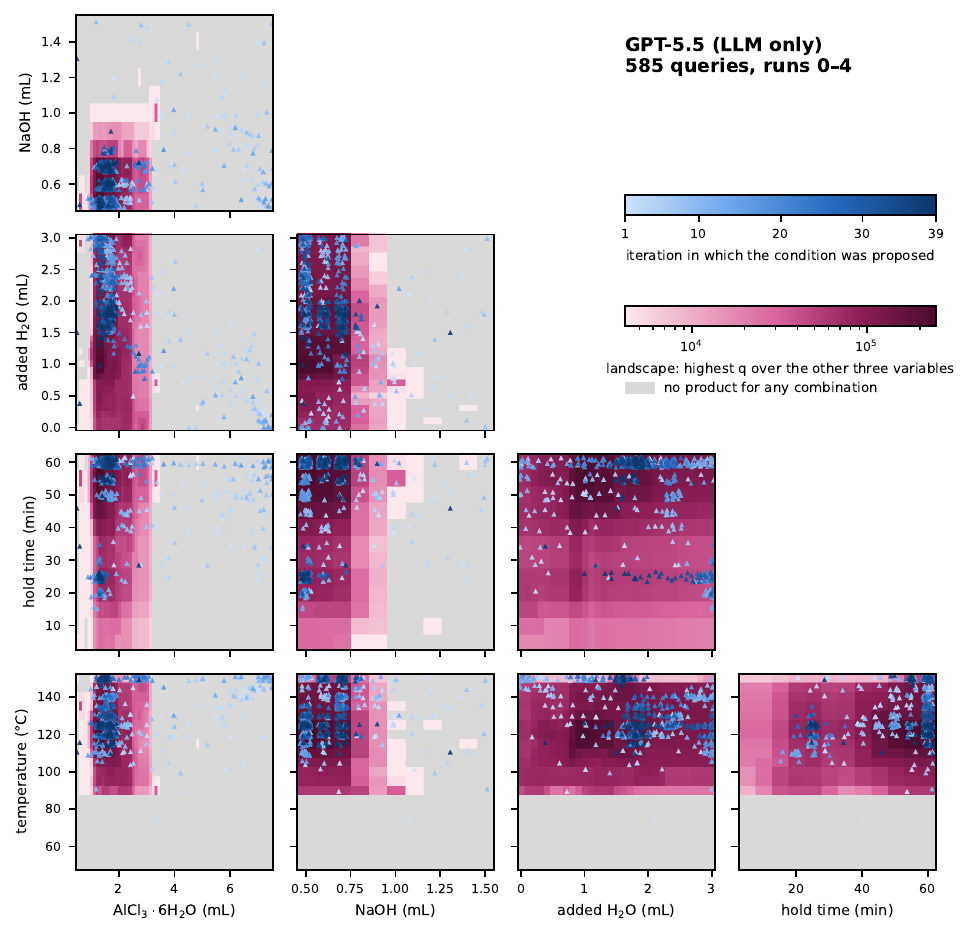}
\caption{Conditions proposed by gpt-5.5 (LLM only), runs 0--4, iterations 1--39, on all ten two-variable projections of the landscape (highest $q$ over the other three variables; gray, no product). Color gives the iteration in which a condition was proposed, light (iteration 1) to dark (iteration 39); later iterations are drawn on top. Points are jittered within their grid cell. The same plot for Mat\'ern-5/2 EI is Figure~\ref{fig:si-gp-proposals}.}
\label{fig:si-llm-vs-gp}
\end{figure}

\paragraph{LLM proposals.} gpt-5.5 concentrates its queries on the productive region (Figure~\ref{fig:si-llm-vs-gp}): of its 585 queries in runs 0--4, 81\% crystallize, 60\% reach the top-1\% tier and 41\% the top-0.1\% tier, compared with 59\%, 39\% and 16\% for Mat\'ern-5/2 EI on the same runs (Figure~\ref{fig:si-gp-proposals}). 87\% of its queries use 0.5--0.7~mL NaOH, 83\% at most 3~mL \ce{AlCl3.6H2O} and nearly all (99.8\%) at least 90~$^{\circ}$C, i.e.\ the known window of Section~\ref{si:landscape}. Its best-so-far nevertheless rises slowly after the first hits (Section~\ref{si:n5-all}, Table~\ref{tab:n5-all}), because it rarely leaves this window to test conditions whose outcome is uncertain.

\FloatBarrier
\subsection{Hybrid GP--LLM Policies}
\label{si:hybrid}

\begin{figure}[htbp]
\centering
\siFig[0.95\linewidth]{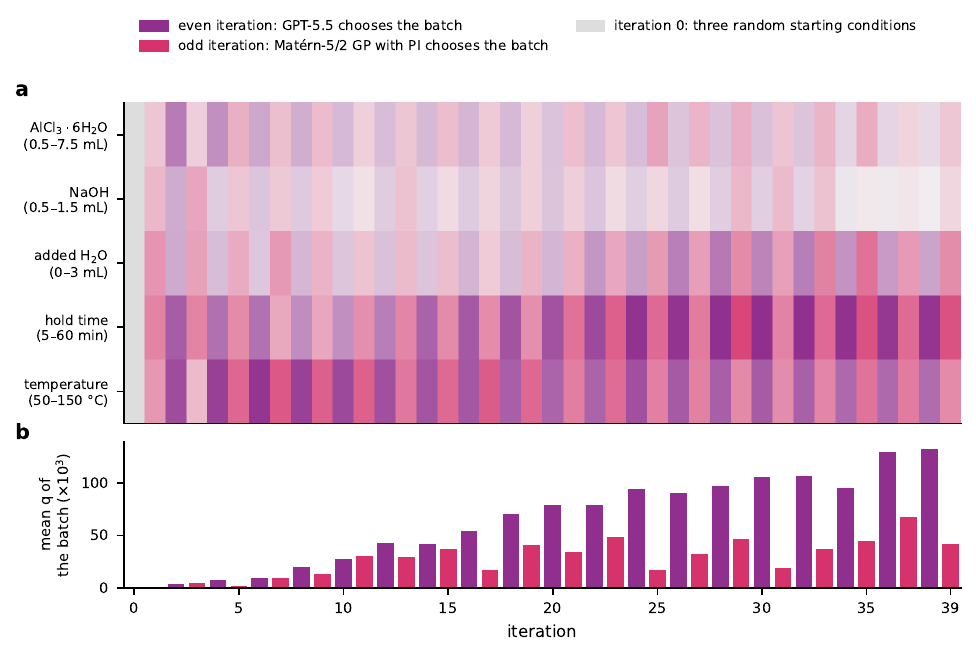}
\caption{How the PI hybrid works. (a) Conditions chosen in each iteration, one row per variable: hue, which decision-maker chose the batch (purple, gpt-5.5, even iterations; pink, GP with PI, odd iterations; gray, random start); lightness, position of the chosen value in the range of the variable (light, low; dark, high), mean over the three conditions and five runs. (b) Mean $q$ of each batch, same colors. Data: five runs of the PI hybrid in which every chosen condition was recorded.}
\label{fig:si-hybrid}
\end{figure}

The hybrid policies alternate between two decision-makers that share one history of experiments (Figure~\ref{fig:si-hybrid}). In odd iterations, a Mat\'ern-5/2 GP is fitted to all experiments of the run so far and picks three conditions from the 15{,}000 candidates with EI or PI, exactly as the GP-only policy. In even iterations, gpt-5.5 receives the same history (its 45 most recent experiments, including those chosen by the GP) with the prompt, window and validation of the LLM-only policy, and writes three conditions. After every iteration, the three conditions and their $q$ are appended to the shared history, so the GP is refitted to the LLM's experiments and the LLM sees the GP's experiments in the next iteration. The alternation was chosen to combine the LLM's prior knowledge of productive conditions with the GP's model of the measured data.

To see which decision-maker moves the search forward, we counted in the logs of the five runs every iteration in which the best-so-far of a run increased, and noted who had chosen the batch of that iteration. The EI hybrid had 55 such improvements in its five runs, 42 in LLM iterations and 13 in GP iterations; the PI hybrid had 58, 38 in LLM iterations and 20 in GP iterations. The final best of the run came from an LLM iteration in all 5 runs of both hybrids. Within this design the LLM contributes most improvements, while the GP iterations mainly add smaller refinements.

\FloatBarrier
\subsection{All Strategies}
\label{si:strategy-table}

Table~\ref{tab:strategies} lists all 39 strategies run on the landscape; 16 of them, chosen as representatives of every class, are compared in the main text. All use the protocol of Table~\ref{tab:protocol}, the target $\log(1+q)$, the same batch construction and, where a pool is used, the same candidate sampling. The metrics of all strategies are in Table~\ref{tab:n5-all}.

\label{si:ensembles}\label{si:rl}Besides GPs, the strategies use other models that return a prediction $\mu$ and an uncertainty $\sigma$ for every candidate; the acquisition rules of Section~\ref{si:acquisition} are then applied in the same way. Two of these models are among the six strategies of Section~\ref{si:leading-n20}. A random forest predicts by averaging many decision trees, each fitted to a resampled copy of the data, and the spread of the trees' predictions, $s_{\mathrm{tree}}$, serves as uncertainty. \emph{BART-like ExtraTrees} (extremely randomized trees; ``ExtraTrees EI'' in the figures and tables) is a variant that chooses the split points of its trees at random; it is used here as a fast stand-in for Bayesian additive regression trees (BART). The \emph{conformal random forest} adds to the tree spread the typical error of the forest on experiments it was not trained on, $\sigma_{\mathrm{conf}} = (s_{\mathrm{tree}}^2 + \hat{r}_{0.9}^2)^{1/2}$, where $\hat{r}_{0.9}$ is the 90\% quantile of these errors. The bandit policies (LinUCB) describe $y$ by a linear function of the five variables; a linear model cannot represent a narrow, closed high-$q$ region, which is consistent with their results (Table~\ref{tab:n5-all}). The settings of all models are given in Table~\ref{tab:strategies} and in the \href{https://github.com/deep-synthesis/ActMOF}{GitHub repository}.

\begin{landscape}
\begingroup
\footnotesize
\setlength{\tabcolsep}{5pt}
\renewcommand{\arraystretch}{1.12}
\begin{longtable}{@{}p{0.20\linewidth}p{0.25\linewidth}p{0.08\linewidth}p{0.15\linewidth}p{0.19\linewidth}p{0.05\linewidth}@{}}
\caption{All 39 benchmarked strategies. Cand., number of candidate conditions scored per iteration (GP iterations only for the hybrids).}\label{tab:strategies}\\
\toprule
Method & Surrogate & Acquisition & Uncertainty & Settings & Cand. \\
\midrule
\endfirsthead
\toprule
Method & Surrogate & Acquisition & Uncertainty & Settings & Cand. \\
\midrule
\endhead
\bottomrule
\endlastfoot
\multicolumn{6}{@{}l}{\textit{GP (kernel and acquisition variants)}} \\
Mat\'ern-3/2 EI & GP, ARD + white noise, Mat\'ern-3/2 & EI & GP posterior s.d. & none & 15{,}000 \\
Mat\'ern-5/2 EI & GP, ARD + white noise, Mat\'ern-5/2 & EI & GP posterior s.d. & none & 15{,}000 \\
RBF EI & GP, ARD + white noise, RBF & EI & GP posterior s.d. & none & 15{,}000 \\
RQ EI & GP, ARD + white noise, rational quadratic & EI & GP posterior s.d. & none & 15{,}000 \\
RobustGaSP EI & RobustGaSP (R), Mat\'ern-5/2, nugget estimated & EI & GP posterior s.d. & max\_\allowbreak{}eval\,=\,60 & 15{,}000 \\
Mat\'ern-5/2 PI & GP, ARD + white noise, Mat\'ern-5/2 & PI & GP posterior s.d. & none & 15{,}000 \\
Mat\'ern-5/2 UCB & GP, ARD + white noise, Mat\'ern-5/2 & UCB & GP posterior s.d. & $\beta$\,=\,2.0 & 15{,}000 \\
Mat\'ern-3/2 PI & GP, ARD + white noise, Mat\'ern-3/2 & PI & GP posterior s.d. & none & 15{,}000 \\
Mat\'ern-3/2 UCB & GP, ARD + white noise, Mat\'ern-3/2 & UCB & GP posterior s.d. & $\beta$\,=\,2.0 & 15{,}000 \\
Mat\'ern-5/2 TS & GP, ARD + white noise, Mat\'ern-5/2 & Thompson & GP posterior s.d. & none & 15{,}000 \\
Mat\'ern-3/2 TS & GP, ARD + white noise, Mat\'ern-3/2 & Thompson & GP posterior s.d. & none & 15{,}000 \\
greedy $\mu$ & GP, ARD + white noise, Mat\'ern-5/2 & greedy $\mu$ & GP posterior s.d. & none & 15{,}000 \\
quantile 0.9 & GP, ARD + white noise, Mat\'ern-5/2 & quantile $\tau$ & GP posterior s.d. & $\tau$\,=\,0.9 & 15{,}000 \\
CVaR 0.9 & GP, ARD + white noise, Mat\'ern-5/2 & CVaR$_{\tau}$ & GP posterior s.d. & $\tau$\,=\,0.9 & 15{,}000 \\
diversity EI ($\lambda = 0.1$) & GP, ARD + white noise, Mat\'ern-5/2 & EI & GP posterior s.d. & $\lambda$\,=\,0.1 & 15{,}000 \\
\multicolumn{6}{@{}l}{\textit{Tree and other ML surrogates}} \\
bootstrap RF EI & bootstrap random forest & EI & s.d.\ across trees & n\_\allowbreak{}estimators\,=\,300 & 15{,}000 \\
conformal RF EI & random forest (300 trees), spread widened by OOB-error quantile & EI & $\sqrt{s_{\mathrm{tree}}^2+\hat{r}_{0.9}^2}$ & n\_\allowbreak{}estimators\,=\,300 & 15{,}000 \\
quantile gradient boosting EI & quantile gradient boosting & EI & 10\%/90\% quantile spread & none & 15{,}000 \\
quantile gradient boosting q0.9 & quantile gradient boosting & quantile $\tau$ & 10\%/90\% quantile spread & $\tau$\,=\,0.9 & 15{,}000 \\
MLP ensemble EI & 8 bootstrapped MLPs (32,16) & EI & s.d.\ across ensemble & n\_\allowbreak{}models\,=\,8; $\alpha$\,=\,1e-3 & 15{,}000 \\
ExtraTrees EI & BART-like ExtraTrees (300 trees) & EI & s.d.\ across trees & n\_\allowbreak{}estimators\,=\,300 & 15{,}000 \\
\multicolumn{6}{@{}l}{\textit{Approximate and linear models}} \\
RFF GP EI & random Fourier features GP & EI & Bayesian linear posterior & n\_\allowbreak{}components\,=\,512 & 15{,}000 \\
Nystr\"om GP EI & Nystr\"om GP & EI & Bayesian linear posterior & n\_\allowbreak{}components\,=\,120 & 15{,}000 \\
laGP ALC EI & local approximate GP (laGP::aGP), ALC design & EI & local GP posterior s.d. & start\,=\,10; end\,=\,60 & 4{,}000 \\
laGP NN EI & local approximate GP (laGP::aGP), nearest neighbours & EI & local GP posterior s.d. & start\,=\,10; end\,=\,60 & 4{,}000 \\
Bayesian linear EI & Bayesian linear model & EI & Bayesian linear posterior & none & 15{,}000 \\
ARD regression on random features EI & ARD regression on RBF random features & EI & ARD posterior & n\_\allowbreak{}components\,=\,256 & 15{,}000 \\
\multicolumn{6}{@{}l}{\textit{Bandits}} \\
LinUCB & contextual LinUCB (linear bandit) & UCB & $\hat{s}\sqrt{\boldsymbol{\phi}^{\top}A^{-1}\boldsymbol{\phi}}$ & $\beta$\,=\,2.0 & 15{,}000 \\
LinUCB q0.9 & contextual LinUCB (linear bandit) & quantile $\tau$ & $\hat{s}\sqrt{\boldsymbol{\phi}^{\top}A^{-1}\boldsymbol{\phi}}$ & $\tau$\,=\,0.9 & 15{,}000 \\
chem. bandit UCB & chemistry-feature bandit, tail-weighted & UCB & $\sqrt{\boldsymbol{\phi}^{\top}A^{-1}\boldsymbol{\phi}}$ & $\beta$\,=\,2.75; ridge\,=\,1.0; success\_\allowbreak{}weight\,=\,2.0; top\_\allowbreak{}weight\,=\,4.0; top\_\allowbreak{}quantile\,=\,0.8 & 15{,}000 \\
chem. bandit TS & chemistry-feature bandit, tail-weighted & Thompson & $\sqrt{\boldsymbol{\phi}^{\top}A^{-1}\boldsymbol{\phi}}$ & ridge\,=\,1.0; success\_\allowbreak{}weight\,=\,2.0; top\_\allowbreak{}weight\,=\,4.0; top\_\allowbreak{}quantile\,=\,0.8 & 15{,}000 \\
chem. bandit q0.9 & chemistry-feature bandit, tail-weighted & quantile $\tau$ & $\sqrt{\boldsymbol{\phi}^{\top}A^{-1}\boldsymbol{\phi}}$ & $\tau$\,=\,0.9; ridge\,=\,1.0; success\_\allowbreak{}weight\,=\,2.0; top\_\allowbreak{}weight\,=\,4.0; top\_\allowbreak{}quantile\,=\,0.8 & 15{,}000 \\
chem. bandit CVaR & chemistry-feature bandit, tail-weighted & CVaR$_{\tau}$ & $\sqrt{\boldsymbol{\phi}^{\top}A^{-1}\boldsymbol{\phi}}$ & $\tau$\,=\,0.9; ridge\,=\,1.0; success\_\allowbreak{}weight\,=\,2.0; top\_\allowbreak{}weight\,=\,4.0; top\_\allowbreak{}quantile\,=\,0.8 & 15{,}000 \\
\multicolumn{6}{@{}l}{\textit{LLM}} \\
GPT-4o & gpt-4o, structured JSON output & LLM proposal & none & temperature\,=\,0.2; random seed per call; window\,=\,45 & none \\
GPT-5.5 & gpt-5.5, structured JSON output & LLM proposal & none & reasoning\_\allowbreak{}effort\,=\,none; random seed per call; window\,=\,45 & none \\
GPT-5-mini & gpt-5-mini, structured JSON output & LLM proposal & none & reasoning\_\allowbreak{}effort\,=\,minimal; random seed per call; window\,=\,45 & none \\
\multicolumn{6}{@{}l}{\textit{LLM--GP hybrid}} \\
Mat\'ern-5/2 EI + GPT-5.5 & odd iterations GP Mat\'ern-5/2 (EI), even iterations gpt-5.5 & EI / LLM & GP s.d.\ (GP iterations) & window\,=\,45 & 15{,}000 \\
Mat\'ern-5/2 PI + GPT-5.5 & odd iterations GP Mat\'ern-5/2 (PI), even iterations gpt-5.5 & PI / LLM & GP s.d.\ (GP iterations) & window\,=\,45 & 15{,}000 \\
\multicolumn{6}{@{}l}{\textit{Random}} \\
random search & none & random & none & none & none \\
\end{longtable}
\endgroup
\end{landscape}

% =====================================================================
\section{Benchmark Results}
\label{si:metrics}
% =====================================================================

All metrics are computed from the best-so-far of each run on the noise-free landscape $q$, over the budget of 120 experiments (iterations 0--39). The comparison proceeds in two steps. First, all 39 strategies are compared with $n=5$ runs (Section~\ref{si:n5-all}). This separates strong from weak strategies but leaves a large group of strong strategies that five runs cannot tell apart. Six strategies of this group are then run 20 times ($n=20$; Section~\ref{si:leading-n20}), and the remaining analyses (Sections~\ref{si:landscape-robustness}--\ref{si:feasibility}) use these six.

\subsection{Metrics}
\label{si:cum-best}
\label{si:hit-rate}
\label{si:early-success}

Table~\ref{tab:metric-defs} collects the quantities used in this section and in Sections~\ref{si:policies} and~\ref{si:transfer}.

\begin{table}[htbp]
\centering
\caption{Quantities used to compare the strategies.}
\label{tab:metric-defs}
\small
\begin{tabular}{@{}>{\raggedright\arraybackslash}p{0.27\linewidth}>{\raggedright\arraybackslash}p{0.66\linewidth}@{}}
\toprule
Quantity & Definition \\
\midrule
Tier thresholds & Top 1\%: $q \ge \qTopOne$; top 0.1\%: $q \ge \qTopPointOne$; quantiles over the 6{,}101{,}077 emulated conditions (Table~\ref{tab:tiers}) \\
Iteration $t$ & Three starting experiments, then three per iteration, so iteration $t$ ends after $3 + 3t$ experiments; $t = 3$ is 12 experiments (the wet-lab budget), $t = 13$ is 42, $t = 39$ is 120 \\
Best-so-far $q^{+}(t)$ & Highest $q$ among all experiments of a run up to iteration $t$ (Eq.~\ref{eq:cum-best-q}); higher is better \\
Hit rate $H_{1\%}(t)$, $H_{0.1\%}(t)$ & Fraction of the runs whose best-so-far has reached the tier by iteration $t$ (Eq.~\ref{eq:hit-rate}); higher is better. Early hit rate: $H_{1\%}(3)$ \\
Normalized regret $R(t)$ & Mean distance of the best-so-far from the highest $q$ on the grid, as a fraction of that value (Eq.~\ref{eq:norm-regret}); 0 if every run has found the best condition; lower is better \\
Query share & Fraction of proposed conditions that give a product or reach a tier (Sections~\ref{si:gp} and~\ref{si:llm}); not a hit rate \\
Leading group & Strategies that beat random search and cannot be distinguished from the best strategy \\
Independent / shared starts & Independent: each strategy and run draws its own three starting conditions (Eq.~\ref{eq:seed}). Shared: all strategies of a run start from the same three conditions \\
Bands in the figures & 95\% confidence interval of the mean over the runs (mean $\pm$ 1.96 standard errors), in all figures \\
$p$ values & Where a $p$ value is given, it comes from a rank test, which compares the order of the values and assumes no particular distribution: Kruskal--Wallis, Dunn and Mann--Whitney tests for independent runs\cite{kruskal1952,dunn1964}, Wilcoxon signed-rank test for shared-start runs. When several pairs are tested, the $p$ values are adjusted by Holm's method\cite{holm1979} \\
Spearman $\rho$ & Correlation between the rank orders of two sets of values (1, same order; 0, unrelated) \\
\bottomrule
\end{tabular}
\end{table}

\paragraph{Best-so-far and hit rate.} The best-so-far after iteration $t$ of run $r$ is
\begin{equation}
q^{+}_{r}(t) = \max_{1 \le i \le 3+3t} q(\mathbf{x}_{r,i}),
\label{eq:cum-best-q}
\end{equation}
including the three starting experiments; iteration 3 corresponds to 12 experiments, the budget of a wet-lab campaign. With the tier thresholds of Table~\ref{tab:tiers}, $q_{1\%} = \qTopOne$ and $q_{0.1\%} = \qTopPointOne$, the hit rate is the fraction of the $n$ runs whose best-so-far has reached the tier,
\begin{equation}
H_{\alpha}(t) = \frac{1}{n}\sum_{r=1}^{n} \mathds{1}\!\left[ q^{+}_{r}(t) \ge q_{\alpha} \right],
\label{eq:hit-rate}
\end{equation}
where $\mathds{1}[\cdot]$ is 1 if the condition in brackets holds and 0 otherwise. The normalized regret is
\begin{equation}
R(t) = \frac{1}{n}\sum_{r=1}^{n}\frac{q_{\max} - q^{+}_{r}(t)}{q_{\max} - q_{\min}},
\label{eq:norm-regret}
\end{equation}
with $q_{\max} = 247{,}107$, the largest $q$ on the grid, and $q_{\min} = 0$. The \emph{early hit rate} is $H_{1\%}(3)$, the hit rate after 12 experiments.

\paragraph{Random baseline.} For reference, 120 random grid points contain a top-0.1\% condition with probability 0.11; the 20 random-search runs gave 0.10.

\FloatBarrier
\subsection{\texorpdfstring{All Strategies, $n=5$}{All Strategies, n = 5}}
\label{si:n5-all}

\begin{figure}[htbp]
\centering
\siFig{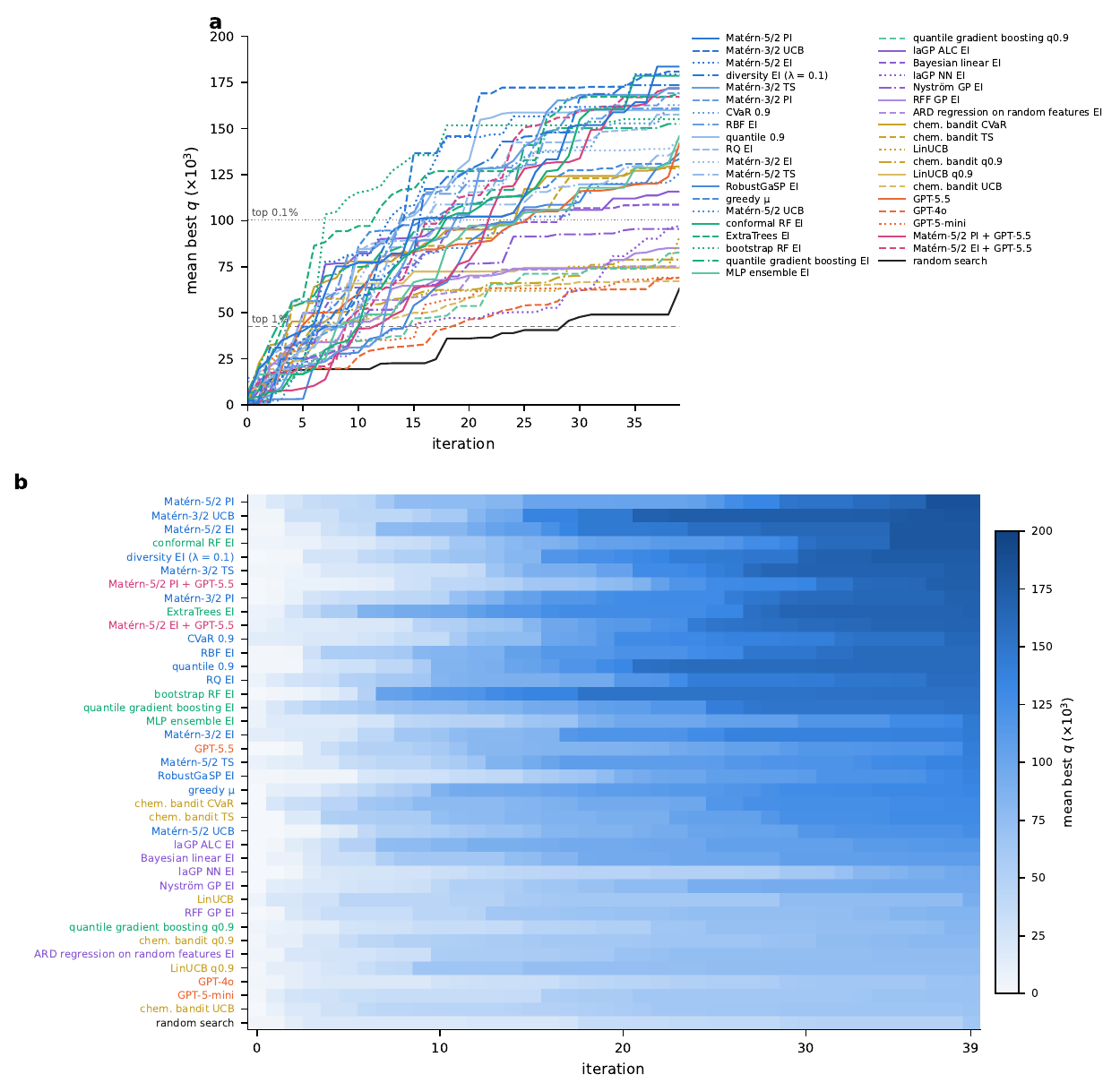}
\caption{All 39 strategies, $n=5$. (a) Mean best-so-far $q$ of every strategy in one plot. Color, class of the strategy (blue, GP; green, tree and other ML; purple, approximate and linear; yellow, bandit; orange, LLM; pink, LLM--GP hybrid; black, random search); tint and line style distinguish the strategies of a class (legend, ordered by class and by final mean). (b) The same means as a heat map, one row per strategy, ordered by mean best $q$ at iteration 39; label color as in (a).}
\label{fig:si-overview}
\end{figure}

\begin{figure}[htbp]
\centering
\siFig{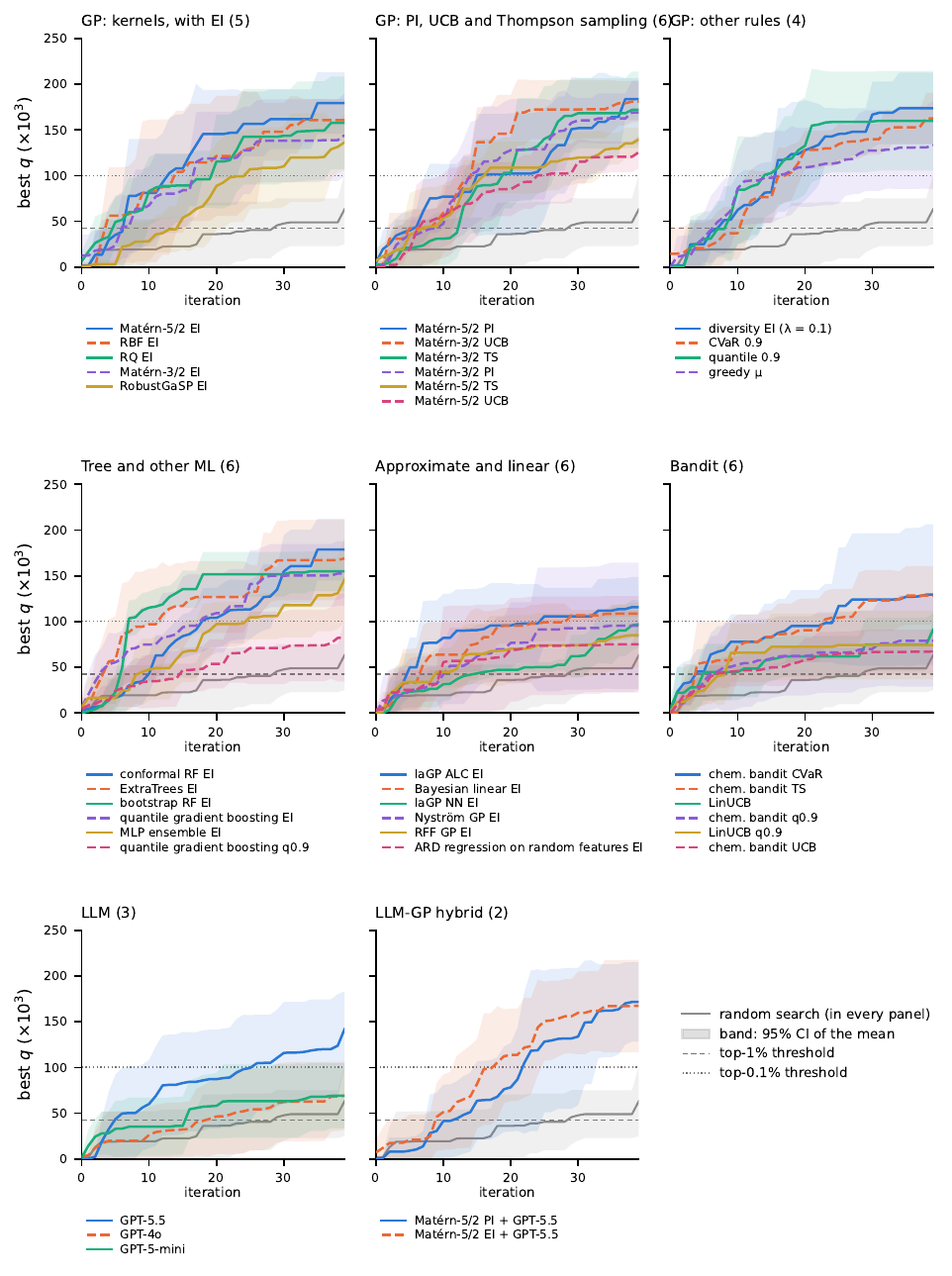}
\caption{The curves of Figure~\ref{fig:si-overview}a in groups, every strategy named (number of strategies in parentheses; the 15 GP strategies in three panels): mean best-so-far $q$ (line) and 95\% CI of the mean (band), $n=5$. Gray, random search, repeated in every panel; dashed and dotted lines, top-1\% and top-0.1\% thresholds.}
\label{fig:si-classes}
\end{figure}

Figure~\ref{fig:si-overview} shows the best-so-far of all 39 strategies together, Figure~\ref{fig:si-classes} shows the same curves in groups, with every strategy named, and Table~\ref{tab:n5-all} lists the metrics; the $n=5$ runs of every other figure and table are the same runs. Most full GPs, the tree ensembles and the two hybrids rise furthest above random search, whereas the bandits, the approximate and linear models and two of the three LLM-only policies stay closer to it. The top-1\% hit rate after 7 iterations is 0.6 for conformal RF and for Mat\'ern-5/2 EI (0.8 for Mat\'ern-5/2 PI), and ExtraTrees reaches the top-1\% tier in a median of 4 iterations.

\begingroup
\scriptsize
\setlength{\tabcolsep}{2.5pt}
\begin{longtable}{@{}llrrccc@{}}
\caption{All 39 strategies, $n=5$. Mean and Median: best $q$ after 120 experiments in thousands (mean $\pm$ s.d.\ over the five runs, and median; higher is better). Hit rates (Eq.~\ref{eq:hit-rate}; higher is better): $H_{1\%}(39)$ and $H_{0.1\%}(39)$, fraction of the five runs that have reached the top-1\% and the top-0.1\% tier after 120 experiments (iteration 39); $H_{0.1\%}(13)$, the same after 42 experiments (iteration 13). Class as in Table~\ref{tab:strategies}. Ordered by median best $q$.}\label{tab:n5-all}\\
\toprule
Strategy & Class & Mean & Median & $H_{1\%}(39)$ & $H_{0.1\%}(13)$ & $H_{0.1\%}(39)$ \\
\midrule
\endfirsthead
\toprule
Strategy & Class & Mean & Median & $H_{1\%}(39)$ & $H_{0.1\%}(13)$ & $H_{0.1\%}(39)$ \\
\midrule
\endhead
\bottomrule
\endlastfoot
conformal RF EI & Tree / ML & 179 $\pm$ 38 & 186 & 1.0 & 0.2 & 1.0 \\
Mat\'ern-3/2 UCB & GP & 181 $\pm$ 26 & 183 & 1.0 & 0.4 & 1.0 \\
Mat\'ern-5/2 PI & GP & 184 $\pm$ 26 & 176 & 1.0 & 0.4 & 1.0 \\
GPT-5.5 & LLM & 142 $\pm$ 47 & 169 & 1.0 & 0.6 & 0.8 \\
quantile gradient boosting EI & Tree / ML & 152 $\pm$ 40 & 169 & 1.0 & 0.4 & 0.8 \\
ExtraTrees EI & Tree / ML & 169 $\pm$ 48 & 166 & 1.0 & 0.8 & 1.0 \\
MLP ensemble EI & Tree / ML & 146 $\pm$ 54 & 165 & 1.0 & 0.2 & 0.8 \\
Mat\'ern-5/2 EI & GP & 179 $\pm$ 38 & 164 & 1.0 & 0.6 & 1.0 \\
quantile 0.9 & GP & 160 $\pm$ 62 & 161 & 1.0 & 0.6 & 0.8 \\
CVaR 0.9 & GP & 163 $\pm$ 32 & 160 & 1.0 & 0.2 & 1.0 \\
Mat\'ern-3/2 EI & GP & 144 $\pm$ 52 & 160 & 1.0 & 0.4 & 0.8 \\
Mat\'ern-3/2 TS & GP & 172 $\pm$ 48 & 159 & 1.0 & 0.2 & 1.0 \\
RBF EI & GP & 161 $\pm$ 26 & 158 & 1.0 & 0.4 & 1.0 \\
greedy $\mu$ & GP & 133 $\pm$ 55 & 156 & 1.0 & 0.6 & 0.6 \\
Mat\'ern-5/2 EI + GPT-5.5 & LLM--GP hybrid & 167 $\pm$ 58 & 155 & 1.0 & 0.4 & 1.0 \\
RQ EI & GP & 158 $\pm$ 58 & 155 & 1.0 & 0.2 & 0.8 \\
Mat\'ern-3/2 PI & GP & 169 $\pm$ 40 & 155 & 1.0 & 0.4 & 1.0 \\
Mat\'ern-5/2 PI + GPT-5.5 & LLM--GP hybrid & 172 $\pm$ 49 & 153 & 1.0 & 0.2 & 1.0 \\
RobustGaSP EI & GP & 136 $\pm$ 35 & 150 & 1.0 & 0.0 & 0.8 \\
bootstrap RF EI & Tree / ML & 155 $\pm$ 25 & 150 & 1.0 & 1.0 & 1.0 \\
Mat\'ern-5/2 TS & GP & 140 $\pm$ 14 & 146 & 1.0 & 0.4 & 1.0 \\
diversity EI ($\lambda$ = 0.1) & GP & 174 $\pm$ 45 & 144 & 1.0 & 0.2 & 1.0 \\
chem. bandit TS & Bandit & 129 $\pm$ 37 & 141 & 1.0 & 0.4 & 0.8 \\
Mat\'ern-5/2 UCB & GP & 126 $\pm$ 20 & 122 & 1.0 & 0.2 & 1.0 \\
laGP ALC EI & Approx. / linear & 116 $\pm$ 38 & 116 & 1.0 & 0.4 & 0.8 \\
Bayesian linear EI & Approx. / linear & 109 $\pm$ 4 & 109 & 1.0 & 0.4 & 1.0 \\
laGP NN EI & Approx. / linear & 97 $\pm$ 30 & 105 & 1.0 & 0.0 & 0.6 \\
LinUCB q0.9 & Bandit & 74 $\pm$ 43 & 101 & 0.6 & 0.2 & 0.6 \\
chem. bandit CVaR & Bandit & 129 $\pm$ 88 & 99 & 1.0 & 0.4 & 0.4 \\
LinUCB & Bandit & 91 $\pm$ 21 & 94 & 1.0 & 0.0 & 0.4 \\
quantile gradient boosting q0.9 & Tree / ML & 83 $\pm$ 55 & 82 & 0.8 & 0.0 & 0.2 \\
GPT-4o & LLM & 69 $\pm$ 42 & 73 & 0.6 & 0.0 & 0.2 \\
random search & Random & 63 $\pm$ 43 & 71 & 0.6 & 0.0 & 0.2 \\
Nystr\"om GP EI & Approx. / linear & 96 $\pm$ 79 & 67 & 0.8 & 0.2 & 0.4 \\
RFF GP EI & Approx. / linear & 85 $\pm$ 40 & 63 & 1.0 & 0.2 & 0.4 \\
GPT-5-mini & LLM & 69 $\pm$ 41 & 58 & 0.8 & 0.0 & 0.4 \\
chem. bandit q0.9 & Bandit & 79 $\pm$ 57 & 53 & 0.6 & 0.2 & 0.4 \\
chem. bandit UCB & Bandit & 67 $\pm$ 28 & 53 & 1.0 & 0.0 & 0.2 \\
ARD regression on random features EI & Approx. / linear & 75 $\pm$ 60 & 49 & 0.8 & 0.2 & 0.2 \\

\end{longtable}
\endgroup

\label{si:variance}
\paragraph{Number of replicate runs.} With five runs, the strategies at the top cannot be ranked: their run-to-run spread is as large as the differences between them (Figure~\ref{fig:si-classes}). Six of them were therefore run 20 times (Section~\ref{si:leading-n20}). They were chosen to cover the two model families of this group: the four Mat\'ern GPs (both kernels, each with EI and PI), the family used in the wet lab, and the two tree ensembles (conformal RF and BART-like ExtraTrees), which share the model family of the emulator and could profit from it (Section~\ref{si:landscape-robustness}). The LLM and hybrid strategies were not extended because of the API cost.

\FloatBarrier
\subsection{\texorpdfstring{Leading Group, $n=20$}{Leading Group, n = 20}}
\label{si:leading-n20}

\begin{figure}[htbp]
\centering
\siFig{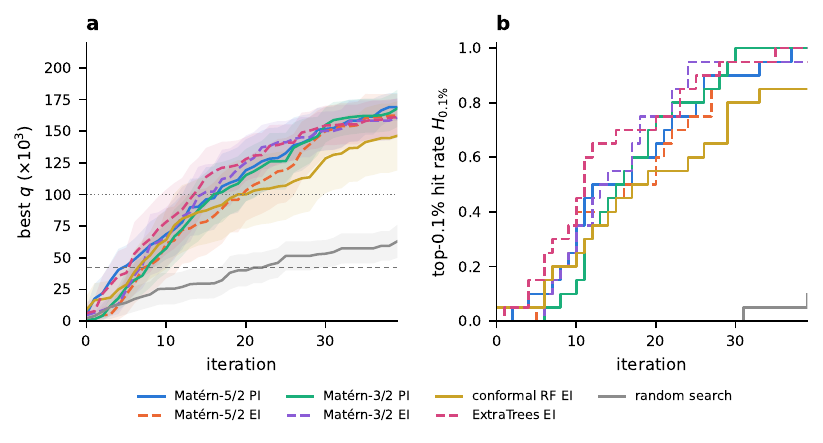}
\caption{The leading group with $n=20$ runs. (a) Best-so-far $q$, mean and 95\% CI; dashed and dotted lines, top-1\% and top-0.1\% thresholds. (b) Top-0.1\% hit rate $H_{0.1\%}(t)$.}
\label{fig:si-leading}
\end{figure}

The six strategies (Mat\'ern-5/2 and Mat\'ern-3/2 GPs with EI and PI, conformal RF and BART-like ExtraTrees) were run 20 times each, together with random search (Figure~\ref{fig:si-leading}, Table~\ref{tab:leading-n20}).

With 20 runs the six strategies still cannot be told apart: their curves overlap throughout (Figure~\ref{fig:si-leading}a), and a rank test finds no difference among them at iteration 13 ($p = 0.72$) or 39 ($p = 0.87$). All six are far above random search (adjusted $p < 0.001$ at iteration 39). After 120 experiments, 85--100\% of their runs have reached the top-0.1\% tier, against 10\% for random search. Their early hit rate $H_{1\%}(3)$ is 0.05--0.35 (random search 0.10), i.e.\ 12 experiments from a random start are rarely enough (Section~\ref{si:warmstart}).

\begin{table}[htbp]
\centering
\caption{The six strategies and random search with $n=20$ runs each. Mean and Median: best $q$ after 120 experiments in thousands (mean $\pm$ s.d.\ over the 20 runs, and median; higher is better). $H_{1\%}(3)$, fraction of the runs in the top-1\% tier after 12 experiments; $H_{0.1\%}(39)$, fraction in the top-0.1\% tier after 120 experiments (higher is better). $R(39)$, normalized regret after 120 experiments (Eq.~\ref{eq:norm-regret}; lower is better).}
\label{tab:leading-n20}
\footnotesize
\setlength{\tabcolsep}{4pt}
\begin{tabular}{@{}lrrccc@{}}
\toprule
Strategy & Mean & Median & $H_{1\%}(3)$ & $H_{0.1\%}(39)$ & $R(39)$ \\
\midrule
Mat\'ern-5/2 PI & 169 $\pm$ 25 & 167 & 0.25 & 1.00 & 0.32 \\
Mat\'ern-5/2 EI & 162 $\pm$ 39 & 155 & 0.05 & 0.95 & 0.34 \\
Mat\'ern-3/2 PI & 168 $\pm$ 34 & 157 & 0.10 & 1.00 & 0.32 \\
Mat\'ern-3/2 EI & 161 $\pm$ 35 & 160 & 0.05 & 0.95 & 0.35 \\
conformal RF EI & 146 $\pm$ 62 & 159 & 0.15 & 0.85 & 0.41 \\
ExtraTrees EI & 161 $\pm$ 33 & 158 & 0.35 & 1.00 & 0.35 \\
Random search & 63 $\pm$ 30 & 61 & 0.10 & 0.10 & 0.75 \\
\bottomrule\end{tabular}
\end{table}

\FloatBarrier
\subsection{Robustness to the Construction of the Landscape}
\label{si:landscape-robustness}

Section~\ref{si:emulator-validation} tests the emulator against measured outcomes. Here we ask whether the strategy comparison would change if the landscape had been built in another reasonable way. We rebuilt it 41 times, each time changing one choice of Section~\ref{si:emulator} and keeping everything else. The changes fall into five groups. (i)~Forest random state (10 landscapes): the random forests were refitted with ten other random seeds, which tests the randomness of the fit itself. (ii)~Anchor resampling (20): the forests were trained on a resampled set of anchors, made by drawing 95 times at random from the 95 anchors and putting each drawn anchor back, so that some anchors appear more than once and others not at all; this tests how much the landscape depends on individual experiments. (iii)~Blending and threshold rules (8): one rule changed at a time, the Rule~1 blend to 60:40 or 100:0 instead of 80:20, the Rule~2 blend to 30:70 or 70:30 instead of 50:50, the intensity threshold to 2000 or 5000 instead of 3000, or the FWHM threshold to 2$^{\circ}$ or 5$^{\circ}$ instead of 3$^{\circ}$. (iv)~Emulator model (2): the random forests replaced by gradient boosting or by $k$-nearest neighbors. (v)~Fewer anchors (1): only the first 60 anchors in recording order. Changing a choice moved the top-0.1\% tier within the productive region but not out of it: 75--100\% of the original top-0.1\% conditions stayed in the top 1\% of every rebuilt landscape.

\begin{figure}[p]
\centering
\siFig{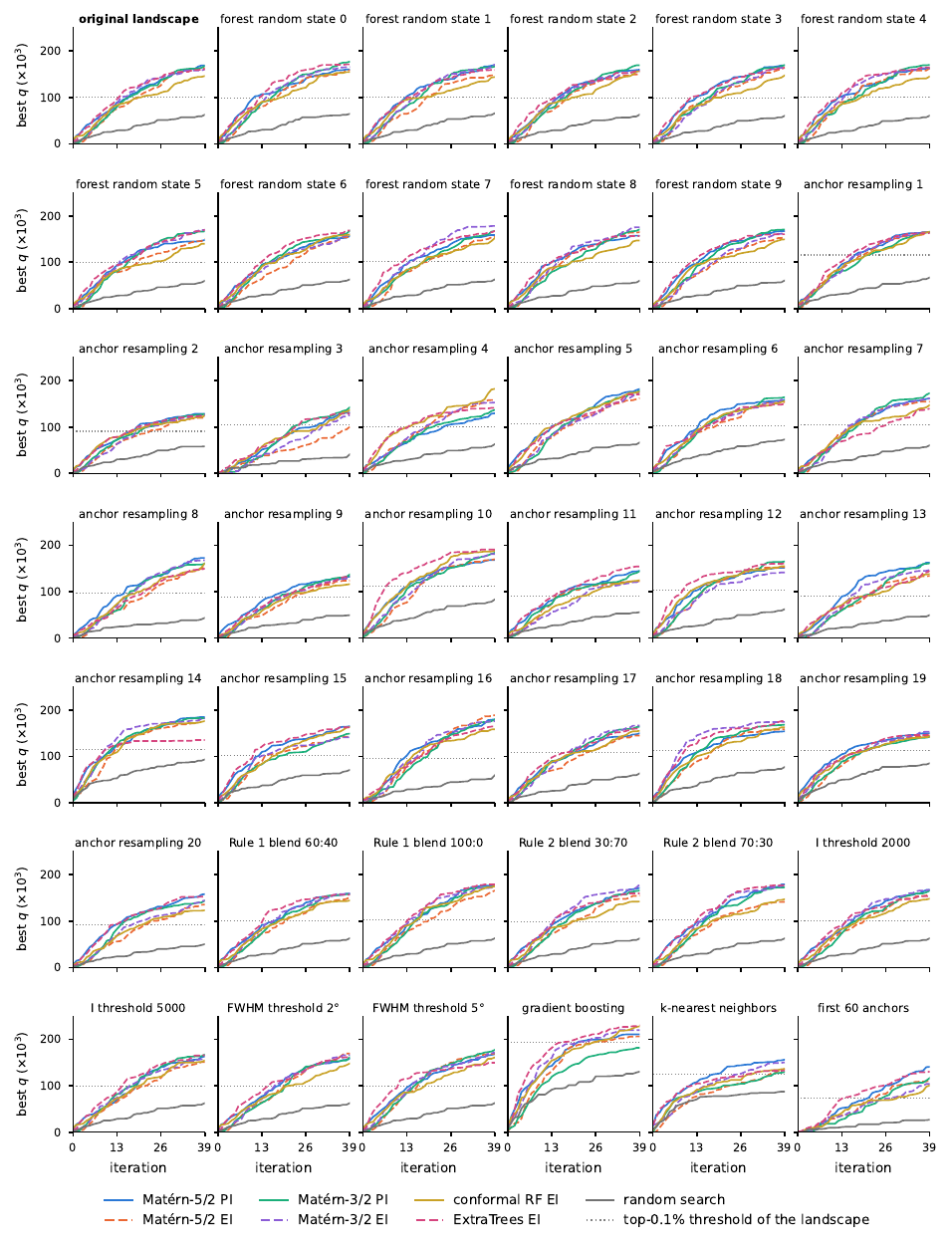}
\caption{Best-so-far $q$ of the six strategies and random search on the original landscape and on each of the 41 rebuilt ones (mean of $n = 20$ runs). Panel titles give the choice that was changed: forest random state 0--9, anchor resampling 1--20, one blending or threshold rule, the emulator model, or the first 60 anchors only. Dotted line, top-0.1\% threshold of the landscape of the panel: every rebuilt landscape has its own $q$ values, so the $q$ that separates its best 0.1\% of conditions differs from landscape to landscape.}
\label{fig:si-robust-grid}
\end{figure}

\begin{figure}[htbp]
\centering
\siFig{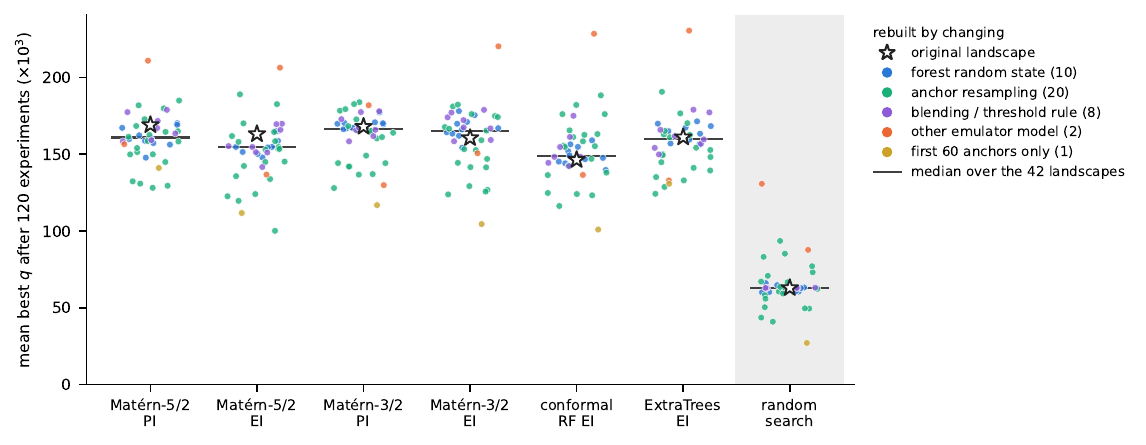}
\caption{Summary of Figure~\ref{fig:si-robust-grid}: mean best $q$ after 120 experiments ($n = 20$ runs) of each strategy on the original landscape (star) and on the 41 rebuilt ones (dots, colored by the group of the change; number of landscapes in parentheses). Line, median over the 42 landscapes; shaded, random search.}
\label{fig:si-landscape-robustness}
\end{figure}

On every landscape, the six strategies and random search were rerun with $n = 20$ (Figure~\ref{fig:si-robust-grid}). Random search was lowest on all 41 landscapes: the mean best $q$ of every one of the six strategies was at least 1.4 times that of random search, and all six beat it on every landscape (rank test, adjusted $p \le 0.033$). The order among the six, however, changed from landscape to landscape, and each of them ranked first on at least one (Figure~\ref{fig:si-landscape-robustness}). The separation from random search is a property of the landscape; the order among the six is not.

\paragraph{Emulator model and search policy.} Because the emulator is a random forest, tree-based policies share its model family and might find the landscape easier than a real synthesis space. The effect exists but is small and runs both ways: the two tree policies ranked first and second on the gradient-boosting map, and two GPs on the $k$-nearest-neighbor map, whereas on the random-forest map used throughout this work the tree policies ranked fourth and sixth. These differences lie within the run-to-run spread, and the GP policies used in the wet lab do not depend on the emulator being a random forest.

\FloatBarrier
\subsection{PI versus EI}
\label{si:pi-vs-ei}

\begin{figure}[htbp]
\centering
\siFig{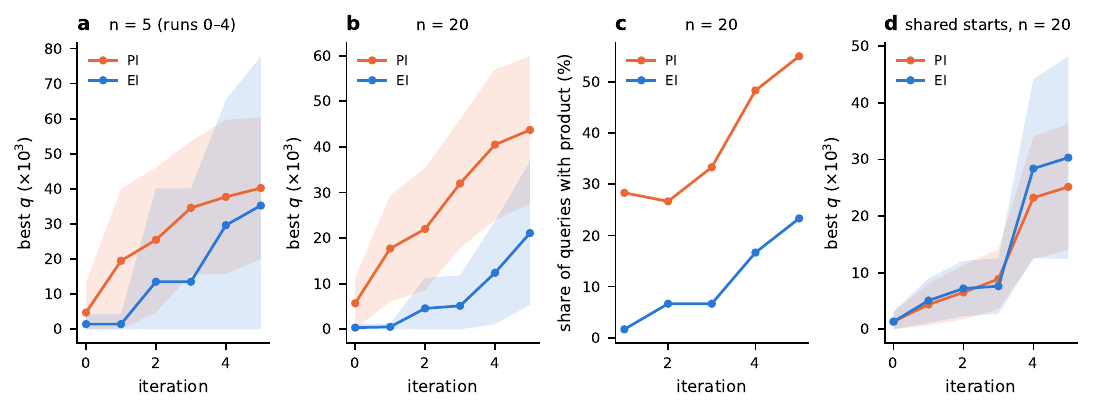}
\caption{Mat\'ern-5/2 PI vs.\ EI in iterations 0--5. (a, b) Mean best-so-far $q$ and 95\% CI of the independent-start runs, $n=5$ (runs 0--4) and $n=20$. (c) Share of queries with a crystalline product, $n=20$. (d) Shared-start rerun: both rules start from the same three experiments, $n=20$.}
\label{fig:si-pi-ei-early}
\end{figure}

\paragraph{Independent-start runs.} In the independent-start runs of Mat\'ern-5/2, PI has a higher mean best-so-far than EI at every iteration from 0 to 5, both with $n=5$ and with $n=20$ (Figure~\ref{fig:si-pi-ei-early}a,b). The lead is present at iteration 0, before either rule has chosen a point, because 7 of the 20 PI starting batches contain a product against 1 of the 20 EI batches. PI also has a larger share of queries that crystallize (panel c; pooled over iterations 1--5, 38\% against 11\%).

\paragraph{Shared-start rerun.} To separate the acquisition rule from the starting batch, we reran both rules with the production code but with a random seed that does not depend on the method, so that PI and EI start from the same three experiments and use the same stream of random numbers ($n=20$). The best-so-far curves then overlap (Figure~\ref{fig:si-pi-ei-early}d): EI leads at iterations 1, 2, 4 and 5 and PI at iteration 3. The early advantage of PI in the independent-start runs is therefore a consequence of its starting batches, not of the acquisition rule.

\begin{figure}[htbp]
\centering
\siFig[0.95\linewidth]{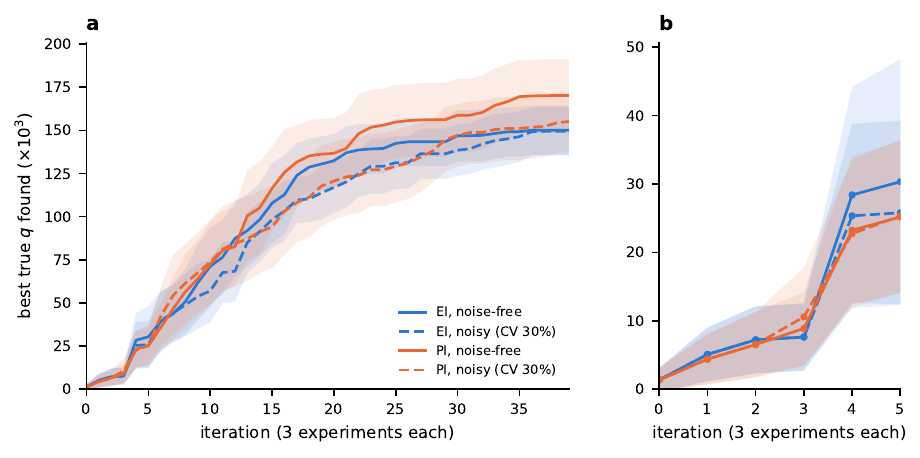}
\caption{Shared-start rerun of Mat\'ern-5/2 PI and EI with a noise-free oracle (solid) and with 30\% log-normal measurement noise (dashed); best-so-far on the true landscape $q$, mean and 95\% CI, $n=20$. (a) Iterations 0--39; (b) iterations 0--5 enlarged.}
\label{fig:si-noise}
\end{figure}

\paragraph{Measurement noise.} In the benchmark, the landscape returns the exact $q$ of a condition. In the wet lab, one powder loaded and measured five times gave $q$ values with a coefficient of variation of 30\% (Section~\ref{si:q-mount}). To see how the two rules cope with this, the shared-start runs of Mat\'ern-5/2 PI and EI were repeated with a noisy landscape. Every time a strategy queries a condition, the landscape value $q$ is multiplied by a random factor $\varepsilon$, drawn anew for every query from a log-normal distribution with mean 1 and a coefficient of variation of 30\%, which reproduces the measured loading-to-loading scatter; a condition without product still returns $q = 0$. The strategy sees and learns from these noisy values only. The curves of Figure~\ref{fig:si-noise}, however, are drawn with the noise-free landscape value of every condition the strategy chose: they show how good the chosen conditions really are, not how high the noisy readings happened to be. Noise lowers the mean best $q$ at iteration 39 by 8.8\% for PI (170{,}212 to 155{,}206) and by 0.3\% for EI (150{,}068 to 149{,}545). PI is the more sensitive rule, as expected from its preference for small predicted gains.

\FloatBarrier
\subsection{Warm Start}
\label{si:warmstart}

\begin{figure}[htbp]
\centering
\siFig[0.86\linewidth]{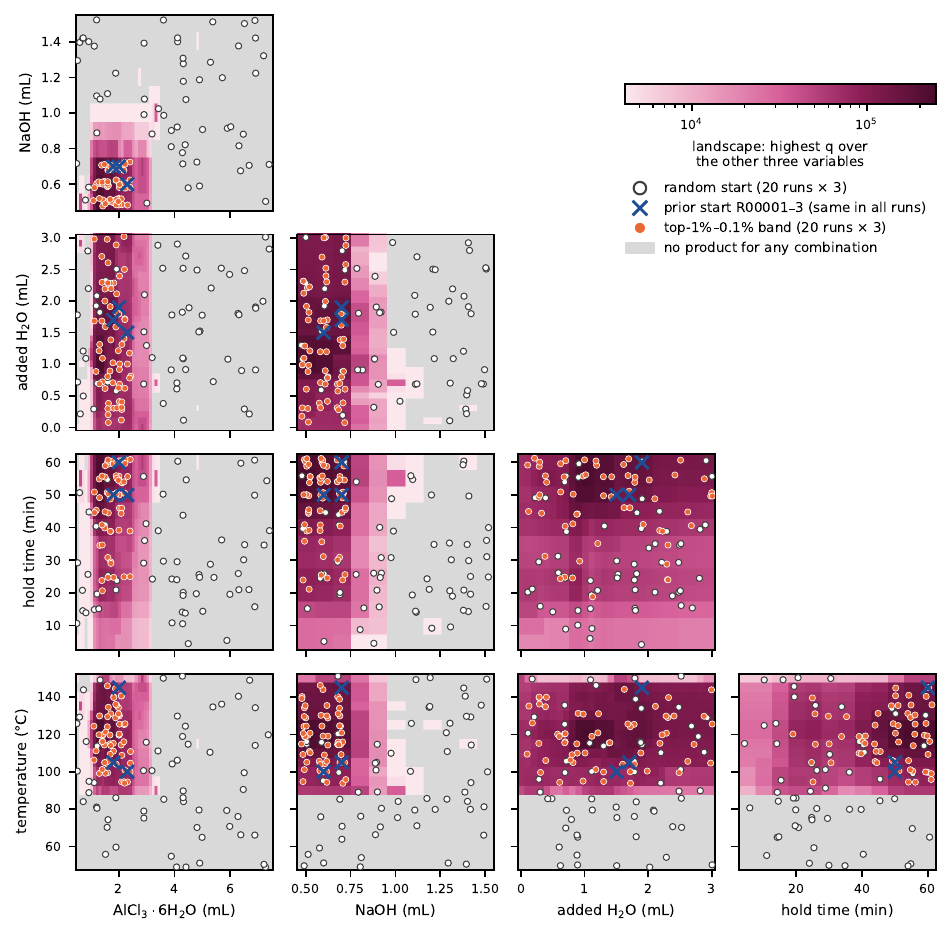}
\caption{Starting conditions of the three start types on all ten two-variable projections of the MOF-321 landscape (highest $q$ over the other three variables; gray, no product). Open circles, random starts (the 20 starting batches of Mat\'ern-5/2 EI, $n=20$); crosses, prior start R00001--R00003 (the same in all runs); orange, starts drawn from the top-1\%--0.1\% band (one batch of three per run). Points are jittered within their grid cell.}
\label{fig:si-warmstart-starts}
\end{figure}

The wet-lab campaigns (Section~\ref{si:transfer}) do not start from random conditions but from the three conditions proposed by the MOF-321 prior (R00001--R00003). These are the first batch that the desktop application proposes before any result of a new MOF has been entered: they are computed from the reference GP fitted to the 95 MOF-321 anchors (Sections~\ref{si:residual-gp} and~\ref{si:calibration}) and are therefore the same for every new MOF to be synthesized. To separate what the starting point contributes from what the search contributes, we reran the six strategies and random search on the MOF-321 landscape with two warm starts: (i) the three prior conditions, identical in all runs; and (ii) three conditions drawn at random from the band between the top-1\% and top-0.1\% thresholds, one draw per run. The top 0.1\% itself is excluded, so the start is never the answer. Figure~\ref{fig:si-warmstart-starts} shows where these starting conditions lie, together with the random starts of Section~\ref{si:leading-n20}. Each arm has 20 runs of 39 iterations, and within a run all strategies share the starting batch and the random seed. The cold-start curves are those of Section~\ref{si:leading-n20}.

\begin{figure}[htbp]
\centering
\siFig{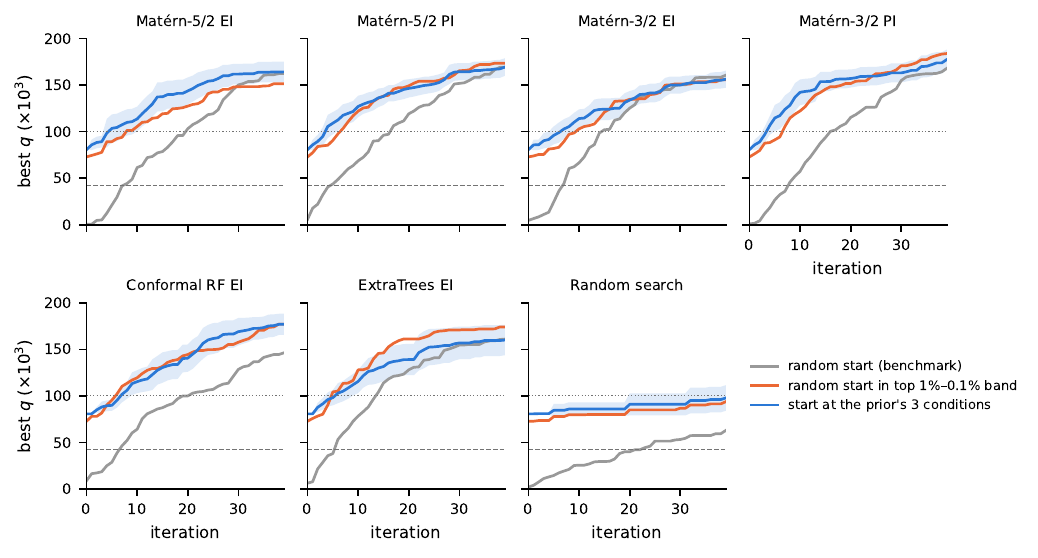}
\caption{Warm start on the MOF-321 landscape. Mean best-so-far $q$ of six strategies and random search from a random start (gray, Section~\ref{si:leading-n20}), from the three conditions proposed by the MOF-321 prior (blue, with 95\% CI) and from three random conditions in the top-1\%--0.1\% band (orange); $n=20$ runs each. Dashed and dotted lines, top-1\% and top-0.1\% thresholds.}
\label{fig:si-warmstart}
\end{figure}

The prior conditions reach $q = 80{,}760$ on the landscape (top 1\%). From there, random search barely improves: its median best $q$ is still 80{,}760 after 39 iterations (Figure~\ref{fig:si-warmstart}), and only 6 of 20 runs reach the top-0.1\% tier (Table~\ref{tab:warmstart}). The six strategies reach the top-0.1\% tier in 65--100\% of runs by iteration 13 and in all runs by iteration 39, and each of them is better than random search (paired rank test, Holm-adjusted $p \le 0.002$ at iterations 13 and 39). The random top-1\%--0.1\% start gives almost the same curves as the prior start. The start therefore decides whether the search begins in the productive region, whatever the particular conditions; the climb from the top-1\% to the top-0.1\% tier is done by the search strategy, and \textbf{this is the part of the campaign that the benchmark evaluates}.

Compared with a cold start, a warm start saves roughly the first 10--15 iterations: with 12 experiments (iteration 3), at most 35\% of cold-start runs reach even the top-1\% tier (Table~\ref{tab:leading-n20}), against 100\% of warm-start runs, of which 10--40\% are already in the top-0.1\% tier. \textbf{This is why the wet-lab campaigns, limited to 12 experiments, use the prior.}

\begingroup
\footnotesize
\setlength{\tabcolsep}{4pt}
\begin{longtable}{@{}llccc@{}}
\caption{Warm start vs.\ cold start ($n=20$ each): top-0.1\% hit rate $H_{0.1\%}(t)$ at iterations $t = 3$, 13 and 39 (12, 42 and 120 experiments; higher is better).}\label{tab:warmstart}\\
\toprule
 & & \multicolumn{3}{c}{$H_{0.1\%}$} \\
\cmidrule(l){3-5}
Strategy & Start & 3 & 13 & 39 \\
\midrule
\endfirsthead
\toprule
Strategy & Start & 3 & 13 & 39 \\
\midrule
\endhead
\bottomrule
\endlastfoot
Mat\'ern-5/2 EI & random & 0.00 & 0.40 & 0.95 \\
 & prior (R00001--3) & 0.15 & 0.85 & 1.00 \\
 & top 1\%--0.1\% band & 0.10 & 0.80 & 1.00 \\
\addlinespace
Mat\'ern-5/2 PI & random & 0.05 & 0.50 & 1.00 \\
 & prior (R00001--3) & 0.20 & 1.00 & 1.00 \\
 & top 1\%--0.1\% band & 0.15 & 0.90 & 1.00 \\
\addlinespace
Mat\'ern-3/2 EI & random & 0.00 & 0.50 & 0.95 \\
 & prior (R00001--3) & 0.20 & 0.80 & 1.00 \\
 & top 1\%--0.1\% band & 0.05 & 0.55 & 1.00 \\
\addlinespace
Mat\'ern-3/2 PI & random & 0.00 & 0.40 & 1.00 \\
 & prior (R00001--3) & 0.40 & 0.95 & 1.00 \\
 & top 1\%--0.1\% band & 0.25 & 0.95 & 1.00 \\
\addlinespace
Conformal RF EI & random & 0.05 & 0.35 & 0.85 \\
 & prior (R00001--3) & 0.10 & 0.65 & 1.00 \\
 & top 1\%--0.1\% band & 0.15 & 0.80 & 1.00 \\
\addlinespace
ExtraTrees EI & random & 0.05 & 0.65 & 1.00 \\
 & prior (R00001--3) & 0.30 & 0.75 & 1.00 \\
 & top 1\%--0.1\% band & 0.10 & 1.00 & 1.00 \\
\addlinespace
Random search & random & 0.00 & 0.00 & 0.10 \\
 & prior (R00001--3) & 0.00 & 0.10 & 0.30 \\
 & top 1\%--0.1\% band & 0.00 & 0.10 & 0.30 \\\end{longtable}
\endgroup

\FloatBarrier
\subsection{Modeling the No-Product Region}
\label{si:feasibility}

In all strategies above, an experiment without product enters the model as an ordinary value, $y = \log(1+0) = 0$, on the same scale as the crystalline results. Because 94\% of the landscape gives no product, one may ask whether the search would be faster if the question ``product or no product'' were modeled separately. This is an established idea in Bayesian optimization, known as optimization with unknown constraints or with failed experiments\cite{gramacy2011,gelbart2014,bachoc2020}, and it has been applied in chemistry and in materials growth\cite{hickman2025,wakabayashi2022}. We tested two policies of this kind against Mat\'ern-5/2 EI.

(i)~\emph{Product probability times EI.} A GP classifier is fitted to all experiments of the run and gives, for every candidate, the probability $p(\mathbf{x})$ that it yields a product. A GP regression is fitted to the crystalline experiments only and gives $\mathrm{EI}_{+}(\mathbf{x})$, the expected improvement if a product forms. Candidates are scored by
\begin{equation}
a(\mathbf{x}) = p(\mathbf{x})\,\mathrm{EI}_{+}(\mathbf{x}),
\label{eq:hurdle-ei}
\end{equation}
which is exactly the expected improvement of a model in which a condition gives no product with probability $1-p$ and a normally distributed $y$ otherwise.

(ii)~\emph{Lowest observed value.} An experiment without product is entered with the lowest $y$ of the products found so far in the run instead of 0, and the usual GP with EI is used\cite{wakabayashi2022}; the failures then no longer lie far below all products. Before the first product, both policies follow the same rule as the baseline (Section~\ref{si:protocol}), and all other settings are those of Table~\ref{tab:protocol}. Each policy was run 20 times with shared starts: in every run, the three policies begin with the same three experiments.

\begin{figure}[htbp]
\centering
\siFig[0.92\linewidth]{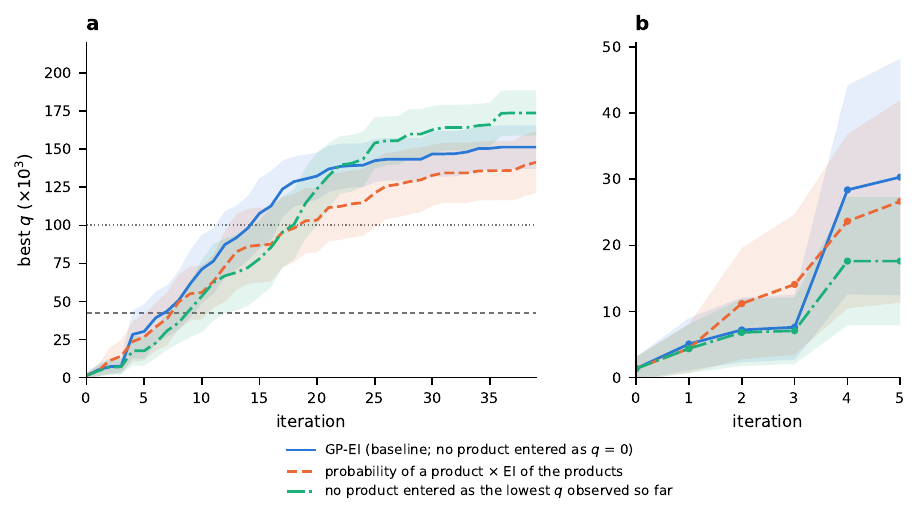}
\caption{Modeling the no-product region: Mat\'ern-5/2 EI (baseline) and the two policies of Section~\ref{si:feasibility}, shared starts, $n=20$. Mean best-so-far $q$ and 95\% CI of the mean. (a) Iterations 0--39; dashed and dotted lines, top-1\% and top-0.1\% thresholds. (b) Iterations 0--5 enlarged.}
\label{fig:si-feasibility}
\end{figure}

Neither policy was faster than the baseline (Figure~\ref{fig:si-feasibility}, Table~\ref{tab:feasibility}). The first product appeared after the same number of experiments in all three (median 12.5). This is so by construction: none of them has a model before a product exists. After 12 experiments the three policies had the same median best $q$. Later, the baseline was nominally ahead after 42 experiments and policy~(ii) after 120, but none of the comparisons with the baseline that we had fixed beforehand (first product, share of productive queries in iterations 1--5, best $q$ after 12 and after 42 experiments; paired rank tests) showed a difference (smallest adjusted $p = 0.34$). With 20 runs, small differences cannot be excluded. Policy~(i) had one drawback: in 10 of its 20 runs it went 10 or more iterations without a new product after its first one (baseline: no run). The regression behind $\mathrm{EI}_{+}$ has seen only products, so its expected improvement is largest in unexplored corners of the grid, and $p$ lowers it there only moderately.

The absence of an early gain has the reason given in Section~\ref{si:protocol}: the early search is limited by finding the first product, and before that a model of the no-product region has nothing to learn from either. In our setting this limit is removed by the prior (Section~\ref{si:warmstart}), not by the model. The plain GP with EI was therefore kept.

\begin{table}[H]
\centering
\caption{Modeling the no-product region, shared starts, $n=20$. First product: median number of experiments until the first product. Median best $q$ in thousands after 12, 42 and 120 experiments (higher is better). $H_{0.1\%}$: runs in the top-0.1\% tier after 42 and 120 experiments (higher is better). Stalled: runs with 10 or more iterations in a row without a new product after the first one (lower is better).}
\label{tab:feasibility}
\small
\begin{tabular}{@{}lccccccc@{}}
\toprule
 & & \multicolumn{3}{c}{Median best $q$} & \multicolumn{2}{c}{$H_{0.1\%}$} & \\
\cmidrule(lr){3-5}\cmidrule(lr){6-7}
Policy & First product & 12 & 42 & 120 & 42 & 120 & Stalled \\
\midrule
Mat\'ern-5/2 EI (baseline) & 12.5 & 1.1 & 92 & 155 & 9/20 & 18/20 & 0/20 \\
(i) product probability $\times$ EI & 12.5 & 1.1 & 78 & 152 & 8/20 & 17/20 & 10/20 \\
(ii) lowest observed value & 12.5 & 1.1 & 55 & 165 & 5/20 & 19/20 & 3/20 \\
\bottomrule
\end{tabular}
\end{table}

\paragraph{Choice of policy for the wet lab.} The six strategies could not be told apart (Section~\ref{si:leading-n20}), and their order changed with the construction of the landscape (Section~\ref{si:landscape-robustness}). The wet-lab campaigns used a Mat\'ern-5/2 GP with EI for two practical reasons. \emph{First, EI is robust to measurement noise}: with 30\% noise it lost 0.3\% of its final best $q$, against 8.8\% for PI (Figure~\ref{fig:si-noise}), and with shared starts it progressed as fast as PI in the first iterations (Figure~\ref{fig:si-pi-ei-early}d). \emph{Second, the GP is easy to adjust}: in the desktop application, the kernel (Mat\'ern-5/2 or Mat\'ern-3/2) and the acquisition function (EI or PI) are each one setting (Section~\ref{si:app}), and the GP also provides the reference and residual models of the transfer prior (Section~\ref{si:residual-gp}).

% =====================================================================
\section{Transfer Learning}
\label{si:transfer}
% =====================================================================

The six validation campaigns (Sections~\ref{si:synthesis}--\ref{si:pxrd}) were run with the ActMOF desktop application (MOF Bayesian Optimization Student App, version 1.0.9). Its settings are those of the benchmark (Section~\ref{si:protocol}) unless stated below.

Bayesian optimization of reaction conditions is established in chemical synthesis\cite{shields2021}, and starting a new optimization from the data of earlier, related ones is known as transfer learning or multi-task Bayesian optimization\cite{swersky2013}. The form used here, a rescaled reference model plus a GP for what the reference does not explain (Section~\ref{si:residual-gp}), follows the model of Kennedy and O'Hagan for combining a cheap, approximate model with a few expensive data\cite{kennedy2000}: the expensive response is written as a constant times the cheap one plus a GP for the difference. Here the MOF-321 model takes the place of the cheap model and the few experiments on the new MOF that of the expensive data. The offset $a$ and the restriction of the slope $b$ described below are our additions for the case of very few experiments.

\subsection{Desktop Application}
\label{si:app}

The search space is the five encoded variables of Section~\ref{si:encoding} (\texttt{metal\_amount}, \texttt{modulator}, \texttt{add\_solvent}, \texttt{reaction\_time}, \texttt{reaction\_temperature}). The user chooses a Mat\'ern-5/2 or Mat\'ern-3/2 kernel and EI or PI; all six campaigns used Mat\'ern-5/2 with EI, 15{,}000 random candidates and the transfer prior with its default schedule. Each iteration proposes three experiments. The user enters the measured PXRD intensity and FWHM of the diagnostic reflection, and the program computes
\begin{equation}
q = \mathrm{round}\!\left(\frac{I}{\mathrm{FWHM}}\right),
\label{eq:app-q}
\end{equation}
with $q=0$ when no reflection is found (entered as FWHM $=30$) or $I \le 0$; $q$ can also be entered directly.

\paragraph{Surrogate.} The program uses its own compact GP implementation (NumPy, Cholesky factorization) instead of scikit-learn, so that it can be distributed as a single executable. Inputs are scaled to $[0,1]$ and targets are $y=\log(1+q)$, standardized before fitting. Unlike the benchmark GPs, the kernel uses one length scale $\ell$ for all five variables (an isotropic Mat\'ern kernel). In a campaign the model is fitted to only 3--12 results of the new MOF; with so few data, five separate length scales cannot be estimated reliably and tend to run to the bounds of the fit, whereas a single length scale is stable. $\ell$ and the noise level are set by trying a fixed list of values and keeping the pair under which the measured data are most probable (highest marginal likelihood); this is repeated after every batch.

\paragraph{Candidates and batch.} As in the benchmark, each iteration scores 15{,}000 untried conditions and picks three of them. Most candidates are drawn at random from the whole grid. Once results exist, up to 2{,}400 of them are instead taken from the neighborhood of the eight best experiments so far (300 around each, with every variable changed by at most three grid steps), so that the surroundings of the good results are always among the candidates. The three conditions are then chosen as in the benchmark, by EI ($\xi = 0.01$) with the distance bonus of Eq.~\ref{eq:diverse-batch} ($\lambda = 0.03$).

\subsection{Reference and Residual GP}
\label{si:residual-gp}

Instead of adding the MOF-321 data to the new project as pseudo-observations, the program uses them through a calibrated transfer prior.

\paragraph{Reference GP.} A GP with the kernel above is fitted to the 95 measured MOF-321 syntheses of Section~\ref{si:seed-data}, built into the program (entries 26 and 31 of Table~\ref{tab:seed-raw} are the same experiment and are counted once), with $y = \log(1+q)$. All 95 experiments are used, including the 28 without product, so the reference GP learns both where MOF-321 crystallizes and where it does not. Its predictive mean $\mu_{\mathrm{ref}}(\mathbf{x})$ and s.d.\ $\sigma_{\mathrm{ref}}(\mathbf{x})$ encode the MOF-321 synthesis trends that also anchor the landscape of Section~\ref{si:landscape}.

\paragraph{Calibration.} Let $n$ be the number of experiments completed so far on the new MOF, with conditions $\mathbf{x}_i$ and results $y_i = \log(1+q_i)$, $i = 1, \dots, n$. The reference prediction is mapped onto the response of the new MOF by a straight line,
\begin{equation}
y_i \approx a + b\,\mu_{\mathrm{ref}}(\mathbf{x}_i).
\label{eq:calibration}
\end{equation}
Before any result ($n=0$), $a=0$ and $b=1$: the reference is used as it is. With one result ($n=1$), only the offset is set, $a = y_1 - \mu_{\mathrm{ref}}(\mathbf{x}_1)$, $b=1$. With $n \ge 2$,
\begin{equation}
(a, b) = \operatorname*{arg\,min}_{a,\,b}\;\Big[\,\sum_{i=1}^{n}\big(y_i - a - b\,\mu_{\mathrm{ref}}(\mathbf{x}_i)\big)^2 + \rho\,(b-1)^2\Big], \qquad \rho = \max(0.05,\, 1/n).
\label{eq:calibration-ridge}
\end{equation}
The first term is an ordinary least-squares fit of the line; the second penalizes slopes far from 1 and so keeps $b$ near 1 while there are few results. In addition, $b$ is limited to $[-2, 3]$. The offset $a$ absorbs differences in the intensity scale (instrument, reflection, framework), and $b$ how strongly the MOF-321 trends carry over: $b \approx 1$ keeps them, $b \approx 0$ switches the prior off, and $b<0$ inverts it.

\paragraph{Residual GP.} The residuals
\begin{equation}
r_i = y_i - \big(a + b\,\mu_{\mathrm{ref}}(\mathbf{x}_i)\big)
\end{equation}
are fitted by a second GP with the same kernel once $n \ge 3$ and the residuals are not all equal. The predictive model of the new MOF is
\begin{equation}
\mu(\mathbf{x}) = a + b\,\mu_{\mathrm{ref}}(\mathbf{x}) + \mu_{r}(\mathbf{x}),
\qquad
\sigma(\mathbf{x}) = \sqrt{\sigma_{r}(\mathbf{x})^2 + \big(0.35\,|b|\,\sigma_{\mathrm{ref}}(\mathbf{x})\big)^2 + \big(0.25\,s_r\big)^2},
\label{eq:residual-gp}
\end{equation}
where $\mu_r$ and $\sigma_r$ are the residual-GP mean and s.d.\ and $s_r$ is the s.d.\ of the residuals (0.35 before any data). The factors 0.35 and 0.25 are fixed settings of the program; they were chosen by us and are not fitted to data. The factor 0.35 reduces the reference uncertainty so that, once results of the new MOF exist, the residual model is the main source of uncertainty; the last term, a quarter of the scatter of the residuals, keeps $\sigma$ from falling to zero next to tested conditions. EI or PI is then computed from $\mu$ and $\sigma$, with the incumbent $y^{+}$ equal to the best $\log(1+q)$ of the project; before the first result, $y^{+}$ is set to the 75th percentile of $\mu$ over the candidates, so that the first batch balances high predicted $q$ and uncertainty.

\subsection{Calibration}
\label{si:calibration}

\paragraph{Duration of the prior.} The calibrated prior is used for the first four batches (the program's default, $M = 4$), i.e.\ for all 12 experiments of each campaign reported here.

\paragraph{Initial batch.} Before any result exists, the proposal depends only on the reference GP and the fixed random seed of the program, so the first batch is the same for every new MOF: R00001 = (18, 7, 17, 10, 21), R00002 = (23, 6, 15, 10, 20) and R00003 = (20, 7, 19, 12, 29), i.e.\ 180/70/170~$\mu$L, 50~min, 105~$^{\circ}$C; 230/60/150~$\mu$L, 50~min, 100~$^{\circ}$C; and 200/70/190~$\mu$L, 60~min, 145~$^{\circ}$C (\ce{AlCl3.6H2O}/NaOH/DI water, at the 0.1~mmol scale of Section~\ref{si:synthesis}). These three conditions lie inside the MOF-321 top-1\% tier (Table~\ref{tab:six-mof-localization}): the campaigns start from the MOF-321 prior, not from random conditions.

\paragraph{Recalibration.} After each batch, $a$, $b$ and the residual GP are refitted to all results of the MOF; the values below are those recorded by the program. For MIL-53-muc, MIL-160 and MTV-MIL-160, $b$ stays between 0.55 and 1.5: the MOF-321 trends carry over and the residual GP adjusts them locally. For MTV-MOF-303, $b$ falls to 0.05 and 0.20 in rounds 2 and 3, i.e.\ the prior is largely switched off and the search relies on the residual GP. For CAU-10 and CAU-23, the six results after round 1 are anti-correlated with $\mu_{\mathrm{ref}}$ and $b$ becomes negative ($-2.0$, the lower bound, and $-0.38$). The prior is then inverted: it predicts the highest $q$ where MOF-321 gives none, and the round-2 batches of both MOFs (the same three conditions, at 60--65~$^{\circ}$C) gave no or almost no product (Tables~\ref{tab:al-cau10} and~\ref{tab:al-cau23}). With these results, $b$ returns to 0.48 and 0.41 in round 3, and the next CAU-10 batch contains its best sample (R00012). The calibration thus corrects itself within one batch, at the cost of one poor batch.

\subsection{Six Campaigns}
\label{si:transfer-outcome}

\begin{figure}[htbp]
\centering
\siFig[0.95\linewidth]{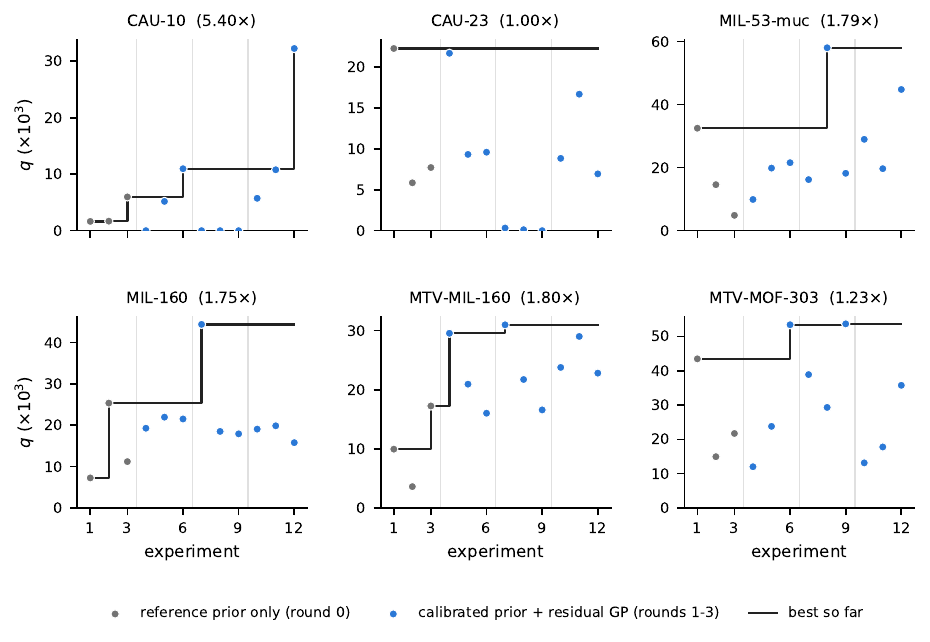}
\caption{Transfer-prior campaigns of the six validation MOFs. Points, measured $q$ of each experiment in synthesis order, first 12 experiments, colored by the model that proposed it: reference GP only (gray, round 0) and calibrated prior with residual GP (blue, rounds 1--3). Line, best so far. Light vertical lines separate the rounds. Titles, best of the first 12 experiments divided by the best of R00001--R00003.}
\label{fig:si-transfer-traj}
\end{figure}

Figure~\ref{fig:si-transfer-traj} shows every measured $q$ and the best-so-far of the six campaigns, and Table~\ref{tab:transfer-summary} compares the best of the three starting experiments with the best of the 12 experiments of each campaign. The best $q$ improved 1.23--5.40-fold for five MOFs; for CAU-23 the first starting condition remained the best. All best samples were found in the calibrated-prior phase (rounds 1--3), except that of CAU-23 (round 0). Of the 72 experiments within the first 12 of each MOF, 67 gave a crystalline product. The five without product are R00004 and R00007--R00009 of CAU-10 and R00009 of CAU-23; the other two round-2 experiments of CAU-23 gave only $q = 321$ and 117.

\begin{table}[htbp]
\centering
\caption{Best $q$ of the three starting experiments and of the first 12 experiments, their ratio, and the best recipe (\ce{AlCl3.6H2O} and NaOH stock solutions and added DI water in $\mu$L, with 0.1~mmol linker; hold time; temperature; Section~\ref{si:synthesis}). Concentrations: amounts of \ce{AlCl3.6H2O} and NaOH from the 100~mg\,mL$^{-1}$ stock solutions ($M$ in Table~\ref{tab:chemicals}) divided by the total volume of the three liquids. \ce{Al^{3+}}:L:\ce{OH^-}, molar ratio per linker (L, 0.1~mmol in total, also for the two multivariate frameworks).}
\label{tab:transfer-summary}
\footnotesize
\setlength{\tabcolsep}{2pt}
\resizebox{\linewidth}{!}{%
\begin{tabular}{@{}lrrccrrrrrr@{}}
\toprule
 & \multicolumn{2}{c}{Best $q$ (ID)} & & \multicolumn{3}{c}{Best recipe} & \multicolumn{4}{c}{Composition} \\
\cmidrule(lr){2-3}\cmidrule(lr){5-7}\cmidrule(l){8-11}
MOF & Start & First 12 & Ratio & \shortstack{Al/NaOH/\ce{H2O}\\($\mu$L)} & \shortstack{$t$\\(min)} & \shortstack{$T$\\($^{\circ}$C)} & \shortstack{{[}\ce{Al^{3+}}{]}\\(mM)} & \shortstack{{[}\ce{OH^-}{]}\\(mM)} & \shortstack{\ce{Al^{3+}}:L:\ce{OH^-}\\(mol)} & \shortstack{\ce{OH^-}:\\\ce{Al^{3+}}} \\
\midrule
CAU-10 & 5{,}978 (R3) & 32{,}259 (R12) & 5.40 & 200/70/180 & 60 & 140 & 184 & 389 & 0.83:1:1.75 & 2.11 \\
CAU-23 & 22{,}254 (R1) & 22{,}254 (R1) & 1.00 & 180/70/170 & 50 & 105 & 178 & 417 & 0.75:1:1.75 & 2.35 \\
MIL-53-muc & 32{,}523 (R1) & 58{,}064 (R8) & 1.79 & 160/60/160 & 50 & 105 & 174 & 395 & 0.66:1:1.50 & 2.26 \\
MIL-160 & 25{,}364 (R2) & 44{,}395 (R7) & 1.75 & 190/60/150 & 55 & 100 & 197 & 375 & 0.79:1:1.50 & 1.91 \\
MTV-MIL-160 & 17{,}258 (R3) & 30{,}990 (R7) & 1.80 & 200/70/180 & 60 & 140 & 184 & 389 & 0.83:1:1.75 & 2.11 \\
MTV-MOF-303 & 43{,}452 (R1) & 53{,}648 (R9) & 1.23 & 190/60/160 & 50 & 110 & 192 & 366 & 0.79:1:1.50 & 1.91 \\
\midrule
\multicolumn{3}{@{}l}{Arithmetic mean / geometric mean / median} & \multicolumn{8}{l}{2.16 / 1.83 / 1.77} \\
\bottomrule
\end{tabular}}
\end{table}

\paragraph{Transferable and framework-specific features of the prior.} The six campaigns suggest that the MOF-321 prior transfers the general crystallization rules of these Al-carboxylate syntheses, and that the optimum of each framework still has to be found by the search. All six best recipes (Table~\ref{tab:transfer-summary}) lie in or next to the productive window of MOF-321 (Section~\ref{si:landscape}). In equivalents relative to the linker, the top-0.1\% MOF-321 anchors use 0.46--0.75~equiv \ce{AlCl3.6H2O} and 1.25--1.75~equiv NaOH (55--60~min, 115--145~$^{\circ}$C), and the six best recipes 0.66--0.83~equiv \ce{AlCl3.6H2O} and 1.5--1.75~equiv NaOH (50--60~min, 100--140~$^{\circ}$C), i.e.\ little metal, little base, long hold times and high temperature. Within this window, however, the optima differ: CAU-10 and MTV-MIL-160 are best at 140~$^{\circ}$C and 60~min, CAU-23, MIL-53-muc and MIL-160 at 100--105~$^{\circ}$C and 50--55~min, and MTV-MOF-303 at 110~$^{\circ}$C and 50~min. None of the best conditions is one of the three starting conditions, except for CAU-23, and none coincides with a MOF-321 optimum: they lie inside the MOF-321 top-1\% tier but 1--2 grid steps from its top-0.1\% tier (Section~\ref{si:six-mof-localization}). The calibration makes the same point: the slope $b$ varies from framework to framework and between rounds (Section~\ref{si:calibration}), so the MOF-321 trends are rescaled, partly switched off (MTV-MOF-303) or temporarily inverted (CAU-10, CAU-23) as results come in, and five of the six best samples were proposed after at least one recalibration.

% =====================================================================
\section{Synthesis of the Validation MOFs}
\label{si:synthesis}

% =====================================================================
\sloppy
\paragraph{General procedure.} All six frameworks were made by the same procedure; only the linker differs. The linker was weighed into a G10 vial, followed by DI water (the volume of the recipe), a small magnetic stir bar and the NaOH stock solution (100~mg\,mL$^{-1}$ in DI water). The mixture was sonicated until the linker had dissolved. The \ce{AlCl3.6H2O} stock solution (100~mg\,mL$^{-1}$ in DI water) was then added, and the mixture was heated in the microwave reactor and worked up as in Section~\ref{si:microwave}. The volumes of the two stock solutions and of DI water, the hold time and the temperature of every synthesis are those proposed by the active-learning campaign; they are listed for each MOF, together with the encoded variables $\mathbf{x}$, in Tables~\ref{tab:al-cau10}--\ref{tab:al-mtvmof303}, and the best recipe of each framework in Table~\ref{tab:transfer-summary}. The validation syntheses were run at one tenth of the scale of Table~\ref{tab:encoding}: 0.1~mmol of linker (for the two multivariate frameworks, 0.05~mmol of each of the two linkers, 1:1), with one step of $x_1$, $x_2$ or $x_3$ equal to 10~$\mu$L instead of 0.1~mL. Molar masses are given in Table~\ref{tab:chemicals}.

\paragraph{(i) CAU-10.}\label{si:syn-cau10} CAU-10 was prepared from \ce{AlCl3.6H2O} and isophthalic acid by the general procedure, with 0.1~mmol (16.6~mg) isophthalic acid; the conditions of the 12 syntheses are in Table~\ref{tab:al-cau10}.

\paragraph{(ii) CAU-23.}\label{si:syn-cau23} CAU-23 was prepared from \ce{AlCl3.6H2O} and thiophene-2,5-dicarboxylic acid by the general procedure, with 0.1~mmol (17.2~mg) thiophene-2,5-dicarboxylic acid; the conditions of the 12 syntheses are in Table~\ref{tab:al-cau23}.

\paragraph{(iii) MIL-53-muc.}\label{si:syn-mil53} MIL-53-muc was prepared from \ce{AlCl3.6H2O} and \textit{trans,trans}-muconic acid by the general procedure, with 0.1~mmol (14.2~mg) \textit{trans,trans}-muconic acid; the conditions of the 12 syntheses are in Table~\ref{tab:al-mil53}.

\paragraph{(iv) MIL-160.}\label{si:syn-mil160} MIL-160 was prepared from \ce{AlCl3.6H2O} and furan-2,5-dicarboxylic acid by the general procedure, with 0.1~mmol (15.6~mg) furan-2,5-dicarboxylic acid; the conditions of the 12 syntheses are in Table~\ref{tab:al-mil160}.

\paragraph{(v) MTV-MIL-160.}\label{si:syn-mtvmil160} MTV-MIL-160 was prepared from \ce{AlCl3.6H2O} and two linkers, furan-2,5-dicarboxylic acid and isophthalic acid by the general procedure, with 0.05~mmol (7.8~mg) furan-2,5-dicarboxylic acid and 0.05~mmol (8.3~mg) isophthalic acid; the conditions of the 12 syntheses are in Table~\ref{tab:al-mtvmil160}. Both linkers were weighed into the vial before the DI water and the NaOH solution were added (feed ratio 1:1).

\paragraph{(vi) MTV-MOF-303.}\label{si:syn-mtvmof303} MTV-MOF-303 was prepared from \ce{AlCl3.6H2O} and two linkers, thiophene-2,5-dicarboxylic acid and 1\textit{H}-pyrazole-3,5-dicarboxylic acid hydrate by the general procedure, with 0.05~mmol (8.6~mg) thiophene-2,5-dicarboxylic acid and 0.05~mmol (8.7~mg) 1\textit{H}-pyrazole-3,5-dicarboxylic acid hydrate; the conditions of the 12 syntheses are in Table~\ref{tab:al-mtvmof303}. Both linkers were weighed into the vial before the DI water and the NaOH solution were added (feed ratio 1:1).

\fussy
% =====================================================================
\section{PXRD and Campaign Trajectories}
\label{si:pxrd}

% =====================================================================

\paragraph{Instrument and data processing.}\label{si:pxrd-instrument} PXRD patterns were collected on a Rigaku MiniFlex diffractometer (Cu K$\alpha$) from 3 to 30$^{\circ}$ $2\theta$ with a step of 0.01$^{\circ}$ at 8$^{\circ}$\,min$^{-1}$. The washed powders were loaded on silicon sample holders and pressed with a spatula as described in Section~\ref{si:q-mount}. The patterns were processed with a Python script (\href{https://github.com/deep-synthesis/ActMOF}{GitHub repository}). In each pattern the diagnostic reflection of the target phase (Table~\ref{tab:q-peaks}) was selected, its maximum intensity was taken as $I$, and the FWHM was taken as the distance between the two $2\theta$ positions at half of this maximum. $I$ and FWHM were entered into the active-learning program, which computes $q = I/\mathrm{FWHM}$ (Eq.~\ref{eq:app-q}); a pattern without a reflection was entered as $I = 0$, FWHM $=30$ ($q = 0$).

The figures and tables below show, for each MOF, the 12 syntheses of its campaign in synthesis order (R00001--R00012), one MOF per page. In each figure, one row of three patterns is one round: round 0, conditions proposed by the reference prior only; rounds 1--3, by the calibrated prior with residual GP (Section~\ref{si:calibration}). The tables give the encoded variables $\mathbf{x} = (x_1, \dots, x_5)$ of Table~\ref{tab:encoding}, the recipe in laboratory units (stock solutions of \ce{AlCl3.6H2O} and NaOH, 100~mg\,mL$^{-1}$ in DI water, and added DI water in $\mu$L; Section~\ref{si:synthesis}) and the measured intensity, FWHM and $q = I/\mathrm{FWHM}$ of the diagnostic reflection; n.d., no reflection detected ($q=0$); bold, best sample. The first three experiments are the same three conditions for every MOF, proposed by the MOF-321 reference prior before any result of that MOF was available. The best $q$ of each campaign and the improvement over the starting experiments are summarized in Table~\ref{tab:transfer-summary} and Figure~\ref{fig:si-transfer-traj}. 

\begin{figure}[H]
\centering
\siFig[0.31\linewidth]{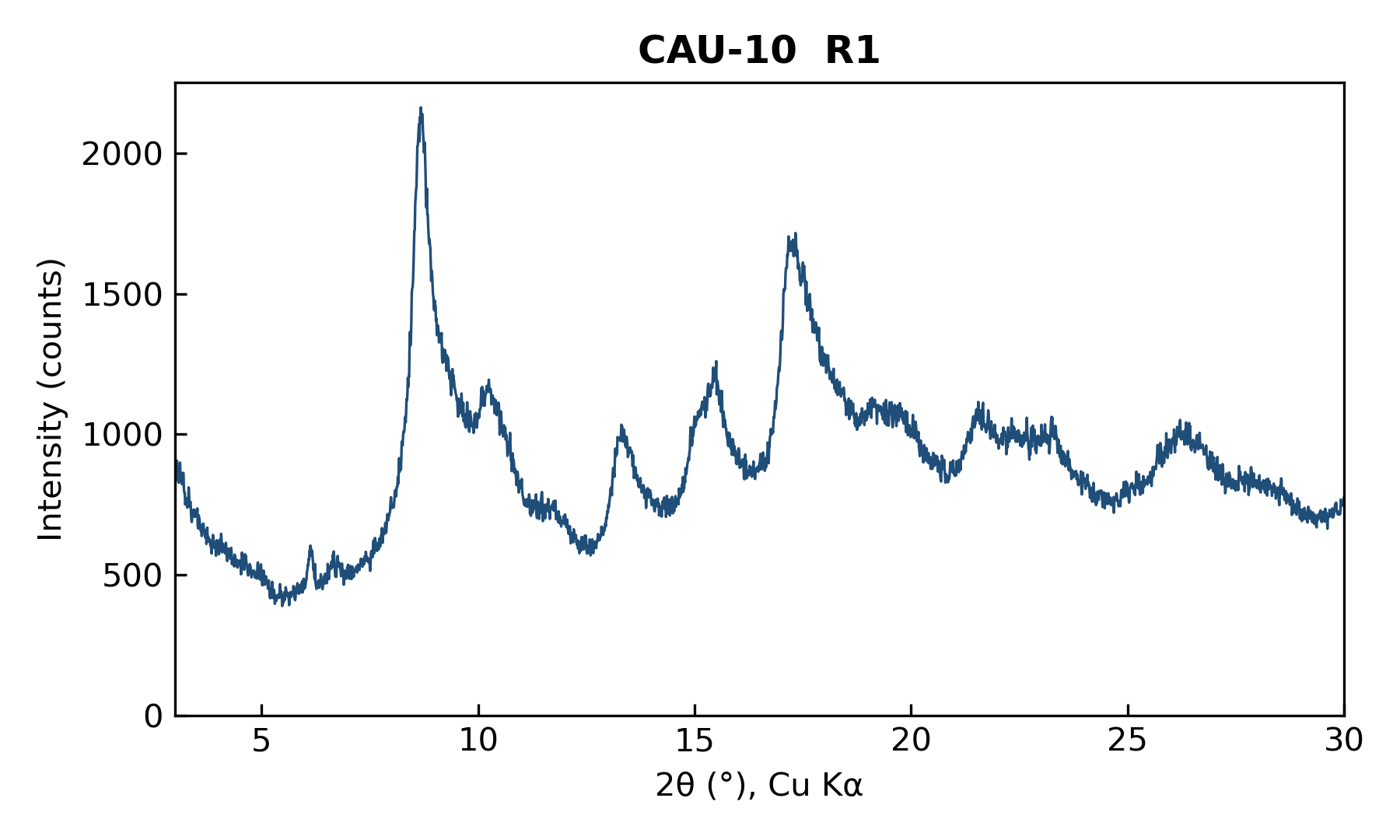}\hfill
\siFig[0.31\linewidth]{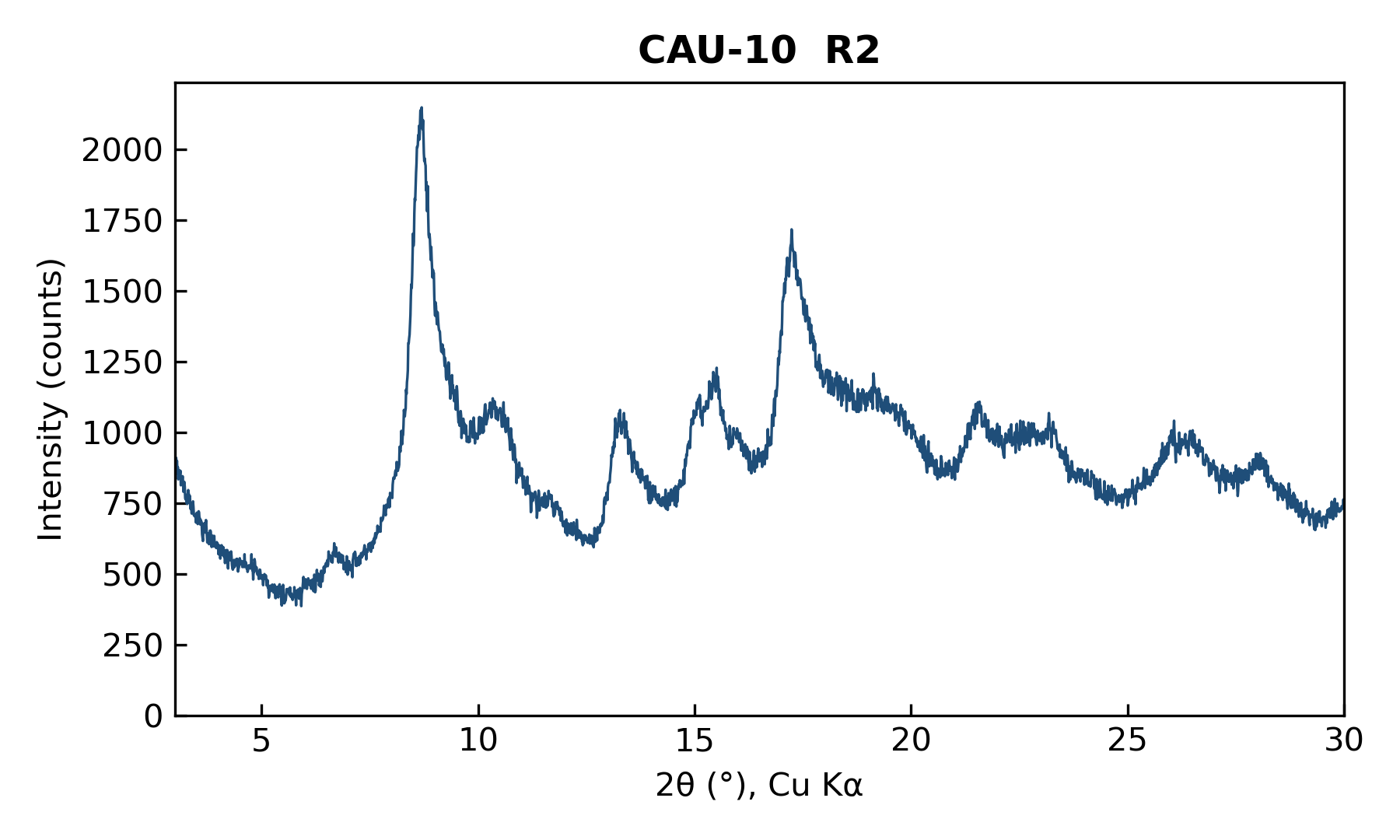}\hfill
\siFig[0.31\linewidth]{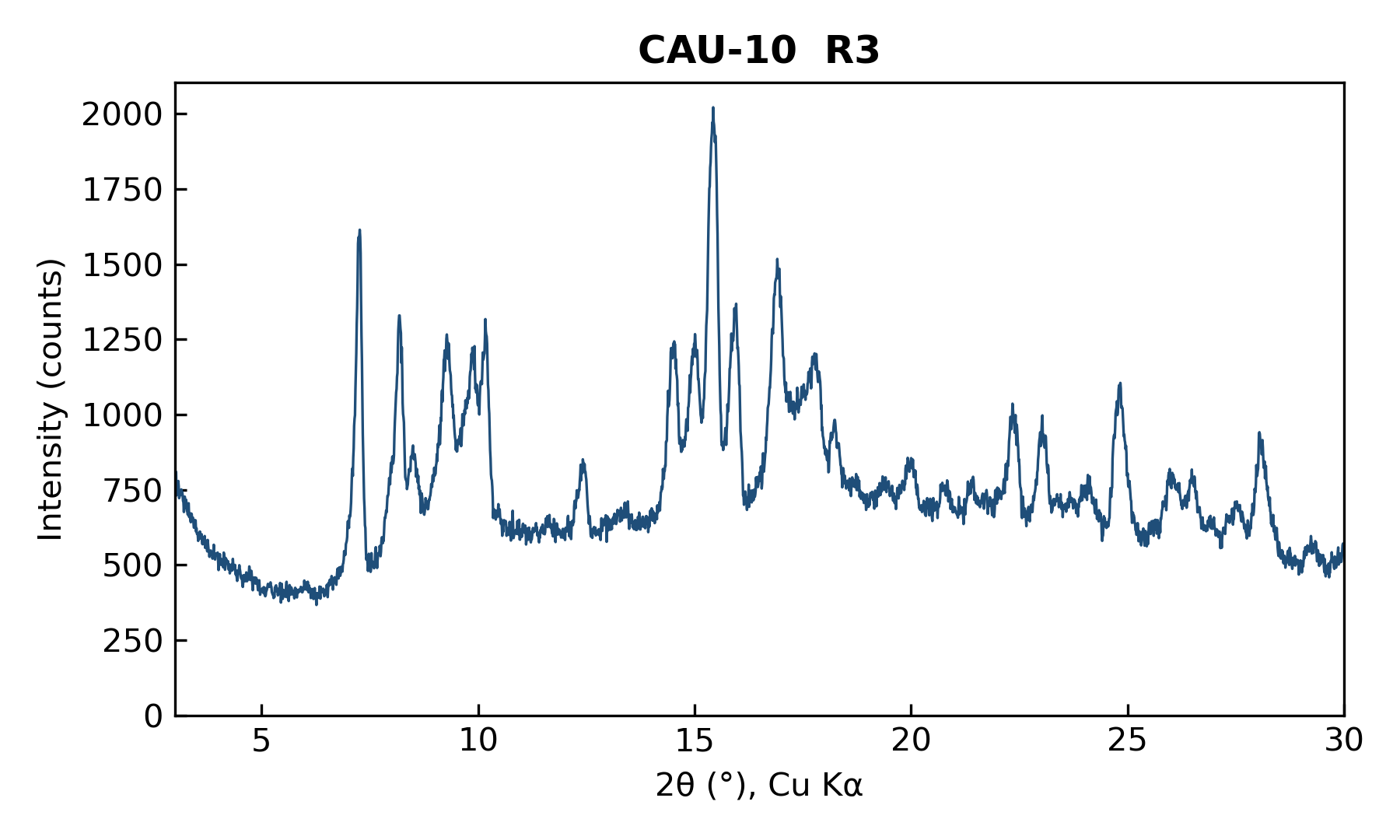}\\[2pt]
\siFig[0.31\linewidth]{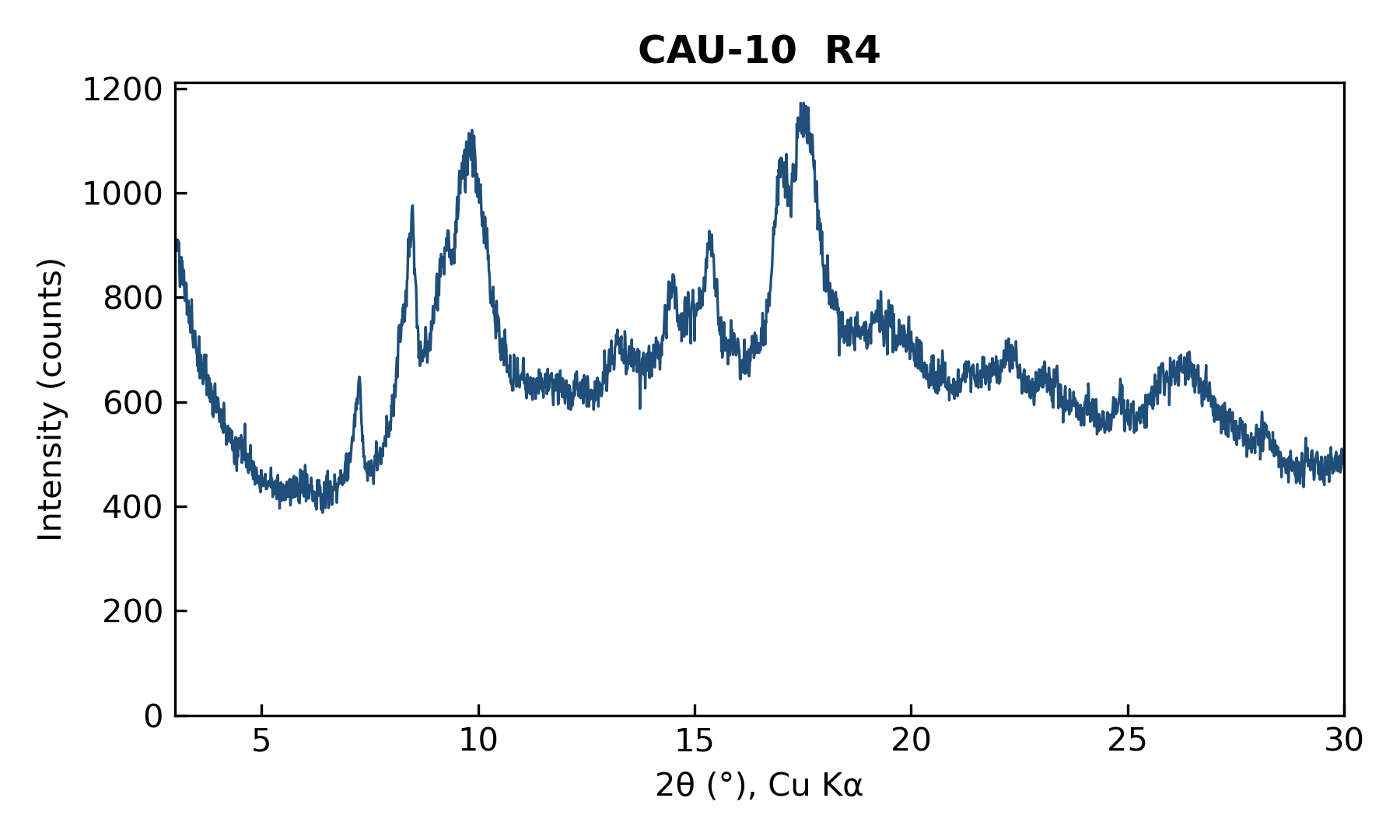}\hfill
\siFig[0.31\linewidth]{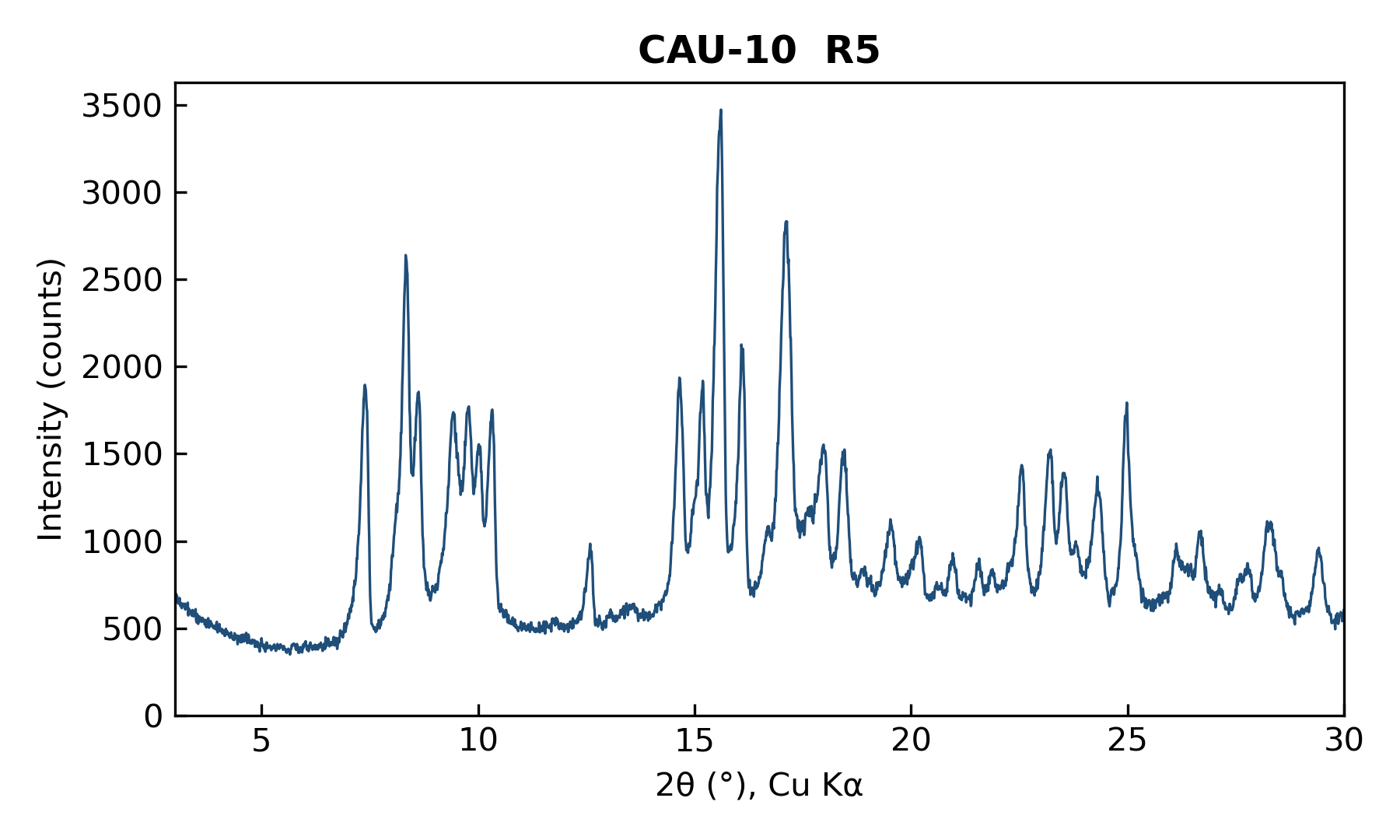}\hfill
\siFig[0.31\linewidth]{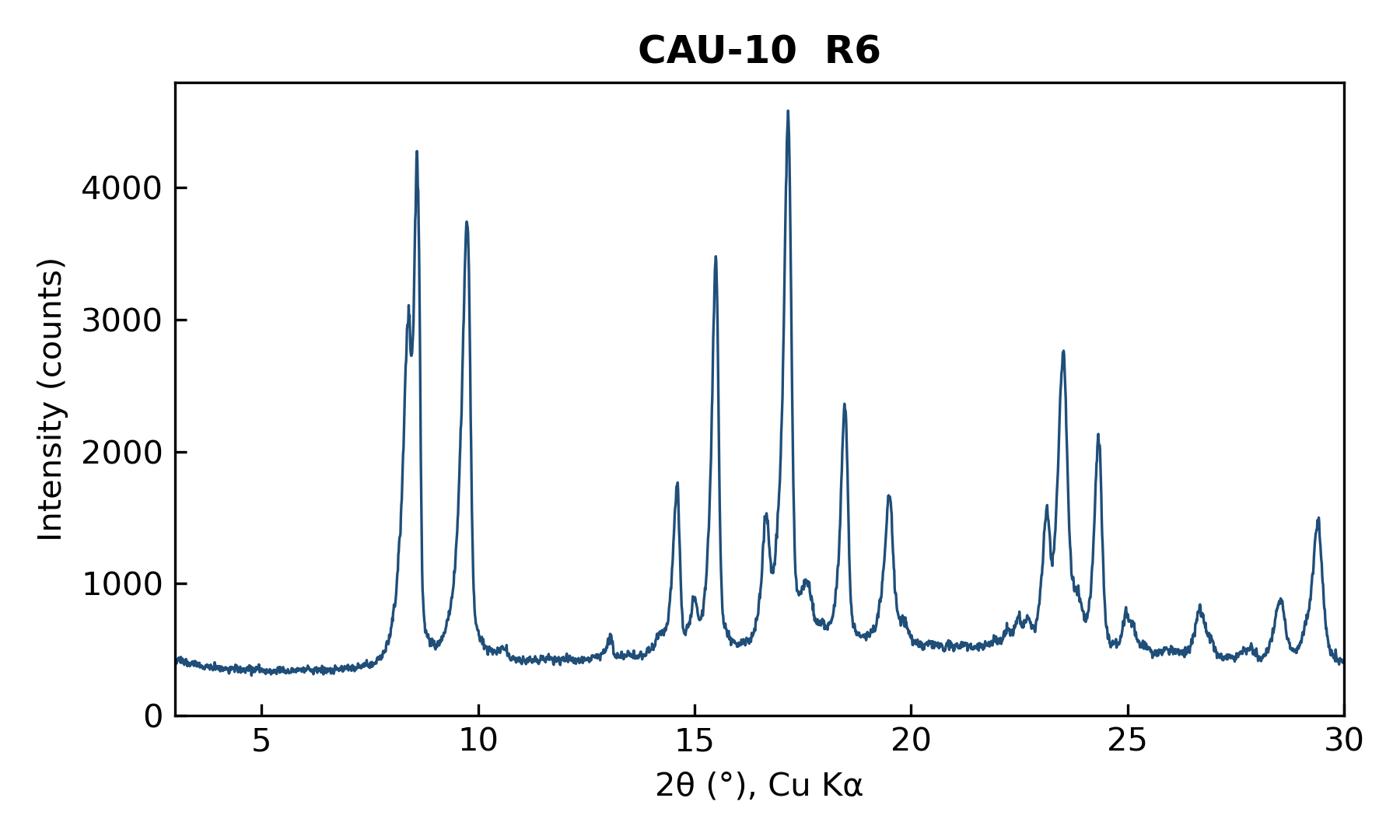}\\[2pt]
\siFig[0.31\linewidth]{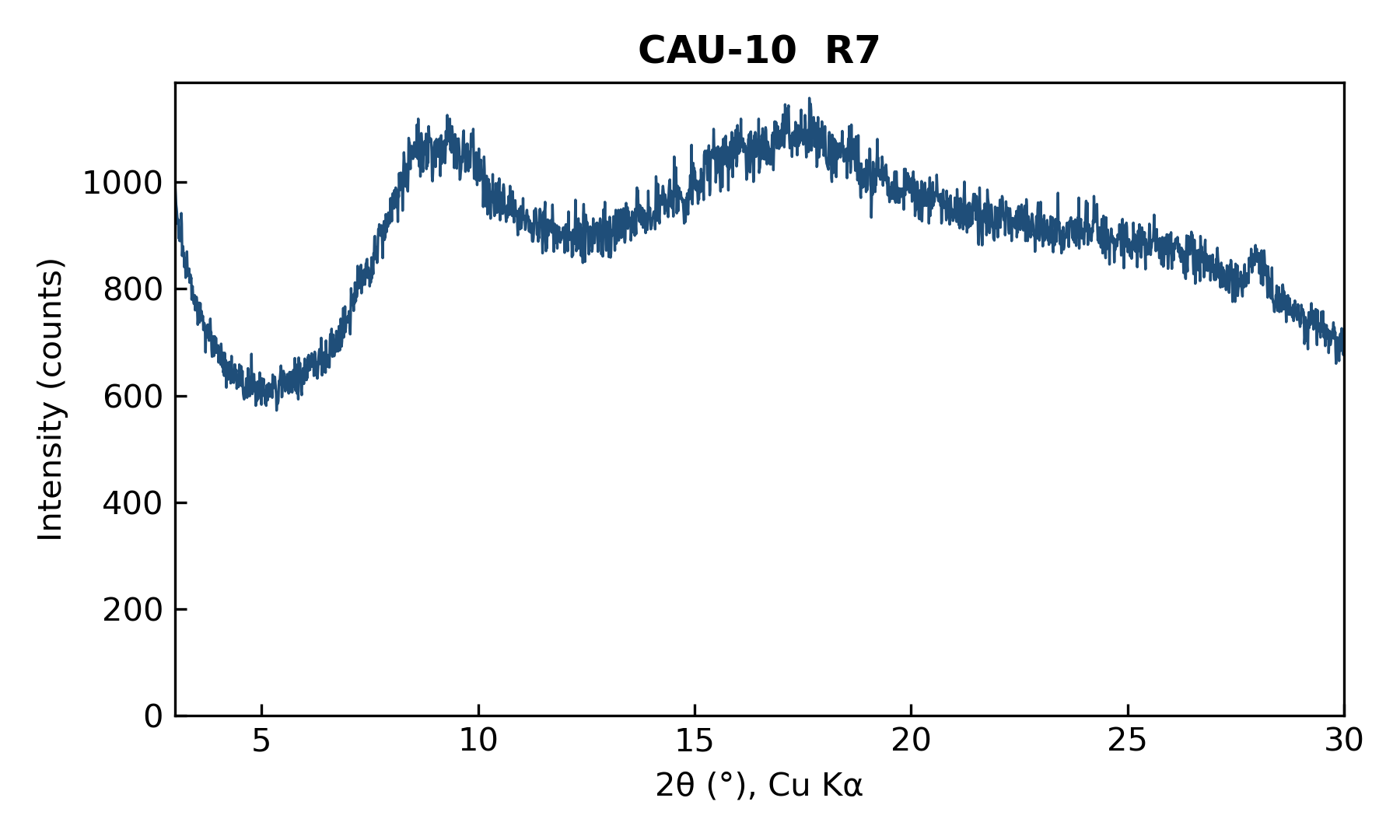}\hfill
\siFig[0.31\linewidth]{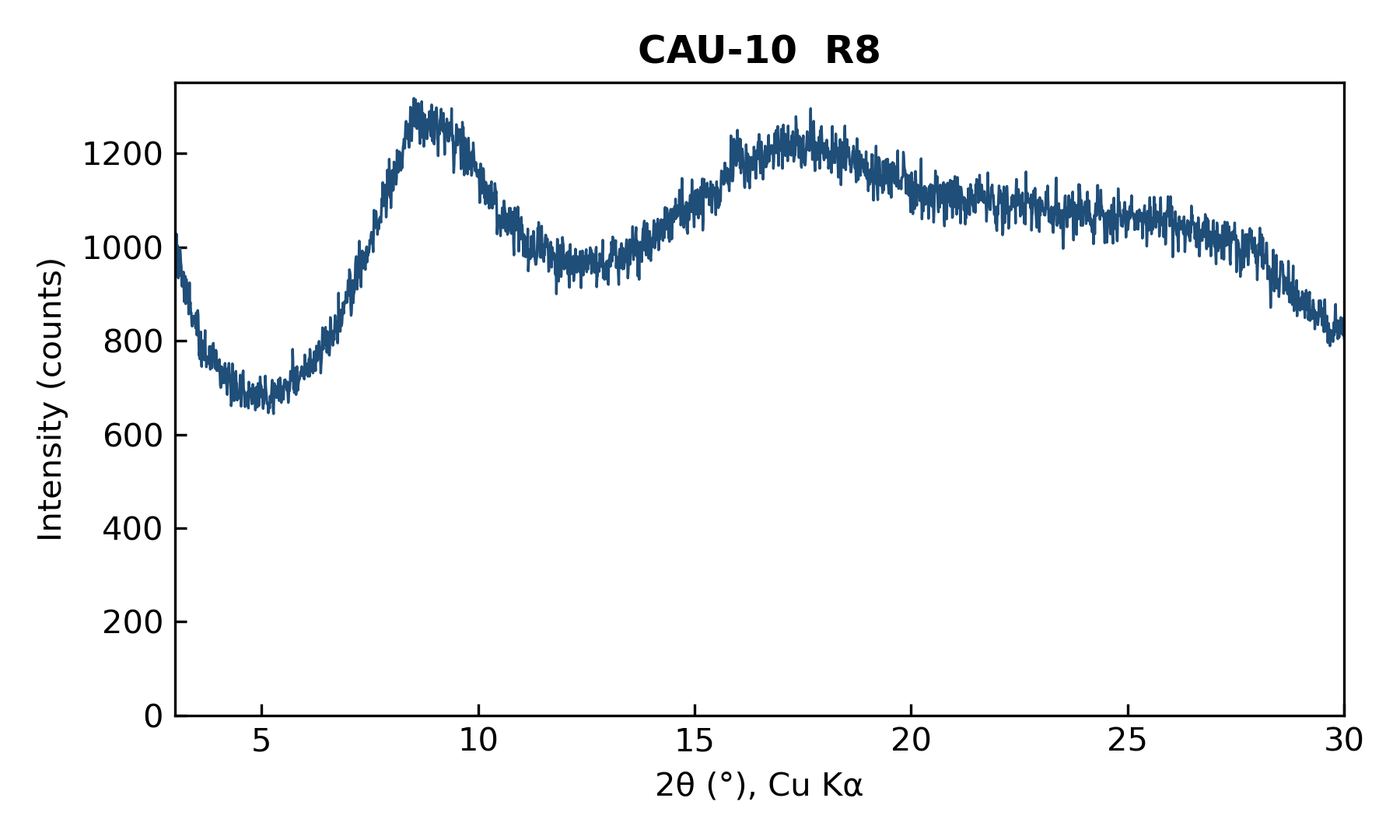}\hfill
\siFig[0.31\linewidth]{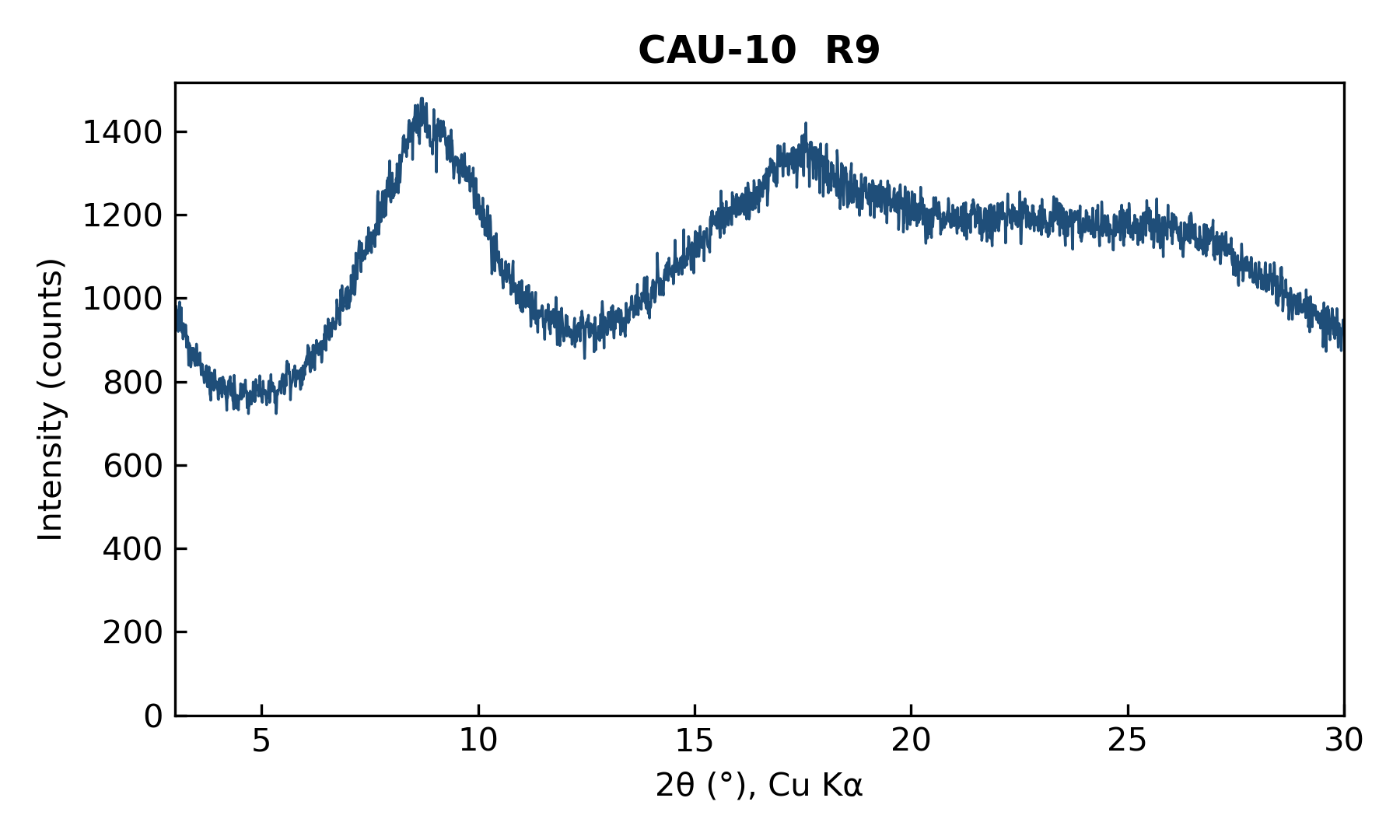}\\[2pt]
\siFig[0.31\linewidth]{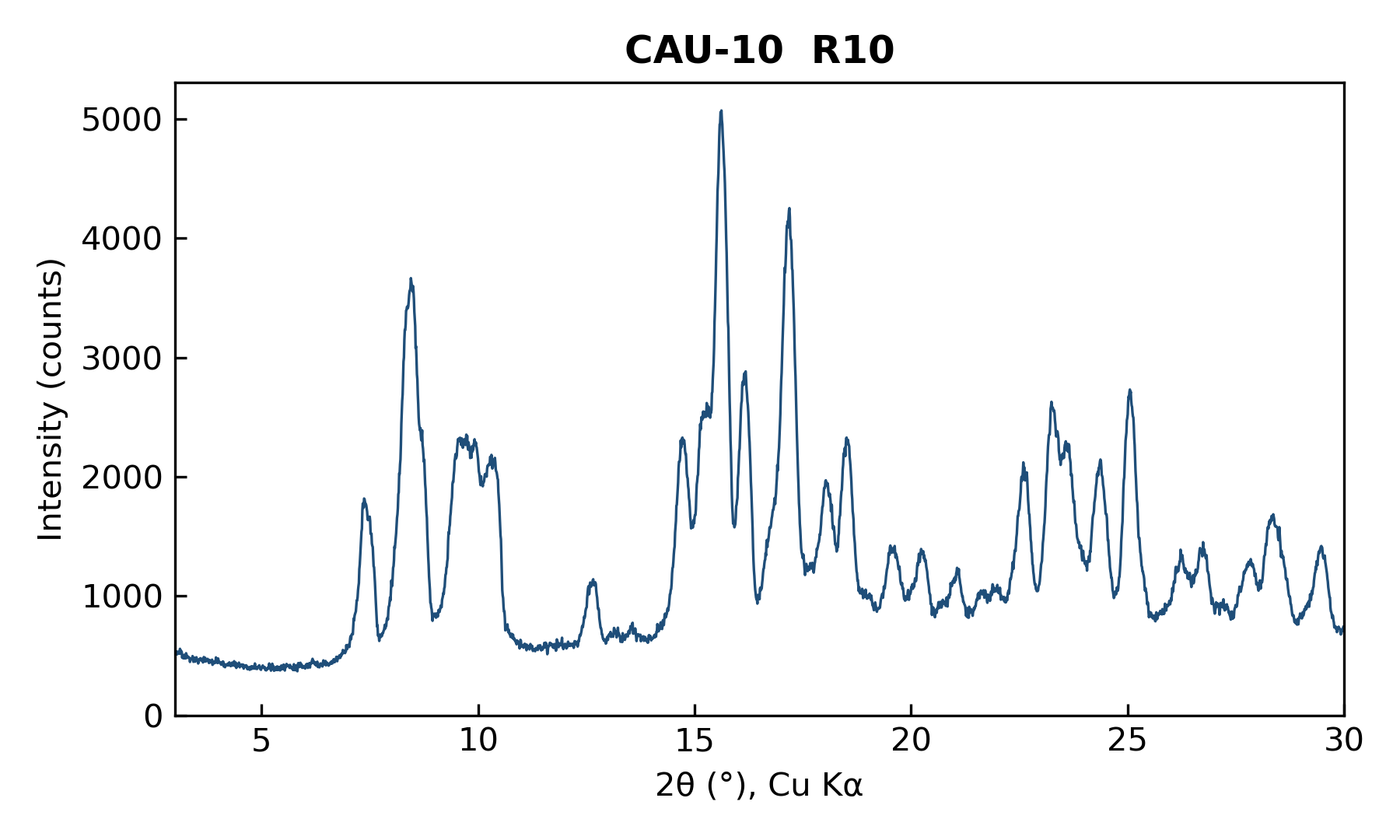}\hfill
\siFig[0.31\linewidth]{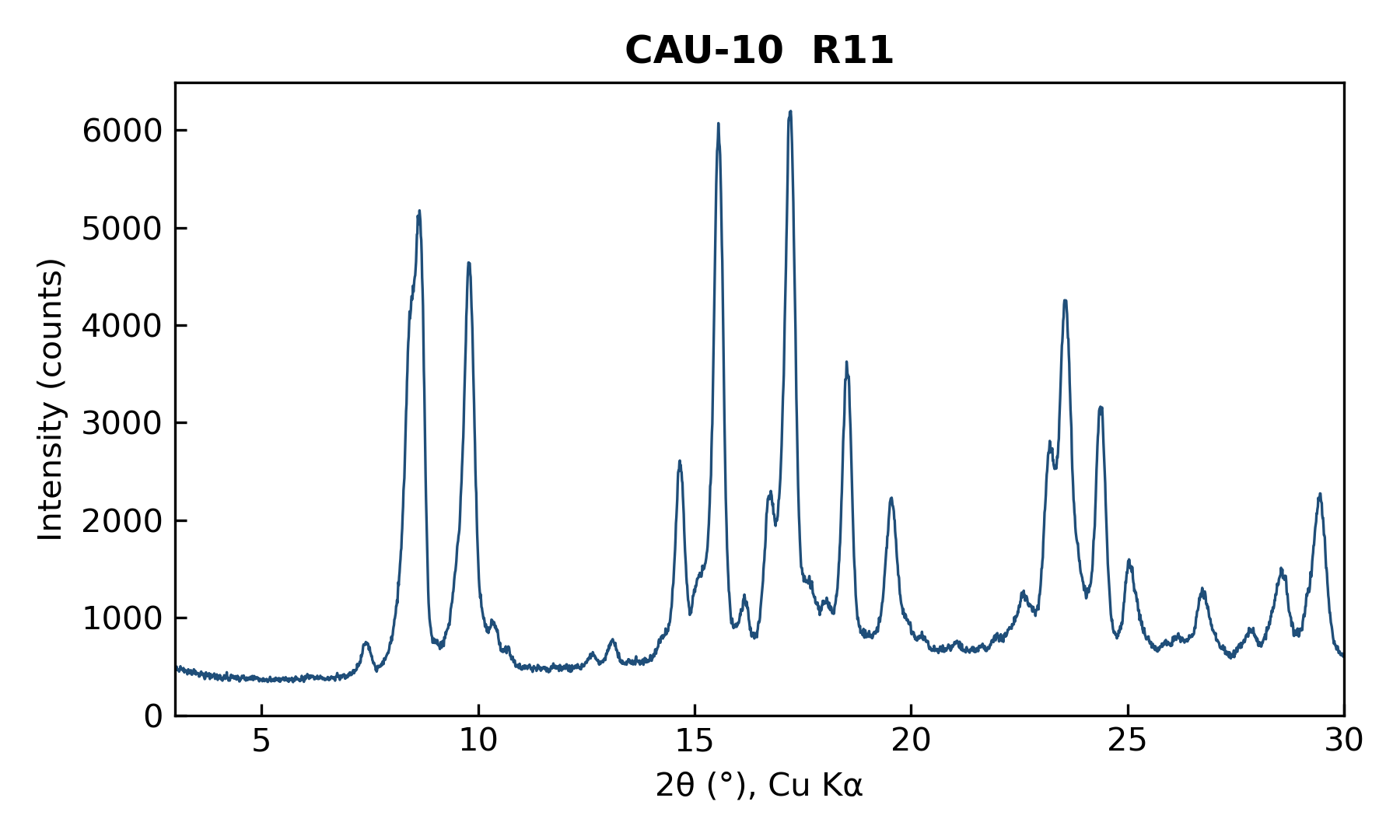}\hfill
\siFig[0.31\linewidth]{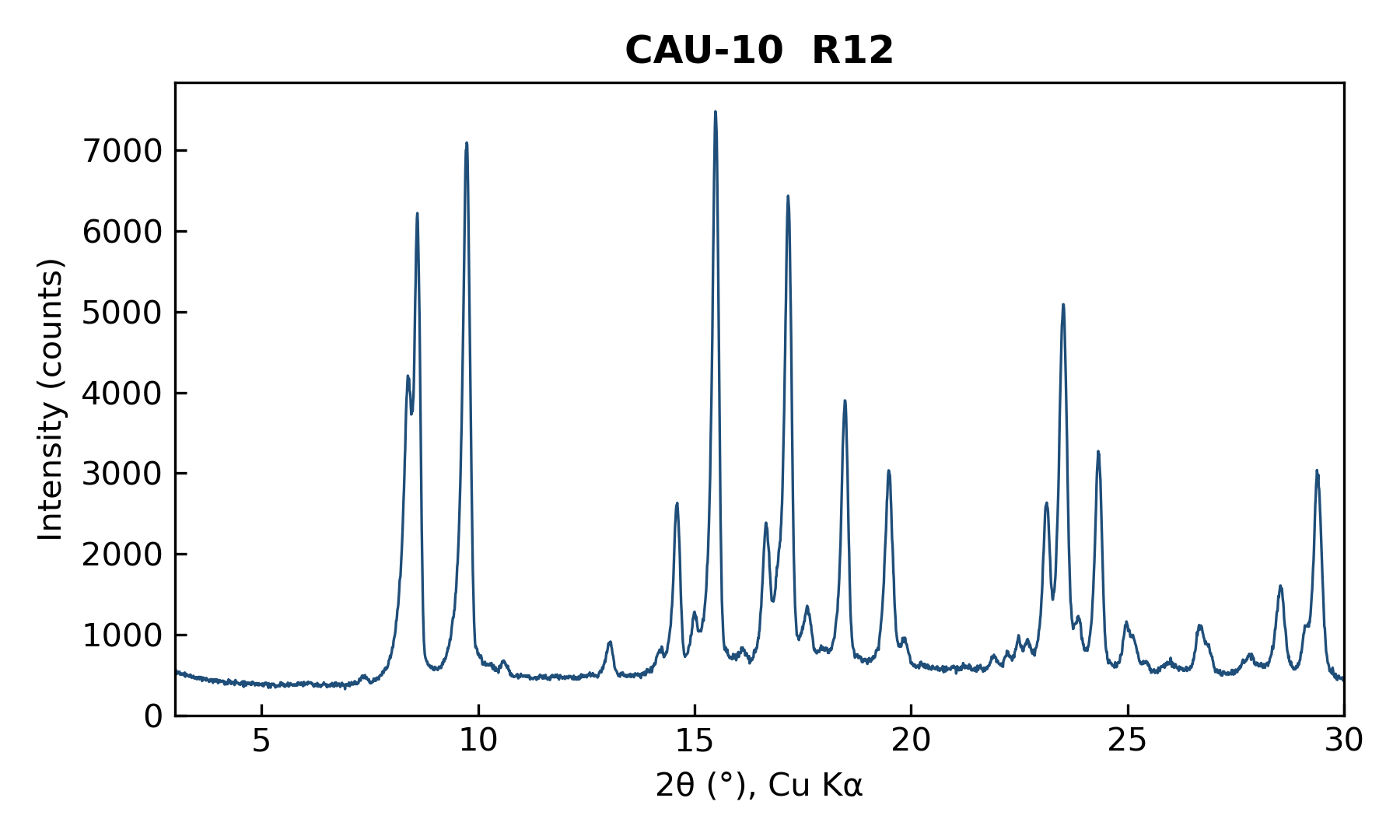}
\caption{PXRD patterns of CAU-10, R00001--R00012 (Table~\ref{tab:al-cau10}).}
\label{fig:pxrd-cau10}
\end{figure}

\clearpage
\paragraph{(i) CAU-10.}\label{si:pxrd-cau10} Recipes and results in Table~\ref{tab:al-cau10}, patterns in Figure~\ref{fig:pxrd-cau10}.
\begin{table}[H]
\centering
\caption{Recipes and PXRD results of CAU-10.}
\label{tab:al-cau10}
\footnotesize
\setlength{\tabcolsep}{3.5pt}
\resizebox{\linewidth}{!}{%
\begin{tabular}{@{}lccrrrrrrrr@{}}
\toprule
 & & & \multicolumn{3}{c}{Stock solution / DI water ($\mu$L)} & & & & & \\
\cmidrule(lr){4-6}
Exp. & Round & Encoded $\mathbf{x}$ & \ce{AlCl3.6H2O} & NaOH & DI \ce{H2O} & Time (min) & $T$ ($^{\circ}$C) & $I$ & FWHM ($^{\circ}$) & $q$ \\
\midrule
R00001 & 0 & (18, 7, 17, 10, 21) & 180 & 70 & 170 & 50 & 105 & 2{,}140 & 1.31 & 1{,}634 \\
R00002 & 0 & (23, 6, 15, 10, 20) & 230 & 60 & 150 & 50 & 100 & 2{,}148 & 1.28 & 1{,}678 \\
R00003 & 0 & (20, 7, 19, 12, 29) & 200 & 70 & 190 & 60 & 145 & 1{,}614 & 0.27 & 5{,}978 \\
R00004 & 1 & (22, 7, 16, 12, 29) & 220 & 70 & 160 & 60 & 145 & 0 & n.d. & 0 \\
R00005 & 1 & (19, 7, 17, 12, 29) & 190 & 70 & 170 & 60 & 145 & 2{,}639 & 0.51 & 5{,}175 \\
R00006 & 1 & (21, 7, 18, 12, 29) & 210 & 70 & 180 & 60 & 145 & 4{,}277 & 0.39 & 10{,}967 \\
R00007 & 2 & (21, 7, 15, 11, 12) & 210 & 70 & 150 & 55 & 60 & 0 & n.d. & 0 \\
R00008 & 2 & (22, 7, 12, 10, 13) & 220 & 70 & 120 & 50 & 65 & 0 & n.d. & 0 \\
R00009 & 2 & (25, 9, 14, 10, 13) & 250 & 90 & 140 & 50 & 65 & 0 & n.d. & 0 \\
R00010 & 3 & (19, 7, 18, 12, 28) & 190 & 70 & 180 & 60 & 140 & 3{,}666 & 0.64 & 5{,}728 \\
R00011 & 3 & (17, 7, 18, 12, 28) & 170 & 70 & 180 & 60 & 140 & 5{,}176 & 0.48 & 10{,}783 \\
\textbf{R00012} & \textbf{3} & \textbf{(20, 7, 18, 12, 28)} & \textbf{200} & \textbf{70} & \textbf{180} & \textbf{60} & \textbf{140} & \textbf{7{,}097} & \textbf{0.22} & \textbf{32{,}259} \\
\bottomrule
\end{tabular}}
\end{table}

\begin{figure}[H]
\centering
\siFig[0.31\linewidth]{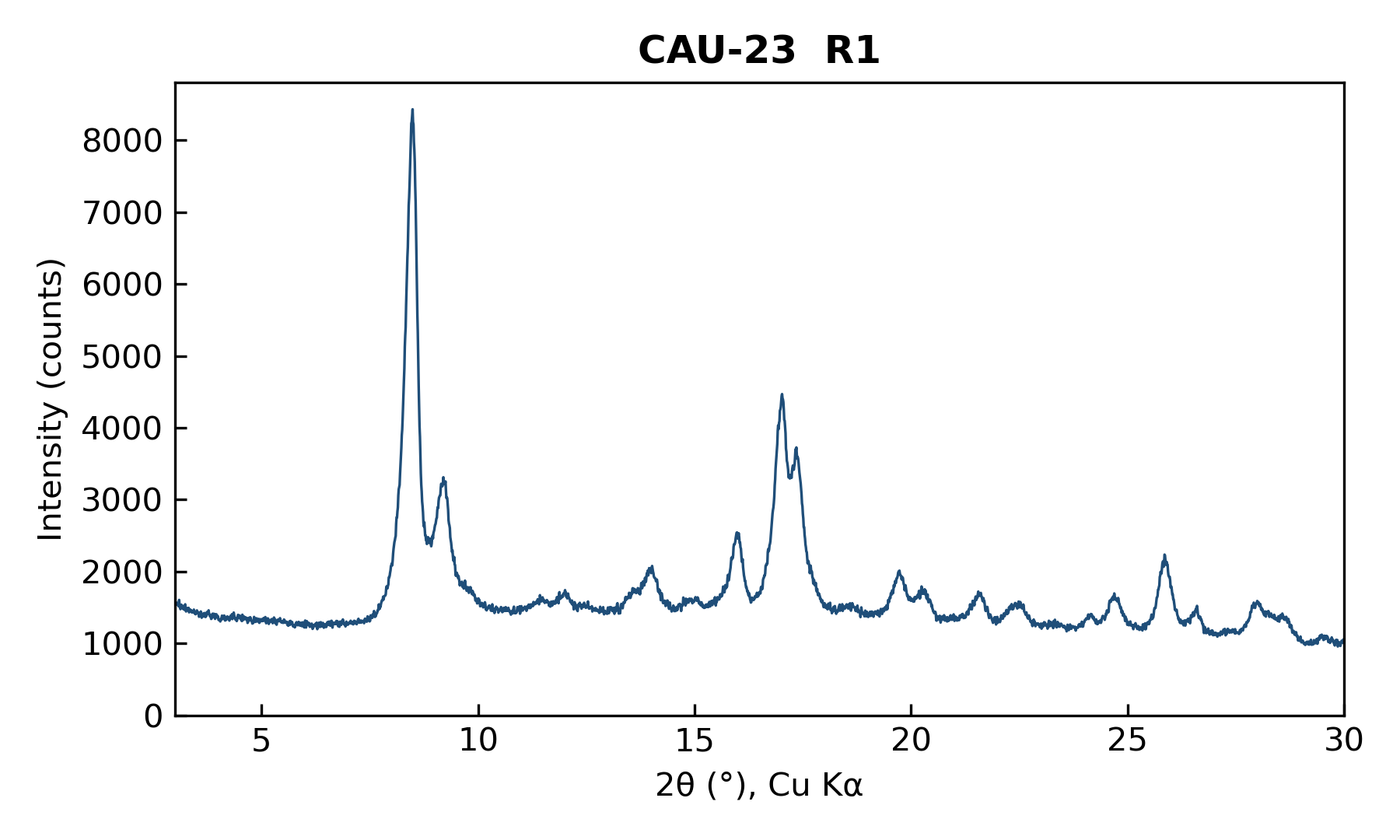}\hfill
\siFig[0.31\linewidth]{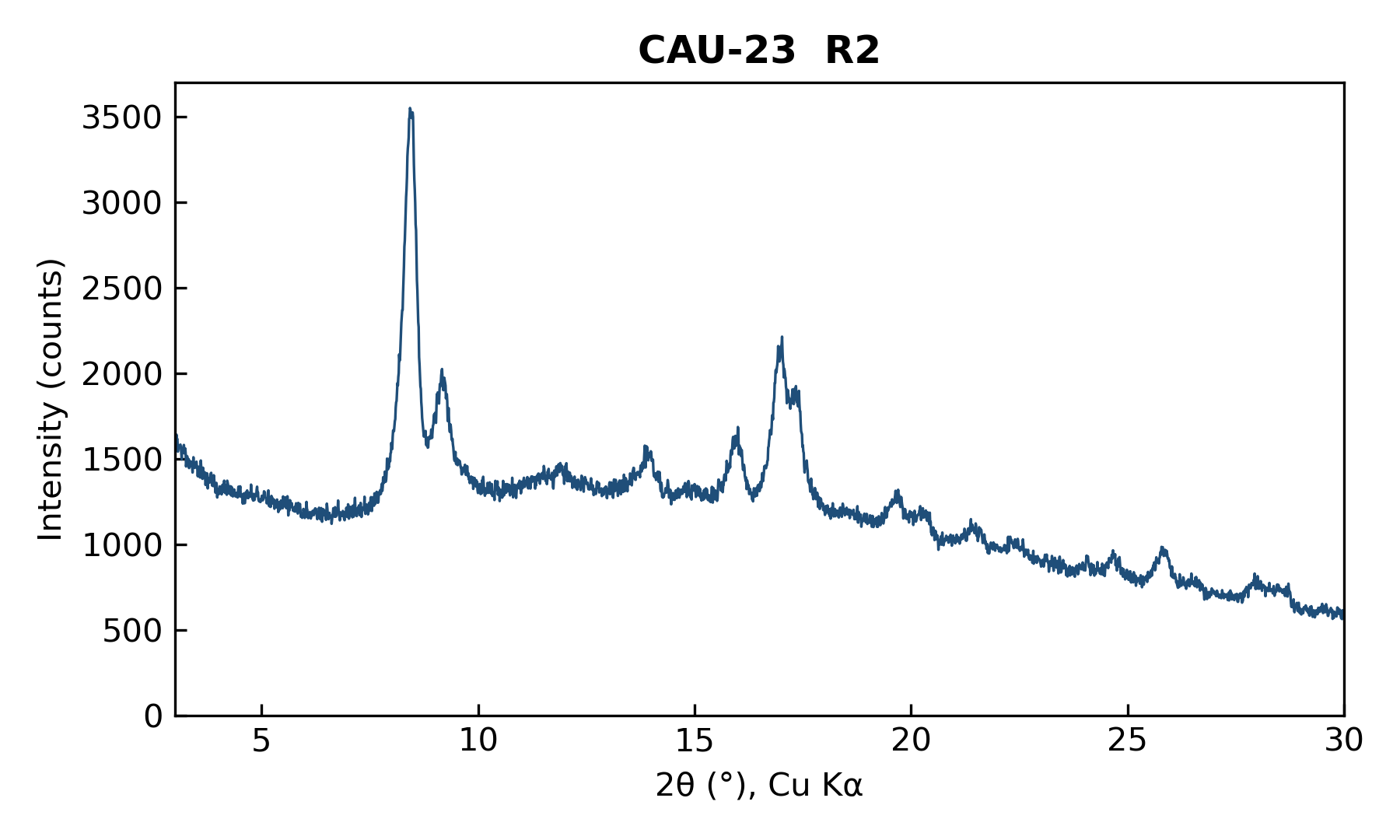}\hfill
\siFig[0.31\linewidth]{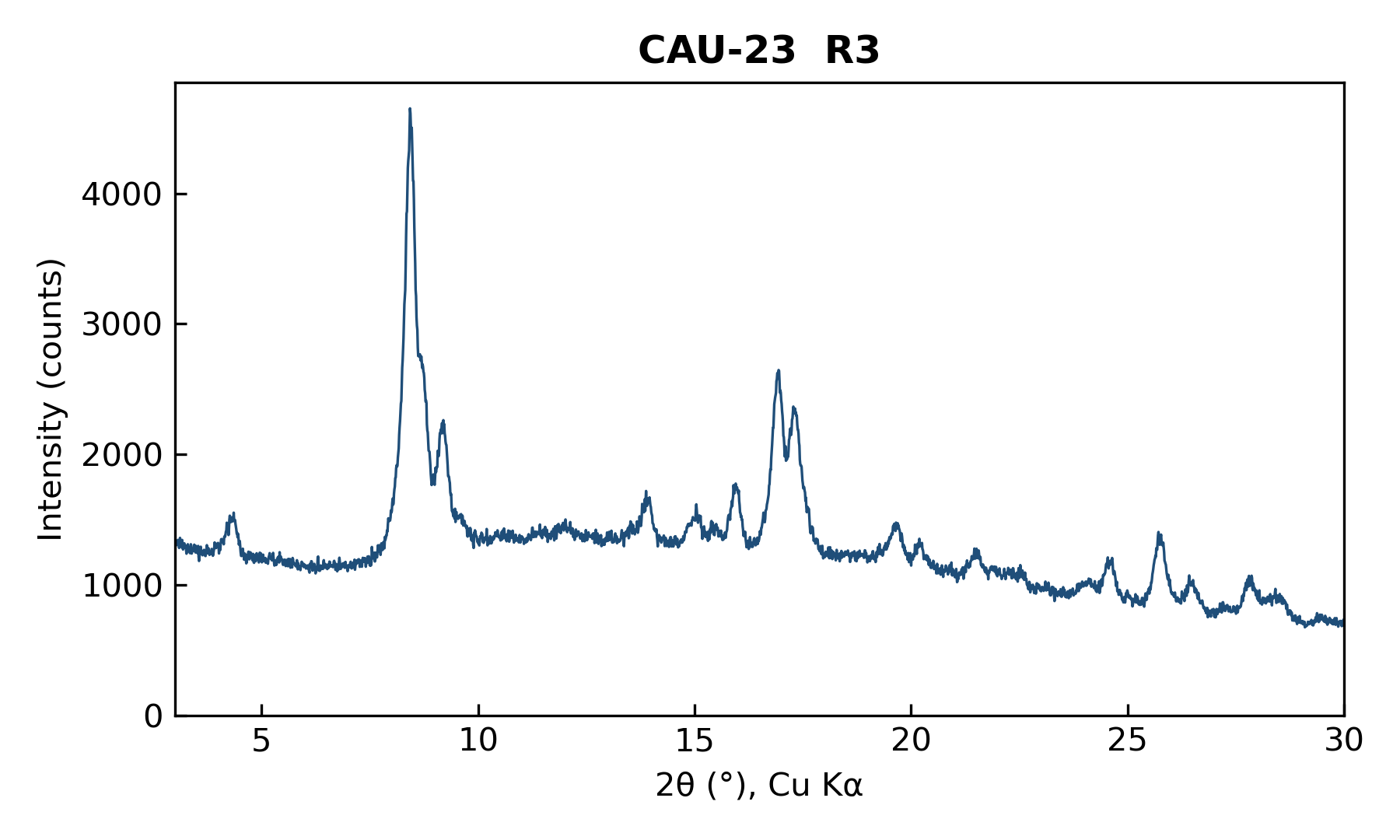}\\[2pt]
\siFig[0.31\linewidth]{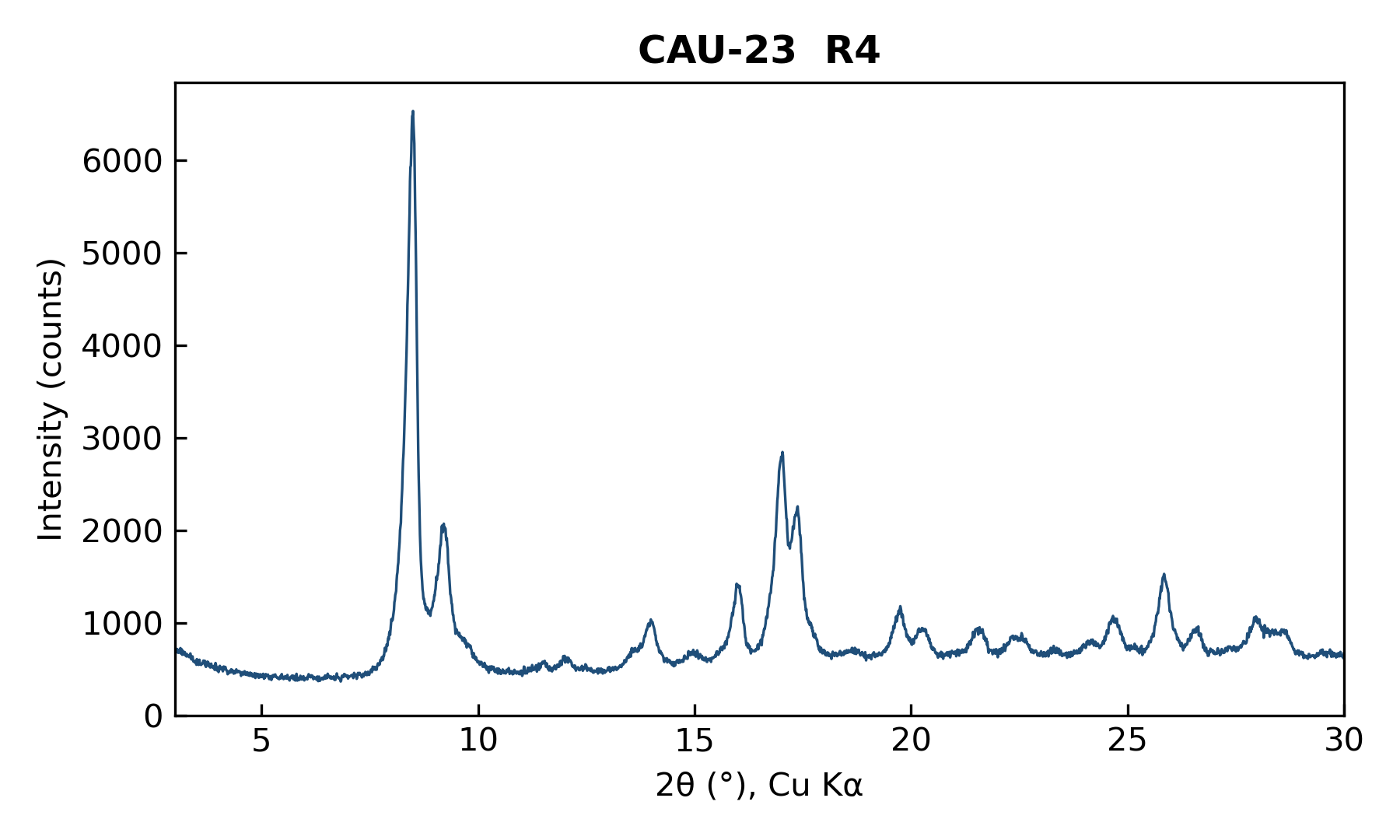}\hfill
\siFig[0.31\linewidth]{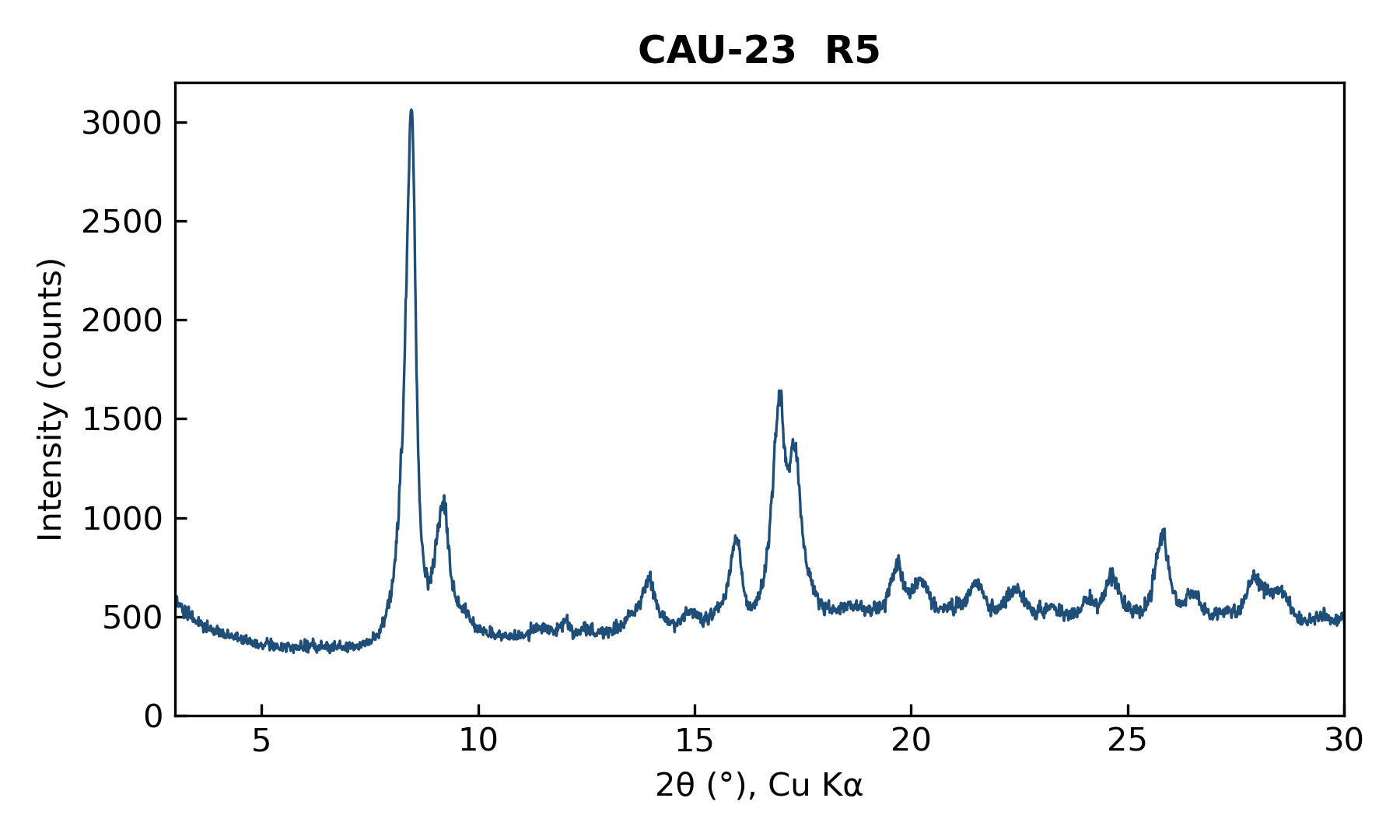}\hfill
\siFig[0.31\linewidth]{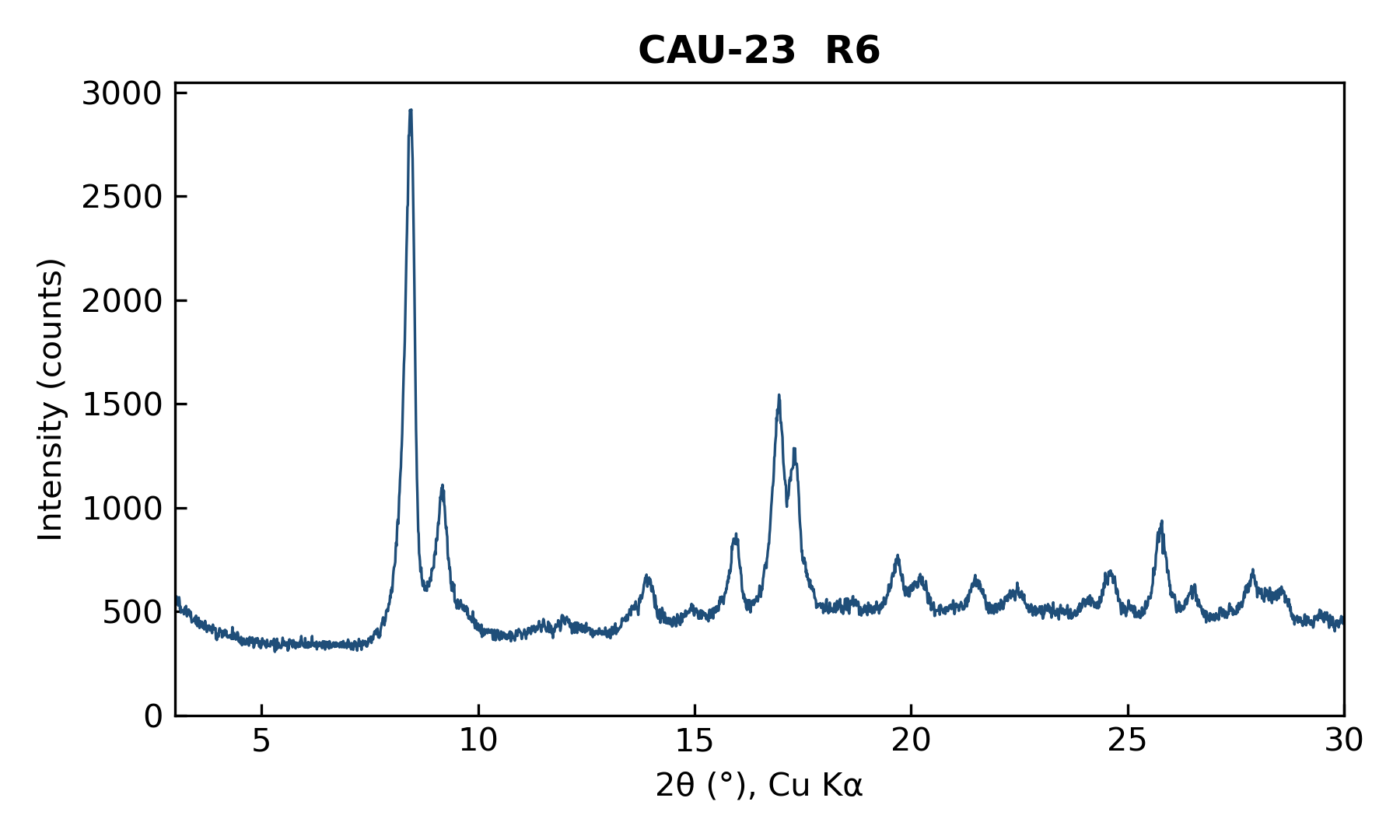}\\[2pt]
\siFig[0.31\linewidth]{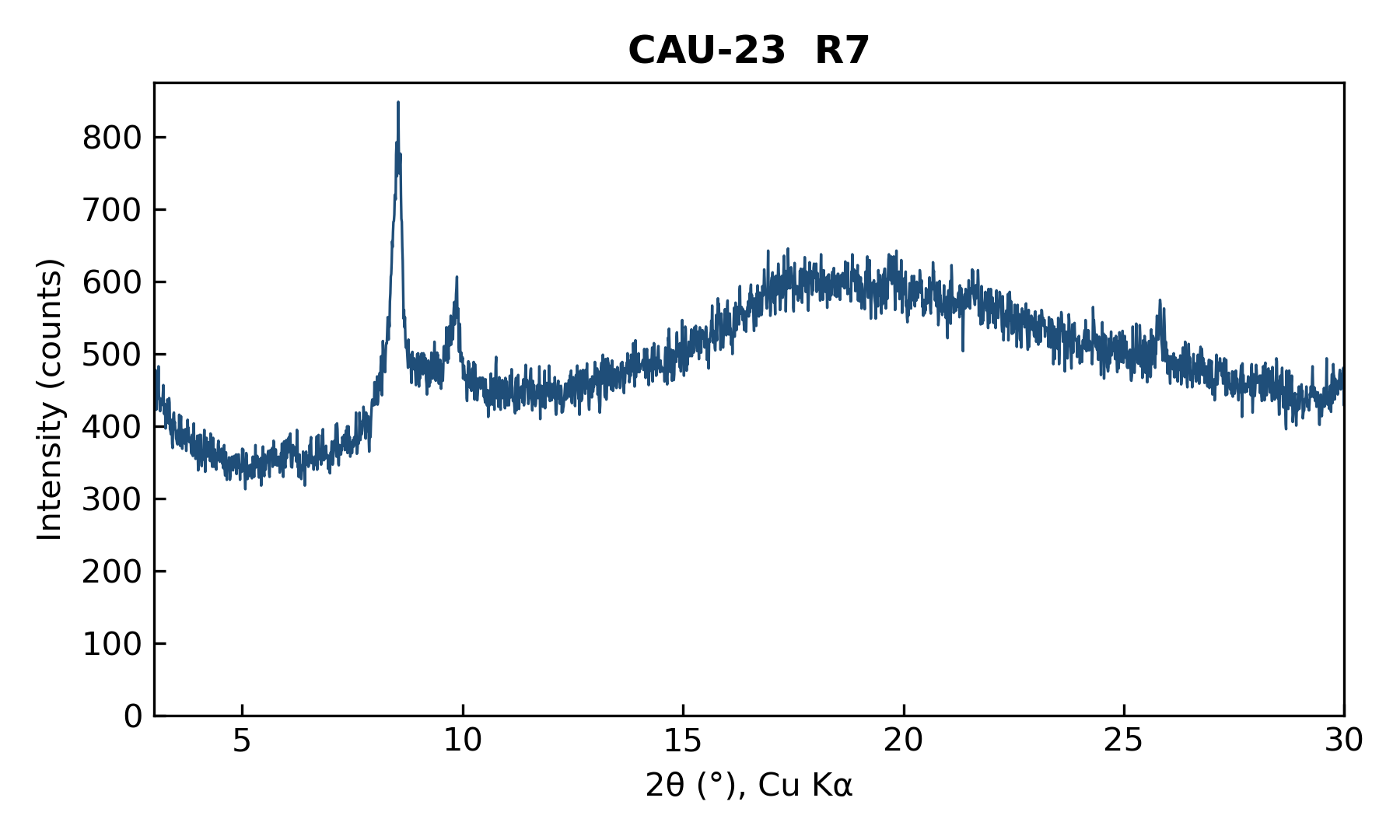}\hfill
\siFig[0.31\linewidth]{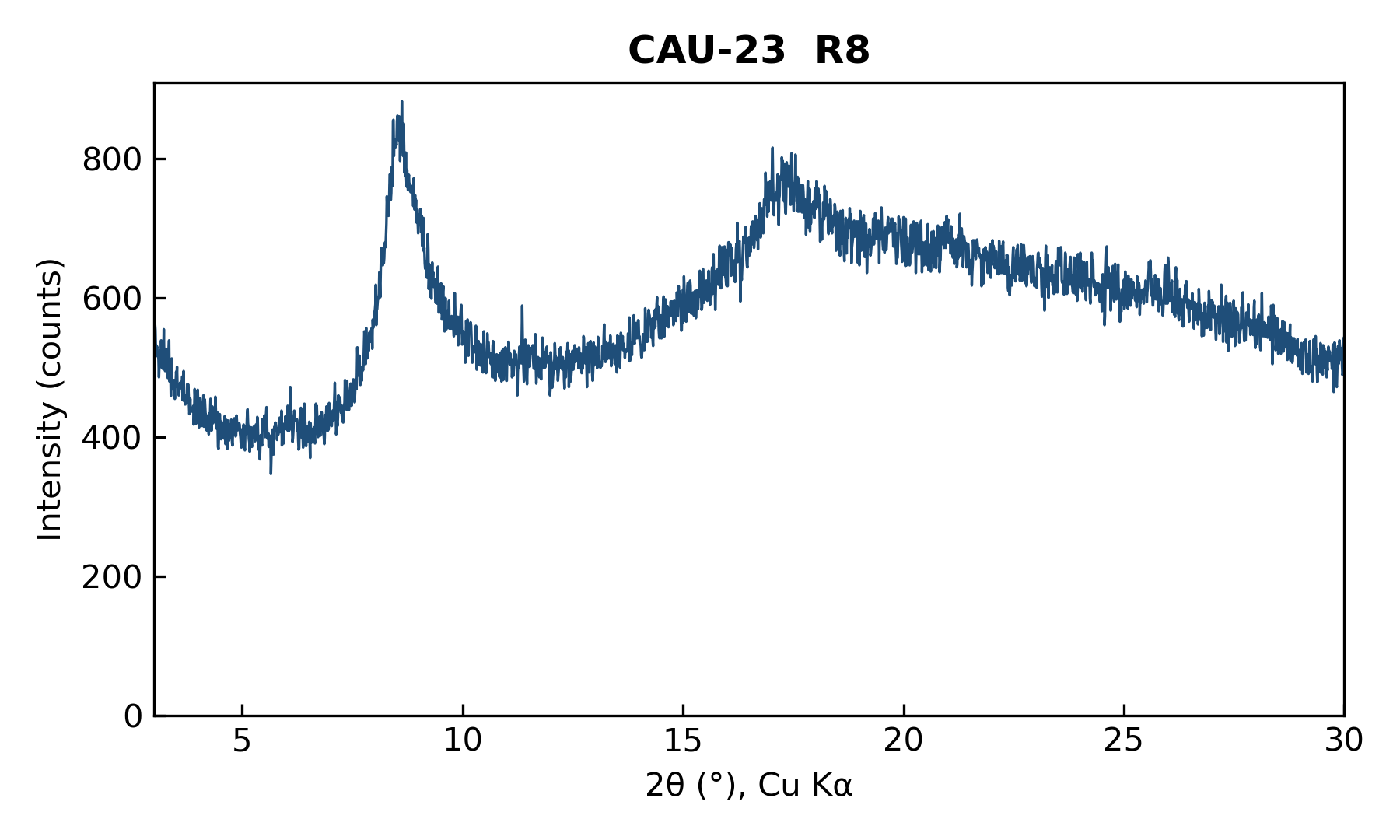}\hfill
\siFig[0.31\linewidth]{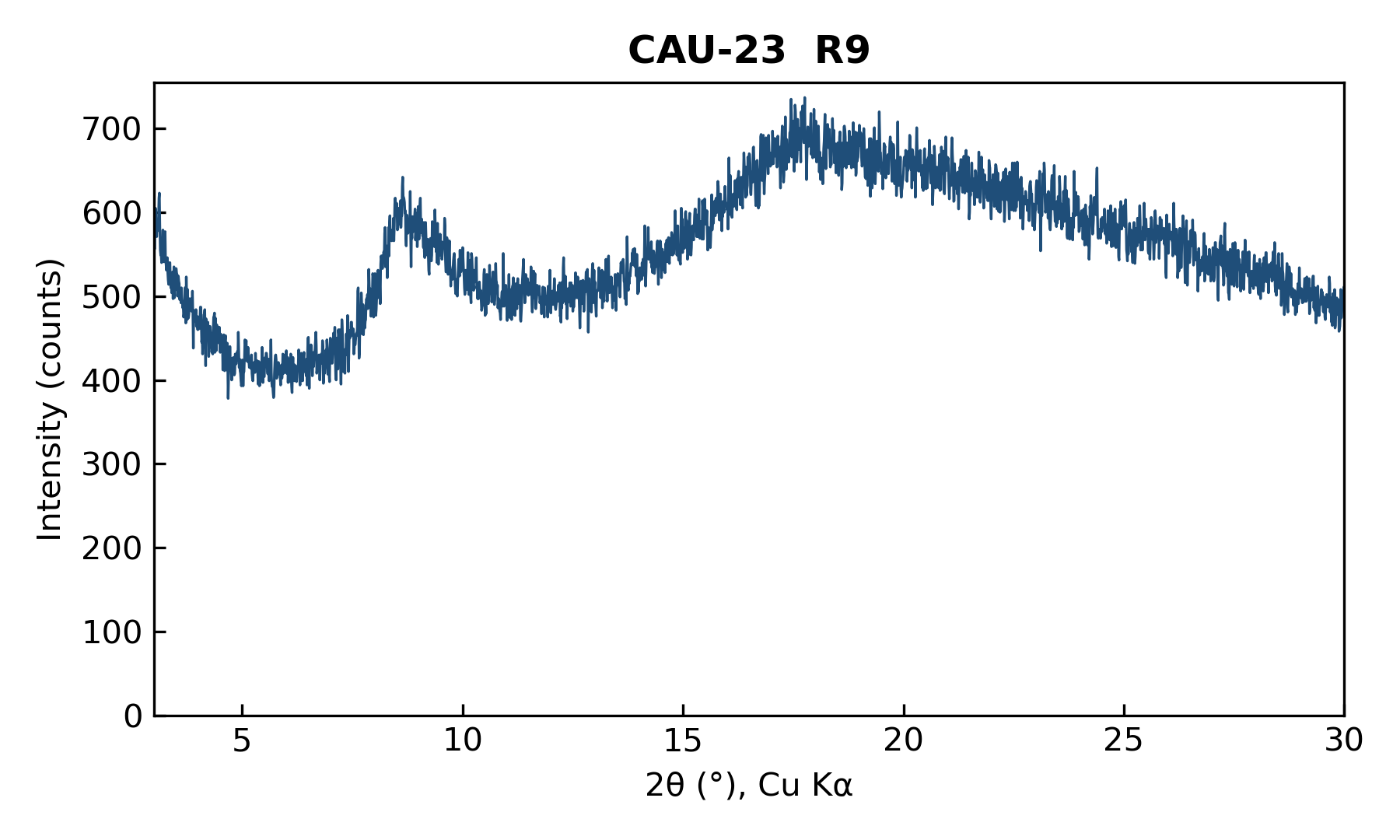}\\[2pt]
\siFig[0.31\linewidth]{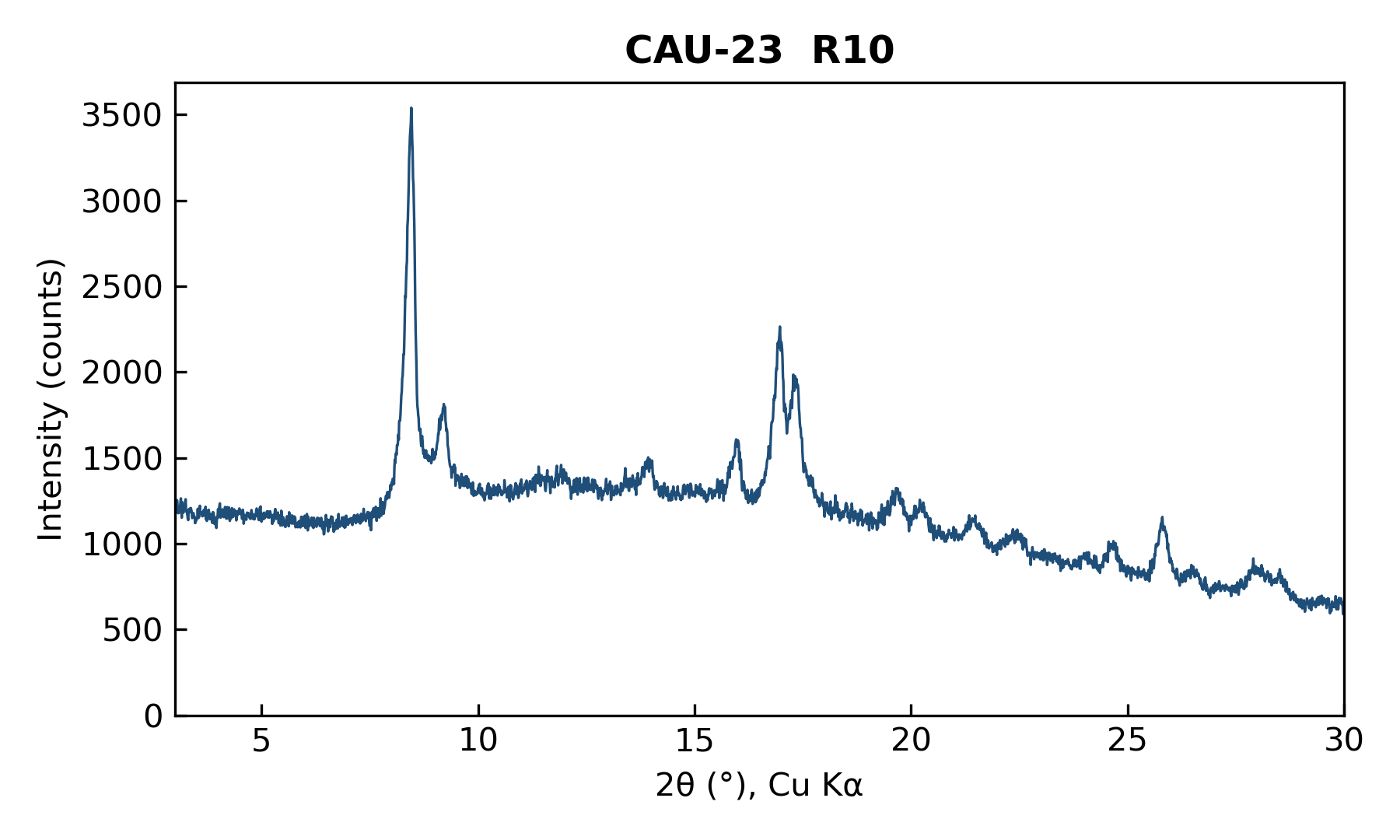}\hfill
\siFig[0.31\linewidth]{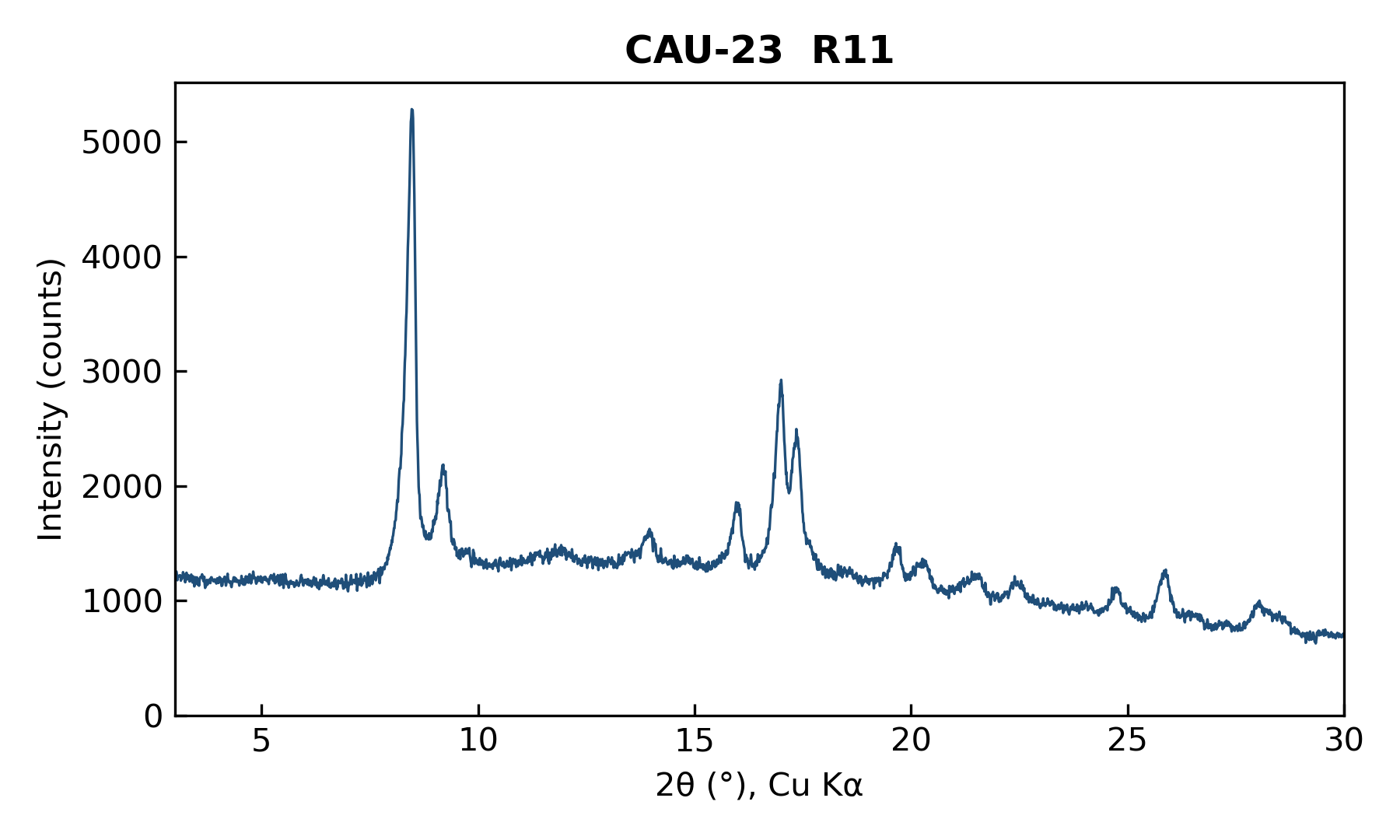}\hfill
\siFig[0.31\linewidth]{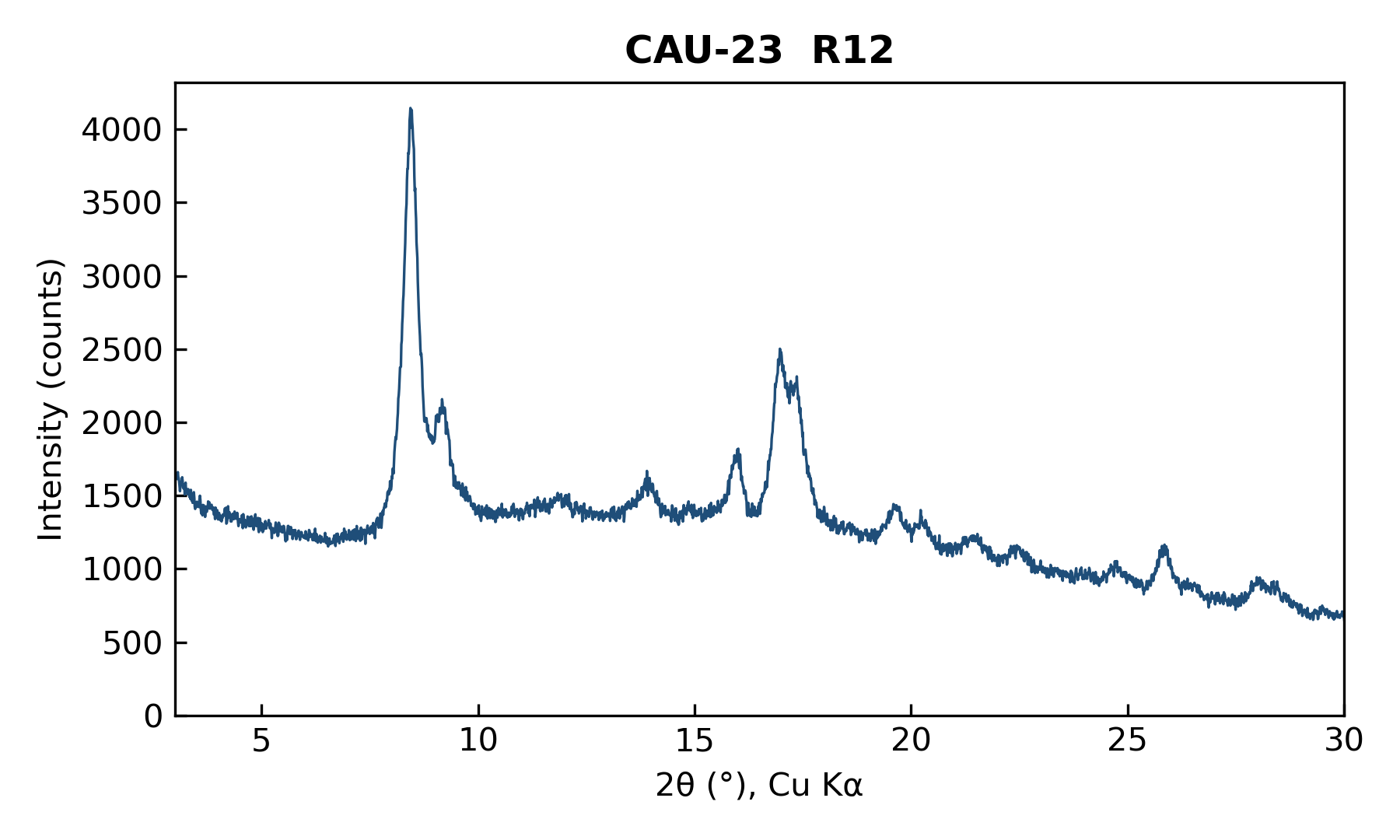}
\caption{PXRD patterns of CAU-23, R00001--R00012 (Table~\ref{tab:al-cau23}).}
\label{fig:pxrd-cau23}
\end{figure}

\clearpage
\paragraph{(ii) CAU-23.}\label{si:pxrd-cau23} Recipes and results in Table~\ref{tab:al-cau23}, patterns in Figure~\ref{fig:pxrd-cau23}.
\begin{table}[H]
\centering
\caption{Recipes and PXRD results of CAU-23.}
\label{tab:al-cau23}
\footnotesize
\setlength{\tabcolsep}{3.5pt}
\resizebox{\linewidth}{!}{%
\begin{tabular}{@{}lccrrrrrrrr@{}}
\toprule
 & & & \multicolumn{3}{c}{Stock solution / DI water ($\mu$L)} & & & & & \\
\cmidrule(lr){4-6}
Exp. & Round & Encoded $\mathbf{x}$ & \ce{AlCl3.6H2O} & NaOH & DI \ce{H2O} & Time (min) & $T$ ($^{\circ}$C) & $I$ & FWHM ($^{\circ}$) & $q$ \\
\midrule
\textbf{R00001} & \textbf{0} & \textbf{(18, 7, 17, 10, 21)} & \textbf{180} & \textbf{70} & \textbf{170} & \textbf{50} & \textbf{105} & \textbf{8{,}430} & \textbf{0.379} & \textbf{22{,}254} \\
R00002 & 0 & (23, 6, 15, 10, 20) & 230 & 60 & 150 & 50 & 100 & 3{,}551 & 0.607 & 5{,}846 \\
R00003$^{\dagger}$ & 0 & (20, 7, 17, 12, 29) & 200 & 70 & 170 & 60 & 145 & 4{,}651 & 0.603 & 7{,}711 \\
R00004 & 1 & (16, 7, 18, 10, 21) & 160 & 70 & 180 & 50 & 105 & 6{,}533 & 0.301 & 21{,}668 \\
R00005 & 1 & (19, 7, 19, 10, 21) & 190 & 70 & 190 & 50 & 105 & 3{,}064 & 0.329 & 9{,}313 \\
R00006 & 1 & (18, 7, 19, 10, 22) & 180 & 70 & 190 & 50 & 110 & 2{,}918 & 0.304 & 9{,}583 \\
R00007 & 2 & (21, 7, 15, 11, 12) & 210 & 70 & 150 & 55 & 60 & 849 & 2.64 & 321 \\
R00008 & 2 & (22, 7, 12, 10, 13) & 220 & 70 & 120 & 50 & 65 & 883 & 7.57 & 117 \\
R00009 & 2 & (25, 9, 14, 10, 13) & 250 & 90 & 140 & 50 & 65 & 0 & n.d. & 0 \\
R00010 & 3 & (15, 6, 16, 10, 22) & 150 & 60 & 160 & 50 & 110 & 3{,}540 & 0.401 & 8{,}830 \\
R00011 & 3 & (14, 7, 16, 10, 21) & 140 & 70 & 160 & 50 & 105 & 5{,}283 & 0.317 & 16{,}671 \\
R00012 & 3 & (17, 6, 17, 11, 20) & 170 & 60 & 170 & 55 & 100 & 4{,}145 & 0.598 & 6{,}934 \\
\bottomrule
\end{tabular}}
\vspace{2pt}\noindent\footnotesize $^{\dagger}$Synthesized with 170~$\mu$L instead of the proposed 190~$\mu$L DI water (pipetting error); the condition actually used is listed and was the one given to the model.
\end{table}

\begin{figure}[H]
\centering
\siFig[0.31\linewidth]{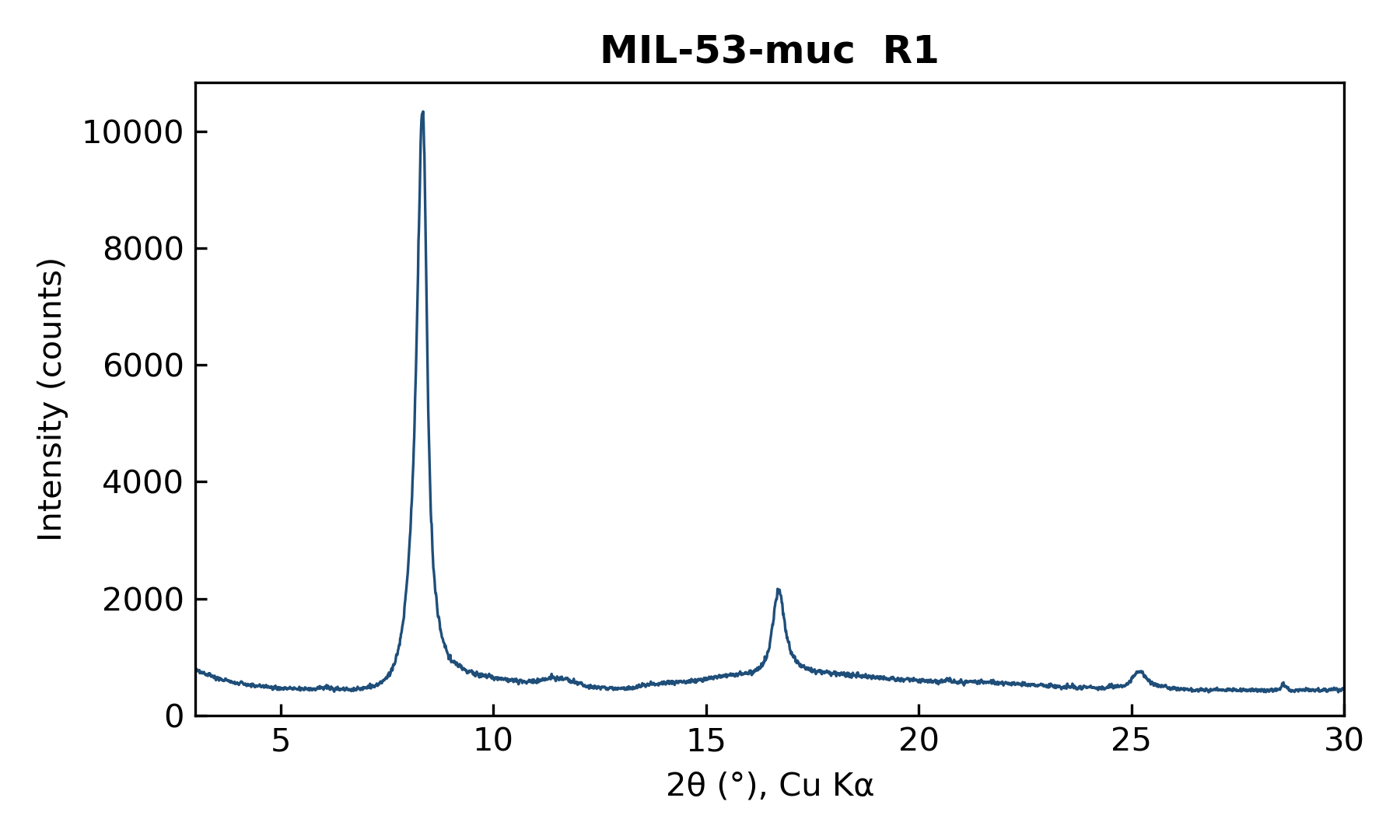}\hfill
\siFig[0.31\linewidth]{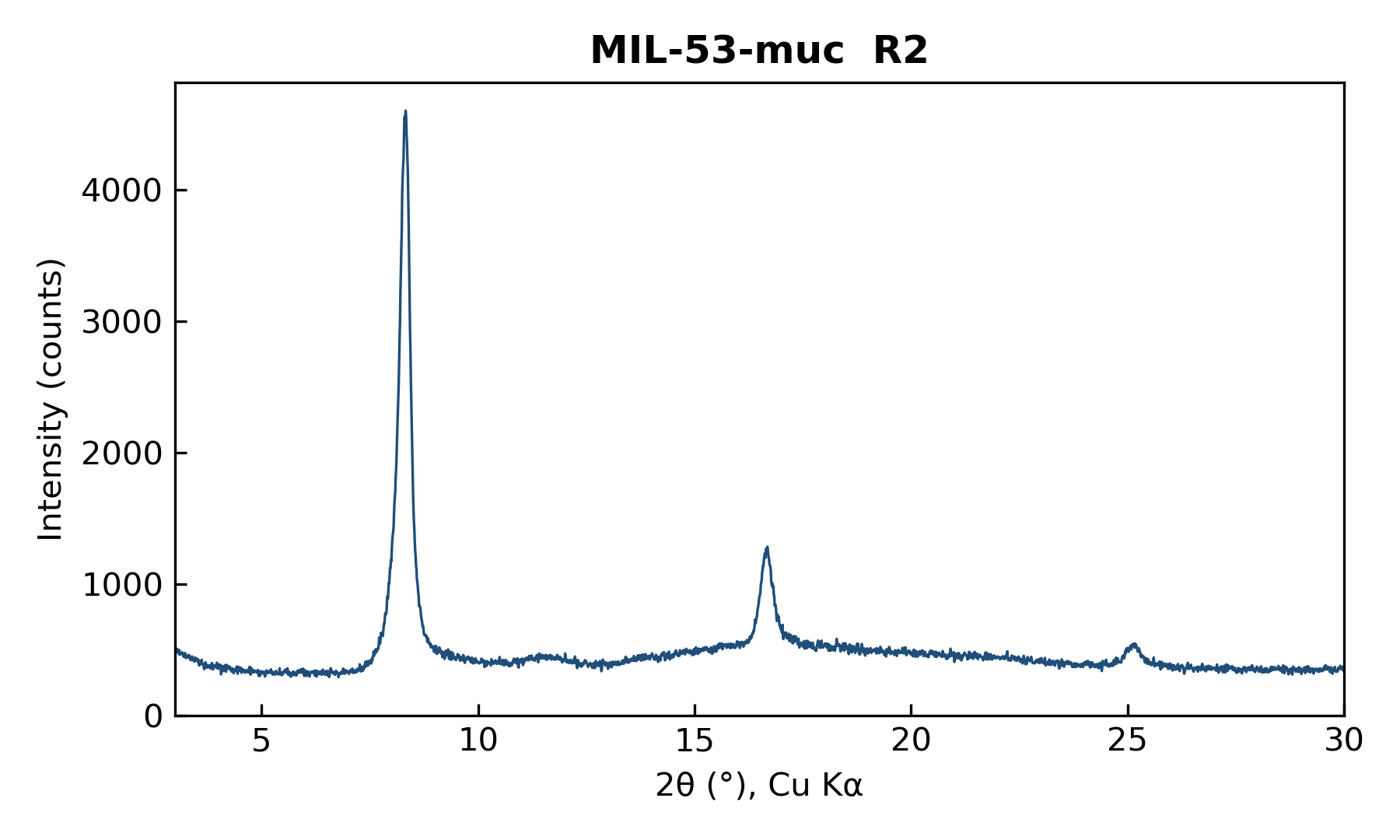}\hfill
\siFig[0.31\linewidth]{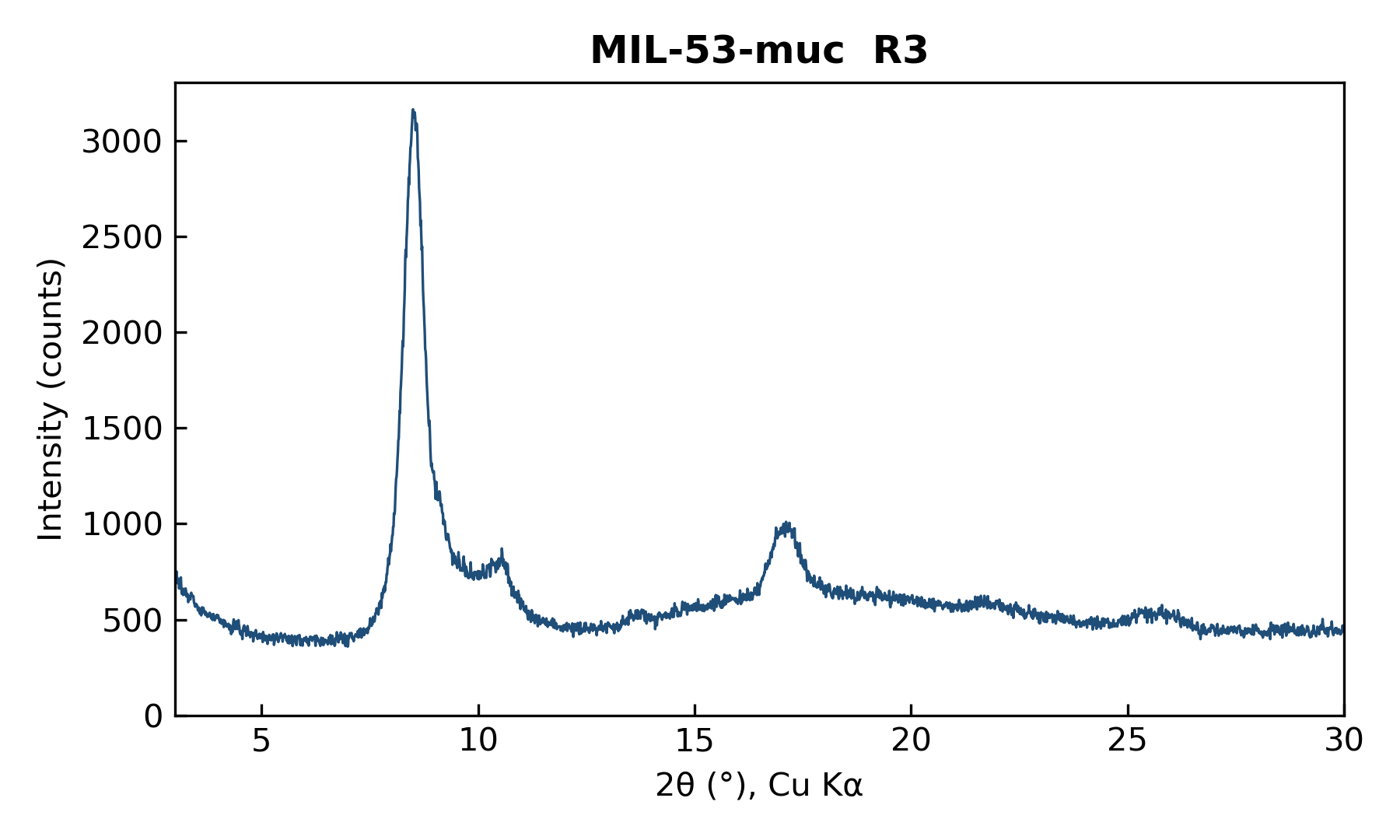}\\[2pt]
\siFig[0.31\linewidth]{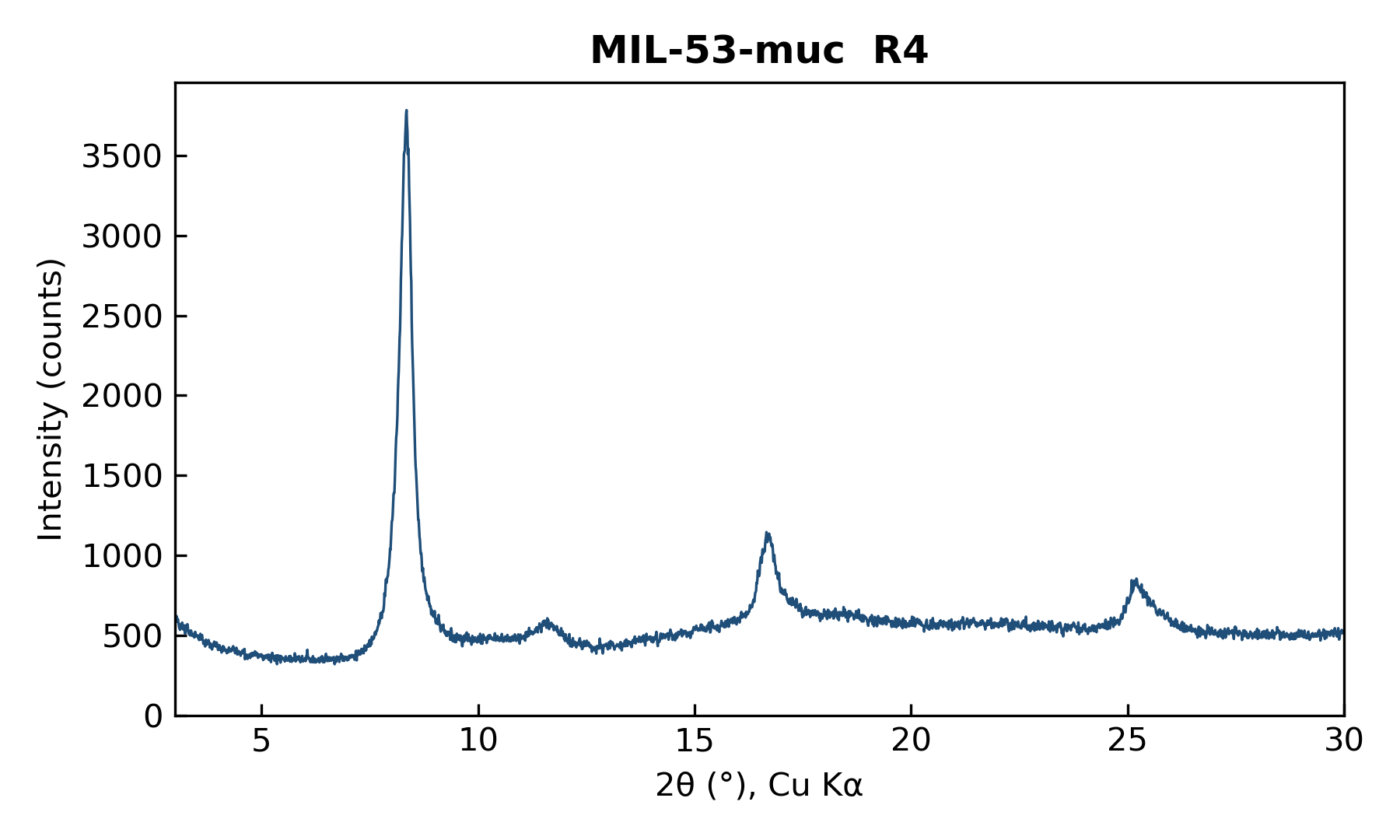}\hfill
\siFig[0.31\linewidth]{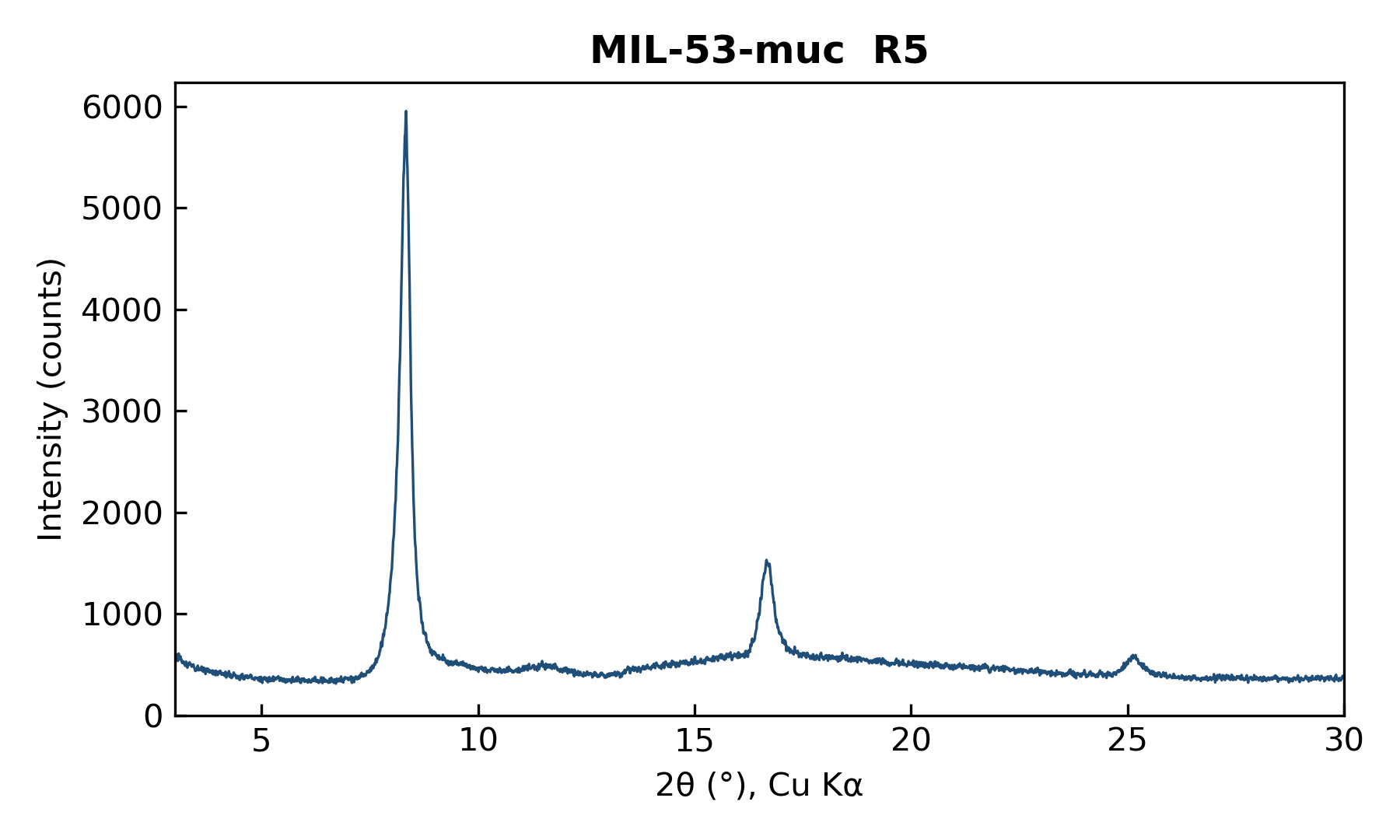}\hfill
\siFig[0.31\linewidth]{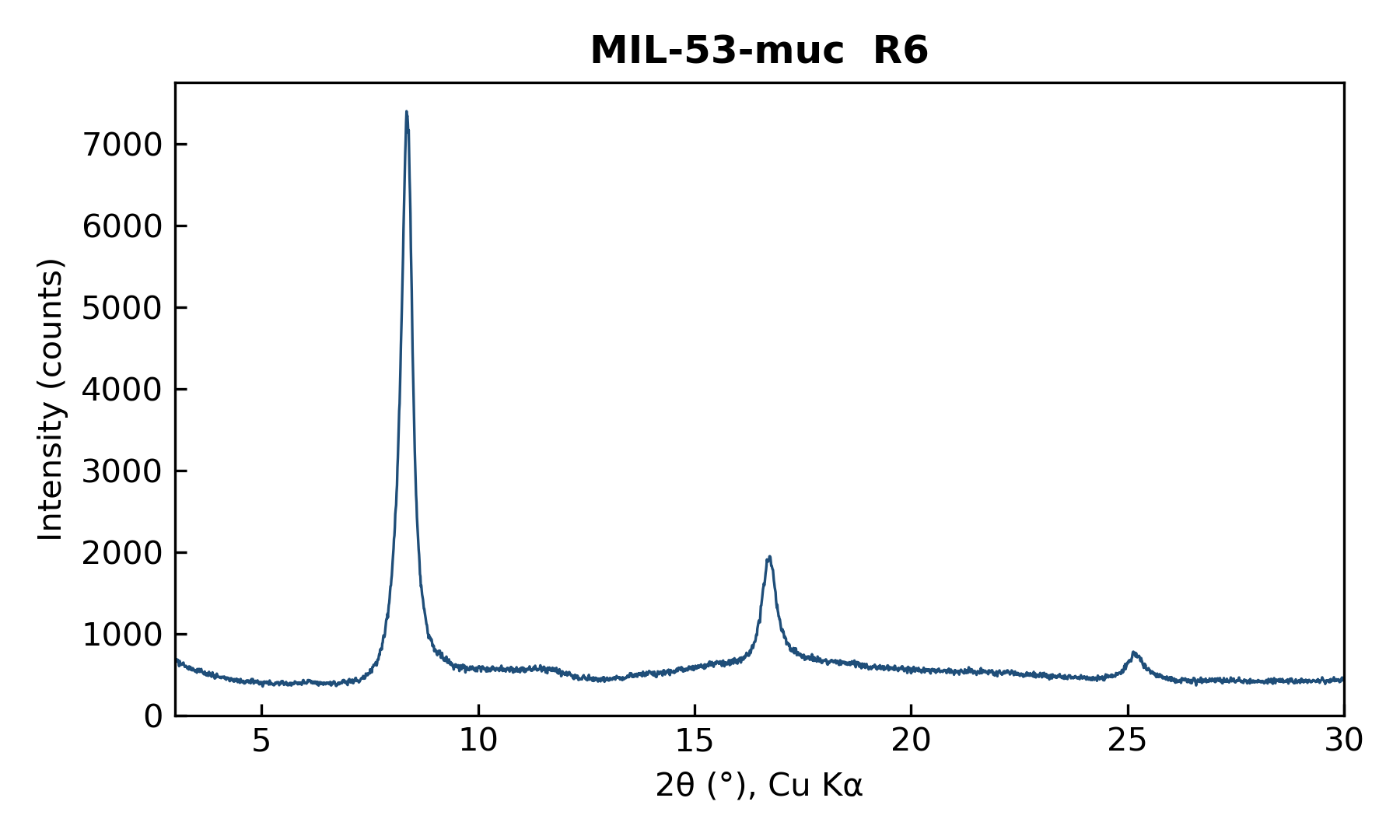}\\[2pt]
\siFig[0.31\linewidth]{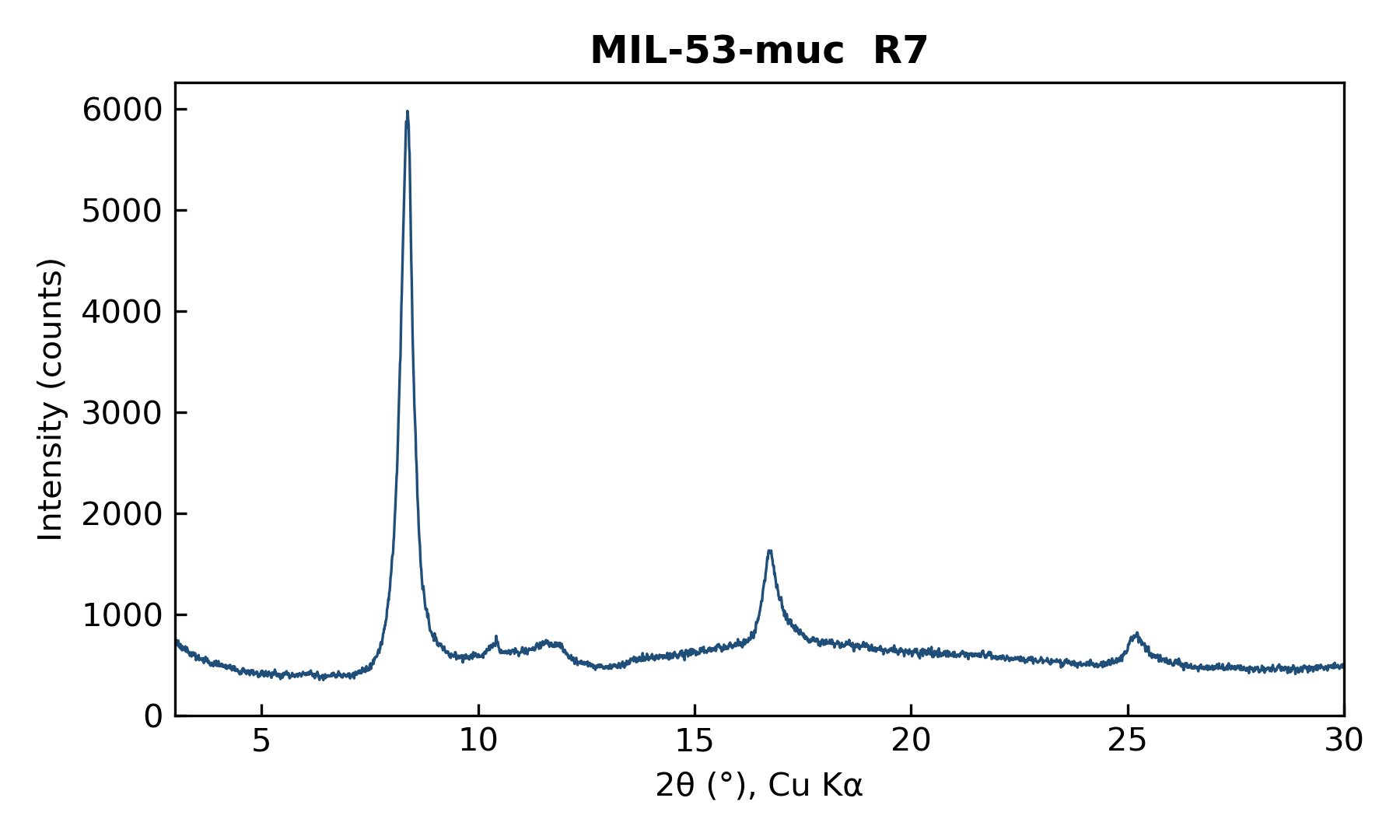}\hfill
\siFig[0.31\linewidth]{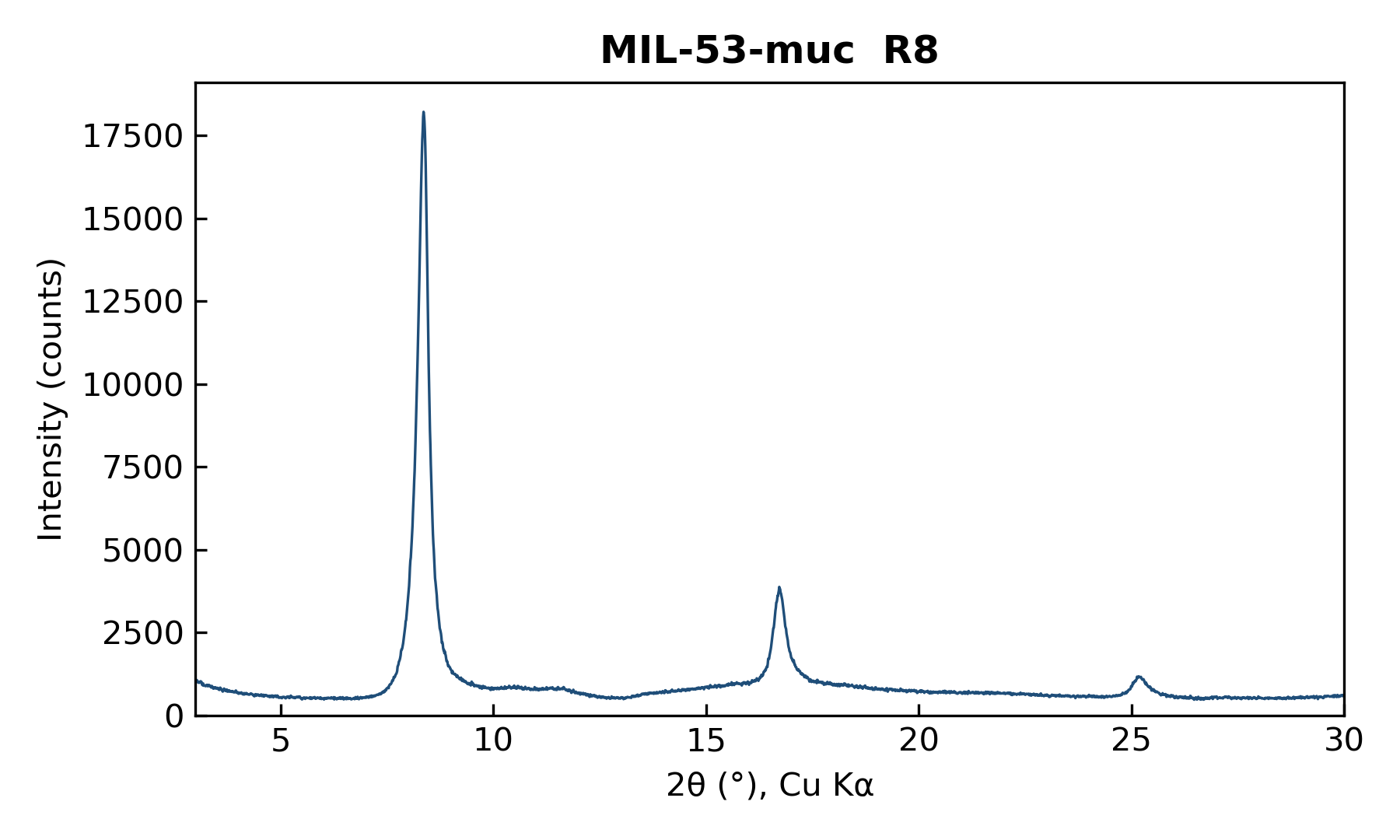}\hfill
\siFig[0.31\linewidth]{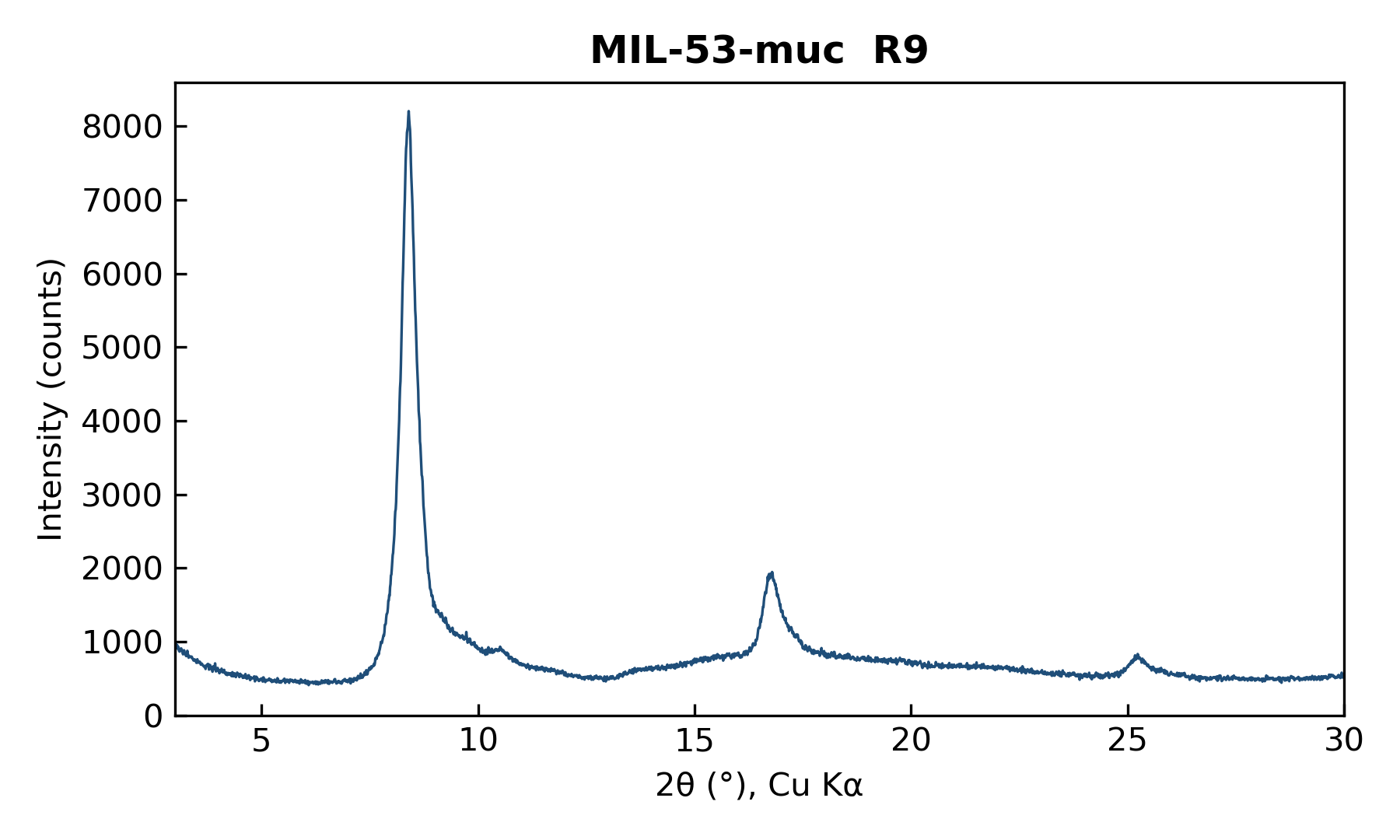}\\[2pt]
\siFig[0.31\linewidth]{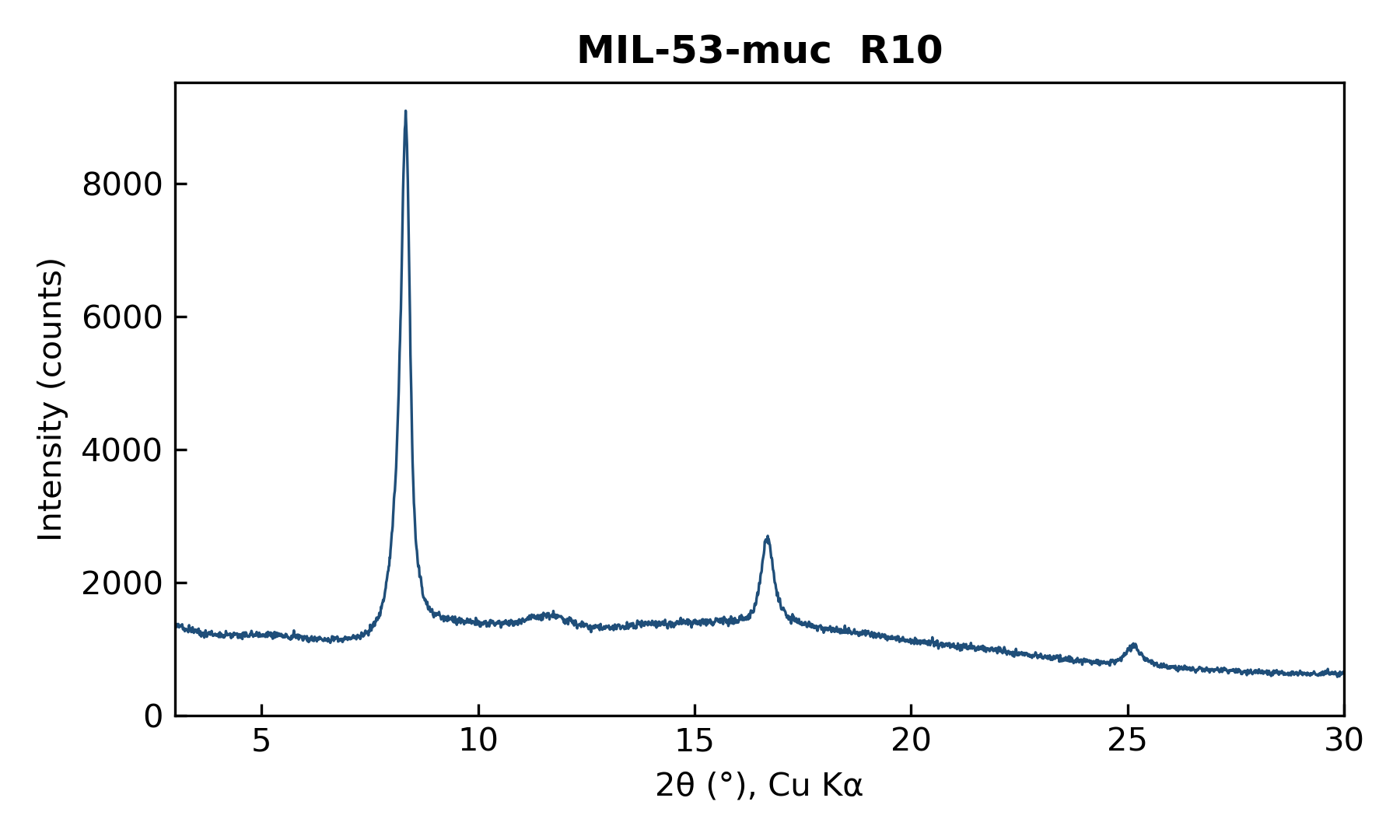}\hfill
\siFig[0.31\linewidth]{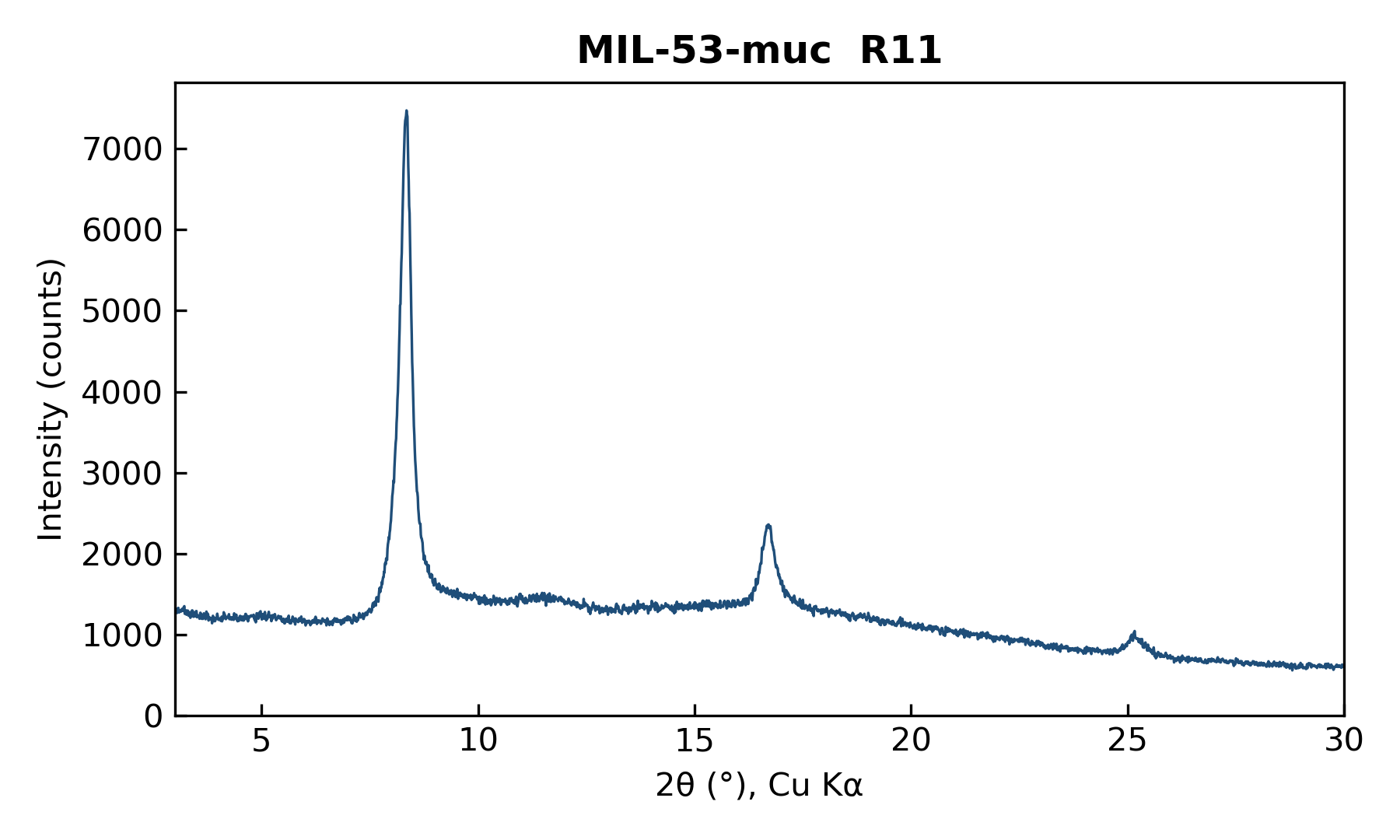}\hfill
\siFig[0.31\linewidth]{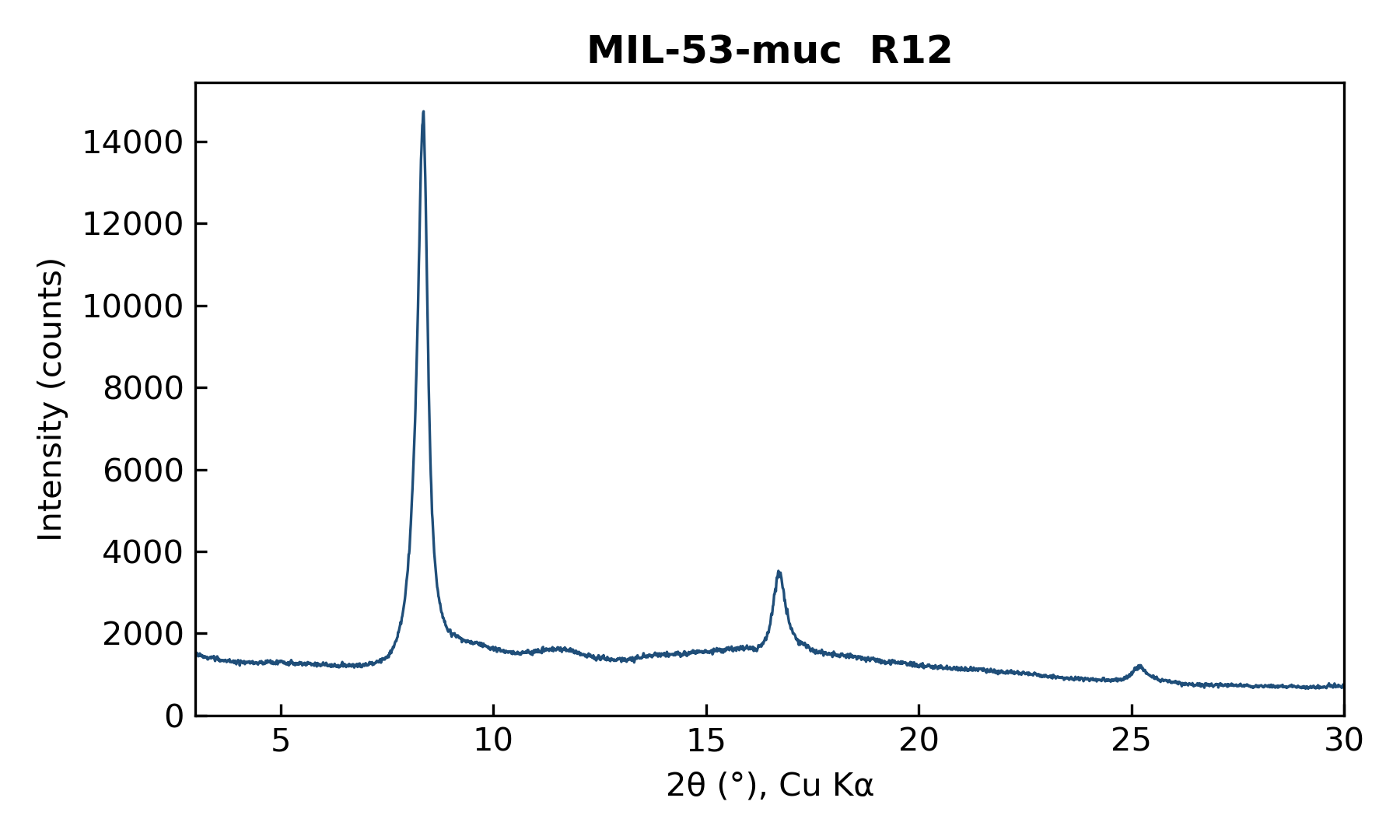}
\caption{PXRD patterns of MIL-53-muc, R00001--R00012 (Table~\ref{tab:al-mil53}).}
\label{fig:pxrd-mil53}
\end{figure}

\clearpage
\paragraph{(iii) MIL-53-muc.}\label{si:pxrd-mil53} Recipes and results in Table~\ref{tab:al-mil53}, patterns in Figure~\ref{fig:pxrd-mil53}.
\begin{table}[H]
\centering
\caption{Recipes and PXRD results of MIL-53-muc.}
\label{tab:al-mil53}
\footnotesize
\setlength{\tabcolsep}{3.5pt}
\resizebox{\linewidth}{!}{%
\begin{tabular}{@{}lccrrrrrrrr@{}}
\toprule
 & & & \multicolumn{3}{c}{Stock solution / DI water ($\mu$L)} & & & & & \\
\cmidrule(lr){4-6}
Exp. & Round & Encoded $\mathbf{x}$ & \ce{AlCl3.6H2O} & NaOH & DI \ce{H2O} & Time (min) & $T$ ($^{\circ}$C) & $I$ & FWHM ($^{\circ}$) & $q$ \\
\midrule
R00001 & 0 & (18, 7, 17, 10, 21) & 180 & 70 & 170 & 50 & 105 & 10{,}339 & 0.318 & 32{,}523 \\
R00002 & 0 & (23, 6, 15, 10, 20) & 230 & 60 & 150 & 50 & 100 & 4{,}600 & 0.315 & 14{,}599 \\
R00003 & 0 & (20, 7, 19, 12, 29) & 200 & 70 & 190 & 60 & 145 & 3{,}163 & 0.651 & 4{,}860 \\
R00004 & 1 & (19, 6, 19, 10, 21) & 190 & 60 & 190 & 50 & 105 & 3{,}784 & 0.382 & 9{,}895 \\
R00005 & 1 & (18, 6, 15, 10, 22) & 180 & 60 & 150 & 50 & 110 & 5{,}953 & 0.3 & 19{,}857 \\
R00006 & 1 & (16, 6, 16, 10, 20) & 160 & 60 & 160 & 50 & 100 & 7{,}402 & 0.343 & 21{,}580 \\
R00007 & 2 & (19, 6, 16, 10, 20) & 190 & 60 & 160 & 50 & 100 & 5{,}981 & 0.369 & 16{,}209 \\
\textbf{R00008} & \textbf{2} & \textbf{(16, 6, 16, 10, 21)} & \textbf{160} & \textbf{60} & \textbf{160} & \textbf{50} & \textbf{105} & \textbf{18{,}209} & \textbf{0.314} & \textbf{58{,}064} \\
R00009 & 2 & (18, 6, 16, 11, 22) & 180 & 60 & 160 & 55 & 110 & 8{,}206 & 0.451 & 18{,}207 \\
R00010 & 3 & (20, 6, 17, 11, 20) & 200 & 60 & 170 & 55 & 100 & 9{,}091 & 0.314 & 28{,}989 \\
R00011 & 3 & (14, 6, 17, 10, 21) & 140 & 60 & 170 & 50 & 105 & 7{,}474 & 0.38 & 19{,}679 \\
R00012 & 3 & (22, 6, 16, 11, 21) & 220 & 60 & 160 & 55 & 105 & 14{,}736 & 0.329 & 44{,}818 \\
\bottomrule
\end{tabular}}
\end{table}

\begin{figure}[H]
\centering
\siFig[0.31\linewidth]{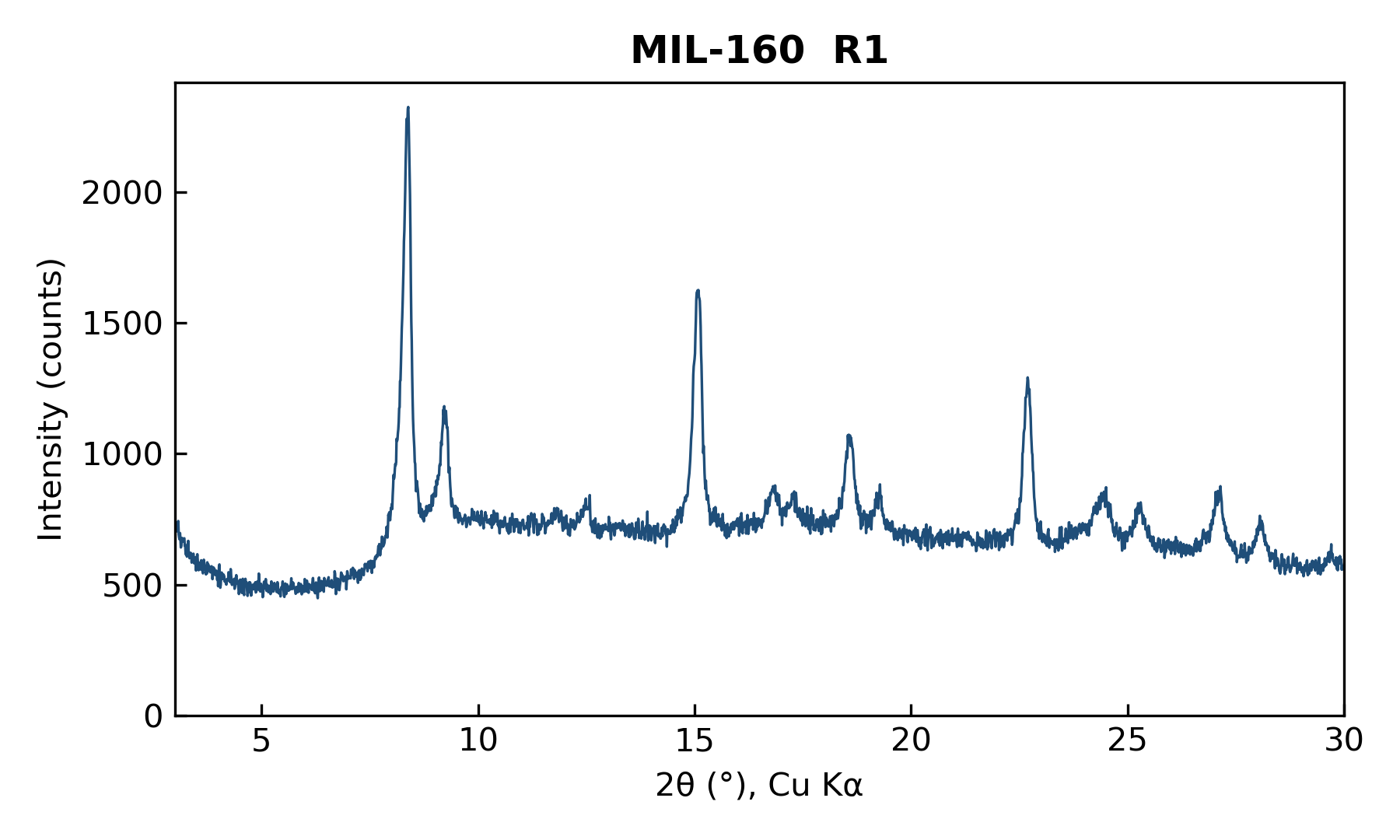}\hfill
\siFig[0.31\linewidth]{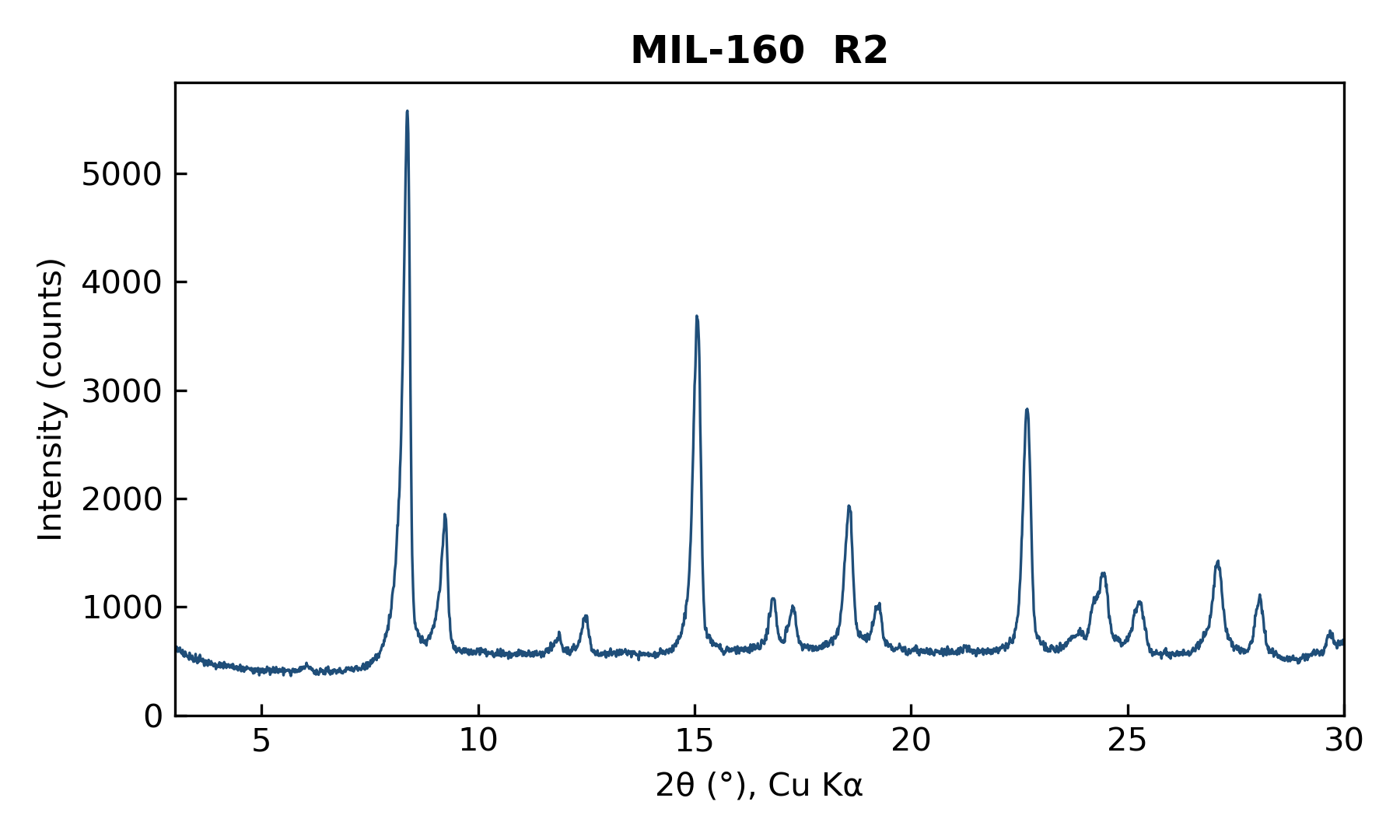}\hfill
\siFig[0.31\linewidth]{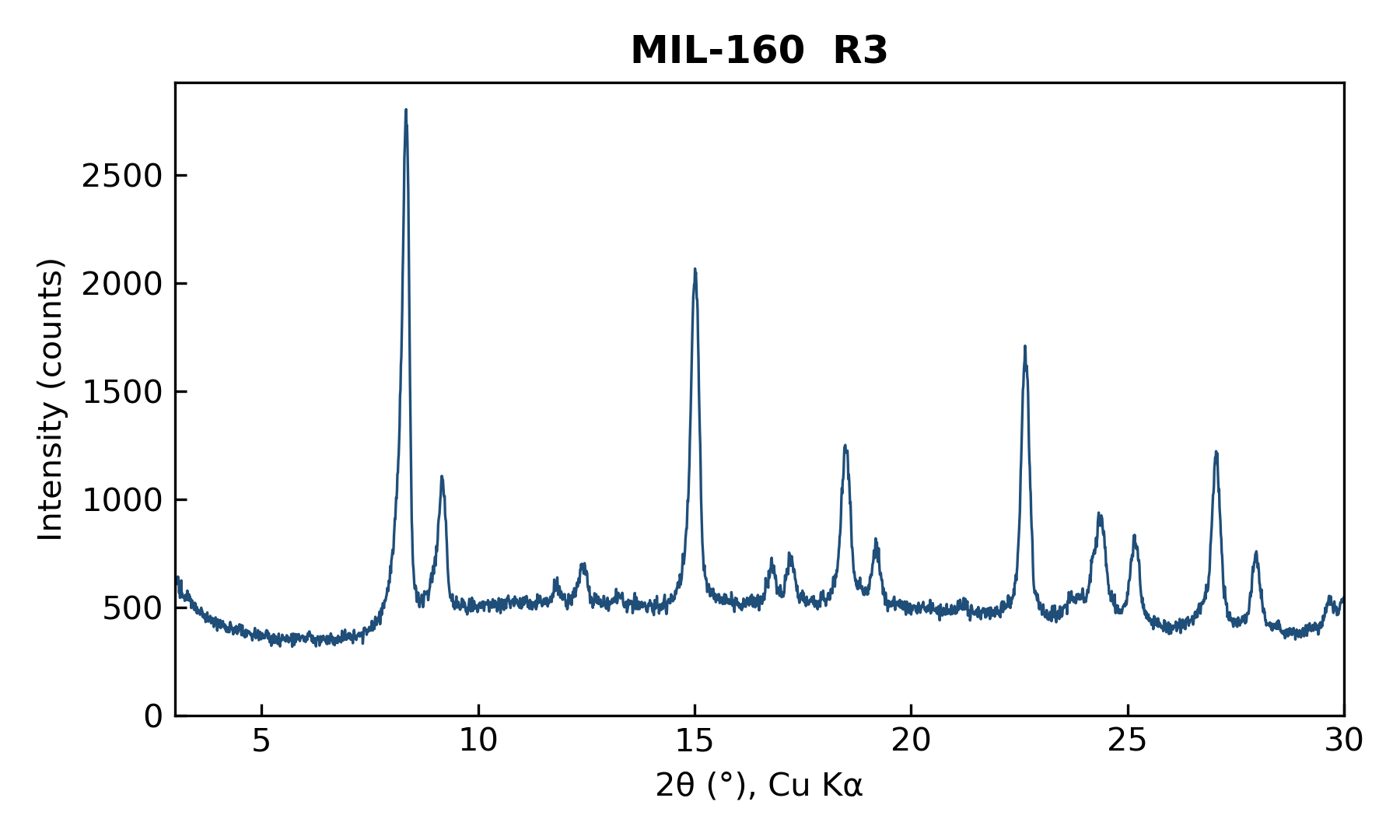}\\[2pt]
\siFig[0.31\linewidth]{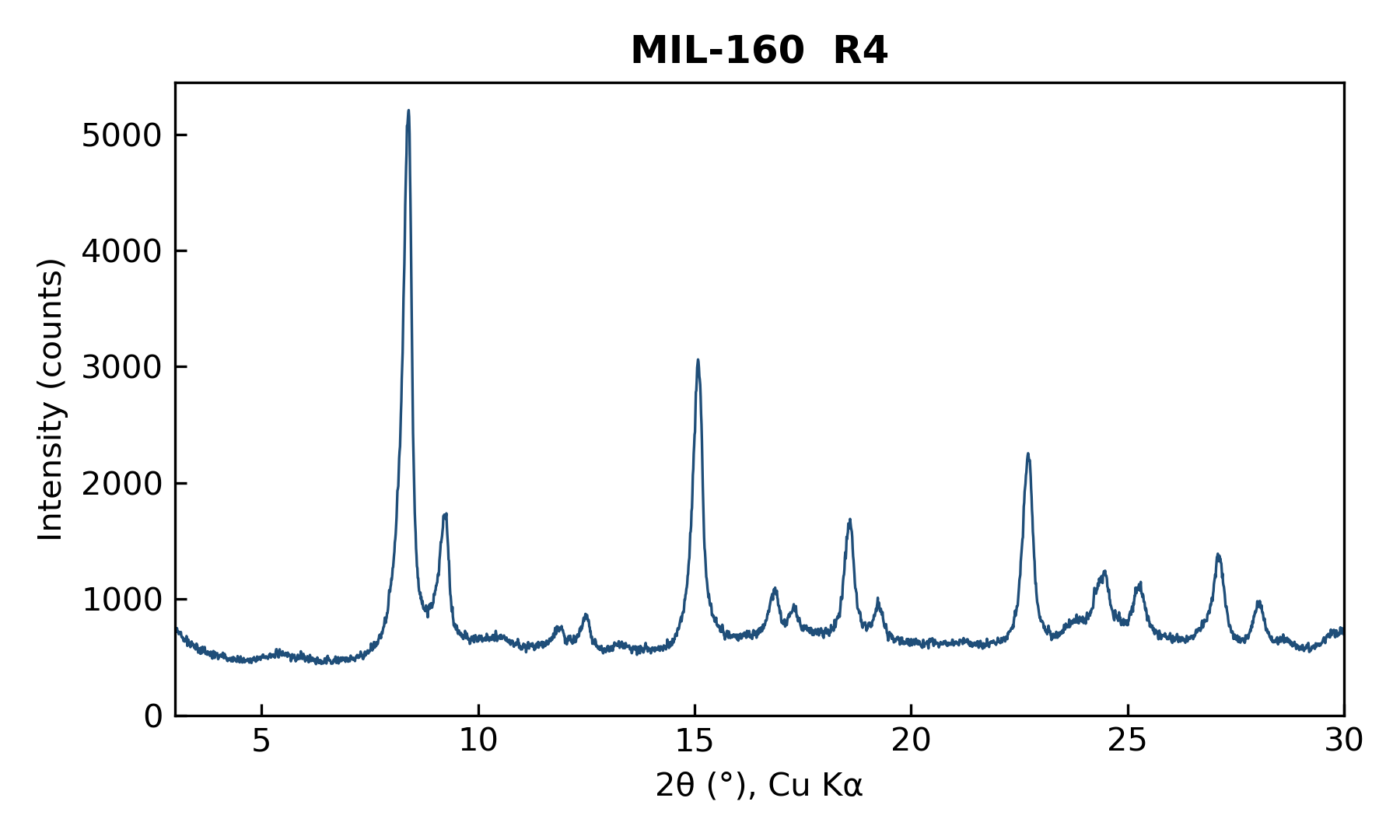}\hfill
\siFig[0.31\linewidth]{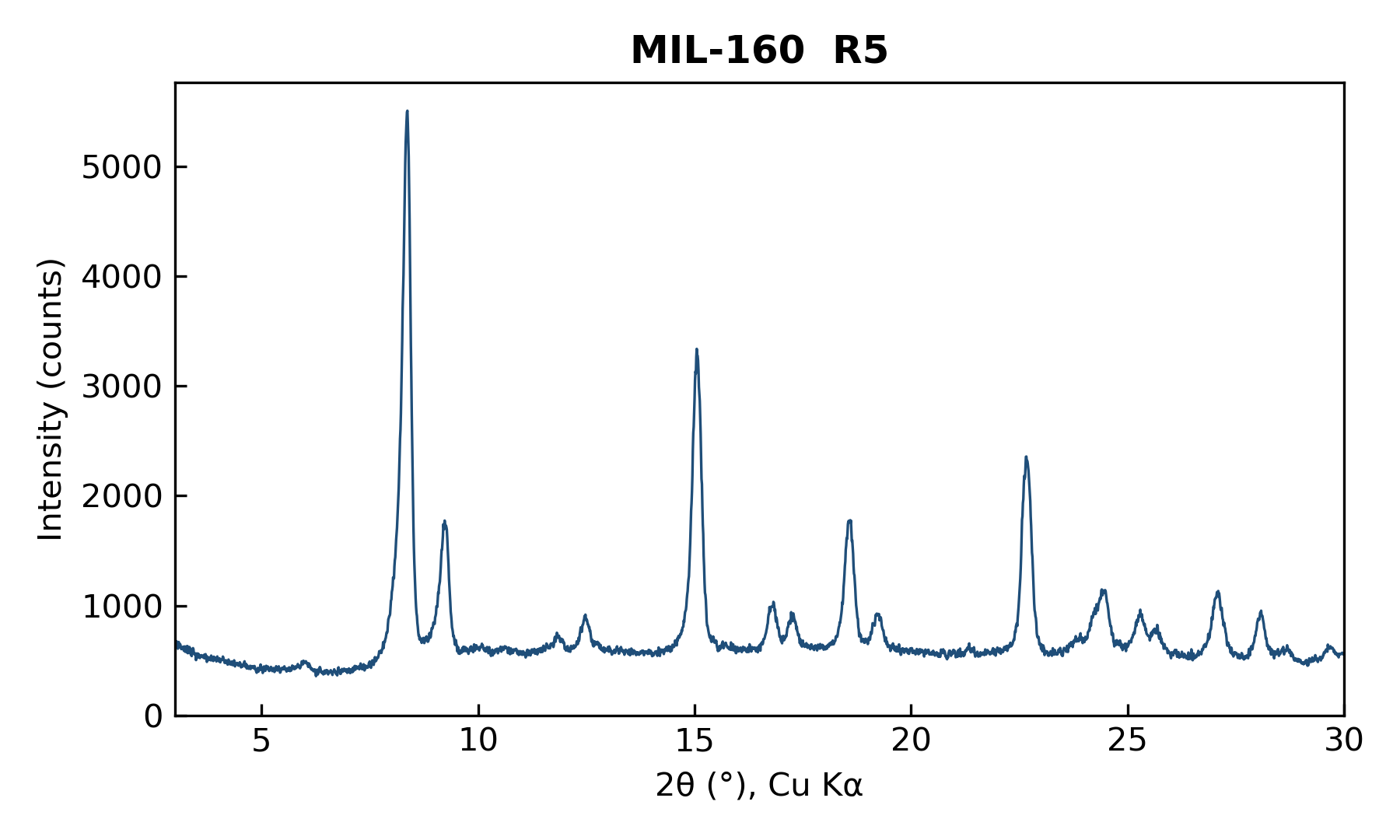}\hfill
\siFig[0.31\linewidth]{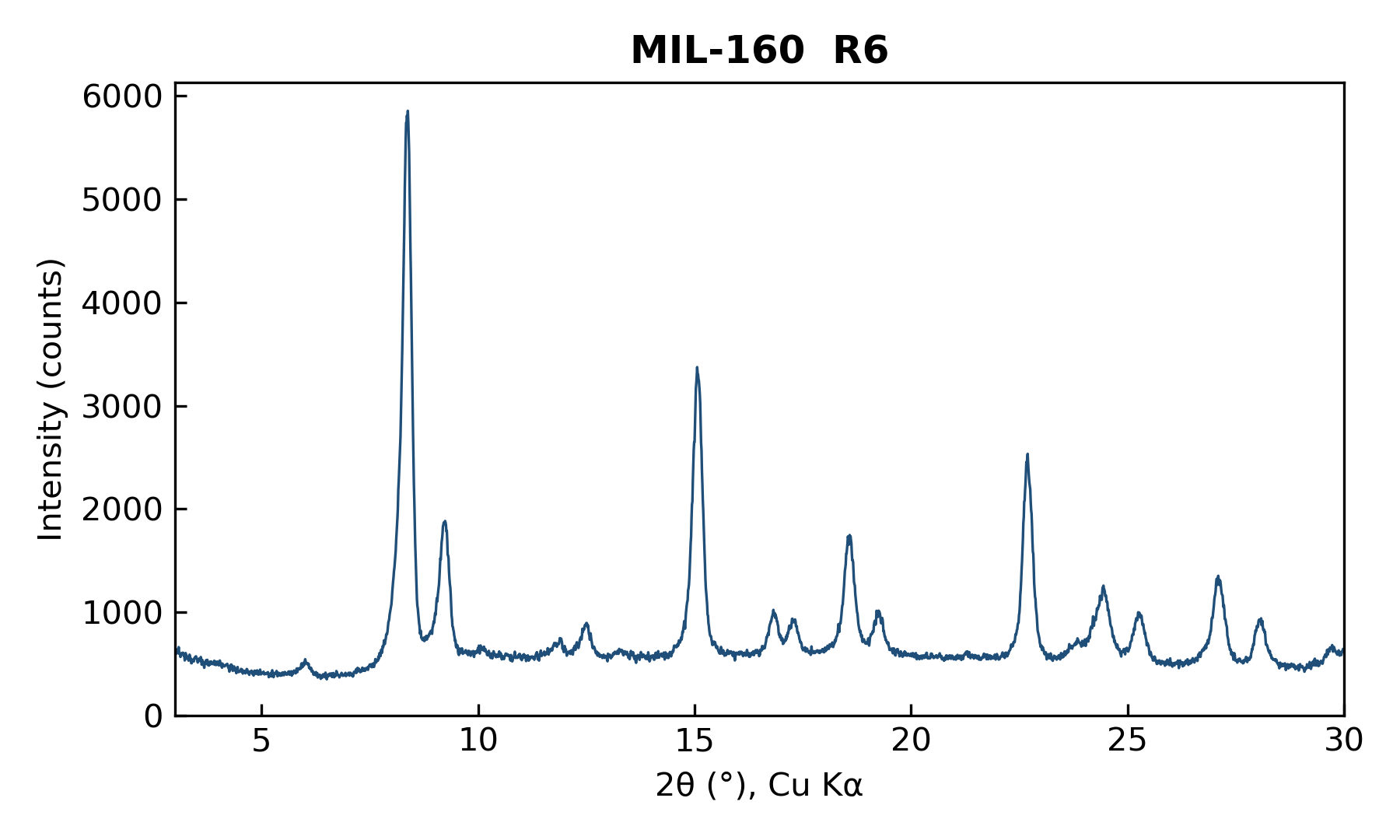}\\[2pt]
\siFig[0.31\linewidth]{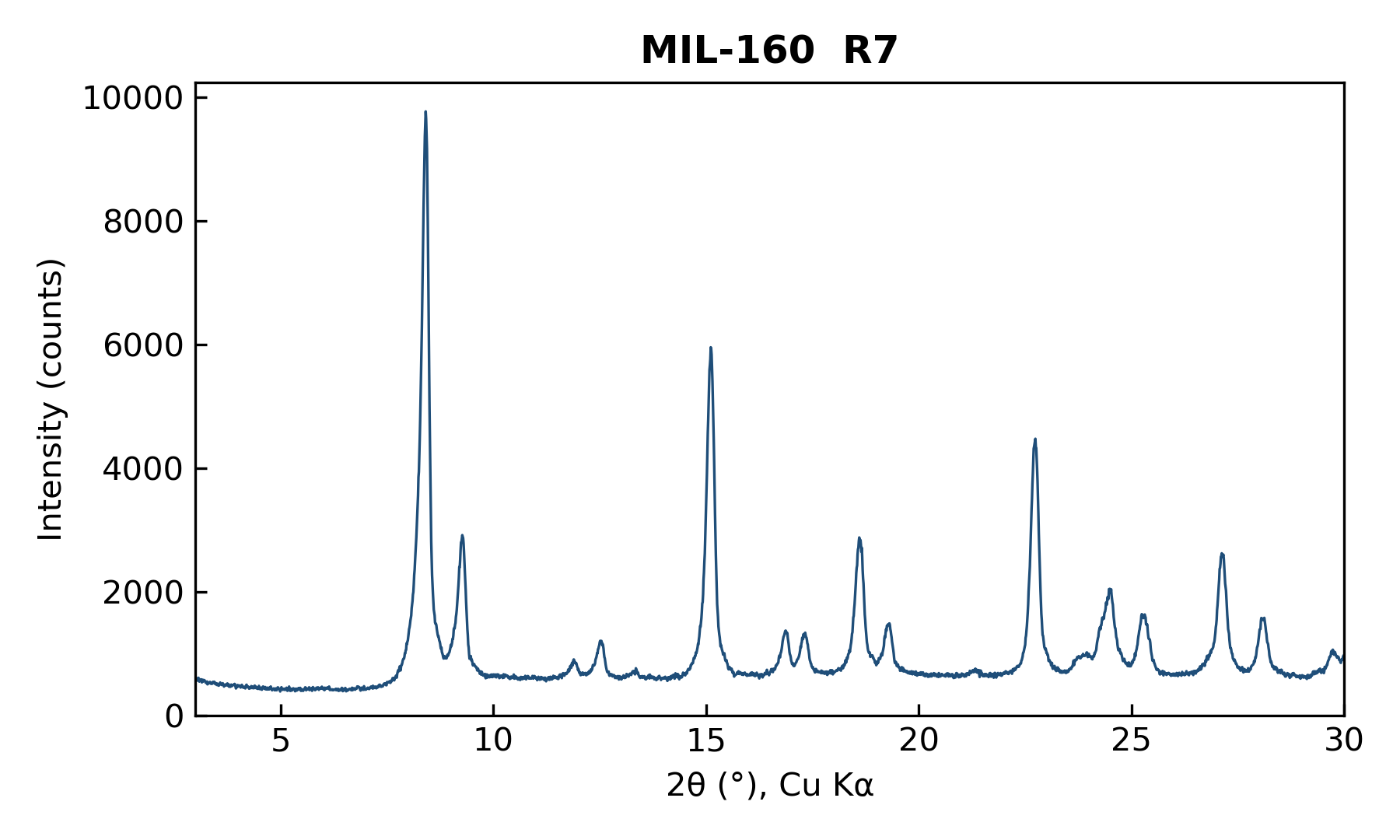}\hfill
\siFig[0.31\linewidth]{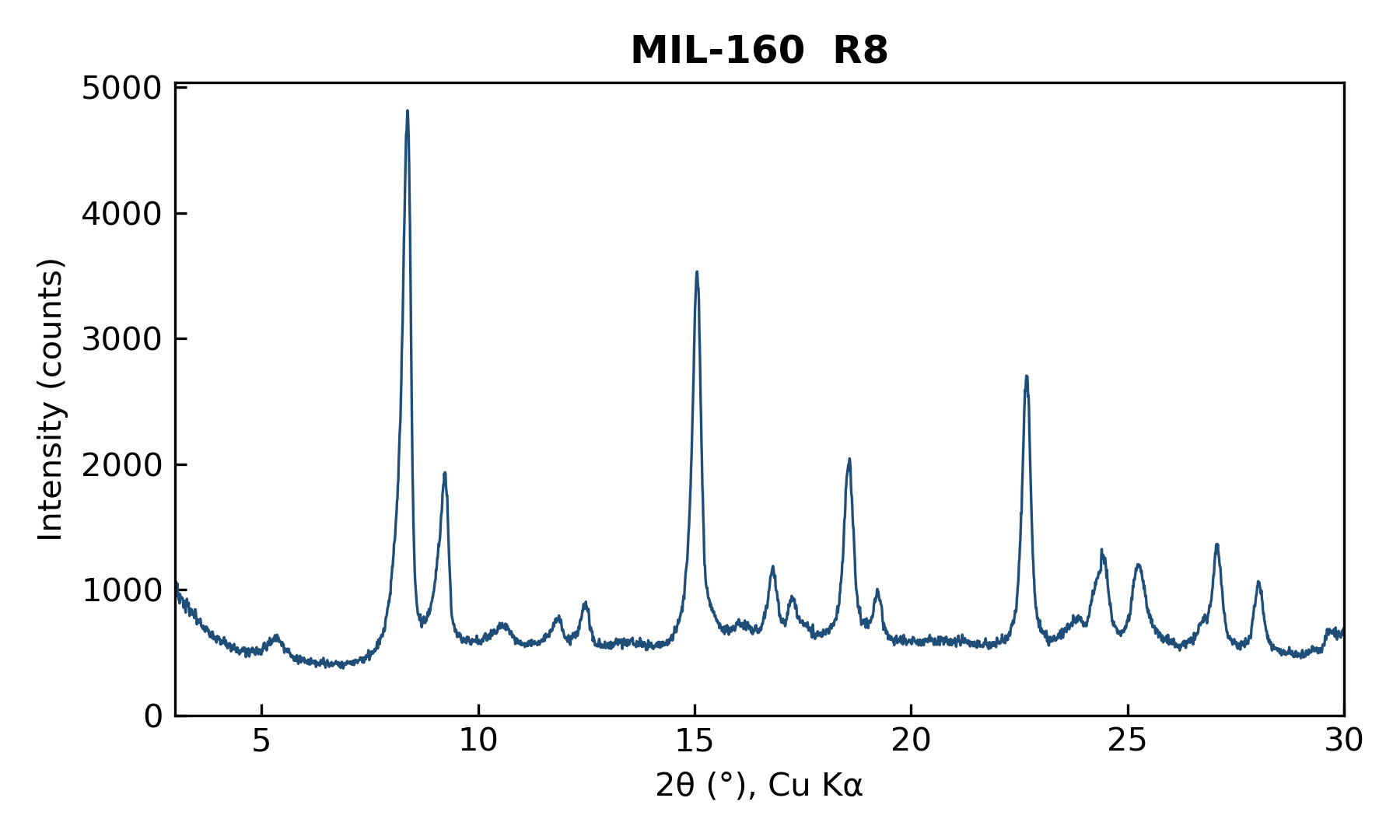}\hfill
\siFig[0.31\linewidth]{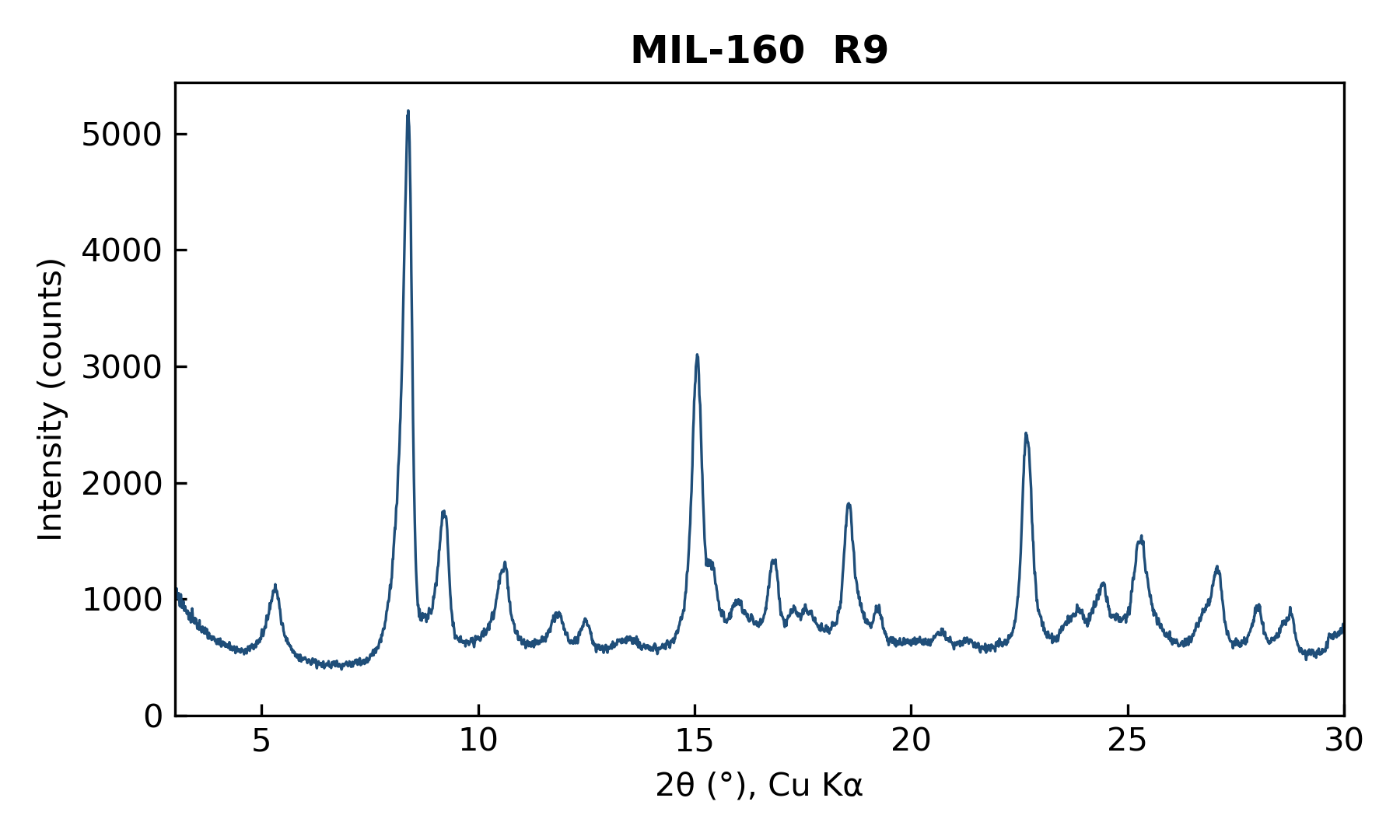}\\[2pt]
\siFig[0.31\linewidth]{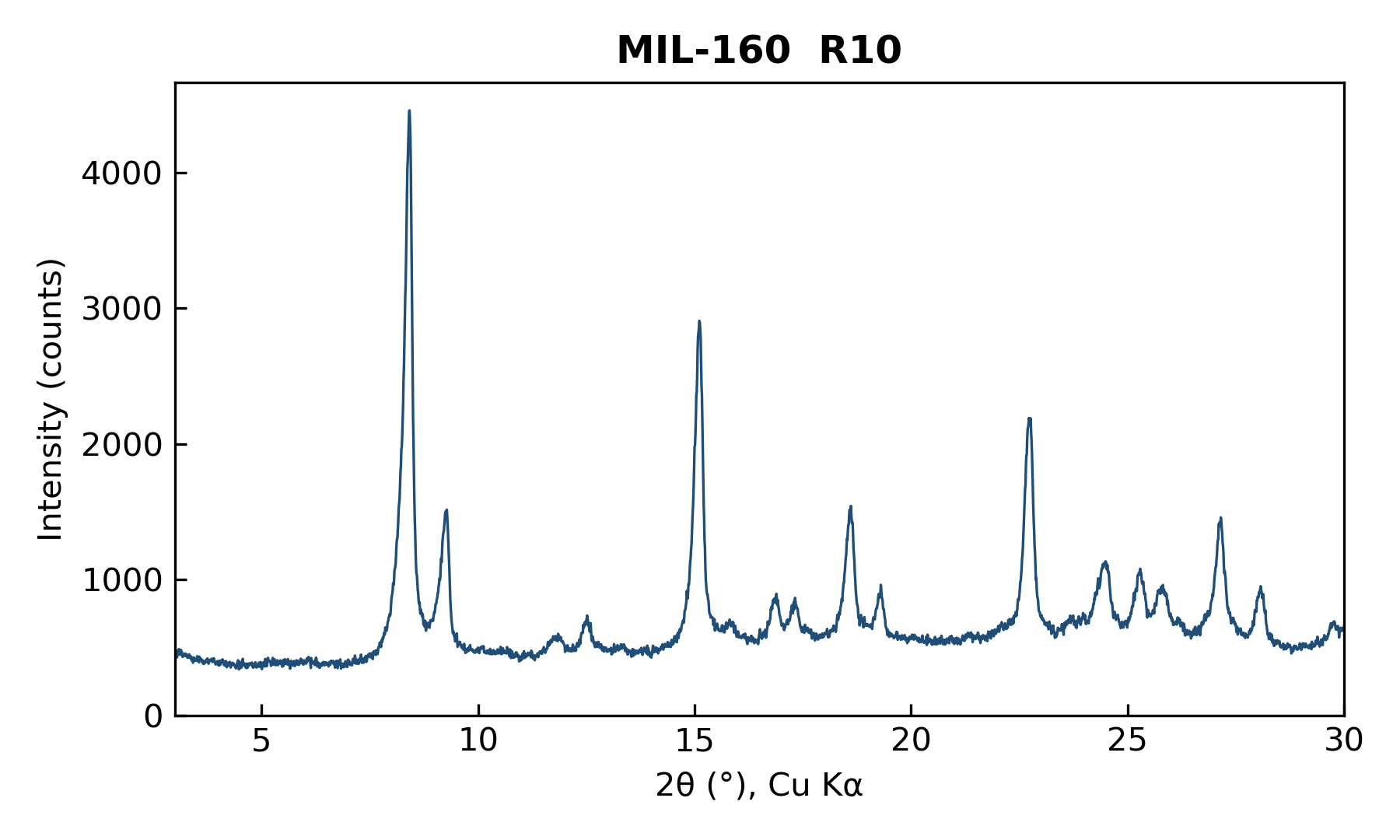}\hfill
\siFig[0.31\linewidth]{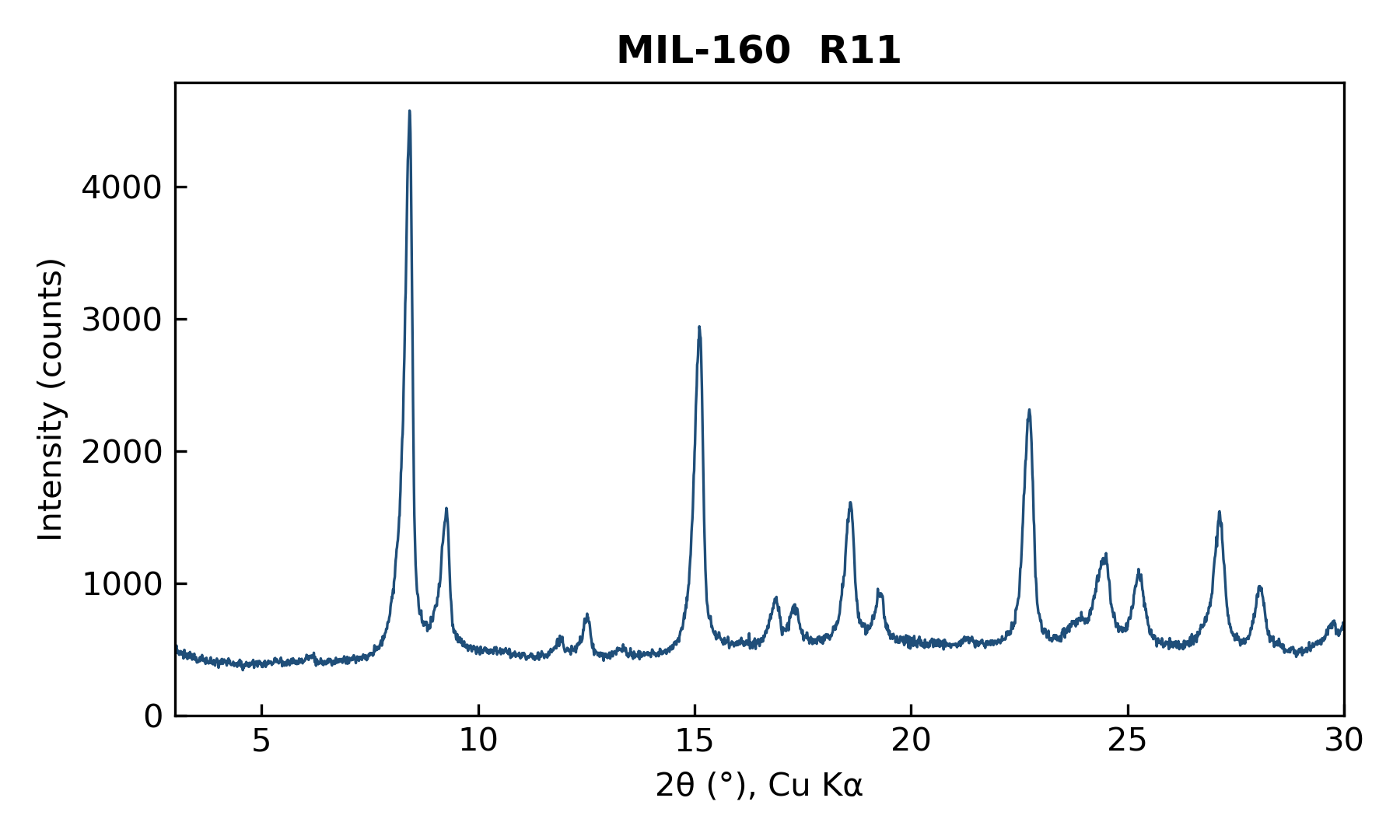}\hfill
\siFig[0.31\linewidth]{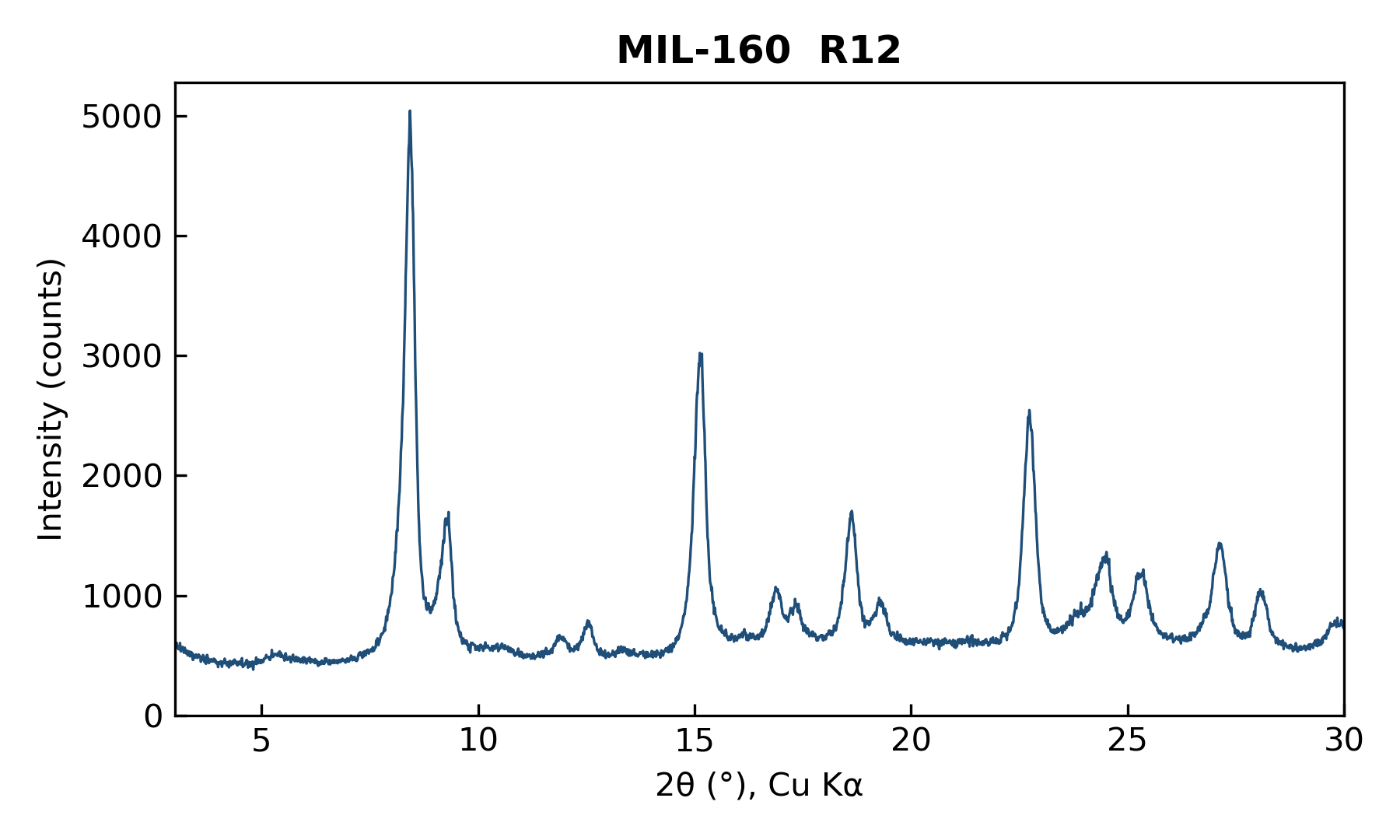}
\caption{PXRD patterns of MIL-160, R00001--R00012 (Table~\ref{tab:al-mil160}).}
\label{fig:pxrd-mil160}
\end{figure}

\clearpage
\paragraph{(iv) MIL-160.}\label{si:pxrd-mil160} Recipes and results in Table~\ref{tab:al-mil160}, patterns in Figure~\ref{fig:pxrd-mil160}.
\begin{table}[H]
\centering
\caption{Recipes and PXRD results of MIL-160.}
\label{tab:al-mil160}
\footnotesize
\setlength{\tabcolsep}{3.5pt}
\resizebox{\linewidth}{!}{%
\begin{tabular}{@{}lccrrrrrrrr@{}}
\toprule
 & & & \multicolumn{3}{c}{Stock solution / DI water ($\mu$L)} & & & & & \\
\cmidrule(lr){4-6}
Exp. & Round & Encoded $\mathbf{x}$ & \ce{AlCl3.6H2O} & NaOH & DI \ce{H2O} & Time (min) & $T$ ($^{\circ}$C) & $I$ & FWHM ($^{\circ}$) & $q$ \\
\midrule
R00001 & 0 & (18, 7, 17, 10, 21) & 180 & 70 & 170 & 50 & 105 & 2{,}326 & 0.32 & 7{,}269 \\
R00002 & 0 & (23, 6, 15, 10, 20) & 230 & 60 & 150 & 50 & 100 & 5{,}580 & 0.22 & 25{,}364 \\
R00003 & 0 & (20, 7, 19, 12, 29) & 200 & 70 & 190 & 60 & 145 & 2{,}804 & 0.25 & 11{,}216 \\
R00004 & 1 & (21, 6, 16, 10, 20) & 210 & 60 & 160 & 50 & 100 & 5{,}207 & 0.27 & 19{,}285 \\
R00005 & 1 & (20, 6, 16, 10, 21) & 200 & 60 & 160 & 50 & 105 & 5{,}491 & 0.25 & 21{,}964 \\
R00006 & 1 & (19, 6, 17, 11, 21) & 190 & 60 & 170 & 55 & 105 & 5{,}811 & 0.27 & 21{,}522 \\
\textbf{R00007} & \textbf{2} & \textbf{(19, 6, 15, 11, 20)} & \textbf{190} & \textbf{60} & \textbf{150} & \textbf{55} & \textbf{100} & \textbf{9{,}767} & \textbf{0.22} & \textbf{44{,}395} \\
R00008 & 2 & (20, 7, 16, 12, 28) & 200 & 70 & 160 & 60 & 140 & 4{,}813 & 0.26 & 18{,}512 \\
R00009 & 2 & (20, 6, 14, 10, 20) & 200 & 60 & 140 & 50 & 100 & 5{,}199 & 0.29 & 17{,}928 \\
R00010 & 3 & (20, 6, 14, 11, 20) & 200 & 60 & 140 & 55 & 100 & 4{,}459 & 0.234 & 19{,}072 \\
R00011 & 3 & (20, 6, 17, 11, 20) & 200 & 60 & 170 & 55 & 100 & 4{,}579 & 0.23 & 19{,}874 \\
R00012 & 3 & (19, 6, 14, 11, 20) & 190 & 60 & 140 & 55 & 100 & 5{,}043 & 0.319 & 15{,}799 \\
\bottomrule
\end{tabular}}
\end{table}

\begin{figure}[H]
\centering
\siFig[0.31\linewidth]{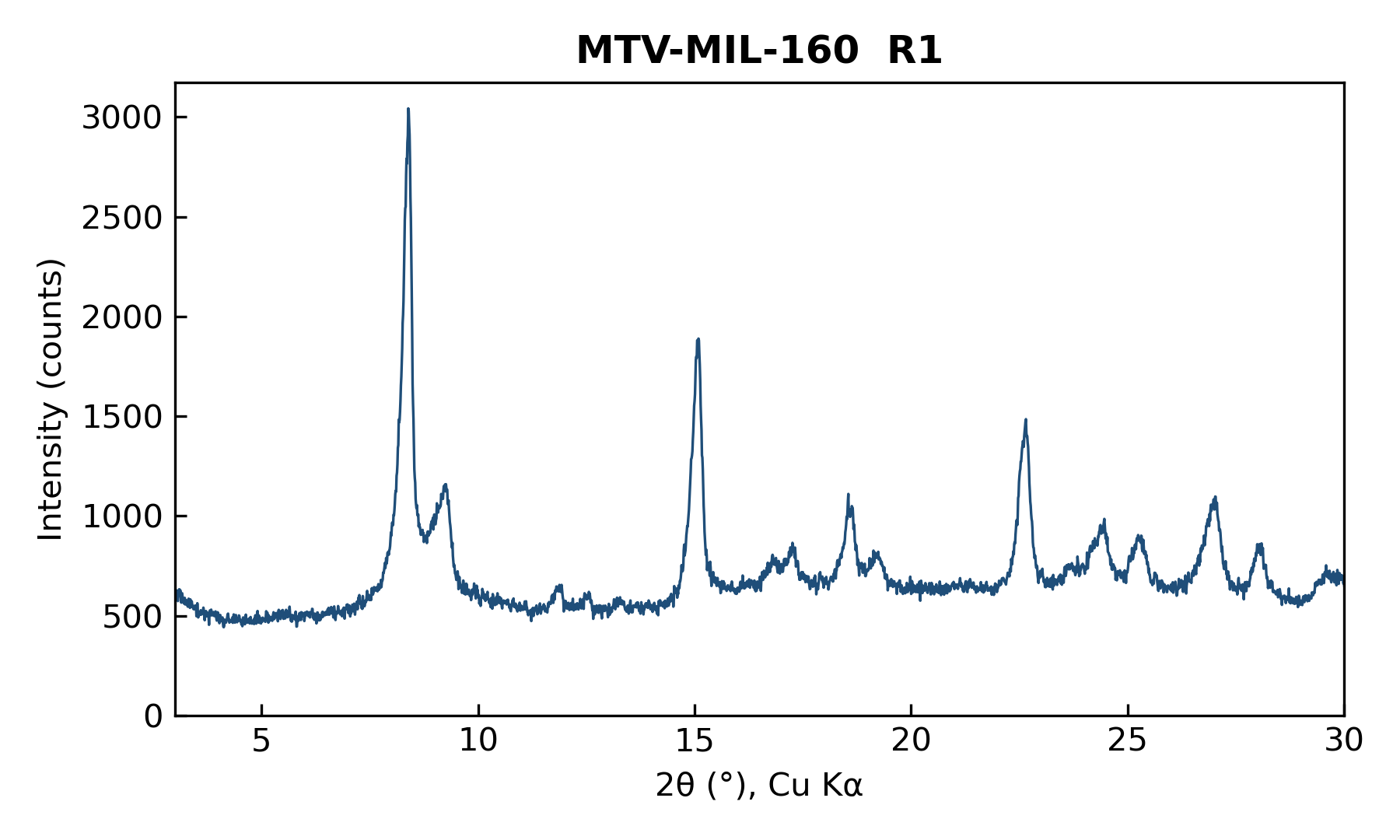}\hfill
\siFig[0.31\linewidth]{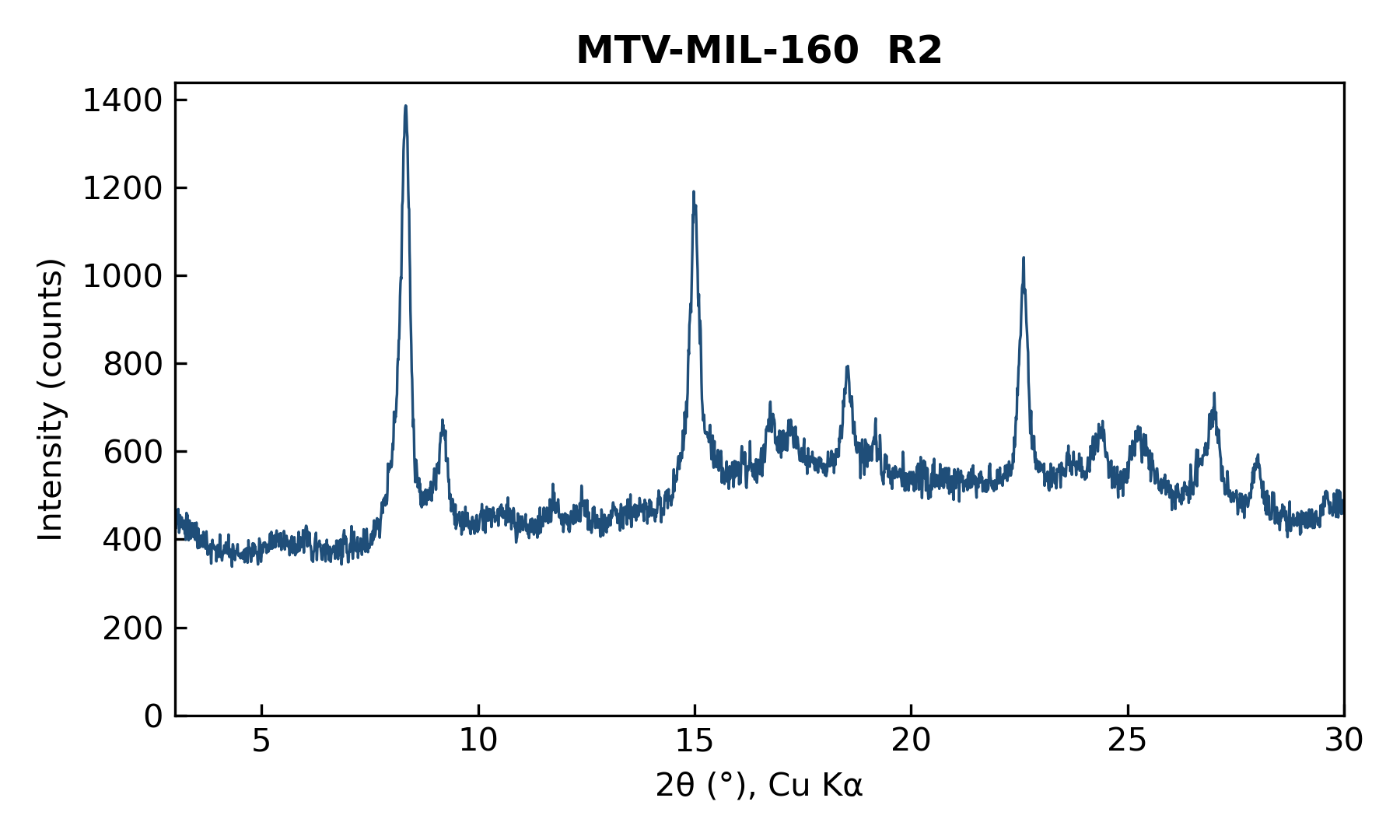}\hfill
\siFig[0.31\linewidth]{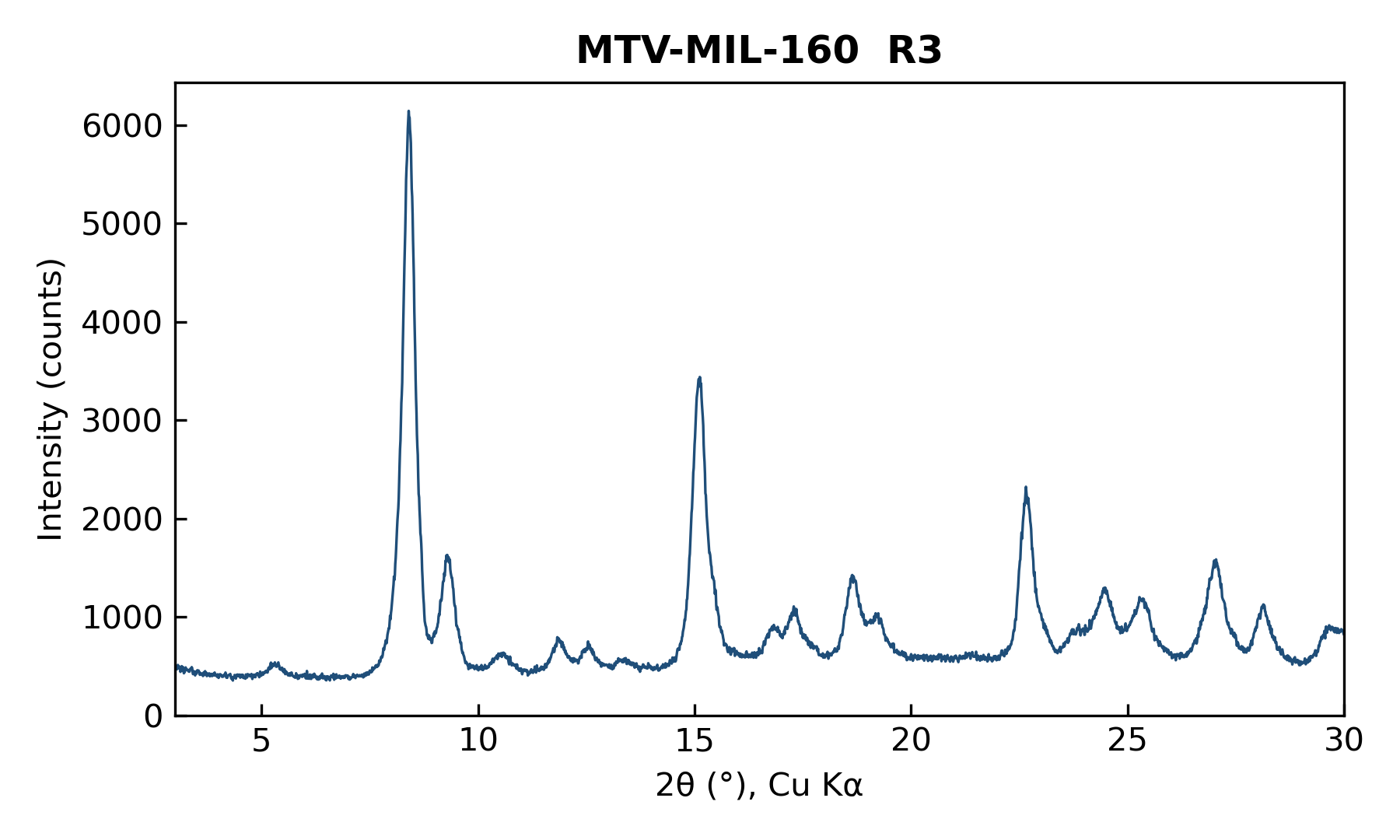}\\[2pt]
\siFig[0.31\linewidth]{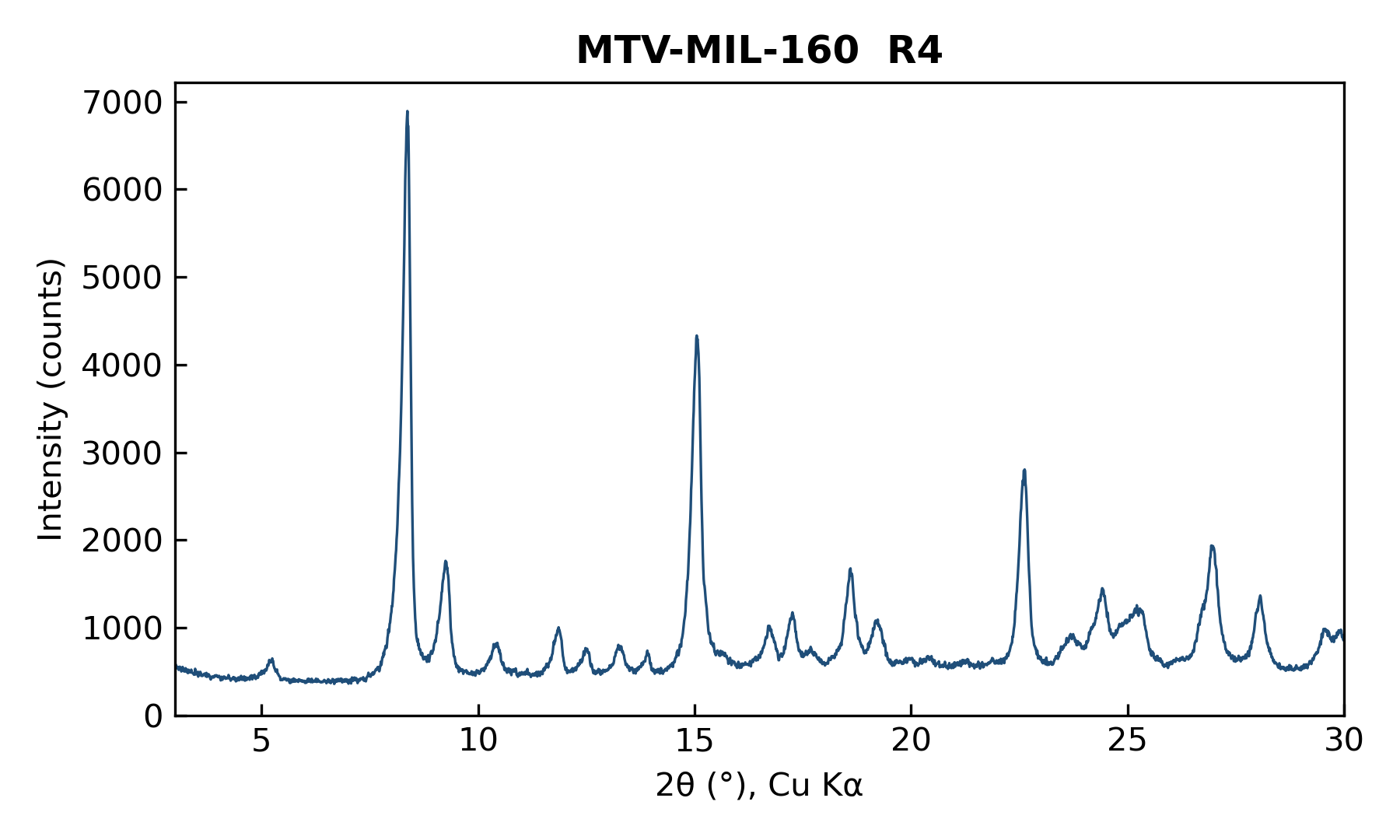}\hfill
\siFig[0.31\linewidth]{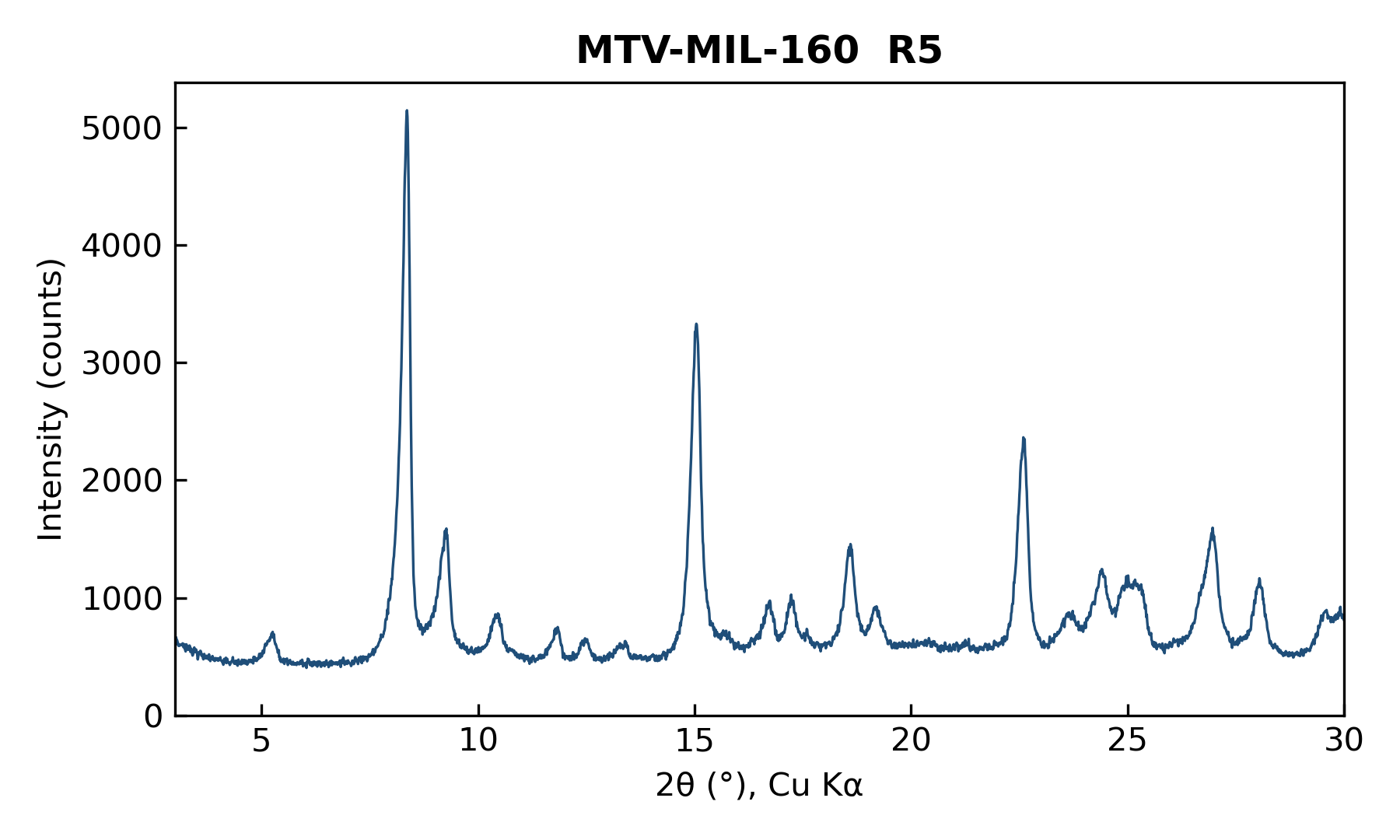}\hfill
\siFig[0.31\linewidth]{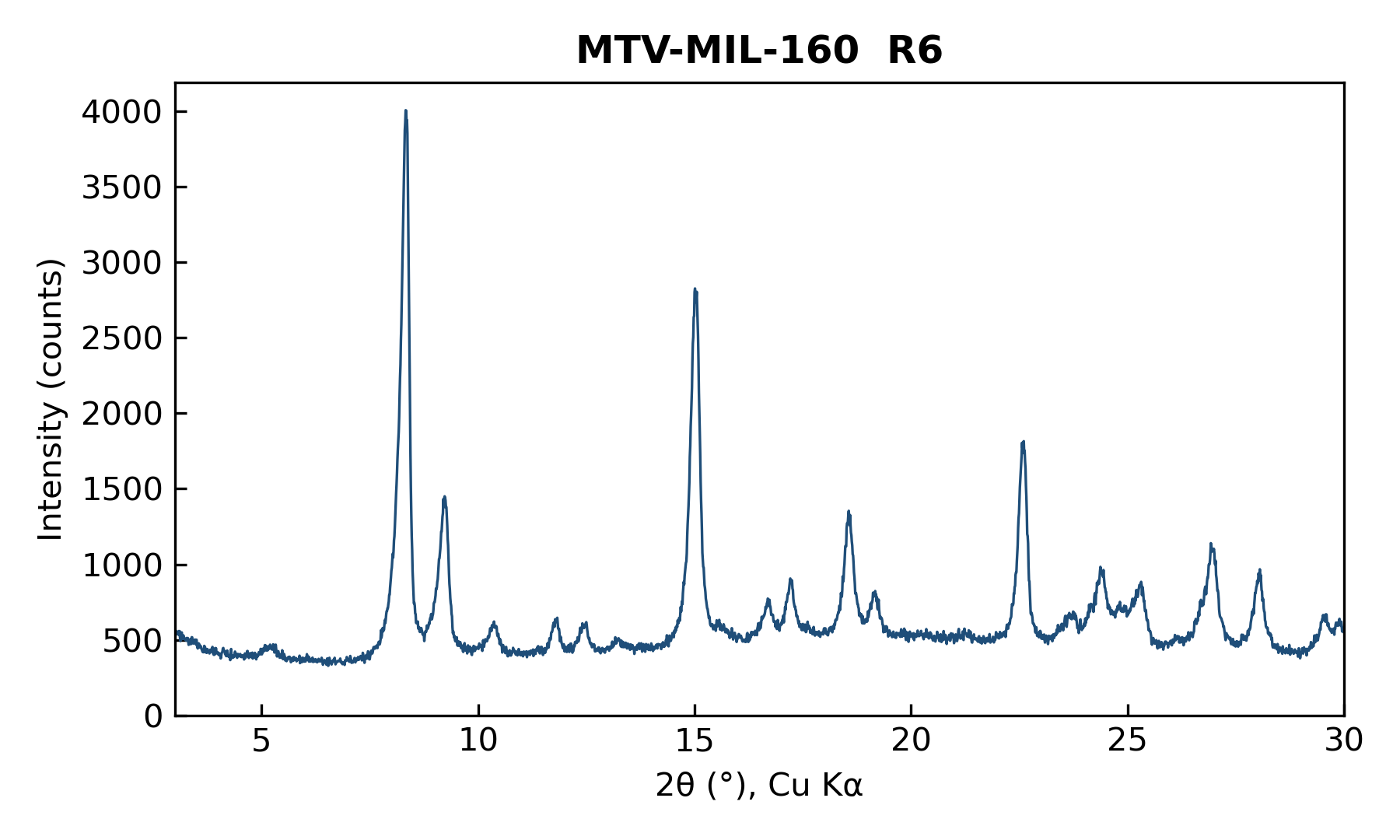}\\[2pt]
\siFig[0.31\linewidth]{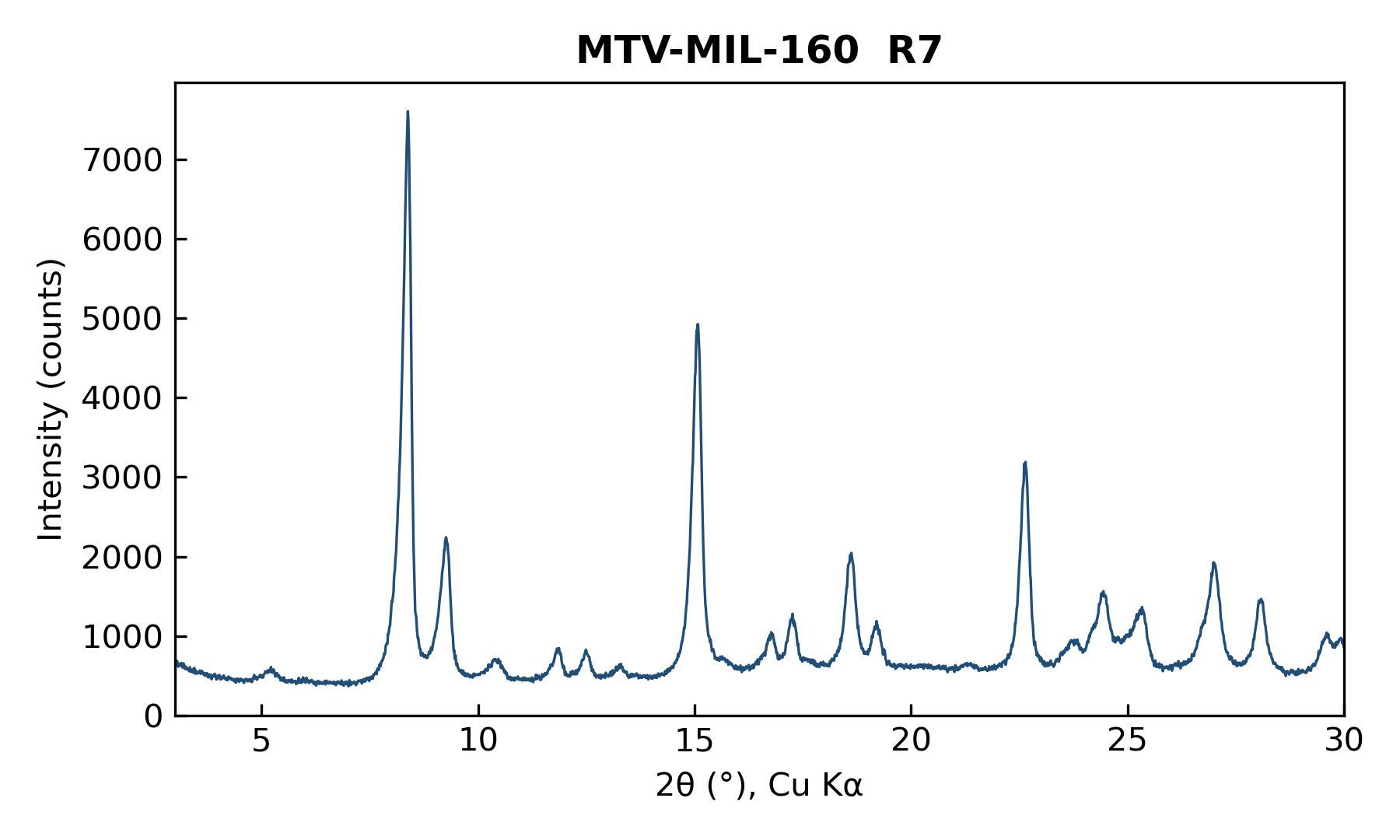}\hfill
\siFig[0.31\linewidth]{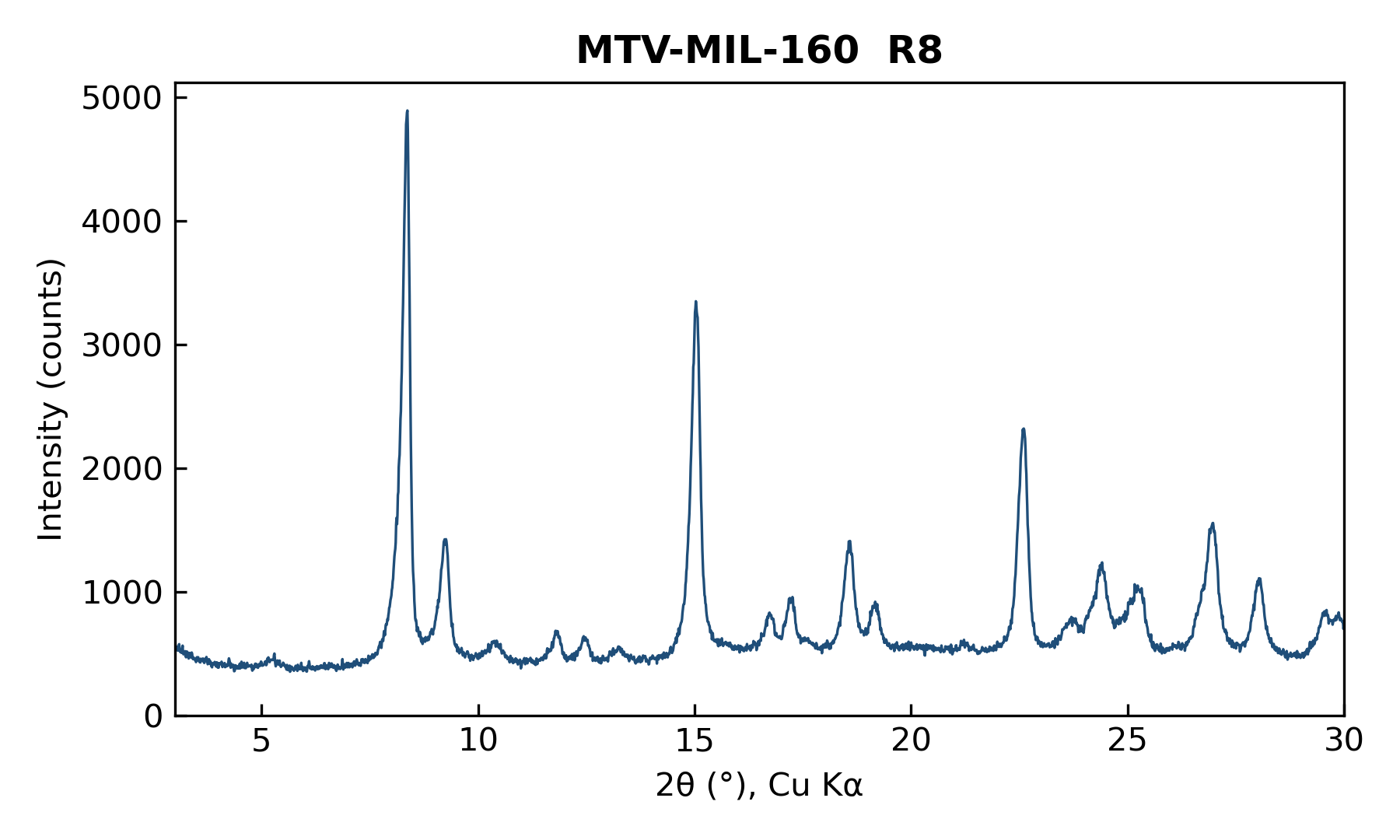}\hfill
\siFig[0.31\linewidth]{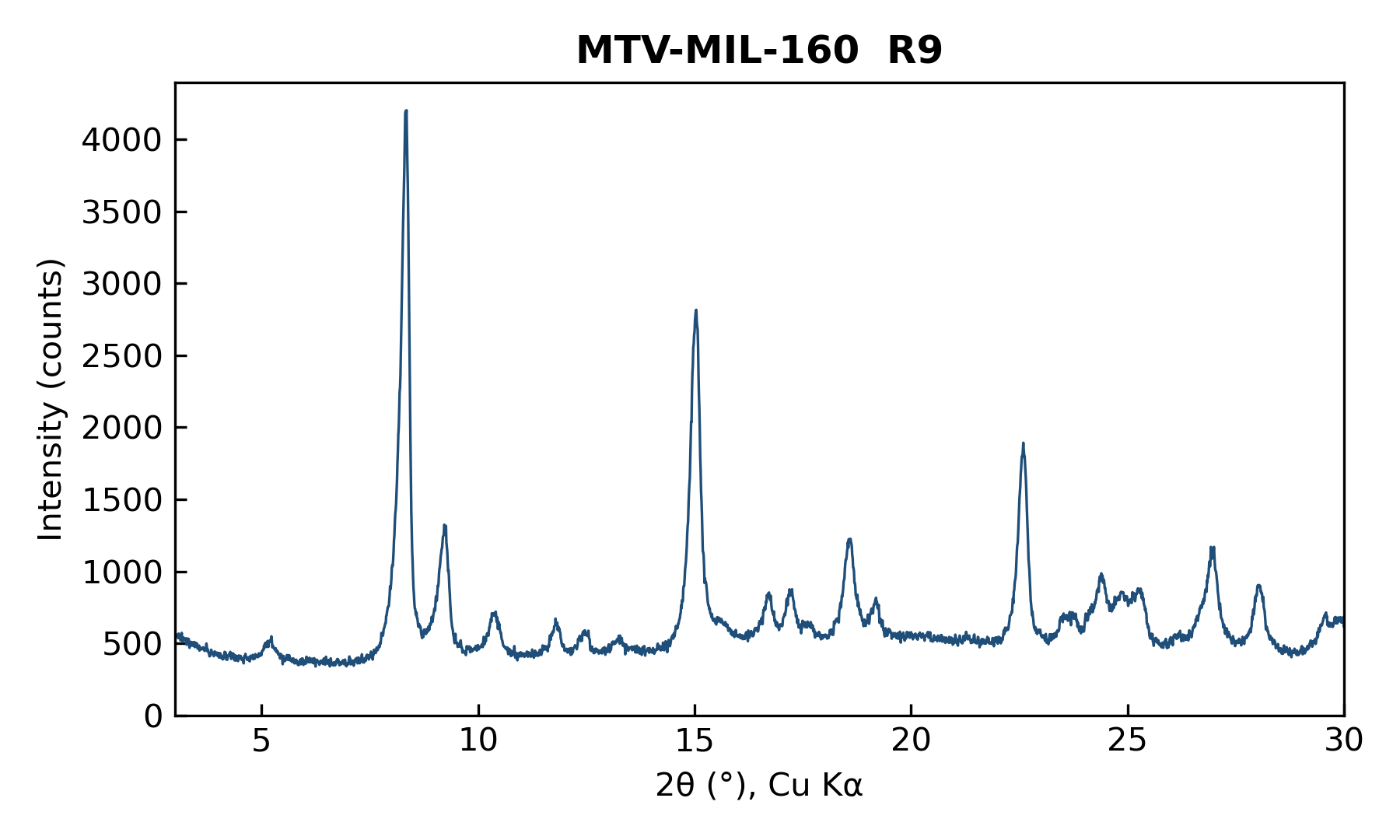}\\[2pt]
\siFig[0.31\linewidth]{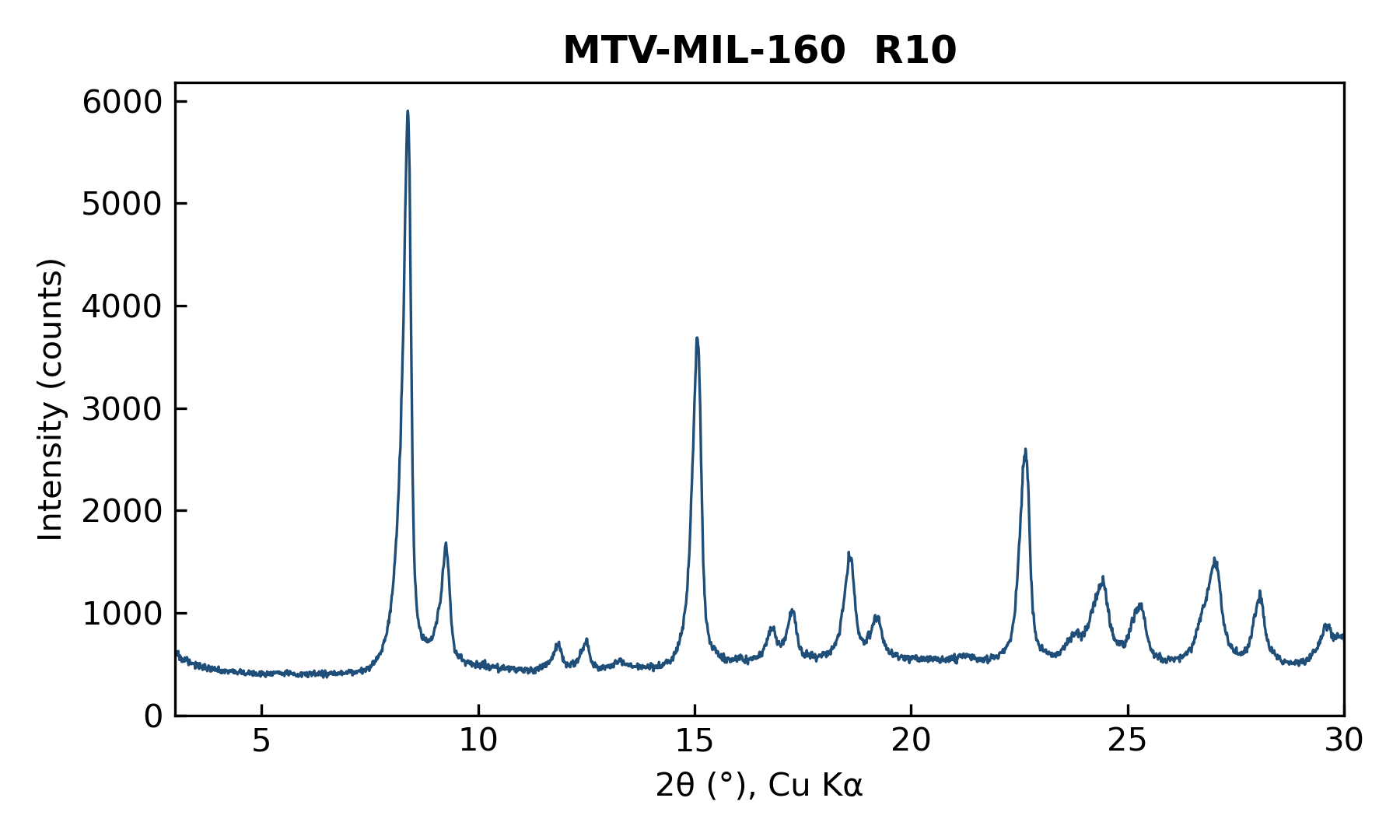}\hfill
\siFig[0.31\linewidth]{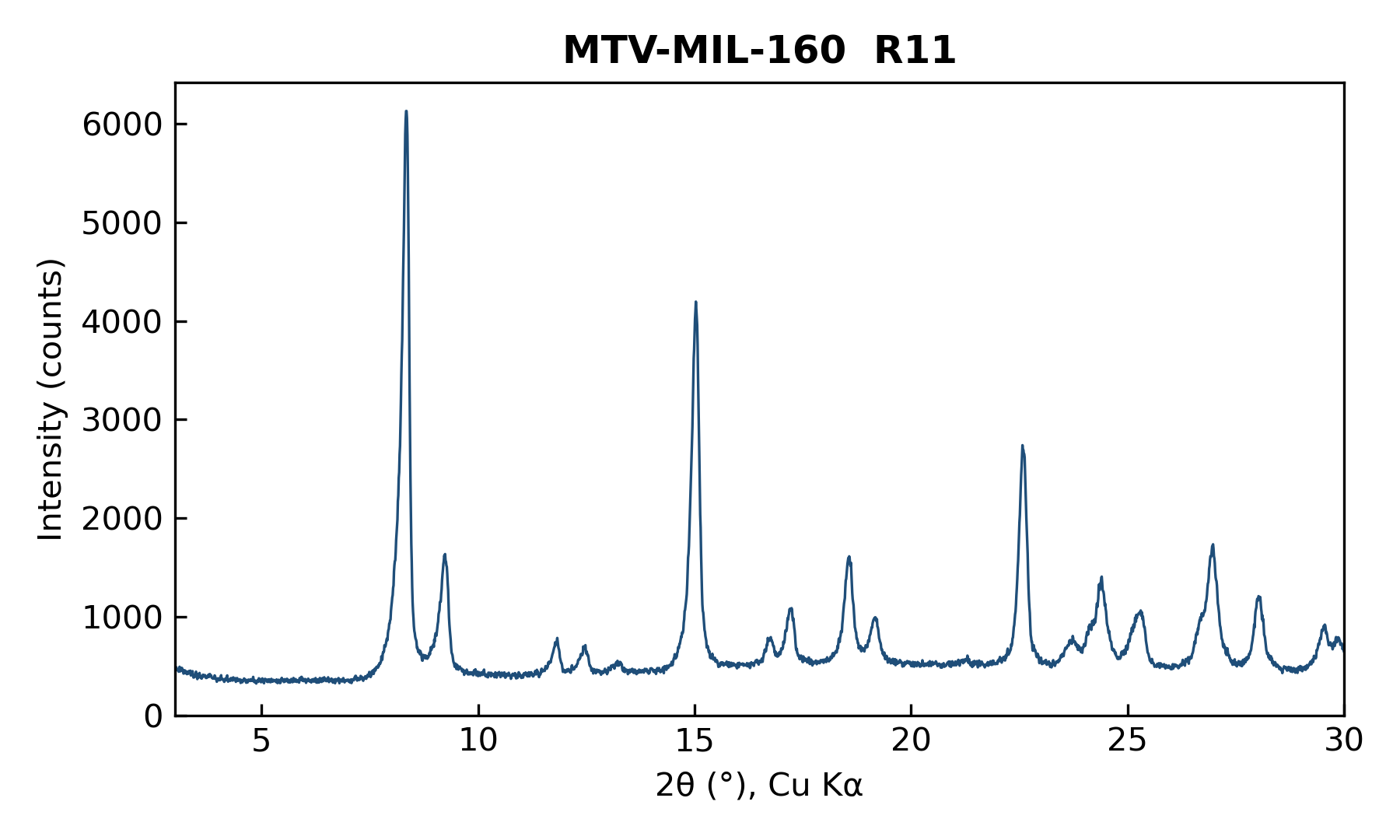}\hfill
\siFig[0.31\linewidth]{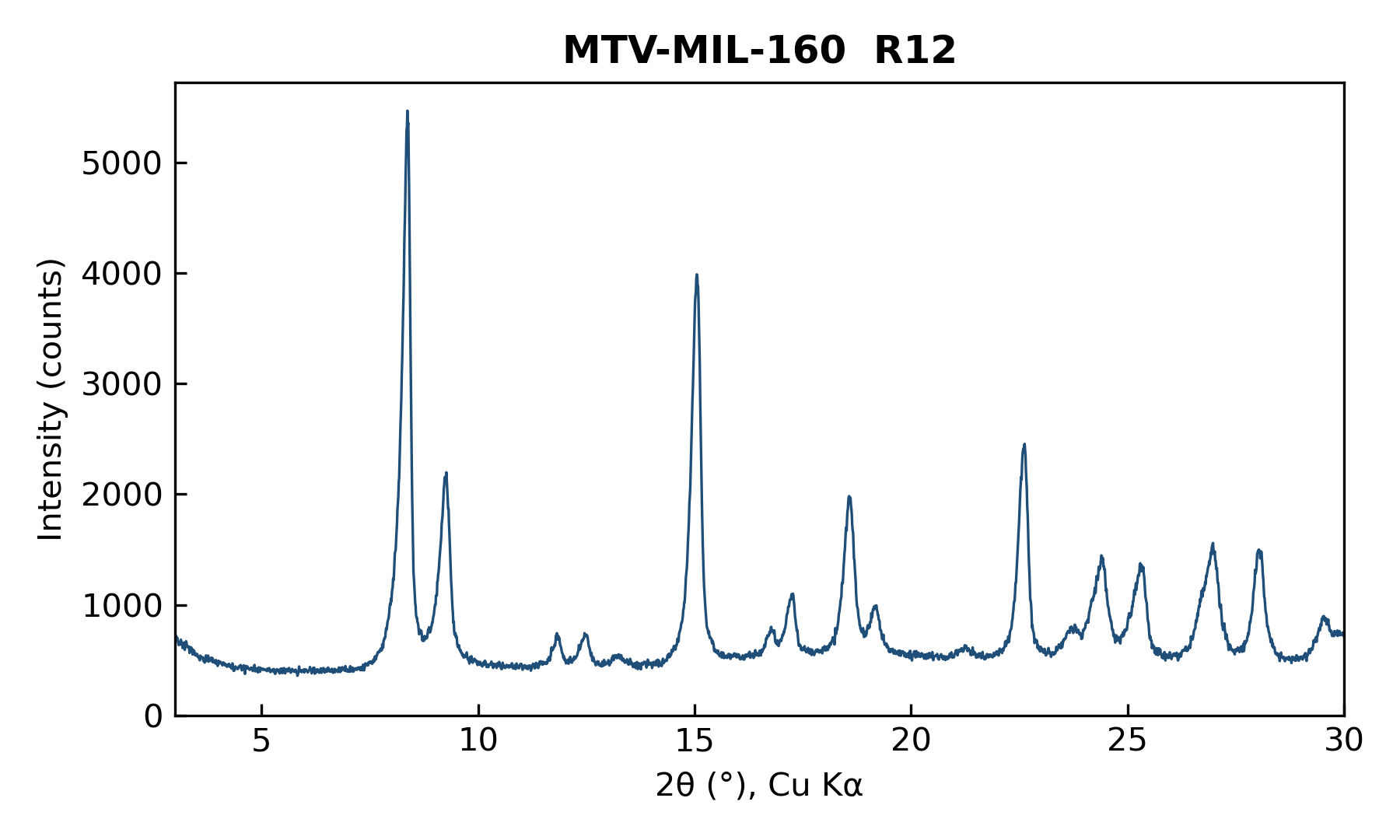}
\caption{PXRD patterns of MTV-MIL-160, R00001--R00012 (Table~\ref{tab:al-mtvmil160}).}
\label{fig:pxrd-mtvmil160}
\end{figure}

\clearpage
\paragraph{(v) MTV-MIL-160.}\label{si:pxrd-mtvmil160} Recipes and results in Table~\ref{tab:al-mtvmil160}, patterns in Figure~\ref{fig:pxrd-mtvmil160}.
\begin{table}[H]
\centering
\caption{Recipes and PXRD results of MTV-MIL-160.}
\label{tab:al-mtvmil160}
\footnotesize
\setlength{\tabcolsep}{3.5pt}
\resizebox{\linewidth}{!}{%
\begin{tabular}{@{}lccrrrrrrrr@{}}
\toprule
 & & & \multicolumn{3}{c}{Stock solution / DI water ($\mu$L)} & & & & & \\
\cmidrule(lr){4-6}
Exp. & Round & Encoded $\mathbf{x}$ & \ce{AlCl3.6H2O} & NaOH & DI \ce{H2O} & Time (min) & $T$ ($^{\circ}$C) & $I$ & FWHM ($^{\circ}$) & $q$ \\
\midrule
R00001 & 0 & (18, 7, 17, 10, 21) & 180 & 70 & 170 & 50 & 105 & 3{,}043 & 0.306 & 9{,}951 \\
R00002 & 0 & (23, 6, 15, 10, 20) & 230 & 60 & 150 & 50 & 100 & 1{,}386 & 0.383 & 3{,}615 \\
R00003 & 0 & (20, 7, 19, 12, 29) & 200 & 70 & 190 & 60 & 145 & 6{,}142 & 0.356 & 17{,}258 \\
R00004 & 1 & (21, 7, 18, 12, 29) & 210 & 70 & 180 & 60 & 145 & 6{,}892 & 0.233 & 29{,}554 \\
R00005 & 1 & (19, 7, 17, 12, 29) & 190 & 70 & 170 & 60 & 145 & 5{,}147 & 0.246 & 20{,}948 \\
R00006 & 1 & (22, 7, 16, 12, 29) & 220 & 70 & 160 & 60 & 145 & 4{,}006 & 0.25 & 16{,}030 \\
\textbf{R00007} & \textbf{2} & \textbf{(20, 7, 18, 12, 28)} & \textbf{200} & \textbf{70} & \textbf{180} & \textbf{60} & \textbf{140} & \textbf{7{,}608} & \textbf{0.245} & \textbf{30{,}990} \\
R00008 & 2 & (20, 7, 17, 12, 28) & 200 & 70 & 170 & 60 & 140 & 4{,}893 & 0.225 & 21{,}737 \\
R00009 & 2 & (22, 7, 17, 12, 29) & 220 & 70 & 170 & 60 & 145 & 4{,}202 & 0.253 & 16{,}589 \\
R00010 & 3 & (20, 6, 18, 10, 21) & 200 & 60 & 180 & 50 & 105 & 5{,}902 & 0.248 & 23{,}779 \\
R00011 & 3 & (20, 7, 18, 12, 29) & 200 & 70 & 180 & 60 & 145 & 6{,}129 & 0.211 & 29{,}020 \\
R00012 & 3 & (22, 7, 18, 12, 29) & 220 & 70 & 180 & 60 & 145 & 5{,}466 & 0.24 & 22{,}813 \\
\bottomrule
\end{tabular}}
\end{table}

\begin{figure}[H]
\centering
\siFig[0.31\linewidth]{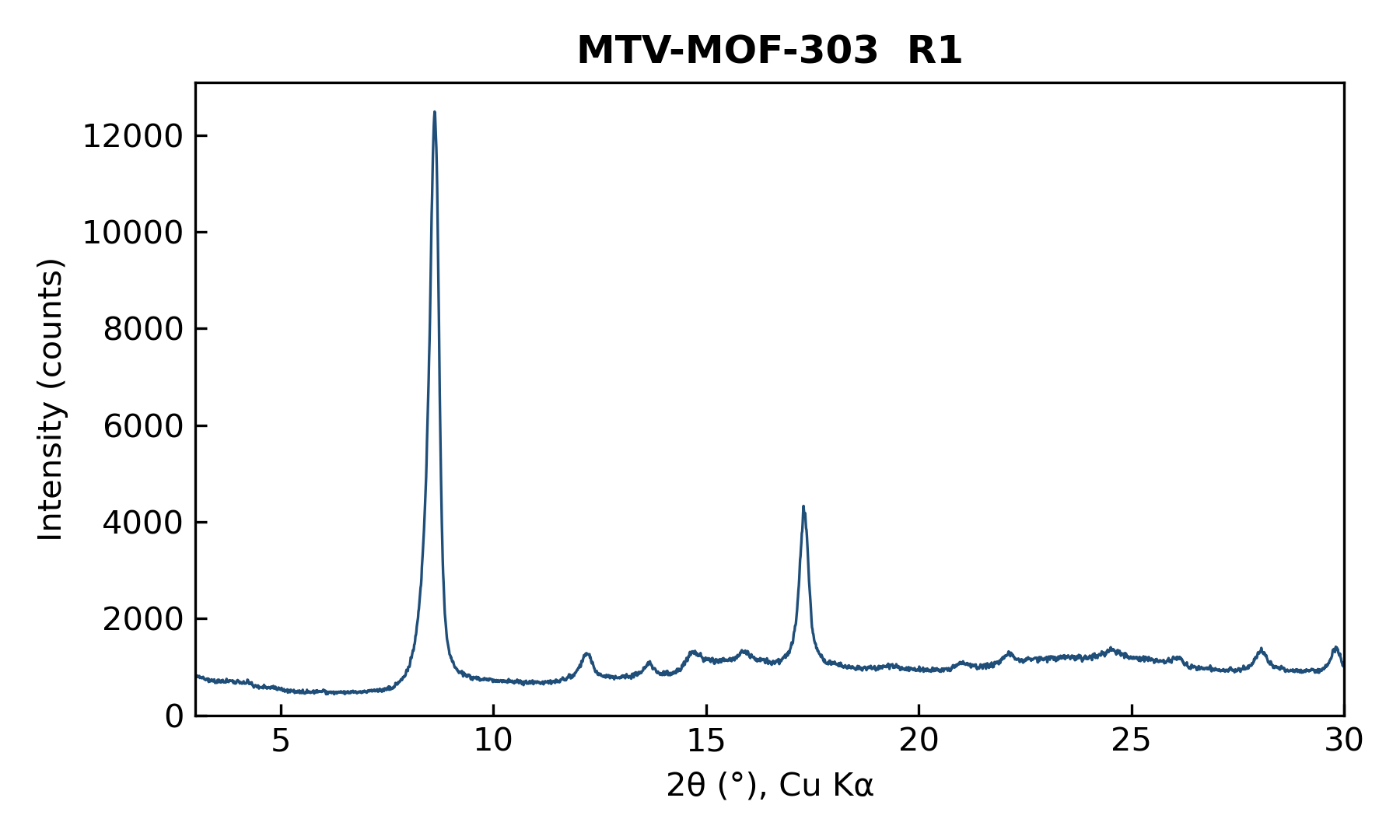}\hfill
\siFig[0.31\linewidth]{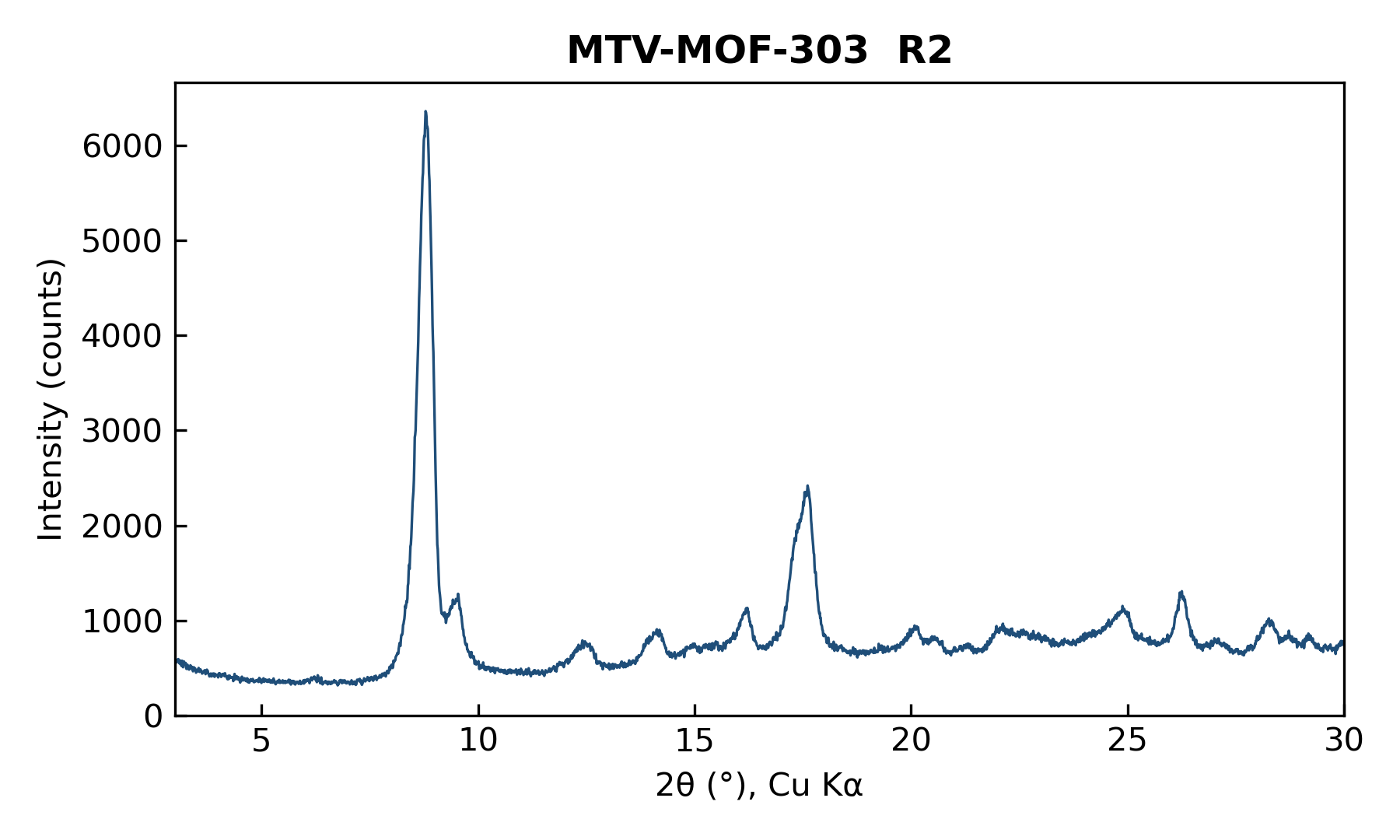}\hfill
\siFig[0.31\linewidth]{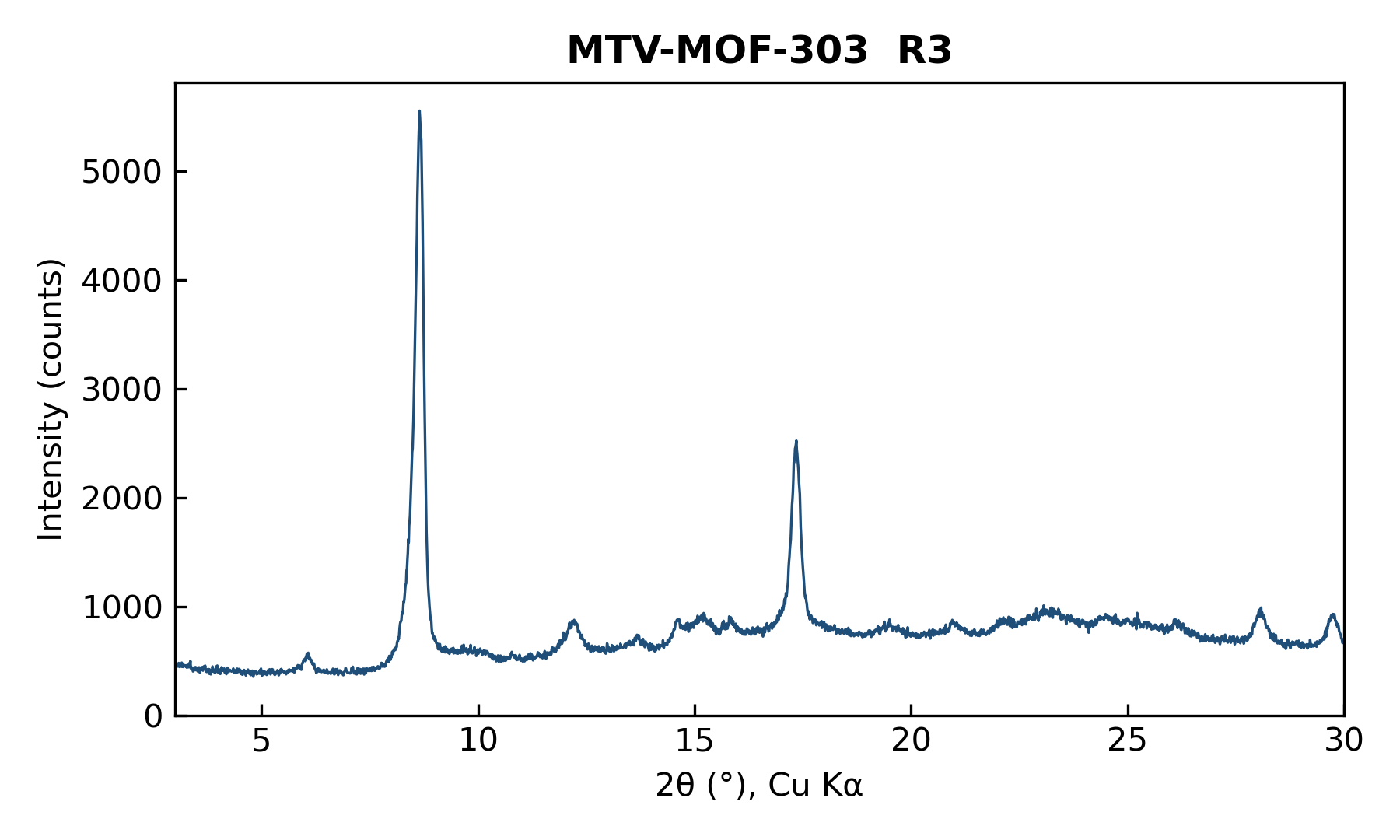}\\[2pt]
\siFig[0.31\linewidth]{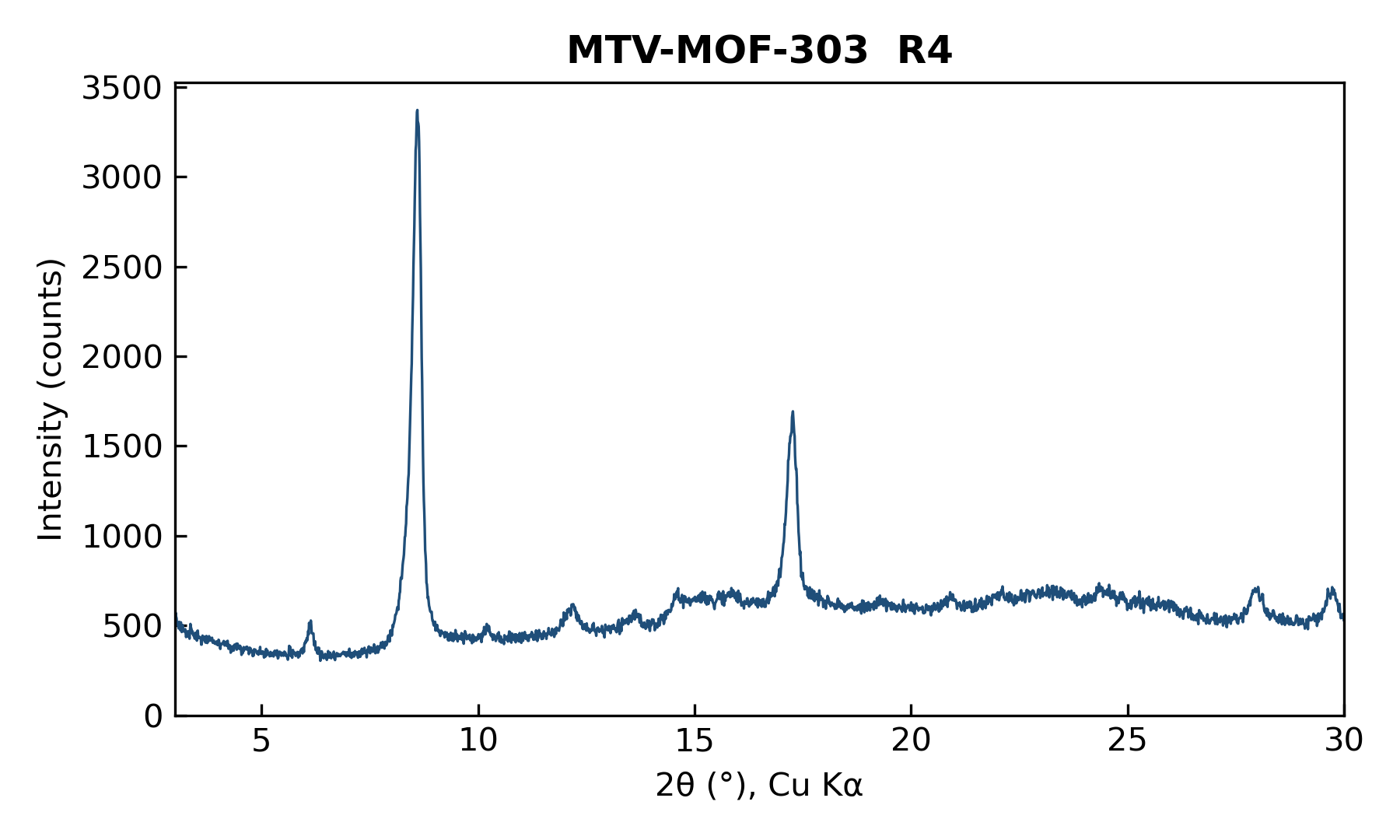}\hfill
\siFig[0.31\linewidth]{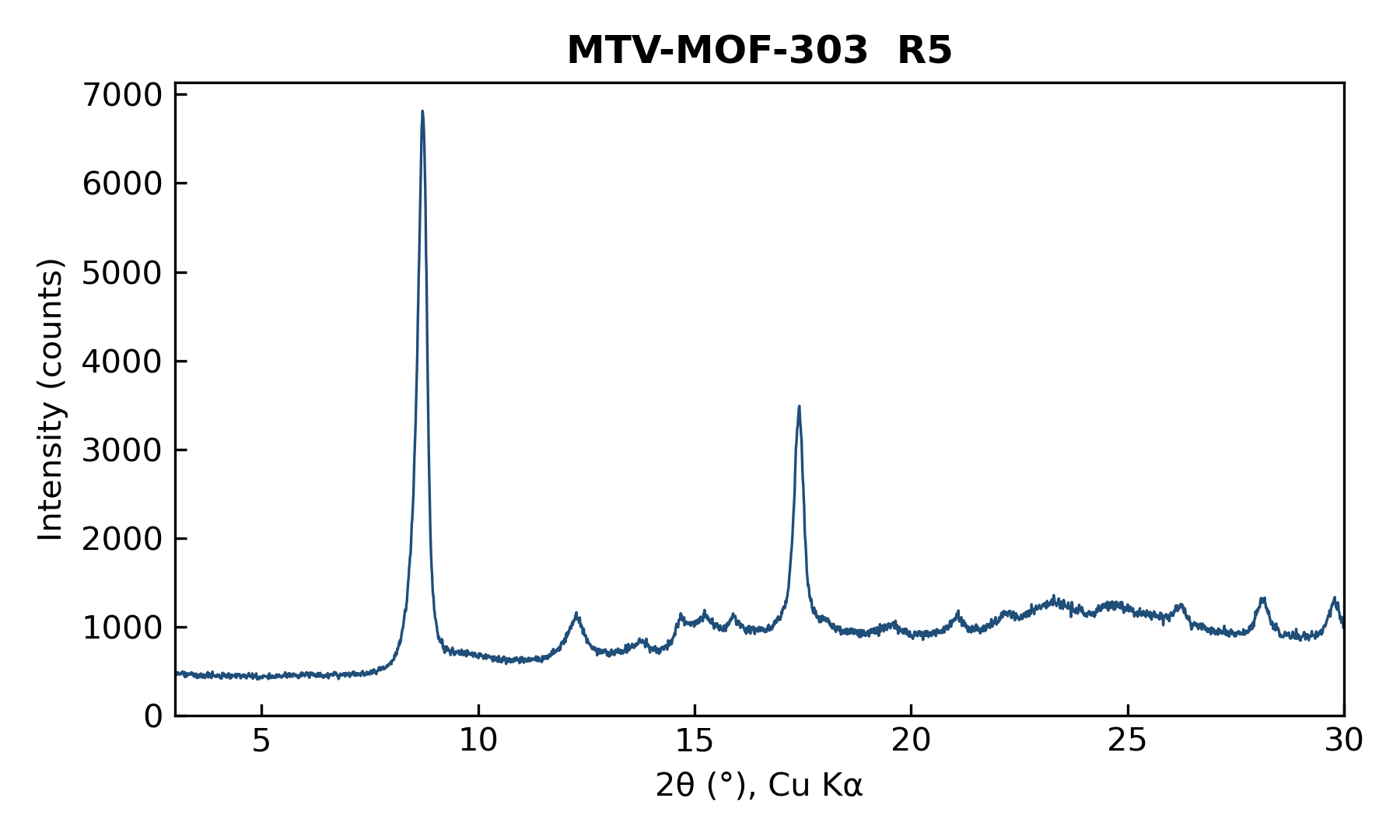}\hfill
\siFig[0.31\linewidth]{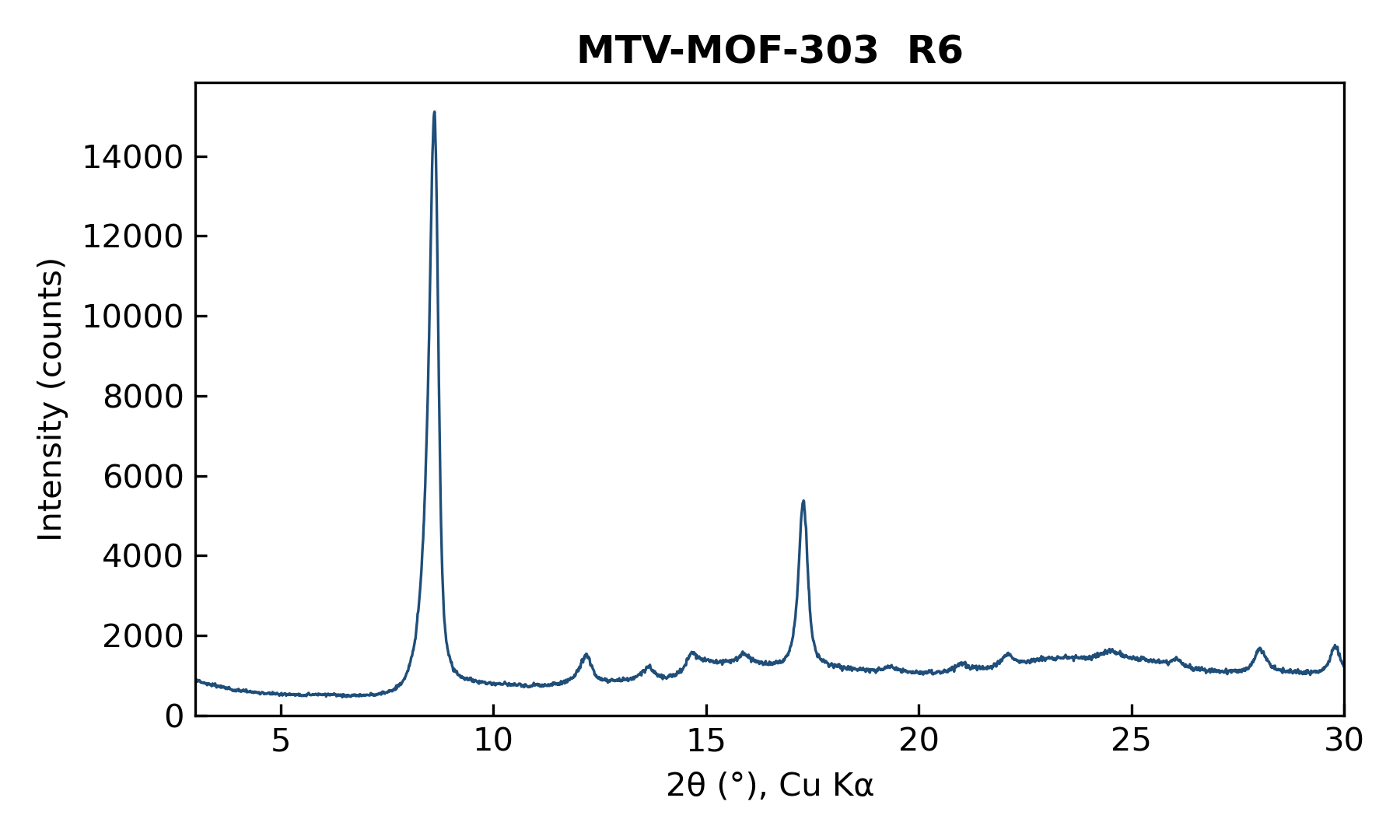}\\[2pt]
\siFig[0.31\linewidth]{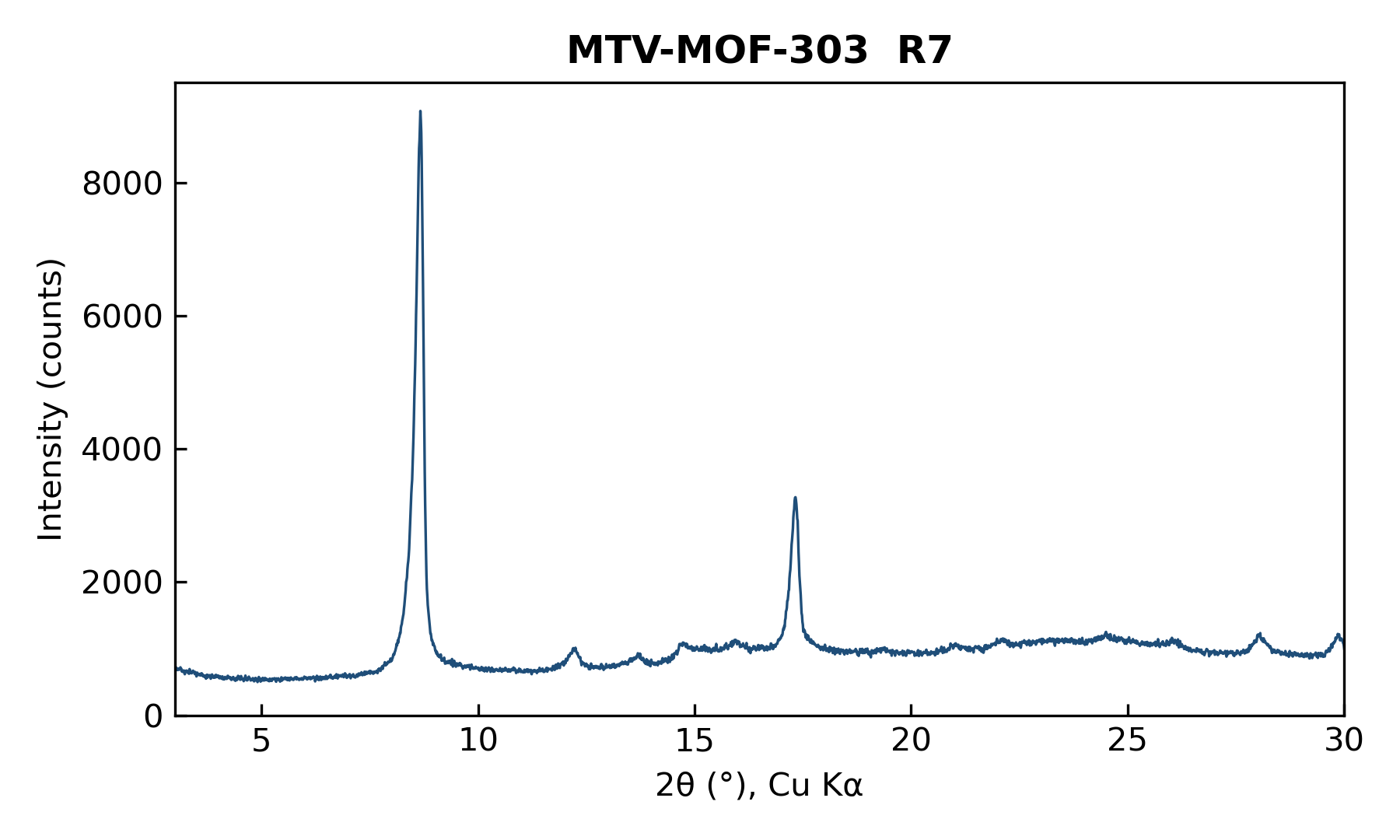}\hfill
\siFig[0.31\linewidth]{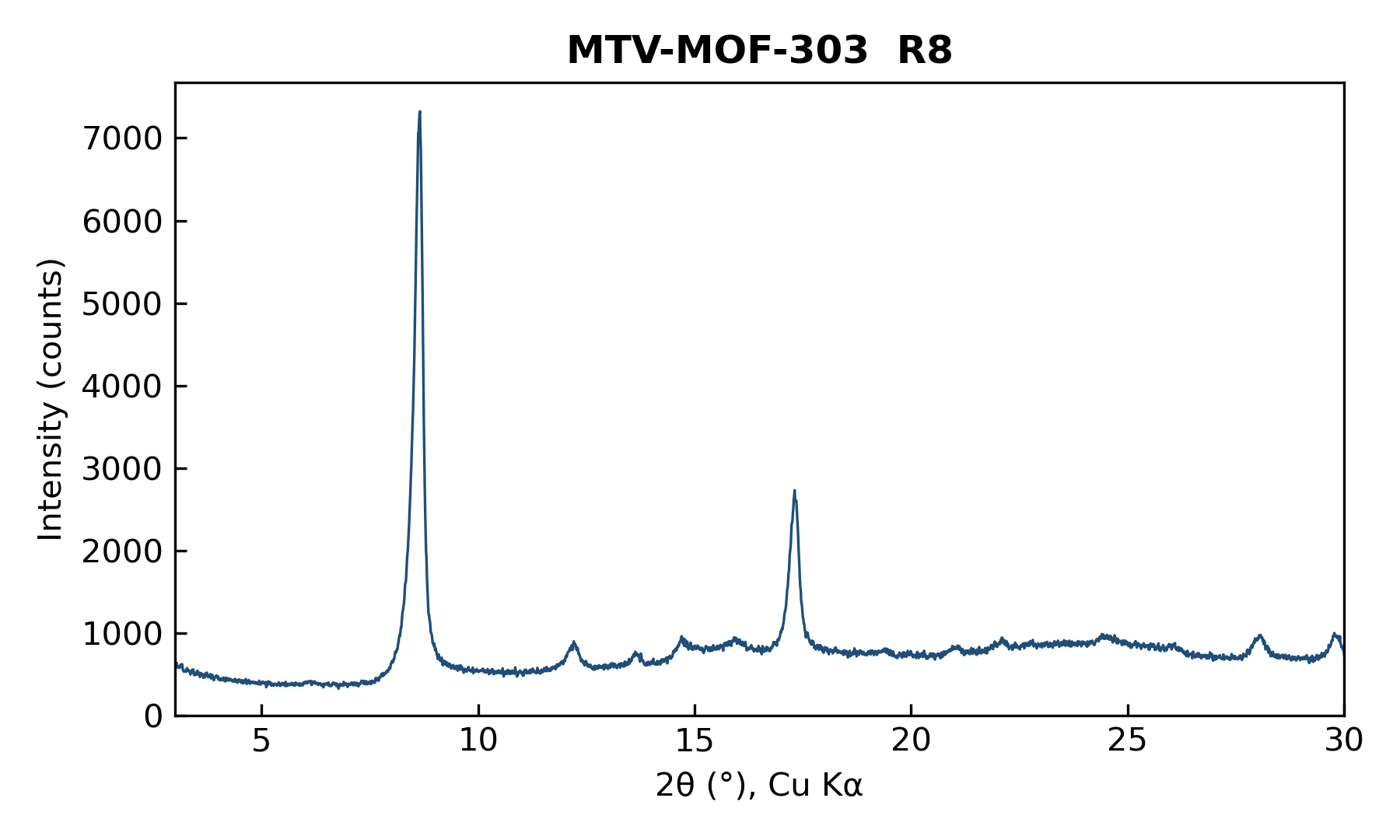}\hfill
\siFig[0.31\linewidth]{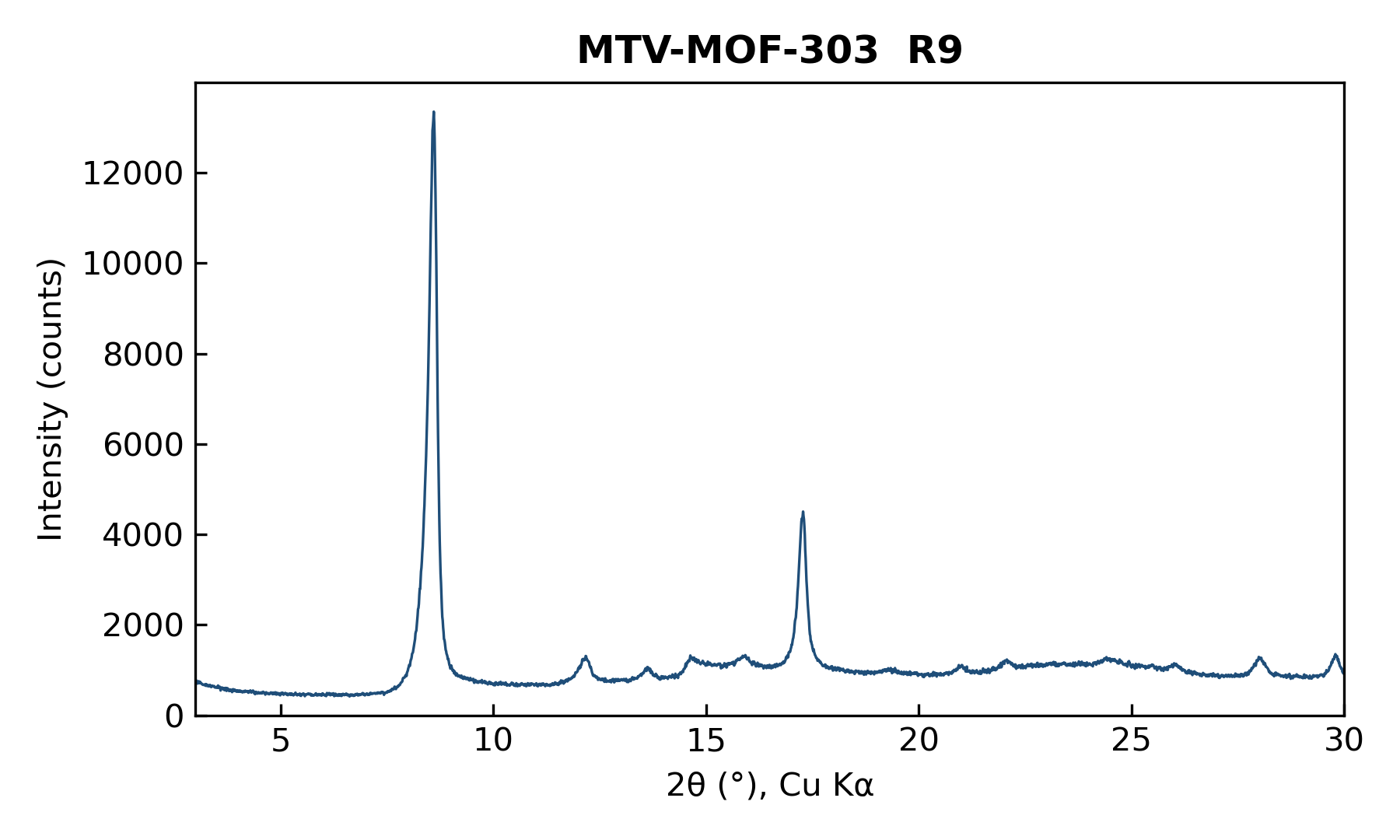}\\[2pt]
\siFig[0.31\linewidth]{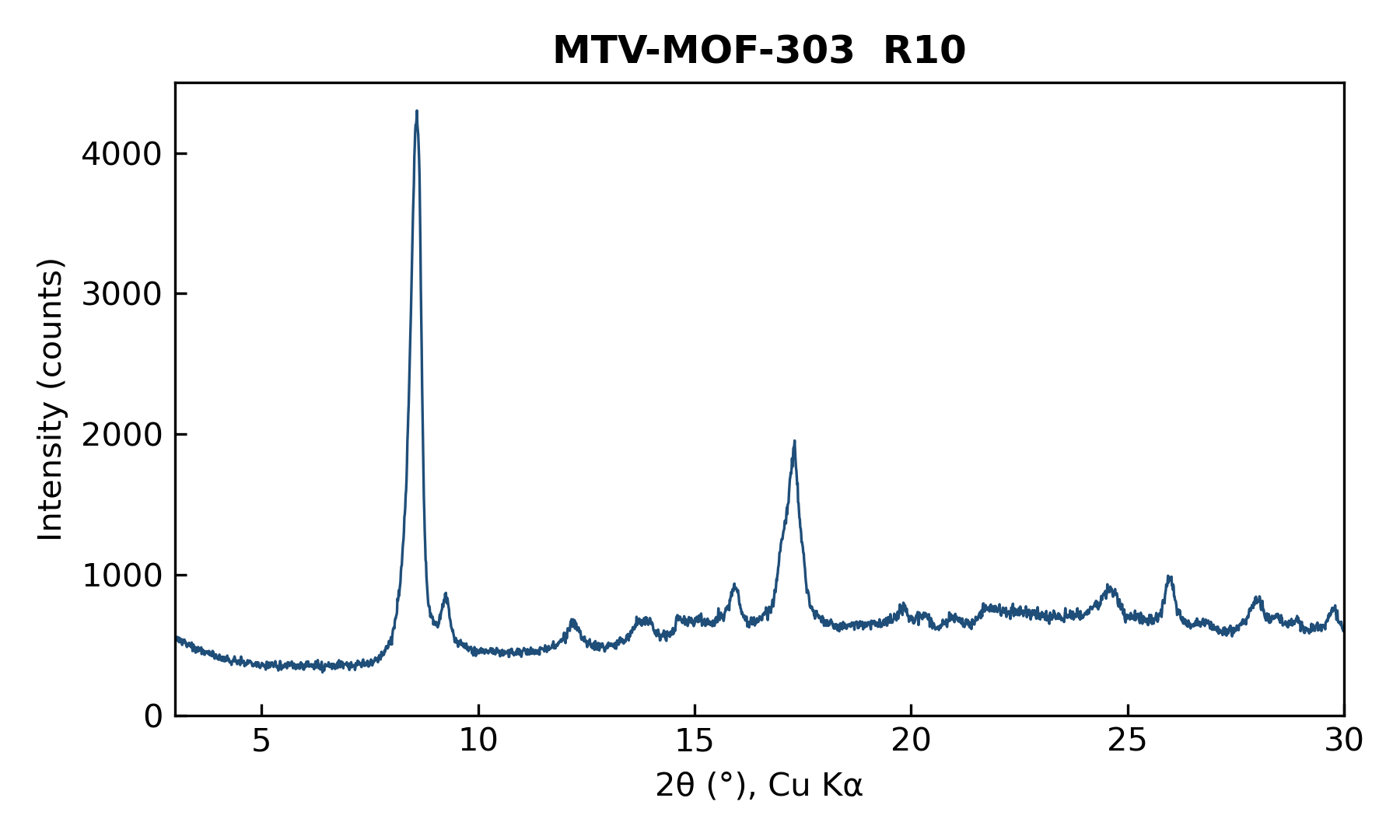}\hfill
\siFig[0.31\linewidth]{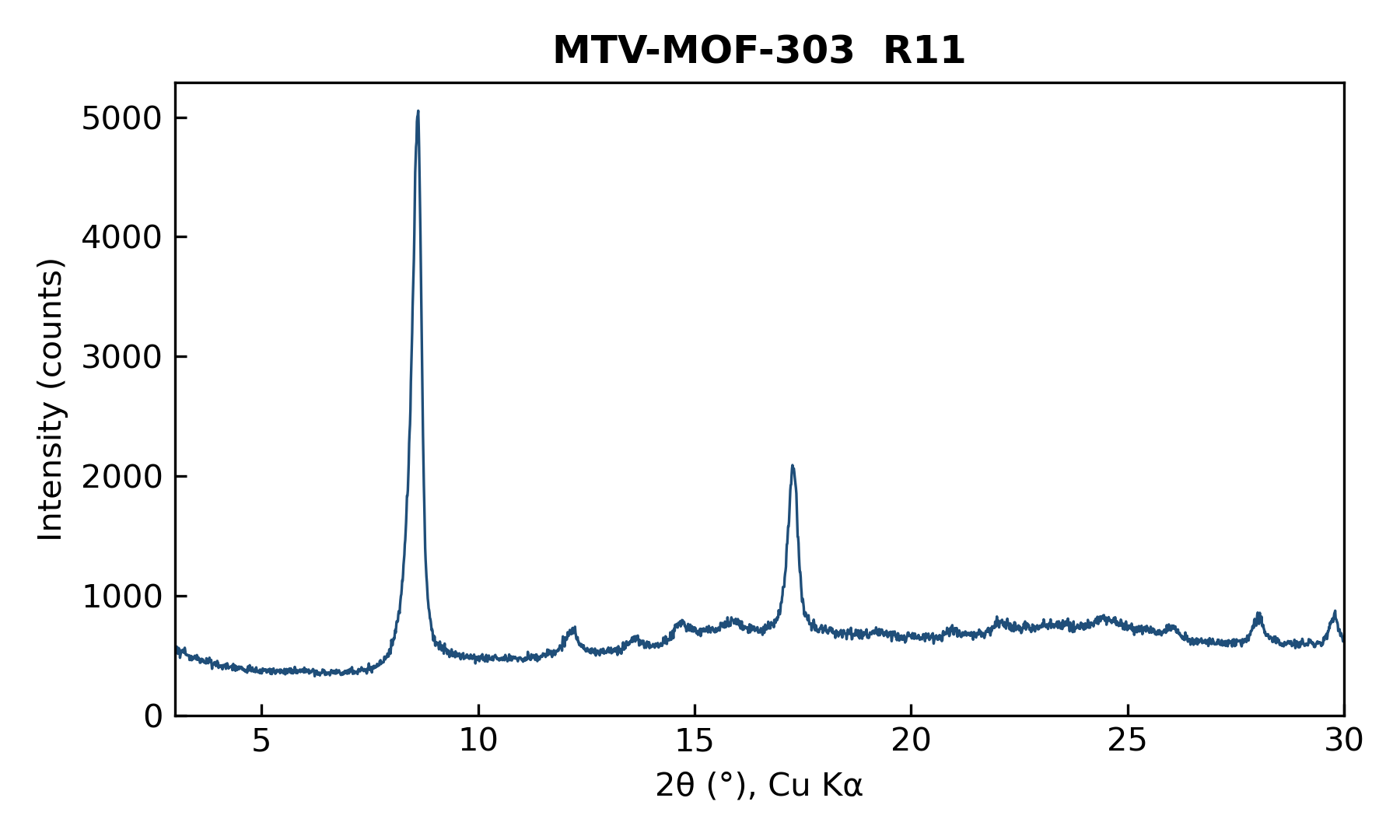}\hfill
\siFig[0.31\linewidth]{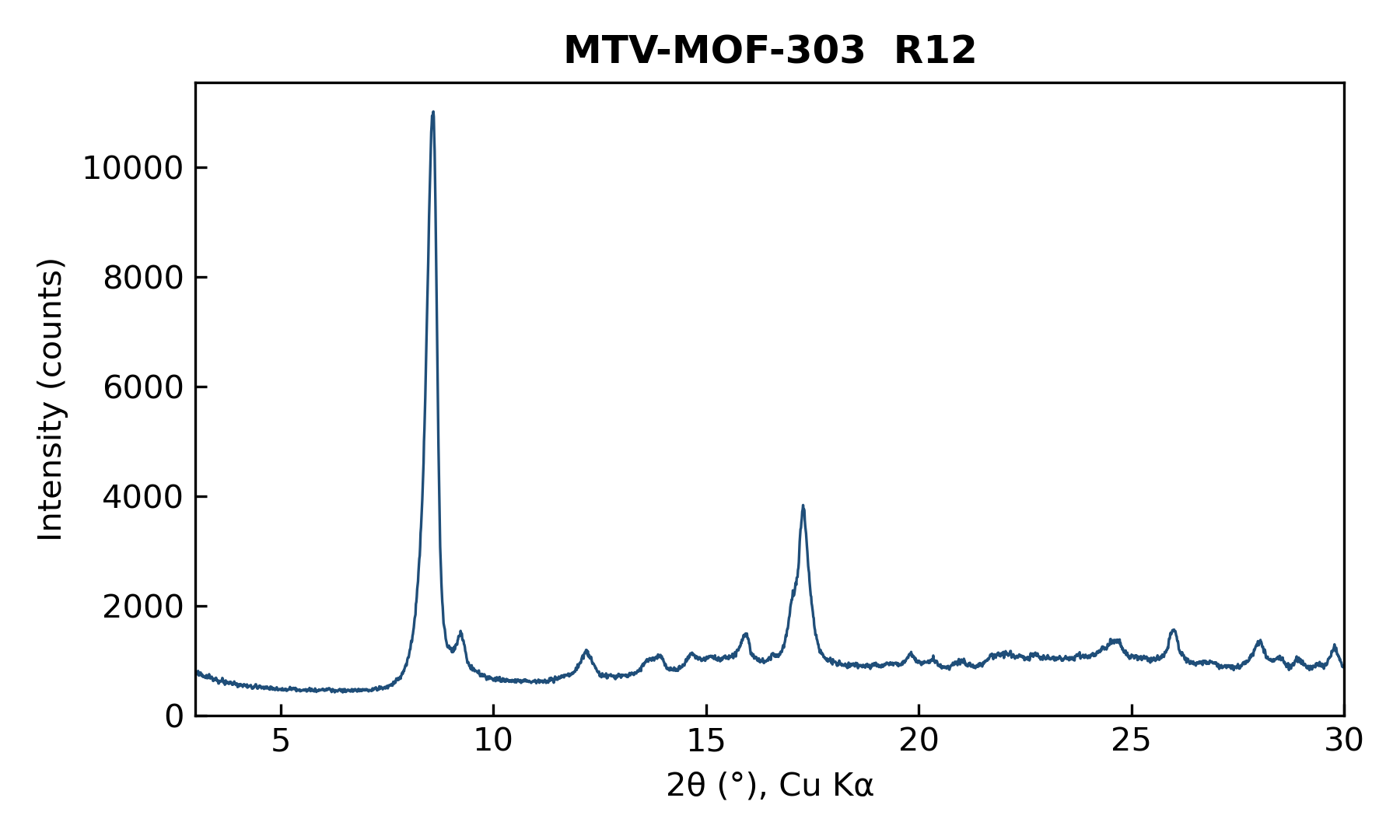}
\caption{PXRD patterns of MTV-MOF-303, R00001--R00012 (Table~\ref{tab:al-mtvmof303}).}
\label{fig:pxrd-mtvmof303}
\end{figure}

\clearpage
\paragraph{(vi) MTV-MOF-303.}\label{si:pxrd-mtvmof303} Recipes and results in Table~\ref{tab:al-mtvmof303}, patterns in Figure~\ref{fig:pxrd-mtvmof303}.
\begin{table}[H]
\centering
\caption{Recipes and PXRD results of MTV-MOF-303.}
\label{tab:al-mtvmof303}
\footnotesize
\setlength{\tabcolsep}{3.5pt}
\resizebox{\linewidth}{!}{%
\begin{tabular}{@{}lccrrrrrrrr@{}}
\toprule
 & & & \multicolumn{3}{c}{Stock solution / DI water ($\mu$L)} & & & & & \\
\cmidrule(lr){4-6}
Exp. & Round & Encoded $\mathbf{x}$ & \ce{AlCl3.6H2O} & NaOH & DI \ce{H2O} & Time (min) & $T$ ($^{\circ}$C) & $I$ & FWHM ($^{\circ}$) & $q$ \\
\midrule
R00001 & 0 & (18, 7, 17, 10, 21) & 180 & 70 & 170 & 50 & 105 & 12{,}488 & 0.287 & 43{,}452 \\
R00002 & 0 & (23, 6, 15, 10, 20) & 230 & 60 & 150 & 50 & 100 & 6{,}362 & 0.426 & 14{,}948 \\
R00003 & 0 & (20, 7, 19, 12, 29) & 200 & 70 & 190 & 60 & 145 & 5{,}557 & 0.256 & 21{,}707 \\
R00004 & 1 & (21, 7, 16, 12, 28) & 210 & 70 & 160 & 60 & 140 & 3{,}371 & 0.281 & 12{,}018 \\
R00005 & 1 & (20, 7, 17, 12, 28) & 200 & 70 & 170 & 60 & 140 & 6{,}815 & 0.287 & 23{,}779 \\
R00006 & 1 & (20, 6, 17, 10, 22) & 200 & 60 & 170 & 50 & 110 & 15{,}110 & 0.283 & 53{,}392 \\
R00007 & 2 & (18, 6, 18, 10, 22) & 180 & 60 & 180 & 50 & 110 & 9{,}073 & 0.233 & 38{,}873 \\
R00008 & 2 & (17, 6, 18, 10, 22) & 170 & 60 & 180 & 50 & 110 & 7{,}324 & 0.25 & 29{,}296 \\
\textbf{R00009} & \textbf{2} & \textbf{(19, 6, 16, 10, 22)} & \textbf{190} & \textbf{60} & \textbf{160} & \textbf{50} & \textbf{110} & \textbf{13{,}353} & \textbf{0.249} & \textbf{53{,}648} \\
R00010 & 3 & (19, 6, 17, 11, 21) & 190 & 60 & 170 & 55 & 105 & 4{,}301 & 0.327 & 13{,}141 \\
R00011 & 3 & (20, 6, 18, 10, 22) & 200 & 60 & 180 & 50 & 110 & 5{,}054 & 0.284 & 17{,}777 \\
R00012 & 3 & (19, 6, 16, 11, 21) & 190 & 60 & 160 & 55 & 105 & 11{,}015 & 0.308 & 35{,}740 \\
\bottomrule
\end{tabular}}
\end{table}

\clearpage
% =====================================================================
\section{Nitrogen Sorption}
\label{si:sorption}

\begin{figure}[htbp]
\centering
\siFig[0.8\linewidth]{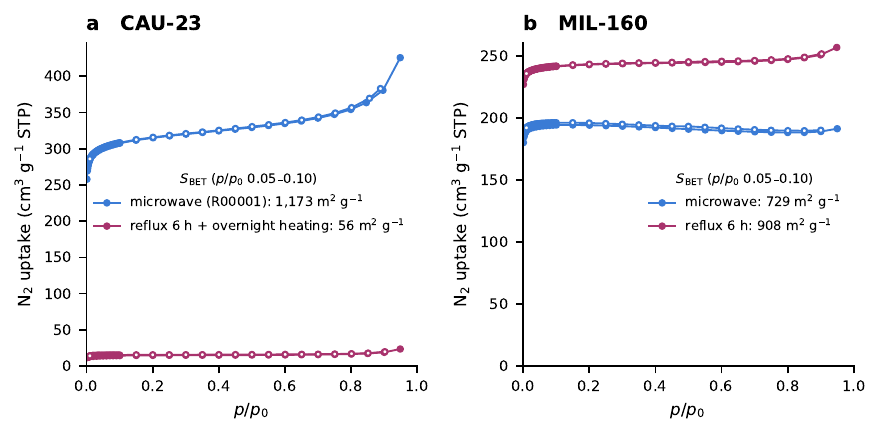}
\caption{N$_2$ isotherms at 77~K of the microwave optima and the 6~h reflux products of (a) CAU-23 (heated overnight after reflux) and (b) MIL-160. Filled: adsorption; open: desorption. Legend: $S_{\mathrm{BET}}$.}
\label{fig:si-bet-iso}
\end{figure}

% =====================================================================
N$_2$ sorption was measured for two representative frameworks, CAU-23 and MIL-160: for each, the microwave optimum of the campaign and a reflux product for comparison. The isotherms were recorded at 77~K on a Micromeritics 3Flex after activation under vacuum at 120~$^{\circ}$C overnight. BET areas were obtained from the adsorption branch by a linear fit of
\begin{equation}
\frac{p/p_0}{n\,(1-p/p_0)} = \frac{1}{n_{\mathrm{m}}C} + \frac{C-1}{n_{\mathrm{m}}C}\,\frac{p}{p_0},
\label{eq:bet}
\end{equation}
where $n$ is the adsorbed amount and $n_{\mathrm{m}}$ the monolayer capacity; $n_{\mathrm{m}}$ was converted to an area with the standard cross-sectional area of an adsorbed N$_2$ molecule, 0.162~nm$^2$\cite{thommes2015}. Langmuir areas were obtained from the same points with the Langmuir equation,
\begin{equation}
\frac{p/p_0}{n} = \frac{1}{n_{\mathrm{m}}K} + \frac{p/p_0}{n_{\mathrm{m}}},
\label{eq:langmuir}
\end{equation}
where $K$ is the Langmuir constant. The same range, $p/p_0=0.05$--0.10 (six points, $R^2\ge0.9997$), was used for every sample and both equations, so that the areas can be compared directly. The microwave samples are the best products of the active-learning campaigns (CAU-23 R00001, MIL-160 R00007); the reflux samples are the 6~h products of Section~\ref{si:q-index} (Table~\ref{tab:bet}, Figure~\ref{fig:si-bet-iso}).

\begin{table}[htbp]
\centering
\caption{BET and Langmuir areas from the adsorption branch at $p/p_0=0.05$--0.10; areas in m$^2$\,g$^{-1}$.}
\label{tab:bet}
\small
\begin{tabular}{@{}llcccc@{}}
\toprule
Framework & Sample & Mass (mg) & $S_{\mathrm{BET}}$ & $S_{\mathrm{Langmuir}}$ & Lit. $S_{\mathrm{BET}}$ \\
\midrule
CAU-23 & microwave, R00001 & 42.6 & 1173 & 1371 & 1250\cite{lenzen2019} \\
       & reflux 6~h, mistakenly heated overnight & 57.2 & 56 & 65 & \\
MIL-160 & microwave, R00007 & 34.3 & 729 & 850 & 1150\cite{permyakova2017} \\
        & reflux, 6~h & 49.2 & 908 & 1058 & \\
\bottomrule
\end{tabular}
\end{table}

\FloatBarrier
\paragraph{Results.} The CAU-23 microwave optimum is microporous (type-I isotherm) with $S_{\mathrm{BET}}=1173$~m$^2$\,g$^{-1}$ ($S_{\mathrm{Langmuir}}=1371$~m$^2$\,g$^{-1}$), 94\% of the literature BET value; the reflux product that was mistakenly heated overnight has converted into a different, nearly non-porous phase (56~m$^2$\,g$^{-1}$). For MIL-160, the microwave synthesis reached in 55~min 80\% of the BET area obtained after 6~h of reflux and measured on the same instrument (729 against 908~m$^2$\,g$^{-1}$; Langmuir, 850 against 1058~m$^2$\,g$^{-1}$). The two microwave optima, which have the highest $q$ of their campaigns, are thus porous frameworks. This supports the use of $q$ as the target of the optimization (Section~\ref{si:q-reflux}, Table~\ref{tab:q-bet}).
% =====================================================================
\section*{Data and Code Availability}
% =====================================================================
% \addcontentsline{toc}{section}{Data and Code Availability}

Code and data are available in the \href{https://github.com/deep-synthesis/ActMOF}{GitHub repository}, \url{https://github.com/deep-synthesis/ActMOF}.

% =====================================================================

\phantomsection

\end{document}